%% file: main.tex
\documentclass[a4,11pt,twoside]{book}
\input{settings.tex}
\title{Aspects of gravitational clustering and\\ structure formation in the Universe}

\date{\today}
\author{Swati Gavas}
\begin{document}

\input{chapters/title}
\frontmatter

\include{chapters/declaration}
\input{chapters/dedication}
\include{chapters/acknowledgement}
\include{chapters/abstract}
\tableofcontents
\include{chapters/papers}
\listoffigures
\listoftables

\mainmatter

\include{chapters/chap1}
\include{chapters/chap2}
\include{chapters/chap3}
\include{chapters/chap4}
\include{chapters/chap5}
\include{chapters/chap6}

\appendix
 \addtocontents{toc}{\protect\setcounter{tocdepth}{0}} 
\input{chapters/appendix1}
\input{chapters/appendix2}
\input{chapters/appendix3}
\input{chapters/appendix4}
\input{chapters/appendix5}
 \addtocontents{toc}{\protect\setcounter{tocdepth}{1}} 

\backmatter
\printbibliography
\end{document}


%% file: settings.tex
\usepackage[a4paper,width=145mm,top=36mm,bottom=36mm,bindingoffset=20mm]{geometry}
\usepackage{graphicx}
\usepackage{braket}
\usepackage{url}
\usepackage{float}
\usepackage[thinc]{esdiff}
\usepackage[dvipsnames]{xcolor}
\usepackage{tensor}
\usepackage{esdiff}
\usepackage{amsmath, amssymb}
\usepackage[linewidth=1pt]{mdframed}
\usepackage[pdfusetitle,bookmarksnumbered=true]{hyperref}
\usepackage{cleveref}
\usepackage{empheq}
\usepackage{siunitx}
\usepackage{emptypage}
\usepackage{commath}
\usepackage[font={small}]{caption}
\usepackage{tabularx}
\usepackage{appendix}
\usepackage{fancyhdr}
\usepackage{bibentry}
\usepackage{booktabs}
\usepackage{dsfont}
\usepackage{multirow}

\fancypagestyle{plain}{ %
  \fancyhf{}

}

\usepackage{titlesec}
\titleformat{\chapter}
{\normalfont\sffamily\LARGE}
{\thechapter}{0.5em}{}

\titleformat{\section}
{\normalfont\sffamily\Large}
{\thesection}{0.5em}{}

\titleformat{\subsection}
{\normalfont\sffamily\large}
{\thesubsection}{0.5em}{}

\titleformat{\subsubsection}
{\normalfont\sffamily\large}
{\thesubsubsection}{0.5em}{}

\titleformat{\paragraph}[runin]
{\normalfont\sffamily\normalsize}
{\theparagraph}{}{}

\crefformat{chapter}{\S#2#1#3}
\crefformat{section}{\S#2#1#3}
\crefformat{subsection}{\S#2#1#3}
\crefformat{subsubsection}{\S#2#1#3}

\usepackage{libertinus}
\usepackage[T1]{fontenc}
\usepackage{titling}

\usepackage{subcaption}
\usepackage[most]{tcolorbox}

\hypersetup{colorlinks = true,
  linkcolor = Black,
  anchorcolor = Black,
  citecolor = Black,
  filecolor = Black,
  urlcolor = Black }

\usepackage[backend=biber,
  sorting=none,
  style=numeric-comp,
  url=false]{biblatex}
\newbibmacro{string+doi}[1]{%
  \iffieldundef{doi}{#1}{\href{http://dx.doi.org/\thefield{doi}}{#1}}}
\DeclareFieldFormat{title}{\usebibmacro{string+doi}{\mkbibemph{#1}}}
\DeclareFieldFormat[article]{title}{\usebibmacro{string+doi}{\mkbibquote{#1}}}

\allowdisplaybreaks

\graphicspath{ {figures/} }

\newcommand{\p}{\partial}

\renewcommand{\d}{\text{d}}

\newcommand{\floor}[1]{\left\lfloor #1 \right\rfloor}


%% file: chapters/title.tex

\begin{titlepage}
   \begin{center}
       \LARGE
       \thetitle
       \vfill
       \Large
       \theauthor

       \vfill

       \large

       \emph{A thesis submitted for the partial fulfillment of\\
         the degree of Doctor of Philosophy}
            
       \vfill
     
       \includegraphics[width=0.225\textwidth]{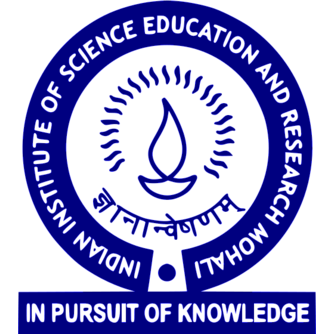}\\

       \vfill
       Department of Physical Sciences\\
       Indian Institute of Science Education and Research Mohali\\
       Knowledge City, Sector 81, SAS Nagar, Manauli PO, Mohali 140306, Punjab,
       India.
       \vfill
       December 2024
            
   \end{center}
\end{titlepage}


%% file: chapters/declaration.tex

\chapter*{}


\begin{center}
  {\Large Declaration}
\end{center}

\noindent
The work presented in this thesis has been carried out by me under the guidance
of Prof. Jasjeet Singh Bagla and Dr. Nishikanta Khandai (NISER Bhubaneswar) at the Indian Institute of Science Education and Research Mohali. This
work has not been submitted in part or in full for a degree, a diploma, or a
fellowship to any other university or institute. Whenever contributions of
others are involved, every effort is made to indicate this clearly, with due
acknowledgement of collaborative research and discussions. This thesis is a bona
fide record of original work done by me and all sources listed within have been
detailed in the bibliography.

\vspace{2cm}
\hfill Swati Gavas

\vspace{2cm}
\noindent
In my capacity as the supervisor of the candidate's thesis work, I certify that
the above statements by the candidate are true to the best of my knowledge.

\vspace{2cm}

\noindent
Prof. Jasjeet Singh Bagla (Supervisor) \hfill Dr. Nishikanta Khandai (Co-supervisor)


%% file: chapters/dedication.tex
\chapter*{}

\begin{center}
\includegraphics[width=0.15\textwidth]{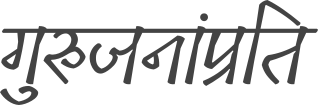}
\end{center}


%% file: chapters/acknowledgement.tex

\chapter*{Acknowledgements}

I sincerely thank Prof. Jasjeet Singh Bagla, my primary supervisor, for his guidance and for providing a valuable learning experience. I am grateful to Dr. Nishikanta Khandai, my co-supervisor, for his crucial feedback and encouragement. I extend my gratitude to Dr. Girish Kulkarni for his suggestions and support as a collaborator. I also thank the anonymous referees for their thorough reviews of my thesis.

I thank my doctoral committee members, Dr. Harvinder Kaur Jassal, Dr. Kinjalk Lochan, Dr. Anosh Joseph, Dr. Ambresh Shivaji and Dr. Pankaj Kushwaha, for their time, constructive feedback, and smooth conduct of my PhD procedures.

I thank IISER Mohali for financial support and for providing research facilities. Thanks to NISER Bhubaneswar and their Computer Center for supporting me with a High-Performance Computing facility and related help. I am grateful to DST-ANRF and IISER Mohali for funding my international travels. Thanks to the hosts of the conferences, schools and workshops for providing an opportunity to present our work and travel support: IIT Indore, IISER Tirupati, KIAS Seoul, IAU, IIT Guwahati, IISc Bangalore, ICTP Trieste, IAP Paris, NCRA-TIFR Pune, ICTS  Bangalore, Observat\'orio Nacional, Rio.

I extend my sincere gratitude to the professors who shared their valuable feedback on our work during various events: Dr. Surhud More, Prof. Thanu Padmanabhan, Dr. Juhan Kim, Dr. Aseem Paranjape, Dr. Shadab Alam, Prof. Arif Babul, Prof. Ravi Sheth, Dr. Fabian Schmidt, Prof. Pierluigi Monaco, Prof. Stephane Colombi.

Throughout my PhD journey, I have been blessed with a wonderful and varied circle of friends and colleagues. I appreciate the Astro Journal Club members, Scholar roommates, batchmates and roommates for all the help and enriching activities we enjoyed together. Thanks are also due to the Gulati family for their kind help during the COVID-19 pandemic and beyond.

Finally, I value having a family.


%% file: chapters/abstract.tex

\chapter{Abstract}

The distribution of galaxies, halo abundance, and peculiar velocities are influenced by non-linear gravitational interactions, making the study of non-linear evolution crucial for accurate cosmological predictions. Thus, it is essential to study the underlying assumptions in these processes to improve predictions like halo abundance and determination of the Hubble-Lema\^itre constant $H_0$. We explore these aspects using N-body simulations. The halo model can be used to make various predictions and interpretation of observations, with the halo mass function as a core ingredient. The theoretical models of mass function can be formulated without referencing the cosmological model and input power spectrum. Mass functions obtained from N-body simulations show systematic deviations of 5-20\% from theoretical predictions. The physical origin of deviations is complex to understand and may result from cosmology, the power spectrum, or both. To investigate the issue, we examine mass function deviations from universality for scale-free power spectra with an Einstein-de Sitter cosmology. We show that the mass function has explicit dependence on the slope of the input power spectrum. We extend our analysis to $\Lambda$CDM cosmologies and show that an effective index of the $\Lambda$CDM model can correspond to the mass function from scale free cosmologies as a first approximation. Our results indicate that an improved analytical theory is required to provide better fits to the mass function. Furthermore, structure formation has led to deviations from homogeneity and isotropy on scales up to at least $100$ Mpc/h, expected to affect measurements of $H_0$. Various efforts have been made to quantify this effect. In our second study, we revisit this issue of the concordance model. We find a correlation between errors in $H_0$ estimates and the density around the observer. Further, our mock observations reveal that deviations of up to 5\% can occur in Milky Way-sized halos. While this finding alone does not fully resolve the Hubble tension, it may account for part of it. We rely on N-body simulations for these studies. Hence, it is essential to understand their limitations to avoid misinterpreting data. We show that the missing power at small scales introduces errors in the root-mean-square fluctuations and, consequently, in the simulated mass function. These errors are expected to diminish as the scale of non-linearity increases. Our analytical calculation indicates that mode coupling between small and large scales depends on resolving collapsed halos. Therefore, accurate mode coupling estimates require sufficient halos in the simulation.


%% file: chapters/papers.tex

\chapter{List of publications}
\label{cha:papers}
\begin{itemize}
\item \fullcite{Gavas_2023}
\item \fullcite{Gavas_2024}
\item \fullcite{Bagla_2024}
\end{itemize}


%% file: chapters/chap1.tex

\chapter{Introduction: Large Scale Structure}

The General theory of relativity describes gravity as the curvature of spacetime resulting from the presence of matter (or energy). This curvature, manifested as the force of gravity, causes matter elements to attract each other, leading to the condensation of matter and the formation of structures such as galaxies, galaxy clusters, and larger cosmic arrangements~\cite{Peebles_1981,  Shandarin_1989, Peacock_1998, padmanabhan_1993, Padmanabhan_2002, Bernardeau_2002, Coil_2013}. These structures, collectively known as the `Large Scale Structure (LSS) of the Universe', encompass various features like halos, filaments, and voids. The study of LSS holds significant importance for several reasons, including understanding cosmic evolution~\cite{Kauffmann_1993, Lacey_1993, Bora_2021}, testing cosmological theories~\cite{Debono_2021, Durrer_2022, Fiorini_2022}, gaining insights into the galactic dynamics~\cite{Benson_2000, Benson_2010, Somerville_2015, Wechsler_2018, Dom_nguez_G_mez_2023}, and cosmography~\cite{HARRISON_1976, Visser_2005, Dalal_2005, Xia_2012, Courtois_2013, Capozziello_2014, Dunsby_2016}. LSS is integral to modern cosmology; its study involves theoretical modeling~\cite{Zeldovich_1970, Gunn_1972, Fillmore_1984,Gurbatov_1989, Matarrese_1992, Brainerd_1993, Bagla_1994, Sahni_1995, Hui_1996,  Bernardeau_2002}, observational astronomy~\cite{Shane_1967, Jarrett_2004, Kirshner_1978, Davis_1982, Colless_2001, York_2000, DESI_ollaboration_2023, lsst_collaboration_2009, Euclid_2011}, and
\emph{simulations}~\cite{Bertschinger_1998, Bagla_1994, Bagla_2002, Springel_2005, Springel_2017, Angulo_2022}.

\section{Observing large scale structure}
\label{sec:obs_lss}

\begin{figure}
  \centering
\includegraphics[width=0.95\textwidth]{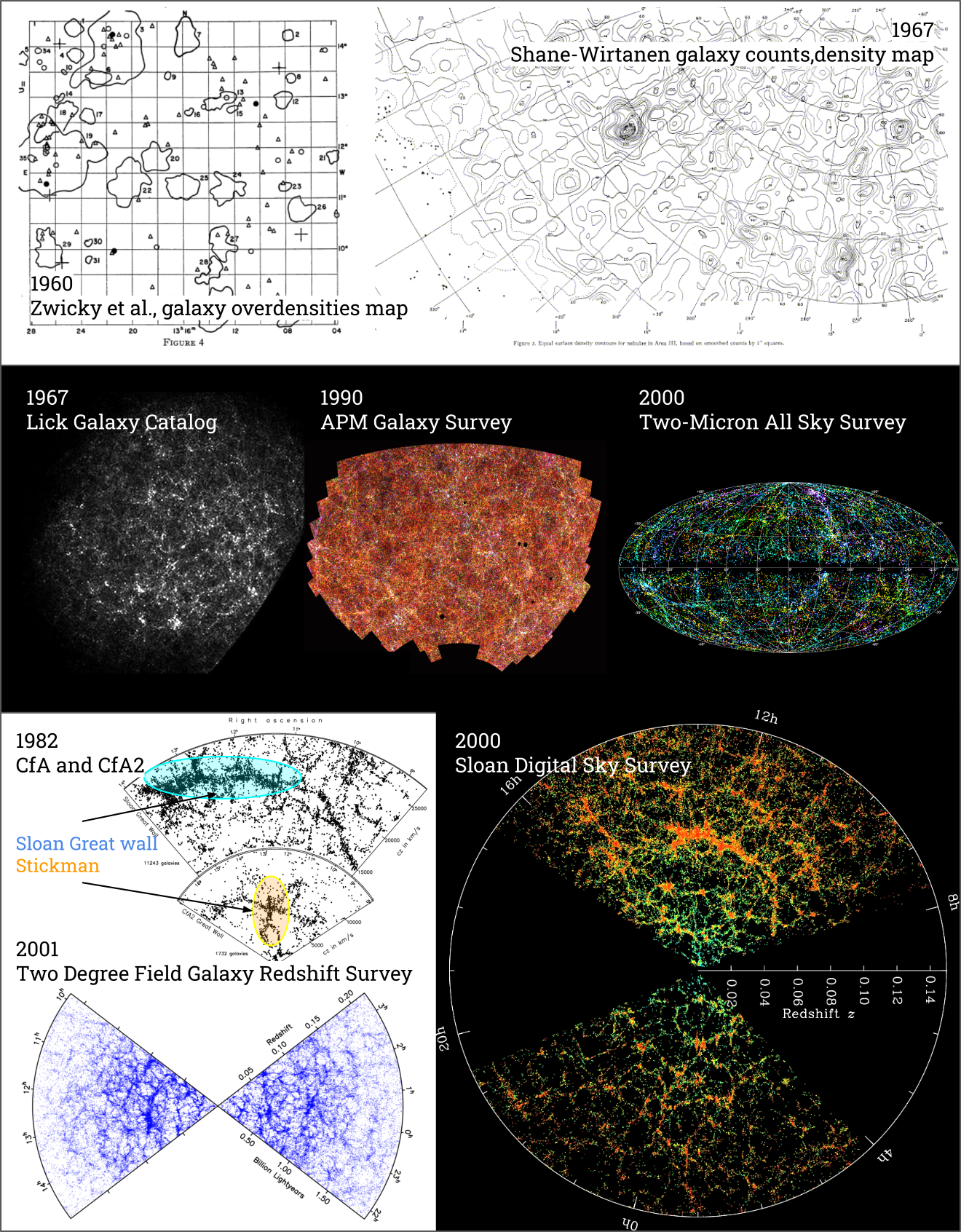}
\caption[Historical overview of observing Large Scale Structure]{\emph{Observing large scale structure:} During the 1930s-50s, Shapley and Zwicky manually mapped the galaxy overdensities using photographic plates. The Lick Galaxy Catalogue($\sim$ million galaxies) revealed the foamy structure of the cosmic web. The automation for reading photographic plates led to the APM galaxy survey ($\sim$2 million galaxies). The Two-Micron All Sky Survey aimed to capture galaxies in the infrared spectrum to attempt photometric redshifts. The CfA ($\sim$9,000 galaxies, $\sim$9,000 deg$^2$) and CfA2 ($\sim$20,000) were among the initial spectroscopic redshift surveys unveiled the famous `Stickman' and `Sloan Great Wall'. Based on the APM data, the Two-Degree Field Galaxy Redshift Survey covered $\sim$220,000 galaxies across $\sim$1500 deg$^2$. The largest galaxy redshift survey is the Sloan Digital Sky Survey, encompassing $\sim$9,30,000 galaxies and covering $\sim$14,555 deg$^2$. Refer~\cref{sec:obs_lss}}
\label{fig:12}
\end{figure}

Notable first attempts to map galaxy distribution can be traced back to Edwin Hubble's examination of 400 `extragalactic nebulae' in 1926~\cite{Hubble_1926}, revealing a general uniformity on large scales. However, further analysis using the larger \emph{Shapley-Ames catalogue} in 1932 indicated \emph{clumpiness} on smaller scales~\cite{Shapley_1932}. In 1934, with a larger statistical sample Hubble found an excess of galaxies within angular scales less than $\sim~10^{\circ}$ compared to a random Poisson distribution~\cite{Hubble_1934}.

Harlow Shapley and Fritz Zwicky led the foundations of constructing galaxy catalogues using photographic plates in the 1930s-50s~\cite{zwicky_1961, Zwicky_1957}. The \emph{Lick Galaxy Catalogue} (1967), produced by Shane $\&$ Wirtanen~\cite{Shane_1956, Shane_1967}, marked a significant advancement, containing data on approximately a million galaxies, revealing a non-uniform distribution in the sky~\cite{Geller_1984}, showing foam-like patterns with possible \emph{walls and filaments}~\cite{Seldner_1977}. Using Hubble's counts of faint galaxies, G\'erard de Vaucouleurs calculated that the total light from all galaxies, including those fainter than counted, would contribute 0.3\%, making it undetectable \cite{Vaucouleurs_1949}. Further, analysis by Peebles and collaborators~\cite{Peebles_1975} indicated that the clustering amplitude decreases for fainter galaxy populations, likely due to larger projection effects along the line of sight, as faint galaxies tend to lie at larger distances. During the 1980s, the emergence of the cosmic web structure was observed through the catalogue of approximately 2 million galaxies generated by the \emph{Automated Plate Measurement (APM) Galaxy Survey}~\cite{Maddox_1990}. It hinted at the presence of a cosmological constant for the first time. In the early 2000s, the \emph{Two-Micron All Sky Survey (2MASS)} mapped the local Universe in the infrared regime~\cite{Jarrett_2004}. It marked the initial attempt to obtain photometric redshifts of galaxies and provided clear evidence for the large scale structure.

These findings are taken further by the emergence of redshift surveys, which use optical spectra to assess redshifts and spatial distributions of galaxy samples. Pioneering work by Gregory $\&$ Thompson (1978) mapped the spatial distribution of 238 galaxies around the Coma/Abell 1367 supercluster, unveiling significant void ($>20 h^{-1}$ Mpc) in the foreground~\cite{Gregory_1978}. Further, Joeveer and collaborators~\cite{Joeveer_1978, Sandage_1975} utilised redshift data to map galaxies across the southern galactic hemisphere, revealing clustered formations, including chains of clusters, now identified as \emph{filaments}. During a similar period, the \emph{KOS (Kirshner, Oemler, Schechter)} survey measured redshifts focusing on 164 brighter galaxies in eight separate fields on the sky, revealing strong clustering in the velocity space~\cite{Kirshner_1978}.

G\'erard de Vaucouleurs and collaborators constructed the earliest large scale galaxy redshift surveys in 1950-1970s~\cite{Vaucouleurs_1964, Vaucouleurs_1976}. The first \emph{CfA (Center for Astrophysics)} survey, completed in 1982, catalogued redshifts for $\sim$9,000 galaxies across both galactic poles, covering$\sim$9000 square degree, aiming to assess the three-dimensional clustering of galaxies~\cite{Davis_1982}. It generated extensive three-dimensional maps of large scale structure, highlighting clusters, voids, and filamentary structures, prompting comparisons with N-body simulations. The survey revealed the famous `Stickman' figure, which is the Coma cluster distorted in redshift space\footnote{Positions are mapped across the celestial sphere, while velocities are measured along the line of sight. Converting these velocities to distances can be distorted due to the peculiar motions of galaxies within clusters.} (\cref{a1:r,a1:rsd}). The second CfA redshift survey, spanning from 1985 to 1995, encompassed spectra for approximately $\sim$20,000 galaxies, unveiling the wide `Sloan Great Wall' supercluster spanning $170 h^{-1}$ Mpc and highlighting widespread underdense voids with a density of about 20\% of the mean density~\cite{Geller_1989}. \emph{Las Campanas Redshift Survey (LCRS)}, conducted from 1991 to 1997, mapped galaxy redshifts over large area out to a redshift of $z = 0.2$ \cite{Shandarin_1998}. It is considered one of the first surveys where the 'Cosmological Principle' of isotropy could be tested. \emph{Canada-France Redshift Survey (CFRS)}, performed from 1989 to 1997, provided the first systematic study of galaxies at redshifts $z>0.5$ \cite{Lilly_1995}.

In the last three decades, redshift surveys have advanced significantly due to multiobject spectrographs and larger telescopes, facilitating deep surveys of nearby and distant galaxies. In late 1990s and early 2000s, massive surveys, such as the \emph{Two Degree Field Galaxy Redshift Survey (2dFGRS)}~\cite{Colless_2001} and the \emph{Sloan Digital Sky Survey (SDSS)}~\cite{York_2000}, provide comprehensive maps of large scale structure, uncovering a sponge-like pattern with voids of approximately 10 Mpc/h and intersecting filaments around galaxy groups and clusters. SDSS has evolved through multiple phases (I, II, III, IV, V), each focusing on different scientific goals. As we now transit to the precision cosmology era, recent Large Scale Structure (LSS) surveys like \emph{Baryon Oscillation Spectroscopic Survey (BOSS)}~\cite{Dawson_2012} (it is an extension of SDSS), \emph{Dark Energy Spectroscopic Instrument (DESI)}~\cite{DESI_ollaboration_2023}, \emph{Large Synoptic Survey Telescope (LSST)}~\cite{lsst_collaboration_2009, Ivezi__2019} and \emph{Euclid Space Telescope}~\cite{Euclid_2011, Euclid_2023} are launched to focus of specific aspects of LSS. These surveys are helping us to constrain cosmological models.

Large scale structure can be explored through a variety of indirect observational probes, such as Weak Gravitational Lensing~\cite{Kaiser_1992, Mellier_1999, Bartelmann_2001}, the Sunyaev-Zel'dovich effect~\cite{Birkinshaw_1999}, Cosmic Microwave Background (CMB) observations~\cite{Sunyaev_1980}, peculier velocity measurements~\cite{kim_2019} and the Lyman-alpha forest~\cite{Weinberg_2003}. Integrating direct and indirect measurements provides insights into the cosmic structure and its evolutionary history.

\section{Theory of large scale structures}
\label{sec:theory_lss}

The current understanding of the theory of Large Scale Structure (LSS) starts with the end of the inflationary phase, $\sim$13.8 billion years ago, where the exponential expansion of the tiny early Universe made quantum fluctuations imprinted on cosmic constituents at that time. This was followed by the nucleosynthesis era, $\sim$3 minutes after the inflation, where protons and neutrons combined to form the first elements. However, the Universe remained too hot for atoms to form due to scatterings of photons and free electrons. Due to the adiabatic nature of the expansion of the Universe, radiation fluid looses energy faster compared to matter particles, leading to matter sector domination in the total energy of the Universe, $\sim$50,000 years after the inflation. At $\sim$380,000 years after the inflation, the Universe became cold enough to sustain atoms, called the recombination phase~\cite{Weinberg_2008, Mbw_2010, Dodelson_2020, Baumann_2022}. The evolution of the matter density field during these times involves intricacies like M\'esz\'aros effect~\cite{Meszaros_1974, Meszaros_1975}, Baryon Acoustic Oscillations (BAO)~\cite{Hu_1996}, Silk damping~\cite{Silk_1968}, integrated Sachs-Wolfe (ISW) effect~\cite{Sachs_1967}, and change in superhorizon scale. All of these can be encapsulated within the transfer function framework \footnote{Sophisticated Boltzmann codes like {\tt CMBFAST}~\cite{Seljak_1996}, {\tt CAMB}~\cite{Lewis_2000} and {\tt CLASS}~\cite{Lesgourgues_2011} are used to calculate transfer function efficiently.}~\cite{Eisenstein_1998, Eisenstein_1999}, it is a mechanism that maps the gravitational potential perturbations field at the end of the inflationary phase to the density and velocity perturbations field for each constituent at the end of recombination. It is possible as the perturbations have a small amplitude at these stages, and the equations governing the evolution are linear. These calculations are consistent with observations to great precision~\cite{Planck_2020}. Recombination is followed by the Dark ages, the formation of the first stars, reionisation, and dark energy domination~\cite{Dodelson_2020, Baumann_2022}. During these phases, the problem of structure formation transitions from a \emph{linear}, \emph{quasi-linear} to the \emph{non-linear regime}\cite{Peebles_1981,Peacock_1998, Mbw_2010}. This section discusses some theoretical modelling in linear and quasi-linear regimes.

\subsection{Homogeneous Universe}
\label{ssec:homogeneous_Universe}

A homogeneous Universe is one where the properties, such as energy density and temperature, are the same at every point in space on large scales. The spacetime geometry of such a universe is given by the Friedmann-Lema\^itre-Robertson-Walker (FLRW) metric~\cite{Friedman_1922, Robertson_1929, Walker_1937, Lematre_1931},

\begin{equation}
  \label{eq:flrw}
  \d s^2 = - \d t ^2 + a^2(t) \left[ \frac{\d x^2}{1- \kappa x^2} + x^2 (\sin^2 \theta \d \phi^2 + \d \theta^2)\right],
\end{equation}
where $a(t)$ is the scale factor normalised to unity at present ($a_0 = 1$). $(x, \theta, \phi)$ is the comoving coordinate of the fundamental observer. Due to the expansion or contraction of the Universe, working with comoving coordinates $\mathbf{x}$ is more convenient than using physical coordinates $\mathbf{r}$. They are related by $\mathbf{r} = a \mathbf{x}$. $\kappa$ is the curvature constant (the positive, zero and negative values of $\kappa$ indicates hyperspherical, flat, and hyperbolic universe respectively). $t$ stands for cosmic time. Alternatively, the conformal time ($\tau$), such that $\d t = a(\tau) \d \tau$, can be used as a time variable. Similarly, redshift defined as $1 + z \equiv a^{-1}$, is often used as alternative to scale factor (\cref{a1:r}). 

The energy density in the Universe governs the evolution of $a(t)$. In order to solve for $a(t)$,  Einstein's field equations~\cite{Einstein_1915} for the FLRW metric need to be solved. Under the assumptions of \emph{homogeneity} and \emph{isotropy}, derived dynamical equations are called Friedmann equations~\cite{Peebles_1981, Kolb_1990, Liddle_2000, Bernardeau_2002, Lesgourgues2004, Trodden2004, Langlois2004, Langlois_2010}. 

\begin{subequations}
\label{eq:Friedmann}
\begin{align}
  \label{eq:Friedmann1}
  \text{Friedmann equations: ~~~~~~~}
  \left( \frac{\dot{a}}{a} \right) ^2 &= \frac{8\pi G}{3} \rho -\frac{\kappa}{a^2} + \frac{\Lambda}{3},\\
  \label{eq:Friedmann2}
  \frac{\ddot{a}}{a} &= - \frac{4\pi G}{3} \left(\rho +3P \right) + \frac{\Lambda}{3},
\end{align}
\end{subequations}
where the dot denotes differentiation with respect to $t$, $\rho$ is the total energy density and $P$ is the pressure and $\Lambda$ is the cosmological constant~\cite{Weinberg_2008, Dodelson_2020, Baumann_2022}. $G$ denotes the gravitational constant, and $c = 1$ unit is adopted. 

Energy conservation in an expanding Universe leads to the evolution of energy density $\rho$, known as the continuity equation.

\begin{equation}
  \label{eq:rho_dot}
  \dot{\rho} = - 3H \left(\rho +P \right),
\end{equation}
where, $H=\dot{a}/a$ is the \emph{Hubble parameter}, its value at present ($a=a_0$) is called the \emph{Hubble-Lema\^itre constant $H_0$}. \Cref{eq:rho_dot} can be solved for perfect fluids with equation of state $P=\omega \rho$. Various  contributors to the total energy include radiation ($\omega_r = 1/3$), non-relativistic matter ($\omega_m \approx 0$), and cosmological constant ($\omega_{\Lambda} = -1$). Let us define \emph{critical density} $\rho_c \equiv 3H^2/8\pi G$, corresponding to the total energy density in a flat Universe. A dimensionless cosmological parameter such as $\Omega = \rho/\rho_c$, can be formulated. From this, it follows that for radiation $\Omega_r \sim a^{-4}$, matter $\Omega_m \sim a^{-3}$ and cosmological constant $\Omega_{\Lambda} \sim a^{0}$. Present values of these ratios are constrained by various observations~\cite{Riess_1998, Alam_2017, Planck_2020}.

\subsection{Dynamics of gravitational instability}
\label{ssec:lss_equations}

The Classical Boltzmann equation (CBE) dictates the fundamental description for particles in the phase space, accounting for advection, external forces, and collisions. For the scope of this thesis, the CBE for a cold dark matter particle field, described as a \emph{collisionless} fluid without external forces, is given as

\begin{equation}
  \label{eq:cbe_cdm}
  \diff{f}{t}  = \diffp{f}{t}  + \mathbf{v} \cdot  \nabla f   - \nabla \Phi \cdot \frac{\p f}{\p \mathbf{v}} = 0,
\end{equation}
where $f(\mathbf{x},\mathbf{v},t)$ is the distribution function, which gives the number density of particles at a specific position $\mathbf{x}$, velocity $\mathbf{v}$, and time $t$ in phase space. $\Phi$ is a gravitational potential field, given by the Poisson equation,

\begin{equation}
  \label{eq:Poisson}
  \nabla^2 \Phi = 4 \pi G \bar{\rho}_m a^2 \delta,
\end{equation}
where, $\bar{\rho}_m$ is average matter densitiy in the Universe. $\delta \equiv \delta(\mathbf{x})$ is the matter density contrast field,

\begin{equation}
  \label{eq:d_contrast}
  \delta(\mathbf{x}) \equiv \frac{\rho_m(\mathbf{x}) - \bar{\rho}_m}{\bar{\rho}_m}.
\end{equation}

The complete analytic solution to the CBE is elusive, but its moment equations can be useful. The zeroth moment of~\cref{eq:cbe_cdm} yields the continuity equation, and the first moment provides the Jeans equation (the collisionless counterpart of the Euler equation in fluid dynamics). Incorporating the expansion of the Universe, they are expressed as 

\begin{subequations}
  \label{eq:cbe_moments}
  \begin{align}
  \label{eq:cbe_moments0}
  \text{Continuity equation: ~~~~~~~~~}
  \diffp{\delta}{t} + \frac{1}{a} \nabla \cdot \left[ (1+\delta) \mathbf{v} \right] &= 0,\\
  \label{eq:cbe_moments1}
  \text{Jeans equation: ~~~~~~~~}
  \diffp{\mathbf{v}}{t} + \frac{\dot{a}}{a} \mathbf{v} +  \frac{1}{a} (\mathbf{v} \cdot \nabla) \mathbf{v}  &= - \frac{\nabla \Phi}{a} -  \frac{\nabla \delta \sigma^2}{a(1+\delta)},
  \end{align}
\end{subequations}
where, $\sigma_{ij} \equiv \left[  \left< v_i v_j \right> -\left< v_i \right> \left< v_j \right> \right]^{1/2} $ is the stress tensor. After the onset of the matter dominant phase of the Universe, the growth of dark matter perturbations ($\delta$) is governed by 

\begin{equation}
  \label{eq:de_delta}
  \diffp[2]{\delta}{t} + 2\frac{\dot{a}}{a} \diffp{\delta}{t} = \frac{1}{a^2} \nabla \cdot (1+\delta) \nabla \Phi+ \frac{1}{a^2} \frac{\p }{\p x_i\p x_j} \left[ (1+\delta) v^i v^j \right] .
\end{equation}
In~\cref{eq:de_delta}, Einstein summation conventions is assumed. It is called the \emph{Vlasov equation}, and is derived using~\cref{eq:cbe_moments}. It is the basic equation for the evolution of density perturbations due to gravitational interaction in a cosmological setting. Detailed derivations can be found in~\cite{Peebles_1981,Dodelson_2020, Mbw_2010}. \Cref{eq:de_delta} can be solved analytically within the linear regime ($|\delta | \ll1$) or by considering specific geometries for the initial perturbations.

\subsection{Eulerian perturbation theory}
\label{ssec:EPT}

It is assumed that matter is slightly perturbed from background values, i.e. $|\delta| \ll 1$ in~\cref{eq:d_contrast}. It holds true for certain epochs in the early Universe. Also, this approximation effectively describes the behaviour of matter on large scales, although there are strong non-linear effects at small scales~\cite{Peebles_1981}. The Vlasov~\cref{eq:de_delta} can be linearised in $\delta$ as

\begin{equation}
  \label{eq:de_1pt}
  \text{$| \delta | \ll1$ ~~~and~~~ $c_s \ll c $ ,~~~~~~~~~~} \diffp[2]{\delta}{t} + 2\frac{\dot{a}}{a} \diffp{\delta}{t} = 4 \pi G \bar{\rho}_m \delta ,
\end{equation}
where $c_s$ is the sound speed and $c$ is speed of light. To arrive at~\cref{eq:de_1pt} a Poisson \cref{eq:Poisson} is used. For structure formation, a \emph{growing mode} solution of~\cref{eq:de_1pt} is relevant, and it has the following form:

\begin{equation}
  \label{eq:del}
  \delta (\mathbf{x}, t) = \mathcal{G} (t) \delta (\mathbf{x},0)  ,
  \text{ where ~~~~~~~~} \mathcal{G} (\mathbf{x}) \propto \frac{H}{H_0} \int_{a(t)}^{\infty} \frac{\d a}{a} \left[ \frac{H_0}{H(a)}\right]^3.
\end{equation}
It shows that the linear matter density field can be scaled in time using the growth solution $\mathcal{G} (t)$. Note, $\mathcal{G} \propto a $ as $\Omega_m \rightarrow 1$.

At higher order, the Eulerian perturbation theory can be implemented by expanding the density and velocity fields,

\begin{subequations}
  \label{dummy}
  \begin{align}
  \label{eq:del_tylor}
  \delta (\mathbf{x}, t)& = \delta^{(1)} (\mathbf{x}, t) + \delta^{(2)} (\mathbf{x}, t) + \dots \text{~~},\\
  v (\mathbf{x}, t) &= v^{(1)} (\mathbf{x}, t) + v^{(2)} (\mathbf{x}, t) + \dots \text{~~},
\end{align}
\end{subequations}
where, $\delta^{(1)} (\mathbf{x}, t)$ is first order $\delta$ same as in~\cref{eq:del}. $\delta^{(2)} (\mathbf{x}, t)$ describes to leading order the non-local evolution of the density field. It is proportional to the \emph{second order growth factor} $\mathcal{G}_2$~\cite{Bouchet1994}. Depending on the cosmological parameters, different expressions for $\mathcal{G}_2$ can be found in the literature~\cite{Bouchet1994, Bernardeau_2002}.

A detailed non-linear Eulerian perturbation theory involves components like kernels, propagators, and vertices. For a comprehensive discussion, refer to ~\cite{Bernardeau_2002}.

\subsection{Lagrangian perturbation theory}
\label{ssec:LPT}

In~\cref{ssec:lss_equations}, fluid approximation for cold dark matter (CDM) is demonstrated. Eulerian perspective, which focuses on specific spatial locations to describe fluid fields, is discussed in~\cref{ssec:EPT}. Alternatively, the field can be described by following the trajectories of particles or fluid elements. It is called the Lagrangian description, which forms the basis of the Lagrangian perturbation theory. It is used to compute the velocities of the particles to generate initial conditions for the simulations.

In this description, the object of interest is the displacement field $\Psi (\mathbf{q})$ which maps the initial comoving particle position $\mathbf{q}$ to its final comoving Eulerian position $\mathbf{x}$~\cite{Buchert_1989, Bouchet1994, Bernardeau_2002}. 

\begin{equation}
  \label{eq:el_trasf}
 \mathbf{x}(\mathbf{q},\tau) \equiv \mathbf{q} + \mathbf{\Psi} (\mathbf{q},\tau).
\end{equation}
Let $J(\mathbf{q},\tau) $ be the Jacobian of the transformation between $\mathbf{q}$ and $\mathbf{x}$,

\begin{equation}
  \label{eq:jacobian}
  J(\mathbf{q},\tau) \equiv \left| \frac{\p \mathbf{x}}{\p \mathbf{q}} \right| = \left| \text{det} \left( \mathcal{I}+ \frac{\p \mathbf{\Psi}}{\p \mathbf{q}} \right) \right|.
\end{equation}
The Jacobian can have the following form if the conservation of mass element is implemented,

\begin{equation}
  \label{eq:jacobian_1}
  J(\mathbf{q},\tau) =\frac{1}{1+\delta(\mathbf{x},\tau)} \text{~~~~~~or~~~~~~} \delta(\mathbf{x},\tau) = J^{-1}(\mathbf{q},\tau) -1.
\end{equation}
When trajectories first cross, fluid elements from different initial positions $\mathbf{q}$ converge to the same Eulerian position $\mathbf{x}$ via mapping in~\cref{eq:el_trasf}. This is called \emph{shell-crossing}, upto it~\cref{eq:jacobian_1} is valid. The Jacobian becomes zero, indicating a singularity and an expected collapse to infinite density. Beyond this point, the dynamics described by the mapping is not valid.

\subsubsection{Equation of motion}
The Vlasov~\cref{eq:de_delta} can be modified to obtain the equation of motion of fluid element with the conformal time as,

\begin{equation}
  \label{eq:jacobian_2}
  J(\mathbf{q},\tau) \nabla_{\mathbf{x}} \cdot \left[ \frac{\p^2 \mathbf{\Psi}}{\p \tau^2} + \mathcal{H}(\tau) \frac{\p \mathbf{\Psi}}{\p \tau} \right] = \frac{3}{2} \Omega_M (\tau) \mathcal{H}^2(\tau)[J^{-1}(\mathbf{q},\tau) -1],
\end{equation}
where $\mathcal{H} = aH$ and $\nabla_{\mathbf{x}}$ is gradient in the Eulerian coordinates. The main challenge of the Lagrangian approach is evident in this equation; the gradient operator needs to be defined with respect to the Eulerian variable $\mathbf{x}$, which is dependent on $\mathbf{q}$. Expressing~\cref{eq:jacobian_2} in Lagrangian coordinates introduces the displacement shears. The resulting non-linear differential equation can then be solved perturbatively, expanding around its linear solution.

\subsubsection{The Zel'dovich approximation}
The Zel'dovich approximation~\cite{Zeldovich_1970, Shandarin_1989} is first order Lagrangian perturbation theory; it takes the linear solution of~\cref{eq:jacobian_2} for the displacement field while keeping the general equation with the Jacobian ~\cref{eq:jacobian_1}. Using linear order relations, $J(\mathbf{q},\tau) \nabla_{\mathbf{x}} \approx \nabla_{\mathbf{q}}$ and $J(\mathbf{q},\tau) \approx 1+ \nabla_{\mathbf{q}} \cdot \mathbf{\Psi}$, ~\cref{eq:jacobian_2} takes the form

\begin{equation}
  \label{eq:1lpt}
  \nabla_{\mathbf{q}} \cdot \mathbf{\Psi}^{(1)} (\mathbf{q},\tau) = - \mathcal{G}(\tau) \delta^{(1)}(\mathbf{q}),
\end{equation}
where $\mathcal{G}(\tau)$ denotes the linear growth factor as in~\cref{eq:del}. Note that the evolution of fluid elements in linear order is local. In linear order, the displacement field is entirely determined by its divergence, i.e., no vorticity. This holds true as initial vorticity diminishes due to the expansion of the Universe, as only the divergence of the velocity field appears in the linear-order fluid equations. Therefore, particles go straight (in comoving coordinates) in the direction set by their initial velocity. 

\subsubsection{Shell-crossing in the Zel'dovich approximation}
Let $\phi(\mathbf{q})$ be Zel'dovich potential such that $\mathbf{\Psi}^{(1)} (\mathbf{q},\tau) = -\mathcal{G}(\tau) \nabla_{\mathbf{q}} \mathbf{\phi}$. If $\lambda_1(\mathbf{q}) \le \lambda_2(\mathbf{q}) \le \lambda_3(\mathbf{q})$ are the local eigenvalues of the Hessian of the Zel'dovich potential, then using~\cref{eq:jacobian_1} the density contrast $\delta$ can be written as~\cite{Bouchet1994, Bernardeau_2002},

\begin{equation}
  \label{eq:za}
  1+\delta(\mathbf{x},\tau) = \frac{1}{[1-\lambda_1(\mathbf{q})\mathcal{G}(\tau)][1-\lambda_2(\mathbf{q})\mathcal{G}(\tau)][1-\lambda_3(\mathbf{q})\mathcal{G}(\tau)]}.
\end{equation}
This equation provides insights into shell-crossing events within the Zel'dovich Approximation (ZA). The perturbation is linear as long as $\mathcal{G}(\tau) \lambda_i \ll 1$. When $\mathcal{G}(\tau) = 1/\lambda_i$ shell crossing happens along the $i^{\text{th}}$ direction of the eigenvector. Therefore, if $\lambda_3 > 0$, and larger than $\lambda_2$ and $\lambda_1$, the ZA predicts a planar collapse to infinite density along this axis, forming a two-dimensional \emph{cosmic pancake}. In cases where at least two eigenvalues are positive, $\lambda_3, \lambda_2 > 0$ and larger than $\lambda_1$, a collapse into a filament occurs. For $\lambda_1, \lambda_2, \lambda_3 > 0$, gravitational collapse happens in all directions, with the largest eigenvalue direction collapsing first, leading to triaxial collapse. For spherical collapse, $\lambda_1 \approx \lambda_2 \approx \lambda_3$ is required near the centre of the spherical symmetric region where the density distribution is isotropic; away from the centre, $\lambda_1 > \lambda_2,\lambda_3$ and $\lambda_2 \approx \lambda_3$, reflecting an axisymmetric collapse. If all eigenvalues are negative, it indicates an evolving underdense region eventually reaching $\delta =-1$. This concept leads to a cosmic web classification algorithm that categorises regions as voids, sheets, filaments, and halos~\cite{Hahn_2007, Lavaux_2010}.

The solution of~\cref{eq:jacobian_2} up to second order in $\phi(\mathbf{q})$ takes into account the fact that
gravitational instability is non-local, i.e. it includes the correction to the ZA displacement due to gravitational
tidal effects~\cite{Melott_1994}. It is referred to as \emph{second-order Lagrangian perturbation theory}, offering a significant enhancement over the Zel'dovich Approximation.

\subsection{Spherical and ellipsoidal collapse}
\label{ssec:collapse}

The spherical collapse model offers a simplified yet insightful depiction of how galaxies or clusters form from initial spherical symmetric perturbations. Such perturbations can be treated as a separate Universe, where the gravitational interaction of matter within the perturbation governs its evolution independently of the surrounding Universe~\cite{Gunn_1972, Fillmore_1984, Bertschinger_1985}. Although the evolution of overdense/underdense regions and background densities ideally requires Friedmann~\cref{eq:Friedmann1,eq:Friedmann2}, often Newtonian mechanics is sufficient to address this problem.

Let $R$ denote the comoving size of the spherical region, then its evolution is given as

\begin{equation}
  \label{eq:dsph}
  \diff[2]{R}{t} = -\frac{4\pi}{3} G\bar{\rho}_m R.
\end{equation}
In general, solving equation~\cref{eq:dsph} reveals the existence of a deterministic relationship between the initial comoving Lagrangian size $R_0$, the linear density of an object $\delta_{\text{lin}}$, and its Eulerian size $R$ at any subsequent time, particularly in an Einstein-de Sitter (EdS) universe,

\begin{equation}
  \label{eq:dsph_sol}
  \frac{R}{R_0} = \frac{1-\cos \theta}{2\bar{\delta}_0},\text{~~~~~~} 
  \frac{t}{t_0} = \frac{3}{4\bar{\delta}_0^{3/2}} \left[ \theta - \sin \theta\right],\text{~~~~~~}
  \delta_{\text{lin}} = \frac{5}{3} \left( \frac{3}{4}\right)^{2/3} \left[ \theta - \sin \theta\right]^{2/3}.
\end{equation}
Initially, overdense region collapse when $\theta = 0$, it \emph{turnaround} at $\theta = \pi$, and collapse completely when $\theta = 2\pi$. Overdense regions collapse in finite time when their extrapolated linear overdensity reaches a threshold of $1.686 \equiv \delta_{sc}$.

Similarly, for evolution of uniform ellipsoid~\cite{White_1979},

\begin{equation}
  \label{eq:dell}
  \diff[2]{R_i}{t} = -\frac{4\pi}{3} G\bar{\rho}_m R_i \left[ 1+ \frac{3}{2} \alpha_i \bar{\delta}\right],
\end{equation}
where, $i$ denotes coordinate axes and $\alpha_i \equiv \alpha_i(a_1,a_2,a_3)$, $a_i$ denoting the half-lengths of the principal axes of the ellipsoid. Unlike spherical collapse where threshold for collapse is constant, ellipsoidal collapse lead to the threshold collapse overdensity $\delta_{ec} \equiv \delta_{ec}(R,t,a_i)$~\cite{Bardeen_1986, Sheth_2001, Sheth_2002}. 

\subsection{Non-linear approximations}
\label{ssec:nla}

In~\cref{ssec:EPT} and \cref{ssec:LPT}, we delve into the theoretical modelling of structure formation, primarily applicable within the linear and quasi-linear regimes. Analysing the non-linear regime analytically proves to be considerably complex. However, several non-linear approximations to the equations of motion have been proposed in the literature to extend analytical calculations into the non-linear regime. These approximations involve substituting one of the dynamic equations (continuity~\cref{eq:cbe_moments0}, Jeans~\cref{eq:cbe_moments1}, and Poisson~\cref{eq:Poisson}) with a different assumption.

The non-linear extension of the Zel'dovich approximation suggests that the relationship between the velocity field and gravitational potential holds beyond just the first order in $\mathbf{\Psi}$~\cite{Bouchet1994, Yoshisato_2006}. It is also possible to formulate local approximations suited for cylindrical or spherical collapse~\cite{Bertschinger_1994, Hui_1996}. Another approach is the `frozen flow' approximation, where the velocity field remains linear while the density field can venture into the non-linear regime~\cite{Coles_1992, Munshi_1994}. Similarly, if a gravitational potential is assumed to be static, it gives a 'frozen potential' approximation \cite{Bagla_1994}. In the `adhesion' approximation, a viscosity term is introduced to the single-stream Euler equation~\cite{Kofman_1988, Gurbatov_1989, Kofman_1992, Valageas_2011}. 

Although all such approximations offer valuable insights, computational methods, particularly N-body simulations are essential for studying gravitational instability for generic density fields.

\section{Challenges in large scale structure studies}
\label{sec:nla}

In the past several decades, progress in cosmology has been remarkable. From complex galaxy distribution patterns to minute variations in the cosmic microwave background (CMB), discoveries have shaped our understanding of the Universe. Using powerful computational tools for simulating structure formation has further paddled cosmology forward, bringing us closer to solving how complex structures in the Universe have emerged from an initially uniform, hot, and dense state. However, these developments have also exposed new questions that need to be addressed, including the validity of the cosmological principles and general relativity, early Universe processes, roles of the dark matter and dark energy, and accurate modelling of astrophysical phenomena such as galaxy formation~\cite{peebles_2003, Steinhardt_2006, Ellis_2009, Gilmore_2011, Colin_2019}. On the other hand, it is an era of precision cosmology, where various cosmological tensions have arisen from inconsistencies between different observational data sets or between observations and theoretical predictions~\cite{Krishnan_2021, Abdalla_2022}. These discrepancies challenge our existing understanding of the Universe. During the last two decades, numerical simulations have become essential in large scale structure studies. Understanding these tools in detail, including how approximations in simulation algorithms impact structure formation, is crucial~\cite{Bagla_2005, Bagla_2006, Bagla_2009}. 

The observed distribution of galaxies, abundance of halos/clusters lensing effects, and peculiar velocities (deviations from the smooth Hubble flow due to local gravitational forces) are influenced by non-linear gravitational interactions. Therefore, Studying the non-linear evolution of large scale structure is essential for accurate cosmological predictions. In the present phase, testing the underlying assumptions and investigating the role of non-linear evolution while making predictions, such as the abundance of halos and the determination of the Hubble-Lema\^itre constant, is necessary. In this thesis, we discuss the case of the abundance of halos and the determination of the Hubble-Lema\^itre constant. We rely on N-body simulations for our studies; therefore, we also discuss the simulation approximations affecting the accuracy of our calculations.

\subsection{Non-universality of halo mass function}
In \cref{cha:3}, we explore the halo mass function in scale-invariant models~\cite{Gavas_2023}. The halo mass function represents the comoving number density of the bound objects, or halos, of mass $m$ at redshift $z$. Theoretical models of the halo mass function are grounded in the assumption that these halos form from regions in the initial density field that eventually collapse due to their sufficient overdensity. The theoretical formulation of mass function can be written in a form that is \emph{universal}, i.e., independent of the cosmology model and initial condition. However, cosmological simulations have shown that this universality is approximate. We investigate the initial power spectrum\footnote{Power spectrum is $\left< \delta(\mathbf {x}) \delta(\mathbf{x})\right>$ in Fourier space. Refer~\cref{ssec:pk}} dependence of halo mass function through N-body simulations of power-law power spectrum initial conditions in an Einstein-de Sitter cosmology (Flat universe with only matter component. Refer~\cref{ssec:scale_free_csm}). This specific choice of cosmology and initial condition ensures the self-similar evolution of the dark matter distribution, allowing us to pinpoint the influence of the power spectrum on the mass function. Our findings reveal a \emph{clear departure from the assumed universality of the mass function}.

\subsection{Determination of the Hubble-Lema\^itre constant}
In \cref{cha:4}, we discuss local measurements of the Hubble-Lema\^itre $H_0$ constant in cosmological N-body simulations. Hubble constant denotes the rate of expansion of the Universe and shows major differences between early and late time measurements. This discrepancy is called `Hubble tension'. Various potential solutions have been proposed, from modifications to the cosmological model to addressing observational systematics~\cite{Di_Valentino_2021}. The peculiar velocities of galaxies, resulting from gravitational clustering, add complexity to  $H_0$ measurement. It introduces a Doppler component to the redshift, affecting the inferred recessional velocity (\cref{a1:rsd}). The effect is significant on scales less than about 100 Mpc/h. In this work, we study the effect of peculiar velocities on the measurement of $H_0$ in the local Universe. \emph{We show a correlation between gravitational clustering and local $H_0$} through simulations. Such correlation can affect the calibration in the local observations, may justify the Hubble tension.

\subsection{Mass discreteness effect in simulations}
In \cref{cha:5}, we study the effect of gravitational clustering at small scales on larger scale by studying mode coupling between virialised halos. The motivation of this study is to understand the impact of finite mass resolution of cosmological N-body simulations on the evolution of perturbations at early times. This difference between the expected evolution and the evolution obtained in cosmological N-body simulations can be quantified using such an estimate. We also explore the impact of a finite shortest scale up to which the desired power spectrum is realised in simulations. Several simulation studies have shown that this effect is small in comparison with the effect of perturbations at large scales on smaller scales~\cite{Peebles_1974, Bagla_1997a}. It is nevertheless important to study this effects and develop a general approach for estimating their magnitude. We provide basic estimates of the magnitude of finite mass resolution effects and their power spectrum dependence.

\subsubsection{}
In this chapter, we trace the history of observing large scale structure (LSS) using galaxy surveys. We explore theoretical models of LSS after establishing the prerequisites. Lastly, we outline general challenges for further exploration in the field and list some of the problems addressed in this thesis. Before delving into these problems, the next chapter introduces some essential tools required for their analysis.


%% file: chapters/chap2.tex

\allowdisplaybreaks

\chapter{Structure formation toolkit}
\label{cha:2}

To interpret the vast datasets from the galaxy surveys and the matter distribution derived from N-body simulations, a mathematical description of their statistical properties is essential. These properties include $n$-point correlation functions, power spectra, cross-correlations, halo mass functions, fractal structure analysis, and Minkowski functionals~\cite{Peebles_1981, Kirkby_1983, Schmalzing1995, Gabrielli_2005}. Further, N-body simulations serve as pivotal computational tool for investigating the formation and evolution of cosmic structures by modelling gravitational interactions among a large number of particles, representing dark matter, gas, and stars~\cite{Bagla_1994, Bagla_2002, Springel_2005, Springel_2017, Angulo_2022}. Thus, structure formation is a multi-faceted field that draws from various disciplines, such as physics, mathematics, and computational science. In this chapter, we review the fundamentals of these tools and key definitions used in the subsequent thesis chapters. We run a suite of simulations required for our studies, described in detail in section~\cref{sec:our_sim}.

\section{Statistical description of the density field}
\label{sec:stat_delta}

This section describes the statistical properties of the density contrast field $\delta(\mathbf{x})$ (\cref{eq:d_contrast}), considering it as a random field. This description is valid for any other random field, such as gravitational potential and velocity divergence field. As discussed in~\cref{sec:theory_lss} the process of generating seed perturbations in $\delta$ is stochastic. For an arbitrary number $n$ of spatial positions $\mathbf{x_i}$, one can define the joint multivariate \emph{Probability Distribution Function} (PDF) $\mathcal{P}$ to have $\delta(\mathbf{x_1})$ between $\delta_1$ and $\delta_1 +\d \delta_1$, $\delta(\mathbf{x_2})$ between $\delta_2$ and $\delta_2 +\d \delta_2$, etc. Such a probability is expressed as

\begin{equation}
  \label{eq:mv_gaussian_pdf}
  \mathcal{P} (\delta_1, \delta_2, \dots , \delta_n) \d \delta_1 \d \delta_2 \dots \d \delta_n.
\end{equation}

\subsection{Average and ergodicity}
The average of a quantity can be defined in two ways. Firstly, one can average by considering numerous realisations drawn from the distribution, all generated in the same manner (e.g. N-body simulations). This is termed the \emph{ensemble average}, for any quantity $X (\delta_1, \delta_2, \dots , \delta_n)$, it is defined as

\begin{equation}
    \label{eq:ens_avg}
    \langle X \rangle \equiv \int X (\delta_1, \delta_2, \dots , \delta_n) \mathcal{P} (\delta_1, \delta_2, \dots , \delta_n) \d \delta_1 \d \delta_2 \dots \d \delta_n.
  \end{equation}
Also, one can compute the average by considering the quantity of interest at various locations within a single realisation of the distribution. This is referred to as the \emph{sample average}. For a volume $V$ in the Universe, the sample average of a quantity $X$ can be expressed as

\begin{equation}
    \label{eq:samp_avg}
    \bar{X}  \equiv \frac{1}{V} \int_V  X(\mathbf{x}) \d^3 \mathbf{x}.
\end{equation}
If the ensemble average of the quantity can be estimated as the sample average, then the system is said to be \emph{ergodic}. In cosmology, ergodicity is frequently assumed if the volume under consideration is sufficiently large. When ergodicity holds in cosmology, the system is considered a \emph{fair sample} of the Universe.

\subsection{Gaussian random field (GRF)}
Gaussian random fields (GRFs) has an extensive application in cosmostatistics~\cite{Lahav_2004, Feigelson_2021}. Evidence suggests that the initial fluctuation field closely resembled a GRF~\cite{Starobinsky_1982, Bardeen_1983, HAWKING_1993}. Perturbation theory based descriptions are used to explain the gravitational clustering originating from Gaussian initial fluctuations~\cite{Bernardeau_2002}. In the late-time structure formation, the large scale distribution of galaxies can be approximately represented as a GRF, especially on scales where gravitational evolution remains well described by linear perturbation theory.

A Gaussian random field is characterised by multivariate Gaussian probability distributions for various $m$-point distributions, such as \cref{eq:mv_gaussian_pdf}. The joint Gaussian probability distribution for random variables $\delta_i$ can be expressed as 

\begin{subequations}
\begin{align}
    \label{eq:mv_gaussian}
    \mathcal{P} (\delta_1, \delta_2, \dots , \delta_n) \d \delta_1 \d \delta_2 \dots \d \delta_n &= \frac{e^{-\mathcal{Q}}}{\left[ (2\pi)^n \text{ Det} (M)\right]^{1/2} }\d \delta_1 \d \delta_2 \dots \d \delta_n, & ~\text{where} \\
    \mathcal{Q} &\equiv \frac{1}{2} \sum \Delta \delta_i (M^{-1})_{i,j} \Delta \delta_j, & \text{~and}\\
    \Delta \delta_i \equiv  \delta_i - \langle \delta_i \rangle, &\text{~~~~~} M_{ij} \equiv \langle \Delta \delta_i \Delta \delta_j \rangle .&\text{~~}
\end{align} 
\end{subequations}
Therefore, the averages of the random variables $\langle \delta_i \rangle$ and covariance matrix $M$ suffice to define the distribution.

The renowned BBKS (Bardeen, Bond, Kaiser, Szalay) paper provides a comprehensive analytical overview of statistics of peaks in Gaussian random fields~\cite{Bardeen_1986}. 
Note that the assumption of Gaussianity for $\delta(\mathbf{x})$ breaks down in the nonlinear regime of structure formation. $\delta_{\text{lin}}(\mathbf{x})$ (denoted as $\delta^{(1)}(\mathbf{x})$ in \cref{eq:1lpt}), is essentially the divergence of the velocity field, represents a Gaussian random field since the initial velocities are Gaussian. The nonlinear density contrast $\delta(\mathbf{x})$  (\cref{eq:za}) is constrained by $1 + \delta(\mathbf{x}) \ge 0$, implying $\delta(\mathbf{x}) \geq -1$. This lower bound and the growth of nonlinear corrections skew the distribution, violating Gaussian symmetry. $\delta_{\text{lin}}(\mathbf{x})$ resembles $\delta(\mathbf{x})$ only when ($|\delta(\mathbf{x})| \ll 1$); thus, the Gaussian assumption holds only in the linear regime, where higher-order contributions are negligible.

\section{Measures of gravitational clustering}
\label{sec:stat_measures}
An enormous variety of clustering statistics is used for large scale structure study. These include correlation functions and their Fourier counterparts, mass functions, clustering/halo models, angular correlations, scaling relations, neighbour statistics, and the Mandelbrot prescription, which are few among all~\cite{Peebles_1981}. This section only focuses on the clustering statistics used for this thesis work.

\subsection{$n$-point correlation functions}
\label{ssec:npcf}
$n$-point correlation functions provide a systematic way to measure the clustering properties of a dataset by examining the spatial relationships between groups of points. These functions quantify the excess probability of finding $n$-points within certain distances of each other, compared to what would be expected in a random distribution. For example, the two-point correlation function measures the excess probability of finding pairs of points at a given distance compared to a random distribution. Higher-order correlation functions, such as the three-point, four-point, or higher, provide information about the clustering beyond pairs of points, capturing more complex spatial arrangements.

The $n$-point correlation function $\xi_n$ of density fluctuations $\delta(\mathbf{x})$ is defined by

\begin{equation}
    \label{eq:npcf}
    \langle \delta(\mathbf{x}_1) \dots \delta(\mathbf{x}_n) \rangle_c \equiv \xi_n (\mathbf{x}_1 \dots \mathbf{x}_n).
\end{equation}
The subscript $c$ stands for connected moments/cumulants\footnote{Connected moments/cumulants are a set of quantities that provide an alternative to the moments of the distribution.}. If all the $\mathbf{x}_i$ are the same, then

\begin{subequations}
    \label{eq:1-4pcf}
\begin{align}
    \langle \delta \rangle_c &= \langle \delta \rangle,\\
    \langle \delta^2 \rangle_c &= \langle \delta^2 \rangle - \langle \delta \rangle_c^2 \equiv \sigma^2,\\
    \langle \delta^3 \rangle_c &= \langle \delta^3 \rangle - 3 \langle \delta^2 \rangle_c \langle \delta \rangle_c - \langle \delta \rangle_c^3,\\
    \langle \delta^4 \rangle_c &= \langle \delta^4 \rangle - 4 \langle \delta^3 \rangle_c \langle \delta \rangle_c - 3 \langle \delta^2 \rangle_c^2 -6 \langle \delta^2 \rangle_c  \langle \delta \rangle_c^2- \langle \delta \rangle_c^4.
\end{align}
\end{subequations}
For the density perturbation field~\cref{eq:d_contrast}, $\langle \delta \rangle =0$. Thus, $\langle \delta^n \rangle_c = \langle \delta^n \rangle$ upto $n=3$. For a Gaussian random field, all higher order moments can be expressed using the two-point correlation function. For fair sample, the two-point correlation function $\xi$ is defined as

\begin{equation}
    \label{eq:2pcf} 
    \xi (\mathbf{r}) = \langle \delta(\mathbf{x}) \delta(\mathbf{x}+\mathbf{r})  \rangle = \frac{1}{V} \int \d^3 \mathbf{x} \delta(\mathbf{x}) \delta(\mathbf{x}+\mathbf{r}).
\end{equation}
Note that the subscript $n$ on $\xi$ is dropped. For discrete $N$ points distribution over volume V, the two-point correlation function can be defined in a way such that,

\begin{equation}
    \label{eq:d2pcf} 
    \delta \mathcal{P} = \bar{n} \delta V \left[ 1 + \xi(\mathbf{r})\right],
\end{equation}
where, $\delta \mathcal{P}$ is probablity of finding neighbour at distance $r=|\mathbf{r}|$ in small volume $\delta V$ with respect to randomly choosen point (Refer section \S 31 of~\cite{Peebles_1981}). Similarly, higher order correlation functions can be formulated. The volume average of the correlation function $\bar{\xi}(r)$ is interpreted as the excess number of neighbours within a distance $r$ of a given point

\begin{equation}
    \label{eq:xi_bar} 
    \bar{\xi}(r) = \frac{3}{r^3} \int_0^r y^2 \xi(y) dy.
\end{equation}

\subsection{Power spectrum}
\label{ssec:pk}

Many of the calculations involved in structure formation studies get simplified considerably in the \emph{Fourier space}. Here, we discuss the power spectrum, which is the Fourier transform of the correlation function.

Let $\delta_{\mathbf{k}}$ be the Fourier transform of $\delta(\mathbf{r})$ (\cref{a2:del}). Thus, 

\begin{equation}
    \label{eq:pk} 
    \xi(\mathbf{r}) = \int \frac{\d^3 \mathbf{k}}{(2\pi)^3} |\delta_{\mathbf{k}}|^2 e^{\iota \mathbf{k}\cdot \mathbf{r}} = \int \frac{\d^3 \mathbf{k}}{(2\pi)^3} P(\mathbf{k}) e^{\iota\mathbf{k}\cdot \mathbf{r}},
\end{equation}
where, $P({\mathbf{k}}) \equiv \langle |\delta_{\mathbf{k}}|^2 \rangle$ is called power spectrum. In an isotropic Universe, the perturbations cannot have a preferred direction; therefore, $\xi(\mathbf{r}) = \xi(r)$ and $P(\mathbf{k}) = P(k)$. This means $\xi$ and $P$ are statistical quantities rather than local quantities like $\delta$. Similarly, higher order quantities like bispectrum and trispectrum can be defined (\cref{a2:cf}). Working with a dimensionless power spectrum is often more convenient. It is defined as

\begin{equation}
    \label{eq:dpk} 
    \Delta^2(k) \equiv \frac{k^3 P(k)}{2\pi^2}.
\end{equation}

\subsection{Mass variance}
\label{ssec:mv}
Mass variance is root-mean-square (rms) fluctuations in mass at a given length scale $r$. A commonly used parameter $\sigma_8$ is the filtered mass variance in spheres of radius $r=8 h^{-1}$ Mpc, roughly corresponding to the scale of massive galaxy clusters. Its more generic form for arbitrary $r$ is given by

\begin{equation}
    \label{eq:sgm} 
    \sigma^2(r) = \frac{\langle [M - \langle M \rangle]^2 \rangle}{\langle M \rangle^2} = \int \frac{\d k}{k} \frac{k^3 P(k)}{2\pi^2} |W(kr)|^2,
\end{equation}
where $W(kr)$ is some window function\footnote{A window function is convolved with density fields for smoothing or filtering to reduce noise and emphasise larger scale features (\cref{a2:wf}).} (\cref{a2:wf}) used to smooth the density field so that $M = f_W r^3 \rho_m$, for a realistic case, we need $f_W = 4\pi/3$, which is satisfied by spherical top-hat window function, however, there are other window functions which in use, some of them and their Fourier transforms are discussed in~\cref{a2:wf}.

\subsection{Halo mass function}
\label{ssec:hmf}
The halo mass function is a statistical distribution that describes the abundance of dark matter halos of different masses at different times. Formally, it is the comoving number density of halos of mass $m$ at redshift $z$,

\begin{equation}
    \label{eq:ndv} 
    \frac{m^2~n(m,z)}{\bar{\rho}_m} \frac{\d m}{m} = \nu f(\nu) \frac{\d \nu}{\nu}, \text{~~~~~} \nu \equiv \frac{\delta_{sc}^2 (z)}{\mathcal{G}^2(z) \sigma^2_{\text{lin}}(m)}.
\end{equation}
where $\delta_{sc}$ is the critical overdensity threshold for spherical collapse, it generally has a weak dependence on cosmology. $\mathcal{G}$ is growth factor. $\sigma_{\text{lin}}$ variance in the initial density perturbation field extrapolated to the present time using linear theory and smoothed with a top hat filter ($W(k,m)$) of scale $r$= $(3m/4\pi\bar{\rho})^{1/3}$. It is calculated as shown in~\cref{eq:sgm}. There are two fundamental models for $\nu f(\nu)$ based on spherical and ellipsoidal collapse (discussed in~\cref{ssec:collapse}), namely Press-Schechter and Sheth-Tormen mass function,

\begin{subequations}
\begin{align}
    \label{eq:ps} 
    \text{Press-Schechter: ~~~~~~~~~~} \nu f(\nu) &= \sqrt{\frac{\nu}{2\pi}}e^{-\nu/2},\\
    \label{eq:st} 
    \text{Sheth-Tormen: ~~~~~~~~~~} \nu f(\nu) &= A(p) [1+(q\nu)^{-p}] \sqrt{\frac{q\nu}{2\pi}}e^{-q\nu/2}.
\end{align}
\end{subequations}
where $p$ and $q$ are termed as the Sheth-Tormen parameters, we revisit them in \cref{cha:3}. When $p = 1/2$ and $q = 1$ in~\cref{eq:st}, the~\cref{eq:ps} is recovered. For $\nu\ll 1$, the mass function scales as $\nu^{0.5-p}$. While the exponential cutoff at $\nu\gg 1$ depends on the value of $q$. The characteristic mass scale denoted as $m_*$, is defined at $\nu = 1$; note that halos more massive than $m_*$ are rare. Both Press-Schechter and Sheth-Tormen formalism assume that all the mass in the Universe is present in the form of halos of some mass, i.e., $\int_0^{\infty} f(\nu)d\nu /\nu = 1$.
This provides the normalisation for the mass functions.
\begin{equation}
    A(p) = \left[1+ \frac{2^{-p} \Gamma(0.5-p)}{\sqrt{\pi}}\right]^{-1}.
    \label{eqn:Ap}
\end{equation}

The derivations of~\cref{eq:ps} in~\cite{Epstein_1983, Peacock_1990, Bond_1991} demonstrate its relationship to a model wherein halos form from spherical collapse. When extended to the ellipsoidal collapse model described by~\cite{Bond_1996}, the same ideas yield equation~\cref{eq:st}~\cite{Sheth_2001, Sheth_2002}, which has been found to offer a satisfactory description of the mass function in numerical simulations. However, improved simulation precision has revealed that the Sheth-Tormen mass function predicts an excess of halos near $m_*$ and underpredicts at both tails. Further, both~\cref{eq:ps,eq:st} exhibit \emph{universality}, implying they are independent of the choice of cosmological model and initial conditions. However, \cref{cha:3} demonstrates this does not hold true.


\section{N-body simulations}
\label{sec:sim} 

When density contrast approaches unity, perturbations enter a non-linear regime, interacting significantly across scales. Solving Vlasov~\cref{eq:de_delta} for perturbations with generic initial conditions becomes challenging. Hence, N-body simulations are indispensable for studying this regime. Additionally, solutions provided by quasi-linear approximations~\cite{Bernardeau_2002, Zeldovich_1970, Gurbatov_1989, Matarrese_1992, Brainerd_1993, Bagla_1994, Sahni_1995, Hui_1996} or scaling relations~\cite{Davis_1977, Hamilton_1991, Nityananda_1994, Peacock_1994, Padmanabhan_1996, Padmanabhan_1996a, Smith_2003} must be validated against simulations before application. Emerging fields like Machine Learning (ML) and Artificial Intelligence (AI) offer an alternative approach to traditional simulations~\cite{He_2019, Meskhidze_2021, Ciprijanovic_2021, Villaescusa-Navarro_2021, Piras_2023, Villaescusa_Navarro_2022}. They are used for emulating, predicting, generating, and detecting patterns and anomalies in large scale structures. However, a substantial amount of training data generated by full N-body simulations is required for these methods.

Over the last three decades, significant progress has been made in cosmological simulations driven by refinement in techniques and increased computing power. State-of-the-art simulations, such as MillenniumTNG~\cite{Pakmor_2023} and EUCLID Flagship Simulation~\cite{Euclid_2020}, involve more than $10^{12}$ particles. Developing new numerical methods, optimisation algorithms, and parallel programming has been crucial in such advancements. These developments have enabled simulations to provide in-depth insights into the structure formation process in the Universe, which is critical for precision cosmology. In this section, we explore fundamental algorithms for N-body simulations, essential concepts in cosmological simulations, and algorithms for structures classification.

\subsection{Equations of motion and Basic algorithm}
In the N-body approach, the continuous matter density field is sampled through a set of discrete identical point-like tracers called `particles', which carry a given mass and only interact via gravity. This is called the \emph{particle approximation} of the Vlasov~\cref{eq:de_delta}. The dynamics of particles is treated in the non-relativistic medium in an expanding cosmological background. It can be studied in the Newtonian limit at much smaller scales than the Hubble radius $d_H =cH_0^{-1}$ (\cref{a1:hs}). For a set of N particles, the system of equations that describe their evolution of trajectories is given by 

\begin{subequations}
\label{eq:nb}
\begin{align}
    \label{eq:nbody} 
    \ddot{\mathbf{r}}_i(t) =  \ddot{a}(t) \mathbf{x}_i (t) +  2\dot{a}(t)\dot{\mathbf{x}}_i(t)  + a(t) \ddot{\mathbf{x}}_i  &= -\nabla [\phi_b + \phi (\mathbf{x}_i)],\\
    \label{eq:nbody_pot} 
    \phi (\mathbf{x}_i) &= -\frac{G}{a^2} \sum_{i \neq j} m_j \frac{1}{|\mathbf{x}_i - \mathbf{x}_j|}, 
\end{align}
\end{subequations}
where $\phi_b = -\ddot{a}r^2 / 2a$ is contribution from the background Universe. $\mathbf{r}_i$ and $\mathbf{x}_i$ are physical and comoving position and $m_j$ is the mass of the $i^{th}$ particle. These are a set of coupled differential equations that are solved in a N-body simulation. \\

The basic algortim to solve equations~\cref{eq:nb} is given in four steps~\cite{Hockney_1988, Bagla_1997},

\begin{enumerate}
    
    \item \emph{Setting up the initial conditions (ICs)}: Initial positions for the particles are distributed according to a Gaussian field using the desired power spectrum for the model of interest~\cite{Buchert_1994, Crocce_2006, Hahn_2011, Michaux_2020, Hahn_2020}. This involves sampling positions from a Gaussian distribution with mean zero and variance determined by the power spectrum. Lagrangian perturbation theory (\cref{ssec:LPT}) is used to compute the initial velocities of the particles. It enables the generation of velocities that that are consistent with the selected initial density field for adiabatic perturbations in the growing mode.

    \item \emph{Force calculations:} Forces acting on the particles are calculated using~\cref{eq:nbody_pot}. Various standard techniques like particle-particle (PM), particle mesh (PM), fast multipole method (FMM), and the Barnes-Hut tree method are used to calculate the forces~\cite{Ewald_1921, Klypin_1983, Efstathiou_1985, Barnes_1986, Hockney_1988, Bagla_2002}. Advanced codes use hybrid methods such as Particle-Particle Particle-Mesh (P$^3$M) and Tree Particle-Mesh (TreePM) for force calculations~\cite{Toukmaji_1996, Bagla_1997, Bagla_2002}.

    \item \emph{Time integration:} Positions and velocities of particles are then updated using a numerical integration scheme such as the leapfrog integration to advance the simulation in time~\cite{Quinn1997}.

    \item \emph{Repeat:} Steps 2 and 3 are repeated multiple times to simulate the evolution of the system over time.
\end{enumerate}

\subsection{Nuances of cosmological N-body simulations}
\label{ssec:Nuances}
The algorithm discussed above is very naive. Working simulation codes implement various techniques for optimisations and modifications because the system is cosmological. We list some of them below.

\subsubsection{Periodic boundary condition (PBC)}
Periodic boundary conditions are used in cosmological simulations to mimic an infinite space by imposing a periodic repetition of the simulation volume. This means that when a particle moves out of one side of the simulation box, it re-enters from the opposite side. The mapping of coordinates (denoted by $\prime$) with PBC can be expressed as follows,

\begin{equation}
    \label{eq:pbc }
    \mathbf{x}^{\prime} = \mathbf{x} - \floor{\frac{\mathbf{x}}{L}}L,
\end{equation}
where, $\floor{\cdot}$ is the floor function and $L$ is the length of periodic box.

\subsubsection{Short-range and long-range force}
Most of the modern simulations, including those utilised for this thesis work, use a hybrid TreePM algorithm for force calculations~\cite{Bagla_1997, Bagla_2002}. This algorithm divides the gravitational force into long-range and short-range components by splitting unity in the Poisson~\cref{eq:Poisson}.

\begin{subequations}
\begin{align}
    \label{eq:pot_l}
    \phi_k^l &= - \frac{4\pi G \rho_k}{k^2} \left[ e^{-k^2 r_s^2}\right],\\
    \label{eq:pot_s}
    \phi_k^s &= - \frac{4\pi G \rho_k}{k^2} \left[ 1 - e^{-k^2 r_s^2}\right],
\end{align}
\end{subequations}
where $\phi_k^s$ and $\phi_k^l$ are the short-range and long-range potentials in the Fourier space $\phi_k = \phi_k^s +\phi_k^l$. And $r_s$ is a \emph{force split scale}. $\phi_k^l$ is computed in the Fourier space with the particle mesh method and involves solving the Poisson equation on the \emph{grid}. $\phi_k^s$  is solved in real space using the Tree method or Fast Multipole Method (FMM).

Short-range forces, dominant at small scales, are calculated with higher computational resolution, ensuring accuracy in capturing local interactions between nearby particles. Long-range forces, significant at larger scales, are efficiently computed using approximation methods such as the particle mesh, enabling simulations to scale to larger computational domains and particle counts while maintaining computational efficiency. This approach optimises computational resources, allowing for accurate and scalable gravitational dynamics simulations across various scales.

\subsubsection{Force softening}
The gravitational potential generated using Poisson~\cref{eq:nbody_pot} approaches infinite values as the distance between particles approaches zero. This creates a problem in simulating collisionless particles affected by the force of gravity. To fix the issue,~\cref{eq:nbody_pot} can be modified to soften the potential at small scales by introducing a small term $\epsilon$, called the \emph{softening length}, in the denominator to regularise the divergences at zero separation, 

\begin{equation}
    \label{eq:pot_soft} 
    \phi (\mathbf{x}_i) = -\frac{G}{a^2} \sum_{i \neq j} m_j \frac{1}{\left[(\mathbf{x}_i - \mathbf{x}_j)^2 + \epsilon^2(\epsilon_0, r)\right]^{1/2}}. 
\end{equation}
The function $\epsilon(r)$ describes a gravitational softening law. Although there are many possible forms for $\epsilon$, in this thesis

\begin{equation}
    \label{eq:soft_fn} 
    \epsilon(\epsilon_0, r) = - \frac{2.8 \epsilon_0}{W_2(r/2.8\epsilon_0)} -r 
\end{equation}
is used. The function $W_2(u)$ is given in~\cite{Springel_2001} (eq. 71).

If $\epsilon$ is excessively small, it cannot effectively prevent infinite forces, resulting in noisy, unstable evolution and longer integration times for the simulation. Conversely, large values diminish spatial resolution, hindering precise calculations on small scale~\cite{Garrison_2016, Garrison_2021}.

\subsubsection{Mass resolution}
Mass resolution in N-body simulations refers to the smallest mass each simulated particle represents ($m_{\text{part}}$). Higher mass resolution implies that each particle represents a smaller mass, allowing for better resolution of small scale structures and gravitational interactions. Similarly, lower mass resolution means that each particle represents a larger mass, producing a coarser representation of the simulated system. Achieving higher mass resolution requires simulating more particles, increasing computational costs~\cite{Kuhlman_1996, Splinter_1998, Joyce_2020, Armijo_2021}. If $V$ is the simulation volume, then

\begin{equation}
    \label{eq:mpart} 
    m_{\text{part}} = \frac{\bar{\rho}_m V}{\text{Number of particles in the simulation}}. 
\end{equation}

\subsubsection{Spatial resolution}
Spatial resolution is related to the size of simulation particles that determines the smallest structures that can be resolved in the simulated Universe~\cite{Yahagi_2001, Joyce_2020}. It can be defined using the grid\footnote{The grid subdivides the simulation volume into discrete cells, facilitating the calculation of quantities at grid points within the volume.} spacing.  

The power spectrum used to generate initial conditions for simulations is restricted at $k_{\text{max}} = \pi/L_{\text{grid}}$, where $L_{\text{grid}}$ is grid spacing. Therefore, the initial positions of particles generated in simulations lack information of the omitted modes ($k > k_{\text{max}} \sim L_{\text{grid}}$). Hence, $L_{\text{grid}}$ can be used to define spatial resolution, below which simulation results are unreliable, especially at early times. The errors due to missing modes propagate further during the evolution of the simulations. This leads to uncertainties in computed quantities at scales comparable to the grid length. We analyse this effect on mass variance and mass function in \cref{cha:5}.

\subsubsection{Finite box size effect}
The lower end of the input power spectrum is truncated at $k_{\text{min}} = 2\pi/L_{\text{box}}$, where $L_{\text{box}}$ represents the length of the simulation box. This truncation introduces notable errors, as modes larger than the box size are disregarded during initial condition generation and subsequent evolution~\cite{Bagla_2006, Bagla_2009, Klypin_2019}. These errors undermine the reliability of certain quantities within a specific range. Therefore, careful consideration of the box size for simulation is crucial, and the guidelines outlined in~\cite{Bagla_2006} offer a prescription for this determination.

\subsection{Structure classification}
Structure classification algorithms aim to categorise and characterise the various components of the large scale structure, such as halos/galaxy clusters, filaments, and voids. These algorithms can be used on observational data from galaxy surveys and simulations. Some standard algorithms for classification are Friends-of-Friends (FoF)~\cite{Davis_1985, More_2011}, density-based algorithms~\cite{Warren_1992, Lacey_1994, Bond_1996, Neyrinck_2008, Cautun_2014}, voronoi tessellation~\cite{Ramella_2001}, watershed algorithm~\cite{Platen_2007, Lin_2016} and topological methods~\cite{Xu_2019, Palomino_2024}.

Halos, gravitationally bound clumps of particles, play a crucial role in understanding the structure formation. While they are not the true observables, galaxies are presumed to form within the potential wells of such halos~\cite{Benson_2010, Somerville_2015, Wechsler_2018}. Therefore, understanding the large scale evolution of galaxies relies on accurately identifying these collapsed structures within the dark matter density field by halo-finder algorithms. Various techniques exist for identifying halos in simulations, each with its advantages and disadvantages, depending on the desired information and the accuracy level required versus the simplicity of the algorithm. Some methods connect neighbouring particles in configuration space, while others incorporate information from the velocity field or identify overdensities in the matter field~\cite{Davis_1985, Warren_1992, Lacey_1994, Bond_1996, Gottloeber1997, Neyrinck_2005, Springel_2001a, Aubert_2004, Tweed_2009}. 

A plethora of halo-finding algorithms are available in the literature~\cite{Springel_2001a, Behroozi_2012, Elahi_2019, Sun_2020, Hadzhiyska_2021}. Here, we discuss the \texttt{FoF-SUBFIND} code, a widely used tool for constructing halo catalogues in cosmological simulations~\cite{Springel_2001a, Springel_2021}. It operates in two main stages. It first identifies Friends-of-Friends (FoF) groups in a given particle distribution. Particles within a certain \emph{linking length} are considered `friends' and grouped into FoF groups based on spatial proximity. Once the FoF groups are identified, the code employs the excursion set algorithm to decompose each FoF object into smaller substructures. This step helps identify subhalos or smaller clumps within larger halos, providing a more detailed characterisation of the halo population.

The {\tt FoF-SUBFIND} employs a spherical overdensity criterion to define halo masses. This implies that halos are identified based on regions of space where the local density exceeds a certain threshold above the background density (e.g. 200 times overdense is $m_{200b}, 18$$\pi^2$ times overdense is $m_{\text{vir}}$). Consequently, the halo masses determined using different overdensity definitions and algorithms may differ.

Note that various factors can affect the accuracy of halo properties obtained from halo finders~\cite{Tinker_2020, Valles-Perez_2022}. For instance, the finite size of the simulation box can lead to an underestimation of halo number density at the high-mass end. In contrast, discreteness noise due to the finite number of simulation particles can affect halo number density at the low-mass end. Careful consideration and correction for these effects are essential to ensure the reliability of halo properties measurements and the validity of subsequent analyses based on these measurements.

\section{Suite of simulations in this thesis}
\label{sec:our_sim}

In this thesis, we use two types of simulations, namely scale-free and $\Lambda$CDM cosmologies. In this section, we highlight their advantages and key features.

\subsection{Scale-free cosmology}
\label{ssec:scale_free_csm}

Scale-free cosmological models are characterised by the Einstein-de Sitter (EdS) background cosmologies featuring a power-law initial spectrum of fluctuations, resulting in a self-similar clustering evolution.

In the EdS universe, the spatial curvature is flat, with no cosmological constant, i.e. $\Omega = \Omega_m = 1$. The Friedmann~\cref{eq:Friedmann} describing the evolution of the scale factor is 

\begin{equation}
    \label{eq:eds} 
    \left( \frac{\dot{a}}{a} \right) ^2 = H_0^2 \Omega_m a^{-3}. 
\end{equation}
The solution of the Friedman~\cref{eq:eds} has the form 

\begin{equation}
    \label{eq:eds_a} 
    a \propto t^{2/3}. 
\end{equation}
Therefore, the scale factor follows a power-law scaling with the coordinate time, and there are no other features in the time scales. When EdS universe is combined with power-law power spectrum as an initial condition of the form 

\begin{equation}
    \label{eq:plpk} 
    P(k) \propto k^n,  
\end{equation}
implies that there are no featured scales in the initial power spectrum either. Note that $4>n\ge -3$; this restriction on the values of $n$ comes from linear theory. $n<-3$ gives divergent density contrast, such as $\xi(r)\rightarrow \infty$ as $r \rightarrow \infty$ at initial times. $n < 4$ assures that non-linear processes at small scales do not affect the behaviour of perturbations at larger scales~\cite{Peebles_1974}.

\subsubsection{Self-similarity}
These models are recognised to have self-similar solutions to the Vlasov~\cref{eq:de_delta}, where any dimensionless statistical quantity $X$ derived from the phase space density can be expressed as

\begin{equation}
    \label{eq:xnl} 
   X(x,t) = X(x/t^{\alpha}),\text{~~~where~} \alpha \in \mathds{R}.
\end{equation}
For instance, the linear power spectrum is given by

\begin{equation}
   P_{\text{lin}}(k,a) = a^2 A k^n,
\end{equation}
where $A$ is the normalisation factor. It gives a dimensionless power spectrum as in~\cref{eq:dpk}

\begin{equation}
   \Delta_{\text{lin}}^2 (k,a) =a^2 \frac{k^3}{2\pi^2} A k^n \propto a^2k^{3+n}.
\end{equation}
We define the self-similar rescaling by the \emph{scale of non-linearity} $r_{nl}$ and corresponding wavenumber $k_{nl}$ such that $\Delta_{\text{lin}}^2(k_{nl},a) \equiv 1$. $k_{nl}$ separates the linear regime ($\Delta^2(k) \ll 1$) from the non-linear regime ($\Delta^2(k) \gg 1$). This definition of non-linearity gives

\begin{equation}
   k_{nl} \propto a^{-2/(3+n)} \propto t^{-4/3(3+n)}.
\end{equation}
This scaling implies that $\Delta_{\text{lin}}^2$ and consequently any dimensionless clustering statistic, is no longer an arbitrary function of two independent variables length $r$ and scale factor $a$, but solely of the rescaled variable $k/k_{\text{nl}}$. Scale-free cosmologies can thus be represented by a single scale, the scale of non-linearity, containing all the information about temporal evolution. If the gravitational clustering evolution does not rely on other scales, it exhibits self-similarity in the rescaled coordinate. A more general version of self-similar rescaling can be given by extending~\cref{eq:xnl} 

\begin{equation}
    \label{eq:exnl} 
   X(x_i,a) = X(x_i/x_{nl,i}(a)),
\end{equation}
where $x_{nl,i}(a)$ is the non-linearity scale with same dimensions as $x_i$. In scale-free simulations, deviations from self-similarity are solely attributable to the shortcomings of N-body description. These departures originate from limitations due to the introduction of unphysical parameters in discretising the density field and the constraints of the simulation volume. Thus, these models simplify the modelling of large scale structures, offering valuable theoretical insights.

\subsubsection{Scaling relations}
As demonstrated above, rescaling of dimensionless power spectrum gives,

\begin{equation}
   k_{nl} \propto a^{-2/(3+n)},
\end{equation}
and corresponding $r_{nl}$ scales as

\begin{equation}
   r_{nl} \propto k_{nl}^{-1} \propto a^{2/(3+n)},
\end{equation}
which leads to the volume averaged 2-point linearly extrapolated correlation as

\begin{equation}
   \bar{\xi}(r) \approx \left[\frac{r}{r_{nl}}\right]^{-(n+3)}.
\end{equation}
Likewise, \emph{mass of non-linearity} defined as the mass contained within a sphere of radius $r_{nl}$ scales as

\begin{equation}
   m_{nl} \propto r_{nl}^3 \propto a^{6/(3+n)},
\end{equation}
leading to linearly extrapolated mass variance of the form

\begin{equation}
    \label{eq:sgm_sf} 
   \sigma(m) = \left[\frac{m}{m_{nl}}\right]^{-\frac{n+3}{6}}.
\end{equation}

When extrapolating scale-free results to $\Lambda$CDM (cosmological constant $\Lambda$ and Cold Dark Matter) cosmologies (described in~\cref{ssec:lcdm_cat}), the transition between scales that emerge during the radiation-dominated era and those that arise during matter domination occurs gradually. Consequently, the change in the slope of the matter power spectrum is slow. Therefore, the Einstein-de Sitter scenario reasonably approximates $\Lambda$CDM cosmologies during the matter domination era. Thus, $\Lambda$CDM epochs can be approximated by interpolating scale-free cosmologies with different spectral indices, effectively capturing the behaviour of the variance at a standard scale. We define the effective spectral index $n_{\text{eff}}$ based on the top-hat variance of density fluctuations as

\begin{equation}
   \frac{\d \log \sigma(m,a)}{\d \log m}\bigg|_{\sigma = 1 }= - \frac{n_{\text{eff}}+3}{6}.
\end{equation}

\subsection{Scale-free simulation catalogue}
\label{ssec:pl_cat}

We conduct a series of dark-matter-only simulations using eight power-law power spectrum initial conditions. The power-law indices $n$ of the power spectrum are listed in column 1 of~\cref{tab:pl_sim}. For simulation runs, we use {\tt GADGET4}~\cite{Springel_2021}, configured in TreePM mode, with initial conditions generated via second-order Lagrangian perturbation theory. As discussed in~\cref{ssec:Nuances}, we set the force split radius to $2$ in grid units. The softening length is set to $\epsilon_0 = 0.05$ grid units across all runs. We employ adaptive time steps with a maximum step size of $0.005$. These parameter selections ensure that the force calculation errors remain well below $1$\%. Halos are identified using {\tt FoF-SUBFIND} code.

\begin{table}
    \begin{center}
	 \begin{tabularx}{\textwidth}
        {|>{\centering\arraybackslash}m{0.7cm} |>
        {\centering\arraybackslash}m{0.7cm} |>
        {\centering\arraybackslash}m{1.4cm} |>
        {\centering\arraybackslash}m{1.4cm} |>
        {\centering\arraybackslash}m{5.2cm} |>
        {\centering\arraybackslash}m{0.7cm} |>
        {\centering\arraybackslash}m{1.4cm}|}

	 \hline
	 $n$ & $z_{\text{start}}$ & $L_{\text{box}}$ & $\sqrt[3]{N_{\text{part}}}$  & $r_{nl~~~(\text{grid~unit})}$ & $r_{nl}^{\text{max}}$ & CPU hours\\ [0.3cm] 
	 \hline
	 0.0 & 453 & 512 ~~~~~[$\times$ 10] & 512 & \raggedright [13]~1.26, 1.63, 2.12, 2.76, 3.58, 4.66, 6.06, 7.88, 10.24, 13.31, 17.30, 22.49, 29.24 &  58.7 & 3,21,020\\ [.1cm]
	 \hline

     & 100 & 512 & 512 & \raggedright [13]~1.26, 1.63, 2.12, 2.76, 3.58, 4.66, 6.06, 7.88, 10.24, 13.31, 17.30, 22.49, 29.24 &  58.7 & 3,20,92 \\ [.1cm]
	 \hline

     -0.5 & 270 & 512 ~~~~~[$\times$ 10] & 512 & \raggedright [14]~1.26, 1.63, 2.12, 2.76, 3.58, 4.66, 6.06, 7.88, 10.24, 13.31, 17.30, 22.49, 29.24, 38.01 &  42.3 & 3,98,608\\ [.1cm]
	 \hline

     -1.0 & 160 & 512 ~~~~~[$\times$ 10] & 512 & \raggedright [12]~1.26, 1.63, 2.12, 2.76, 3.58, 4.66, 6.06, 7.88, 10.24, 13.31, 17.30, 22.49&  27.4 & 2,30,320\\ [.1cm]
	 \hline

      & 160 & 768 & 768 & \raggedright [14]~1.26, 1.63, 2.12, 2.76, 3.58, 4.66, 6.06, 7.88, 10.24, 13.31, 17.30, 22.49, 29.24, 38.01 &  41.1 & 91,776\\ [.1cm]
      \hline

      & 160 & 1024 & 1024 & \raggedright [15]~1.26, 1.63, 2.12, 2.76, 3.58, 4.66, 6.06, 7.88, 10.24, 13.31, 17.30, 22.49, 29.24, 38.01, 49.41 &  54.8 & 192,482\\ [.1cm]
      \hline

      -1.5 & 96 & 1024 & 1024 & \raggedright [12]~1.26, 1.63, 2.12, 2.76, 3.58, 4.66, 6.06, 7.88, 10.24, 13.31, 17.30, 22.49 &  27.8 & 1,18,800\\ [.1cm]
      \hline

      -1.8 & 70 & 1024 & 1024 & \raggedright [10]~1.26, 1.63, 2.12, 2.76, 3.58, 4.66, 6.06, 7.88, 10.24, 13.31 &  14.5 & 88,998\\ [.1cm]
      \hline

      -2.0 & 57 & 1536 & 1536 & \raggedright [16]~0.2, 0.26, 0.34, 0.44, 0.57, 0.74, 0.97, 1.26, 1.63, 2.12, 2.76, 3.58, 4.66, 6.06, 7.88, 10.24 &  11.4 & 1,30,743\\ [.1cm]
      \hline

      -2.2 & 46 & 1536 & 1536 & \raggedright [12]~0.2, 0.26, 0.34, 0.44, 0.57, 0.74, 0.97, 1.26, 1.63, 2.12, 2.76, 3.58 &  4.4 & 1,53,520\\ [.1cm]
      \hline

      -2.5 & 34 & 1536 & 1536 & \raggedright [3]~0.2, 0.26, 0.34 &  0.4 & 1,24,058\\[.1cm]
	 \hline
	 \end{tabularx}
	 \caption[Scale-free simulation catalogue]{\emph{Scale-free simulation catalogue: } Column 1: Power-law power spectrum index for the model, Column 2: Initial redshift used to start the simulation, Column 3: Side length of the cubical simulation box, Column 4: Cube root of the total number of particles put in the simulation, Column 5: Range of the scale of non-linearity($r_{nl}$) covered in the simulation, Column 6: Maximum limit on $r_{nl}$ considering the finite box size effect, Column 7: Number of CPU hours taken by the simulation.}
     \label{tab:pl_sim} 
    \end{center}
\end{table}

These models necessitate the adoption of Einstein-de Sitter (EdS) cosmology as the background cosmology, where the growth factor of the growing mode scales with the scale factor. As previously mentioned, the only scale involved in these models is the scale of non-linearity, $r_{nl}$, induced by gravity. It is defined as the scale at which the linearly extrapolated value of the mass variance at a given epoch, $\sigma_{\text{lin}}(a, r_{nl})$, equals unity. Consequently, we can denote epochs in these models in terms of $r_{nl} \propto a^{2/(n+3)}$, where the normalisation of the power spectrum determines the proportionality constant. All power spectra are normalised such that $\sigma_{\text{lin}}(a=1, r_{nl}=8) = 1$.

The initial redshift $z_{\text{start}}$ for the simulation is determined by the epoch at which the mass variance at grid length reaches approximately $10^{-2}$, as listed in column 2 of~\cref{tab:pl_sim}. This ensures a common starting point with respect to the perturbation amplitude for all eight models used in this thesis. However, it is essential to acknowledge that systematic errors in simulations can arise from initial conditions, as indicated by~\cite{Michaux_2020}. Hence, we have confirmed that the results presented using these simulations remain unaffected by the chosen starting redshift.

At the beginning of the simulation, the average inter-particle separation is set at a unit grid size, wherein the number of particles in the simulations, denoted as $N_{\text{part}}$, is equal to the volume of the simulation in grid units. Column 3 of~\cref{tab:pl_sim} displays the size of the simulation box, denoted as $L_{\text{box}}$, which is also equal to the cube root of the number of particles, as displayed in column 4. We conduct ten realisations of smaller boxes to account for sample variance for power-law indices $n=0.0$, $-0.5$, and $-1.0$. A single realisation of larger boxes is employed for other indices to cover a wider range of halo masses.

We output particle positions and velocities at different $r_{nl}$ values listed in column 5 of~\cref{tab:pl_sim}. Values inside square brackets in column 5 denote number of $r_{nl}$s for perticular run. The epochs are chosen such that increments in $\log (r_{nl})$ are constant. We configure the spacing to the minimum necessary to ensure that outputs are considered independent.

\begin{figure}
    \centering
  \includegraphics[width=0.75\textwidth]{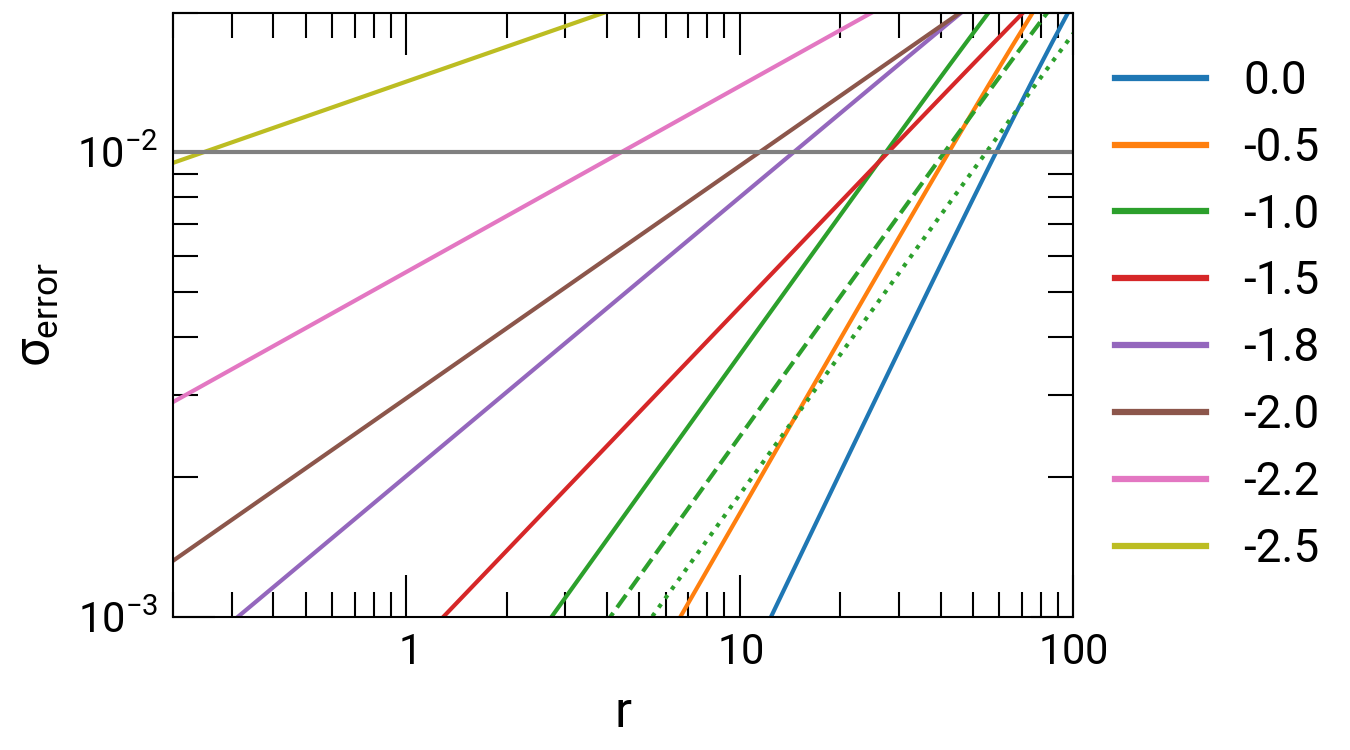}
  \caption[Finite box size effect]{\emph{Finite box size effect:} The plot shows the fractional error in the mass variance due to the finite box size used in models listed in~\cref{tab:pl_sim} as a function of length scale in the grid units. Different coloured solid lines represent the models with different power-law power spectrum index $n$. For $n=-1$, three variants of $L_{\text{box}}$ are shown, solid $512$, dashed $768$ and dotted $1024$. The grey line represents the $1\%$ mark.}
  \label{fig:sgm_err}
\end{figure}

We account for finite box size effects while determining the appropriate box size and the range of $r_{nl}$ where results can be considered reliable. This is discussed in section~\cref{ssec:Nuances}. We set a tolerance limit of $1\%$ error in the mass variance at the scale of non-linearity. \Cref{fig:sgm_err} illustrates the fractional error in the linearly extrapolated mass variance as a function of length scale in grid units, computed using the aforementioned criterion proposed by~\cite{Bagla_2005, Bagla_2006}. The various coloured solid lines represent the models listed in~\cref{tab:pl_sim}. Larger box sizes are required for models with more negative indices to maintain errors below $1\%$. We have verified the stability of our results against variations in the box size. Column 6 of~\cref{tab:pl_sim} provides the maximum limit on the scale of non-linearity within our criteria.

Column 7 lists the CPU time in hours taken by simulations to run. In total, the simulations have taken over 2 million CPU hours.

\begin{figure}
\centering
  \includegraphics[width=0.95\textwidth]{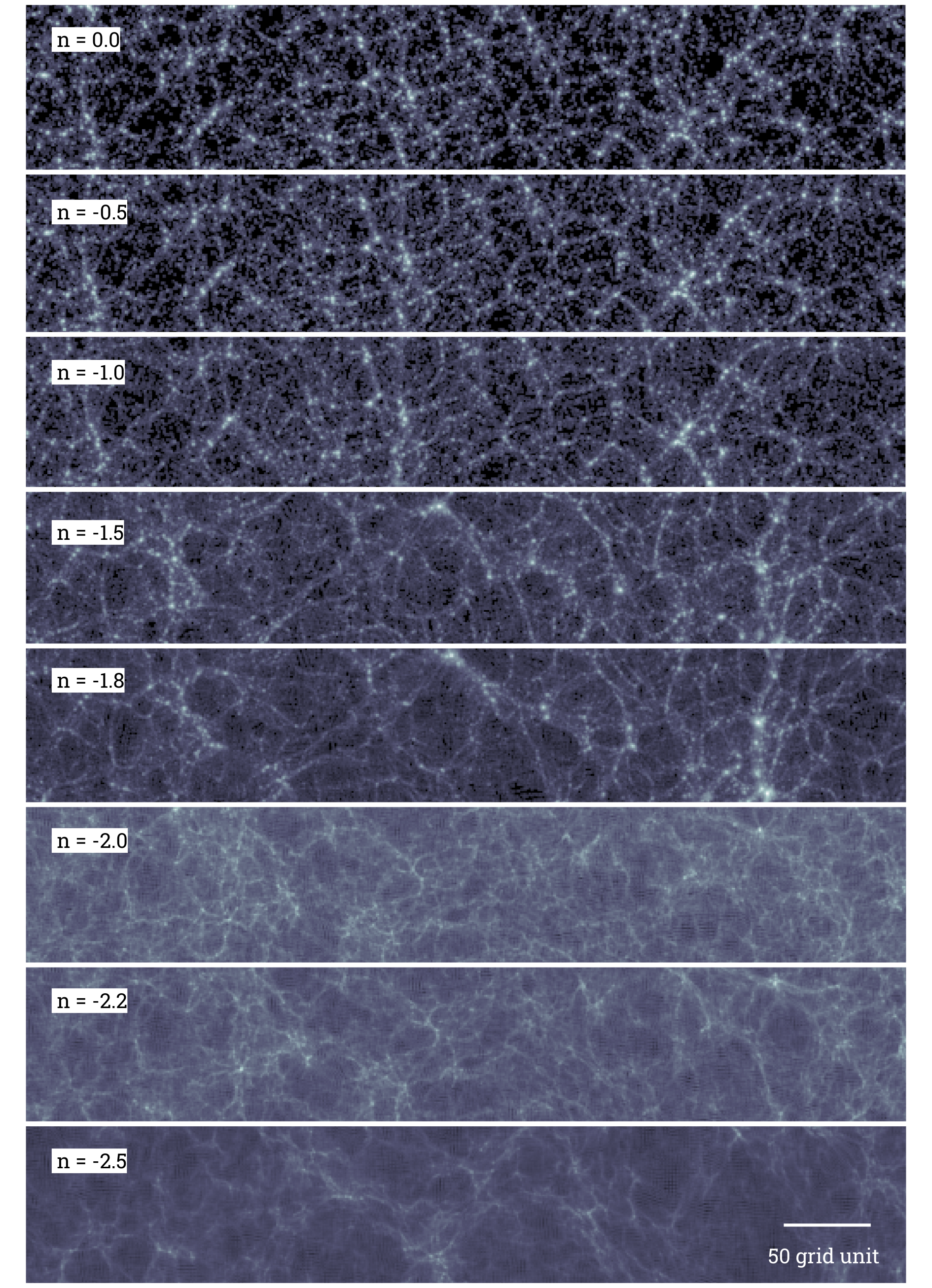}
  \caption[Large scale structure in scale-free cosmologies]{\emph{Large scale structure in scale-free cosmologies:} A 512 $\times$ 96 grid units slices of simulation boxes listed~\cref{tab:pl_sim}. Each slice spans ten grid units in thickness. All slices uses $r_{nl} = 10.24$. Observable differences in clustering are evident across different models and scales. Further quantification of these variations can be achieved through clustering statistics. Refer~\cref{ssec:pl_cat}}
  \label{fig:lss_pl}
\end{figure}

\begin{figure}
\centering
  \includegraphics[width=0.95\textwidth]{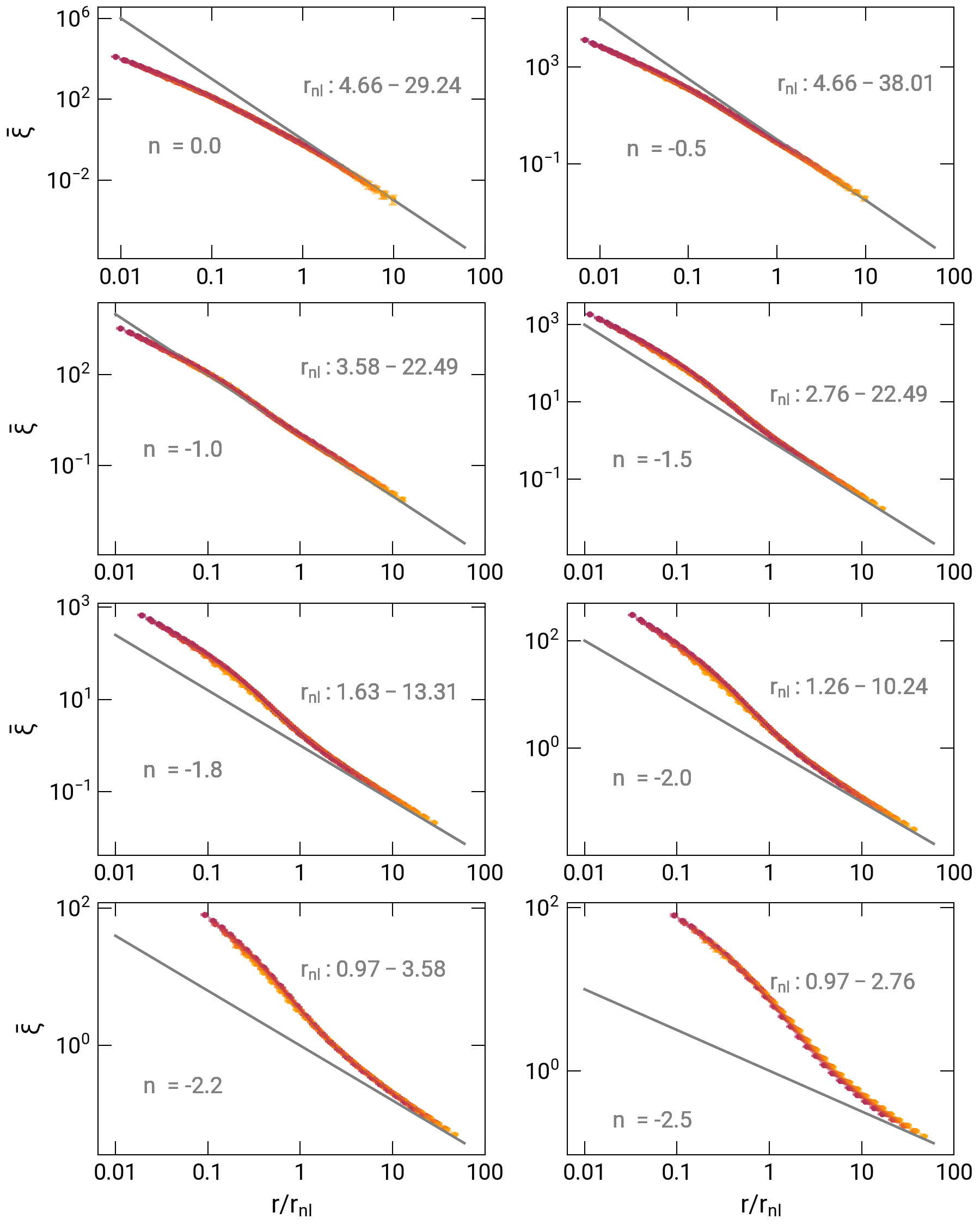}
  \caption[Demonstration of self-similarity in scale-free models]{\emph{Self-similarity of $\bar{\xi}$:} Figure shows the volume averaged two-point correlation function $\bar{\xi}$ scaled with non-linearity scale ($r_{nl}$) for all models listed in~\cref{tab:pl_sim}. Power-law power spectrum index n is displayed on each panel to identify the models. Different coloured solid lines represent $\bar{\xi}$ at different epochs, and $r_{nl}$ range is displayed on the panels. The grey line represents the linear theory prediction for $\bar{\xi}$. The self-similar nature of these models is evident through this figure, as scaled $\bar{\xi}$ from different epochs converge into a single curve. Further, it shows a characteristic linear, quasi-linear, and non-linear regime. Refer~\cref{ssec:pl_cat}}
  \label{fig:corr}
\end{figure}

\Cref{fig:lss_pl} displays a slice of $512 \times 96 $ grid units, spanning $10$ grids in thickness, from N-body simulations representing scale-free models with varying power-law power spectrum indices. $r_{nl} = 10.24$ is chosen for this visualisation \footnote{Note that $r_{nl} = 10.24$ is greater than the $r_{nl}^{\text{max}}$ for  $n=-2.2$, $-2.5$.}. Observable differences in clustering are evident across different models and scales. Further quantification of these variations can be achieved through clustering statistics. We show the average correlation function, $\bar{\xi}$, in~\cref{fig:corr}.

When mode coupling becomes important, resulting in power transfer from large to small scales. This effect becomes significant as the spectral index approaches $-3$, where all modes possess equal power. To mitigate errors involved in such effects, we employ self-similar models which allow us to validate the simulations~\cite{Maleubre_2023}. Thus, validating self-similar clustering evolution across epochs in these simulations is essential for the analysis presented in this thesis work. For this, we use the volume-averaged two-point correlation function, $\bar{\xi}$. We compute $\bar{\xi}$ by randomly sampling particles from the simulation output. Utilising the {\tt Python} module {\tt Corrfunc}~\cite{Sinha_2019}, we calculate the correlation function and show the results in the~\cref{fig:corr}. Different coloured solid lines represent $\bar{\xi}$ scaled with $r_{nl}$, while the solid grey line denotes the linear theory prediction. In these plots, we exclude epochs where $r_{nl} > r_{nl}^{\text{max}}$ due to the significant errors arising from finite box size effects, generating unreliable data. We omit a few initial epochs to account for the initial transient evolution in simulations. The $r_{nl}$ range corresponding to the plots is specified in the panels.

The self-similar nature of these models is evident as $\bar{\xi}$ plotted for various epochs converge into a single curve as a function of $r/r_{nl}$. The $\bar{\xi}$ curves exhibit three main regimes, linear ($\bar{\xi} \gg 1$), quasi-linear ($\bar{\xi} \approx 1$), and non-linear ($\bar{\xi} \ll 1$). During the evolution of the matter density field, small-scale collapses initially enter the quasi-linear regime before transitioning into the non-linear regime. The growth rate in the quasi-linear regime closely mirrors the linear rate for $n = -1$, faster for $n < -1$, and slower for $n > -1$. The slope of $\bar{\xi}$ ultimately approaches a $-2$ in the non-linear regime for all indices, the same as the linear slope for $n = -1$~\cite{Bagla_1997a}.

\subsection{$\Lambda$CDM simulation suite}
\label{ssec:lcdm_cat}

The $\Lambda$CDM (Lambda Cold Dark Matter) cosmological model is the concordant model of cosmology, taking into account the presence of dark energy (represented by $\Lambda$) and cold dark matter (CDM). In this model, the late time expansion of the Universe is primarily driven by dark energy, while cold dark matter provides the gravitational skeleton necessary for the formation of large scale structures such as galaxies and galaxy clusters. A wide range of observational data supports this model, including Cosmic Microwave Background Radiation (CMBR), large scale structure surveys, and observations of distant supernovae~\cite{Planck_2020, Alam_2017, Dawson_2012, DESI_ollaboration_2023, Riess_1998}.

\begin{table}
    \begin{center}
	 \begin{tabular}{|>{\centering\arraybackslash}m{1.5cm} |>{\centering\arraybackslash}m{1.5cm} |>{\centering\arraybackslash}m{1.5cm} |>{\centering\arraybackslash}m{1.8cm}  |>{\centering\arraybackslash}m{2cm} |}
	 \hline
	$L_{\text{box}}$ & $\sqrt[3]{N_{\text{part}}}$ & $\epsilon_0$ & $m_{\text{part}}$ & CPU hours\\
	$_{(\text{Mpc/h})}$ &  & $_{(\text{Kpc/h})}$ & $_{(M_{\odot}/\text{h})}$ & \\
	 \hline
	 150 & 1024 & 5 & 2.71 $\times 10^{8}$  & 67,500\\ [.1cm]
	 \hline
	 300 & 1024 & 10 & 2.17 $\times 10^{9}$  & 63,870\\ [.1cm]
	 \hline
     500 & 1024 & 16.6 & 1.01 $\times 10^{10}$ & 42,306\\ [.1cm]
	 \hline
     750 & 1024 & 24.4  & 3.39$\times 10^{10}$  & 36,426\\ [.1cm]
	 \hline
     1000 & 1024 & 32.2 & 8.04 $\times 10^{10}$  & 25,578\\ [.1cm]
	 \hline
	 \end{tabular}
	 \caption[$\Lambda$CDM simulation suite]{\emph{$\Lambda$CDM simulation suite:} Column 1: Side length of the cubical simulation box, Column 2: Cube root of the total number of particles put in the simulation, Column 3: Softening law parameter as in~\cref{eq:soft_fn}, Column 4: Mass of a single N-body particle in the simulation, Column 5: Number of CPU hours taken by the simulation.}
     \label{tab:lcdm_sim}
    \end{center}
    \vspace{-0.5cm}
\end{table}

\begin{figure}
    \centering
  \includegraphics[width=0.95\textwidth]{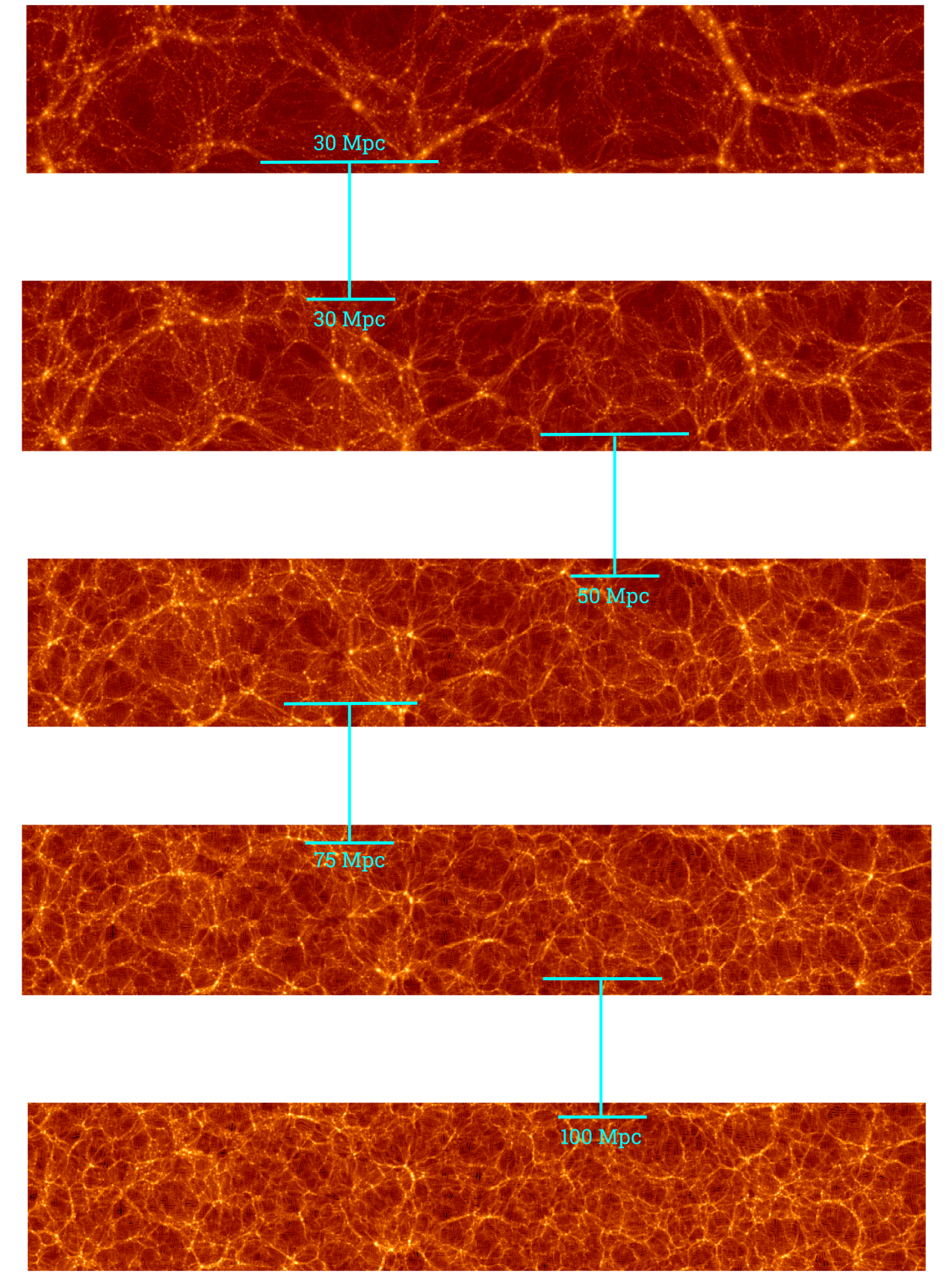}
  \caption[Large scale structure in $\Lambda$CDM cosmology]{\emph{Large scale structure in $\Lambda$CDM cosmology:} Each slice shows a $10$ Mpc/h slice of different simulation boxes listed in~\cref{tab:lcdm_sim}. The cyan markers denote different length scales in various simulation boxes. This illustration shows the hierarchical structure formation and clustering across different scales in the $\Lambda$CDM cosmology.}
  \label{fig:lcdm_sim}
\end{figure}

We run five simulations based on the Planck Cosmology parameters ($\Omega_{m0}= 0.3111$, $\Omega_{\Lambda0}= 0.6889$, $\Omega_{b0} = 0.0490$, $H_0 = 67.66$, $\sigma_8 =0.8102$, $n_s=0.965$, where the subscript $0$ represents values at $a=a_0=1$) ~\cite{Planck_2020}. The simulations feature different box sizes of $150$, $300$, $500$, $750$, and $1000$ Mpc/h, each containing $1024^3$ particles, as listed in~\cref{tab:lcdm_sim}. The smaller box simulations offer a better mass resolution, which is crucial for capturing low-mass halos and probing early epochs where massive halos are rare. Conversely, larger box simulations allow for the exploration of high-mass halos. An intermediate-sized box bridges the two ends of the halo mass spectrum. 

We use {\tt GADGET4}\cite{Springel_2021}, operating in TreePM mode, to perform simulations. The initial conditions are generated using second-order Lagrangian perturbation theory. Similar to the scale-free simulations, we employ a force split radius of $r_s = 2$ and adaptive time steps with a maximum step size of 0.005. The softening length $\epsilon_0$ varies across different boxes, as indicated in~\cref{tab:lcdm_sim}. To construct the halo catalogue, we use the built-in {\tt FoF-SUBFIND} algorithm~\cite{Springel_2001a} of {\tt GADGET4}, using a linking length of $0.2$ along with its default parameters.

The simulations are initialised at $z_{\text{start}} = 150$, and we record particle positions and velocities at $20$ different scale factors: $0.059$, $0.067$, $0.077$, $0.083$, $0.091$, $0.1$, $0.111$, $0.125$, $0.143$, $0.167$, $0.2$, $0.25$, $0.333$, $0.4$, $0.5$, $0.571$, $0.667$, $0.8$, $0.909$, and $1.0$. 

The~\cref{fig:lcdm_sim} depicts the large scale structure in these simulations, each showcasing a $10$ Mpc/h slice. The cyan markers denote different length scales in various simulation boxes. This illustration shows the hierarchical structure formation and clustering across different scales in the $\Lambda$CDM cosmology. As resolution increases, structures at smaller scales become more discernible. Nonetheless, small box simulations are more susceptible to errors arising from cosmic variance for high mass halos and larger structures.

\vspace{0.5cm}
This chapter begins with a review of some statistical tools essential for large scale structure studies. Following that, we delve into the fundamentals of N-body simulations. Further, we explore the advantages of scale-free and $\Lambda$CDM cosmologies and provide a catalogue of simulations for both. These tools are the foundation for addressing problems in structure formation studies. We discuss these problems in subsequent chapters \ref{cha:3}, \ref{cha:4} and \ref{cha:5}.


%% file: chapters/chap3.tex

\chapter[Halo mass function in scale-invariant models]{Halo mass function in scale-invariant models}
\label{cha:3}

This chapter is based on the following paper: \fullcite{Gavas_2023}. \\

Understanding the theoretical foundations of the large scale structure mapped by galaxies and galaxy clusters with precision has become crucial with the advent of state-of-the-art observations. The halo model \cite{Peacock_2000, Seljak_2000, COORAY_2002} asserts that all mass in the Universe is contained within virialised dark matter halos. This model is developed by quantifying the spatial distribution of halos and the dark matter distribution within each halo. It has been effectively used to make a wide range of predictions \cite{Mandelbaum_2005, Mead_2015} and to interpret observations \cite{Skibba_2009, Alam_2020, Greig_2024}.

The spatial distribution of halos is quantified using the excursion set formalism, which also underpins the theory of the halo mass function. These are essential components in various studies, such as semi-analytical theories of galaxy formation \cite{White_1991, Somerville_2015}, constraining cosmological parameters using galaxy cluster abundance \cite{Majumdar_2003, Allen_2011, Planck_2020, To_2021}, calculating halo merger rates \cite{Lacey_1993, Lacey_1994, Cohn_2001, Giocoli_2008, Yacine_2017, Saeed_2021}, gravitational lensing \cite{Bartelmann_1997, Bartelmann_2010, Massey_2010}, and constraining non-Gaussianity in the primordial power spectrum of matter perturbations \cite{Maggiore_2010, Dionysios_2018, Shirasaki_2021}. \cite{Shimin_2023} investigates the effects of variations in the halo mass function on the modelling of matter power spectra and the structure growth parameter $\sigma_8$, finding that varying halo abundance may provide an alternative solution to the $\sigma_8$ tension. Therefore, understanding the theory of mass functions and its validity is critical for precision cosmology. 

Press-Schechter presented the first detailed theory of the mass function of collapsed objects \cite{Press_1974}. This model identifies spherical regions in initial linear Gaussian random fields representing peaks in the density field dense enough to eventually collapse into non-linear structures. The criterion for collapse is given by the spherical collapse mechanism \cite{Gunn_1972}. \cite{Bardeen_1986} established a formal correspondence between the number density of peaks in the matter density field and the number density of halos, providing a quantitative connection. However, their theory underestimates the mass function by a factor of two because it does not account for peaks that are part of bigger peaks. Initial numerical results were consistent with the Press-Schechter mass function theory \cite{Efstathiou_1988, Carlberg_1989, Lacey_1994}. \cite{Peacock_1990} discussed conceptual problems with the Press-Schechter mass function, such as not accounting for mass in underdense regions, the assumption that the formalism is independent of the choice of the filter function, and the lack of detail on how overdense regions relate to bound objects. Several attempts were made to revisit the derivation of the mass function, but none resolved these conceptual issues satisfactorily. Larger and more accurate simulations have shown clear discrepancies when compared with theory, particularly the over-prediction of halos near the characteristic mass ($m_*$) and under-prediction at both lower and higher masses \cite{Monaco_1997, Audit_1998, Lee_1999, Sheth_1999}.

\cite{Bond_1991} adopted an approach based on excursion set formalism to calculate the mass function. This method employs Markovian random walks of mass elements and a spherical up-crossing absorbing barrier to determine the number density of halos. These calculations, initially in Lagrangian space, were later transformed into Eulerian space by \cite{Mo_1996} to address issues related to the clustering of halos. This approach naturally resolves the factor-of-two discrepancy in earlier theories by utilizing a top-hat filter in k-space \cite{Paranjape_2012}.

The non-zero tidal field around a collapsing region naturally results in an ellipsoidal collapse rather than a spherical one. Additionally, the shape of equal-density contours around peaks in the density field is often not spherical. These concepts were investigated to modify the Press-Schechter mass function \cite{Monaco_1997, Lee_1998, Sheth_2001}. Sheth-Tormen \cite{Sheth_2001} provided a functional form for the ellipsoidal collapse barrier required in the excursion set formalism and derived the mass function from it.

It is possible to develop the Press-Schechter formalism without explicitly referencing the cosmological model, the power spectrum of matter perturbations, or the epoch. Sheth-Tormen preserves this universality of the mass function. However, these are simple theoretical predictions that do not capture the intricacies of halo formation. The accurate mass function can be obtained using cosmological N-body simulations. Studies using simulations have revealed a 5-20\% systematic deviation from theoretical predictions \cite{Jenkins_2001, Reed_2006, Courtin_2011, Diemer_2020}. Further, attempts to examine the universality of the mass function in various settings—such as different cosmologies \cite{Bhattacharya_2011}, power spectra \cite{Ondaro_2022}, epochs \cite{Watson_2013}, mass definitions \cite{Despali_2015}, and mass intervals \cite{Reed_2006}—have shown a systematic trend in each case. This indicates that a complete understanding requires us to go beyond the excursion set formalism \cite{Delos_2024}.

Understanding the physical origin of non-universality in the mass function is a complex issue that may involve factors such as cosmology, the power spectrum, or a combination of both. The threshold density for collapse introduces cosmology dependence \cite{Barrow_1993}. The Cold Dark Matter (CDM) class of power spectra has a gradually varying slope; the spectral index $n(k)$ decreases with decreasing scale (increasing $k$). As perturbations at smaller scales collapse earlier, the effective index of the power spectrum is smaller at earlier times and increases towards later times. Therefore, variations in the mass function and deviations from a universal mass function may arise due to the changing slope of the power spectrum, cosmology, or both. The alternative is to use fitting functions for each of these quantities derived from N-body simulations \cite{Jenkins_2001, Reed_2003, Warren_2006, Reed_2006, Tinker_2008, Crocce_2010, Bhattacharya_2011, Courtin_2011, Angulo_2012, Watson_2013, Bocquet_2015, Despali_2015, Comparat_2017, McClintock_2019, Nishimichi_2019, Bocquet_2020, Diemer_2020, Seppi_2021}. The effect of baryons on the mass function is observed in large cosmological hydrodynamic simulations \cite{Khandai_2015, Euclid_2024}. \cite{Guo_2024} use a deep-learning model to study partial non-universality of the halo mass function due to cosmological parameters.

Since CDM spectra lack the simplicity of scale-free spectra, we approach the problem of non-universality differently. We examine departures from non-universality in the mass function for scale-free power spectra of initial fluctuations with an Einstein-de Sitter (EdS) background. This approach helps us determine if non-universal behaviour can be attributed to spectrum dependence. Our choice of cosmology does not introduce any scale into the problem, and the threshold overdensity remains constant over time in a straightforward manner. Thus, we can isolate the non-universality of the mass function that arises from the slope of the power spectrum. We provide spectrum-dependent fits for the parameters in the Sheth-Tormen mass function and demonstrate that this allows for a better fit to simulation data.

\cite{Bagla_2009} explored this idea using a similar approach. The key results of their work overlap with our findings, although, as we will demonstrate, our results are robust across many variations in the analysis. A similar analysis was also conducted by \cite{Euclid_2023}, who observed a similar departure from the universality of the mass function.

In \cref{sec:mf_method}, we summarise the basic framework of mass function theory and our simulations setup. In \cref{sec:results}, we present results from our data analysis. We extend this work to $\Lambda$CDM models in \cref{sec:mf_lcdm}. Finally, we conclude our findings in \cref{sec:mf_conclusion}.

\section{Methodology}
\label{sec:mf_method}

\subsection{Mass function}
The form of equations thought to be universal \cite{Lacey_1994} for Press-Schechter and Sheth-Tormen mass function is given by \cref{eq:ps,eq:st}. In \cref{eq:st} $p \approx 0.3$ and $q \approx 0.75$ are the standard values of parameters proposed by Sheth-Tormen. These mass functions can be connected to the number density of halos of a given mass (per unit comoving volume),
\begin{equation}
    \frac{\d n}{\d \ln m} = \frac{\bar{\rho}}{m} \frac{\d \ln \sigma^{-1}}{\d \ln m} f(\nu).
    \label{eqn:dn}
\end{equation}
The dependence of the mass function on cosmology comes mainly through the growth factor $\mathcal{G}$ and small dependence from threshold density $\delta_{sc}$. All the power spectrum dependence lies in mass variance $\sigma(m)$ (see \cref{ssec:hmf},and \cref{ssec:mv}). Both of these dependencies are absorbed in the definition of $\nu$. Thus \cref{eq:ps,eq:st} have a universal form. The Press-Schechter mass function has only one parameter, $\delta_{sc}$, and this is fixed by theoretical considerations. The Sheth-Tormen mass function has two more free parameters $p$ and $q$. We aim to study whether these parameters are independent of the slope of the power spectrum. If the theory of mass function can be constructed in a truly universal fashion, then these should be independent of the power spectrum.

\subsection{Simulations}
\label{subsec:simulations}

\begin{table}
	\begin{center}
	 \begin{tabular}{|c| c| c| c| c| c| c|} 
	 \hline 
    $n$ & $z_{\text{start}}$ & $N_{\text{box}}$ & $\sqrt[3]{N_{\text{part}}}$ &realisations & $r_{nl}^{mf}$ & $r_{nl}^{\text{max}}$\\ [0.3cm] 
	 \hline
	 0.0 & 453 & 512 & 512 & 10 &7.87-29.2 & 58.7    \\ [.1cm]
	 \hline
	 -0.5 & 270 & 512 & 512 & 10  &6.06-38.01 & 42.3 \\[.1cm]
	 \hline
	 -1.0 & 160 & 512 & 512 & 10  & 4.66-22.49 & 27.4\\[.1cm]
	 \hline
	 -1.5 & 96 & 1024 & 1024 & 1  &2.76-22.49 & 27.8 \\ [.1cm]
	 \hline
	 -1.8 & 70 & 1024 & 1024 & 1  &2.12-13.30 & 14.5 \\[.1cm]
	 \hline
	 -2.0 & 57 & 1536 & 1536 & 1  &1.25-7.88 & 11.4  \\[.1cm]
	 \hline
	 -2.2 & 46 & 1536 & 1536 & 1  & 0.96-3.58& 4.4   \\
	 \hline

	 \end{tabular}
	 \caption[Halo mass function simulation setup]{\emph{ Simulation Setup:} Column 1: power-law power spectrum index for the model, Column 2: Initial redshift used to start the simulation, Column 3: Side length of the cubical simulation box, Column 4: Cube root of the total number of particles put in the simulation, Column 5: Number of realisations run for the model, Column 8: Range of $r_{nl}$ used to compute the mass function, Column 7: Maximum limit on $r_{nl}$ considering the finite box size effect.}
     \label{tab:sim_mf}
	\end{center}
	\end{table}
	
We use a suite of dark-matter-only simulations for seven power-law power spectrum initial conditions. Power-law indices ($n$) of the power spectrum ($P(k)=Ak^n$) are shown in column 1 of \cref{tab:sim_mf}. We use \textsc{gadget4} \cite{Springel_2021} to run the simulation suite. These simulations are discussed in detail in \cref{ssec:pl_cat}.

Einstein-de-Sitter (EdS) is chosen as the background cosmology, it is the simplest model where scale factor grows with single slope and threshold collapse overdensity is constant. We identify epochs in terms of scale of non-linearity $r_{nl}$ ($r_{nl} \propto a^{2/(n+3)}$) as discussed in \cref{ssec:pl_cat}.

The initial redshift $z_{\text{start}}$ for the simulation is estimated from the epoch where the mass variance per unit length is about $10^{-2}$, as discussed in \cref{ssec:pl_cat}. However, as per \cite{Michaux_2020}, systematic errors can arise in simulations due to initial conditions. Therefore, we have verified that the results presented here are not affected by the chosen $z_{\text{start}}$; analysis is presented in \cref{a3:1}.

Column 5 of \cref{tab:sim_mf} lists the number of realisations run for a given model. For the first three models, we use multiple realisations of smaller boxes to account for sample variance. For others, we use a single realisation of larger boxes to cover a broad mass range. 

Finite volume in N-body simulations causes significant errors, as modes greater than the box size are not considered while generating initial conditions and during the evolution. The errors in the mass variance can become arbitrarily large as the power spectrum slope approaches $-3.0$. Furthermore, as demonstrated by \cite{Klypin_2019}, the finite box size can propagate errors in the mass function for $\Lambda$CDM cosmologies. These factors should be considered when determining the appropriate box size and the regime where the results can be regarded as reliable (see \cref{ssec:Nuances}). We take it into account using the prescription provided by \cite{Bagla_2006} for deciding the box size and the range of $r_{nl}$. \Cref{fig:sgm_err} demonstrate that as we move towards more steep slope models, larger box sizes must be selected to keep errors below $1$\%. We have confirmed the stability of our results against variations in the box size and shown in \cref{a3:2}.

Column 6 of table \ref{tab:sim_mf} lists the $r_{nl}$ range covered to compute the mass function. Column 7  lists the maximum limit on the scale of non-linearity within our criteria. Refer to \cref{ssec:pl_cat} for details.

Finite box size effects induce error in non-linear spectra, where Fourier mode coupling becomes dominant. During this process, power is transferred from large to small scales; the effect becomes significant as the spectral index approaches $-3$, as there is equal power in all modes in this limit. We are able to avoid errors arising from such effects by working with self-similar (power-law) models and EdS background. This approach has been used in many simulation studies to ensure that results are free from errors arising from finite box size or other effects, e.g., \cite{Maleubre_2023}. We require a similar self-evolution of clustering across epochs for the analysis in this study. We check for self-similarity using the volume-averaged two-point correlation function $\bar{\xi}$, demonstrated using \cref{fig:corr}. 

\subsection{Finding halos}
\label{subsec:halo_finding}

The inbuilt \textsc{fof-subfind} \cite{Springel_2001a} algorithm of \textsc{gadget4} is used to construct the halo catalogue with a linking length equal to $0.2$ and all its default parameters. It first identifies Friends-Of-Friends (FOF) \cite{Davis_1985} groups in a given distribution and then decomposes each found object into substructures using the excursion set algorithm. We only consider isolated halos when computing the halo mass function. The halo-finding approach uses spherical overdensity (SO) to define halo masses. Hence, SO masses may differ from the FOF masses, and the finite box size effect and discreteness noise can affect halo masses at the high and low mass ends, respectively. Care is required to ensure that such effects do not creep into our analysis. We describe our approach in the following section.

\section{Results}
\label{sec:results}

Here, we present the halo mass function from our simulations of power-law models and compare them with theory predictions. First, we bin the available halo catalogue with mass to compute the halo mass function. We construct adaptive bins in mass such that each bin contains a fixed number of halos. This gives us a halo count per mass bin, $\d n/\d\ln m$. With this, \cref{eqn:dn} can now be rewritten as,

\begin{equation}
    f(\nu) = \frac{6}{n+3} \frac{m}{\bar{\rho}_m} \frac{\d n}{\d\ln m}.
    \label{eqn:fnu}
\end{equation}
Note that $\bar{\rho}_m =1$ in units used here. \Cref{eq:sgm_sf} is used to arrive at \cref{eqn:fnu}.

We fit the Sheth-Tormen mass function to our data using $\chi^2$ minimisation with $p$ and $q$ as free parameters. To determine the errors in our analysis, we considered the Poisson errors associated with the number of halos in each bin. Furthermore, we assessed the suitability of our fits by employing Jackknife and Sample Variance errors, discussed in \cref{a3:3}. We verify that using Poisson errors is a conservative approach and that the outcomes obtained using this method are equally robust as those obtained using other error types. In subsequent sections, we discuss mass function plots, tolerance to the filtering criteria, the dependence of Sheth-Tormen parameters on power-law $n$, and epoch dependence.

\subsection{Mass function}
\label{sec:mf_result}

\begin{figure}
    \centering
    \includegraphics[width=0.99\textwidth]{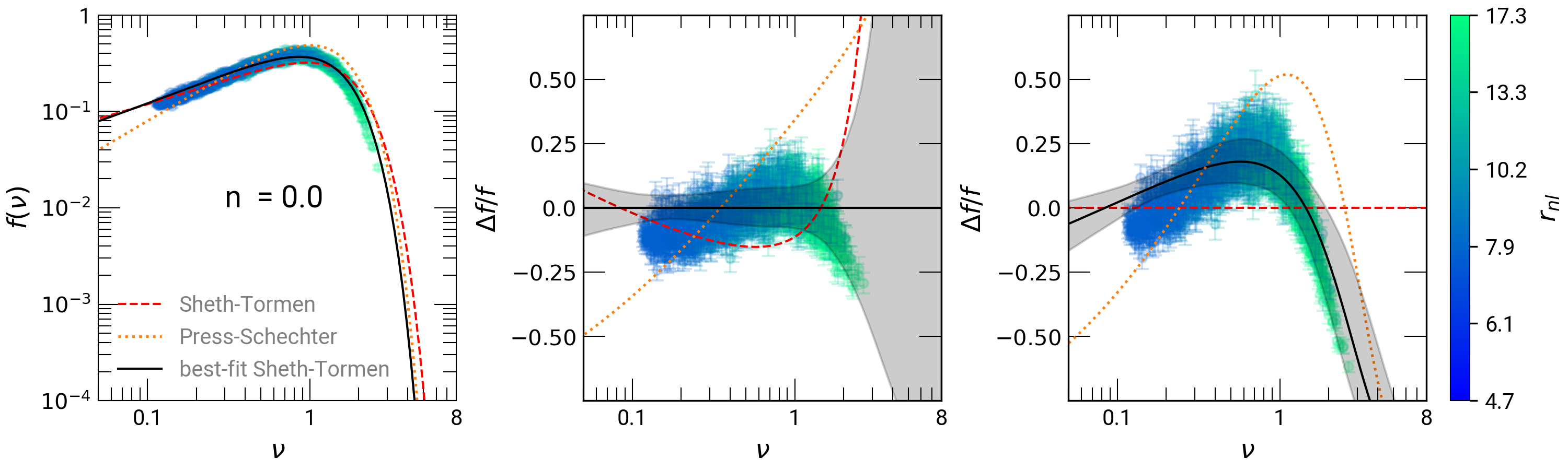}
    \includegraphics[width=0.99\textwidth]{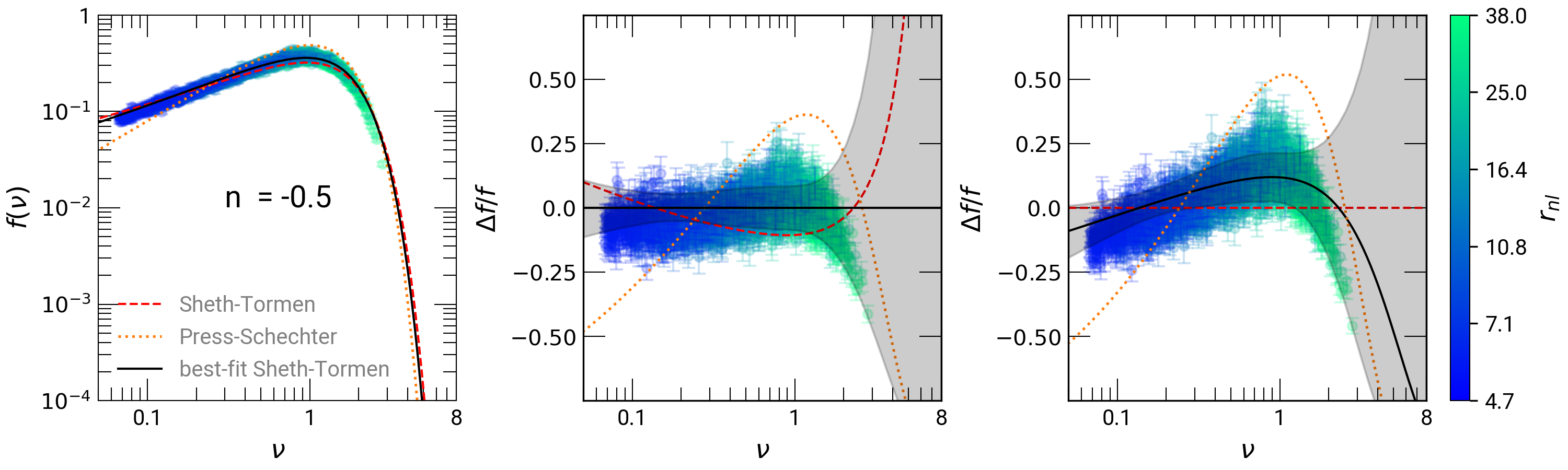}
    \includegraphics[width=0.99\textwidth]{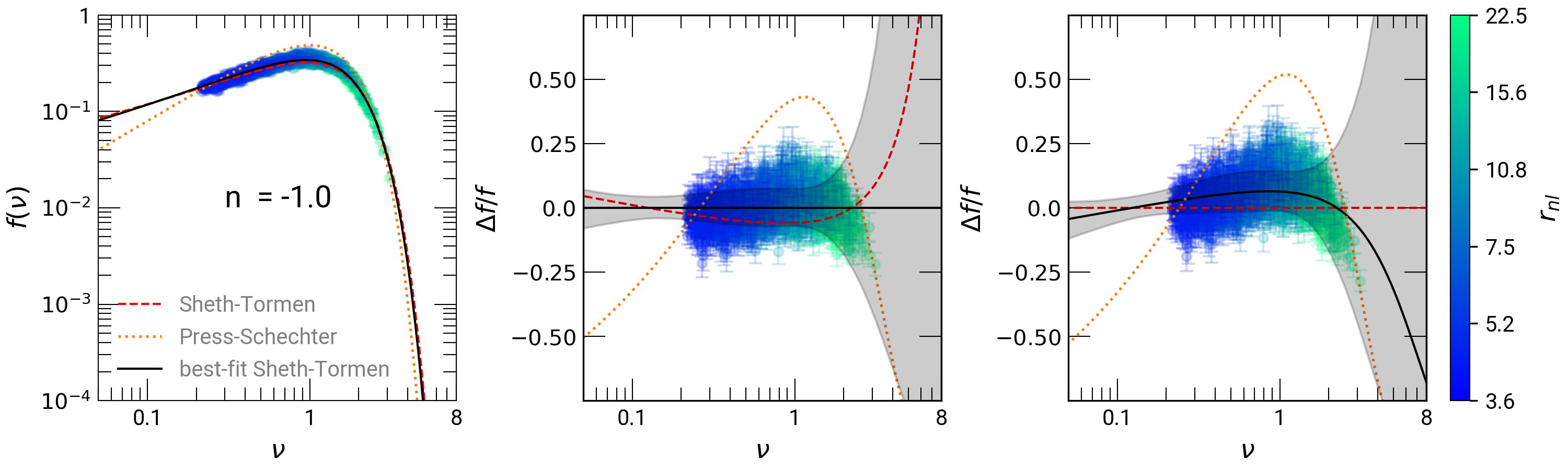}
    \includegraphics[width=0.99\textwidth]{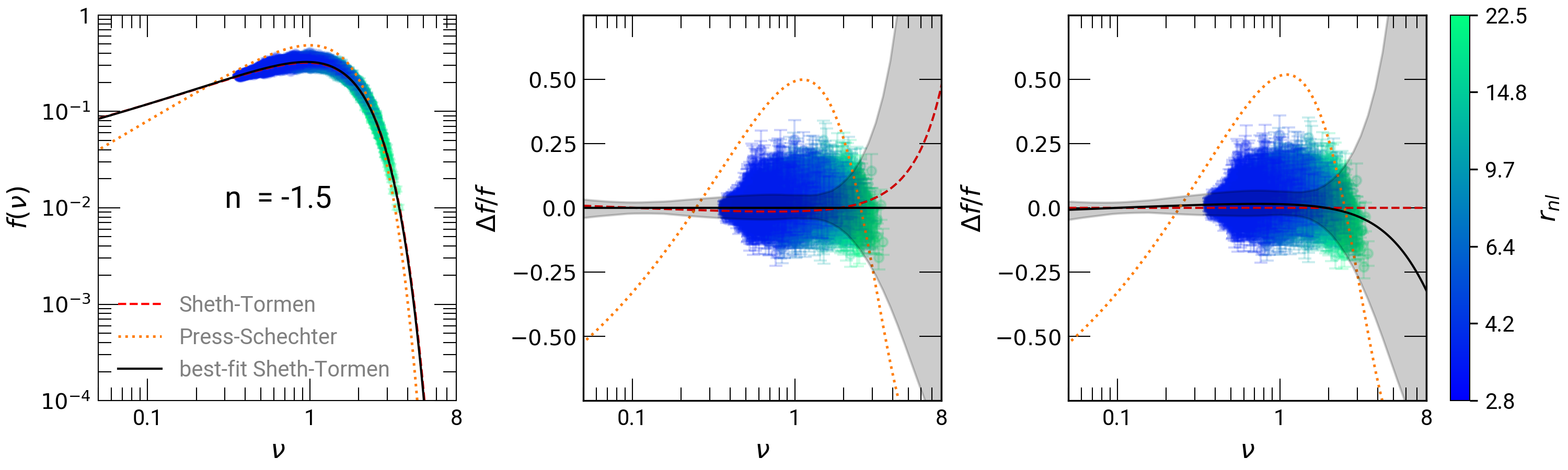}
    \caption[Mass function in scale-invariant models ($n=0.0,-0.5,-1.0,-1.5$)]{\emph{Mass function:} The mass function from the simulations (left column), their residuals after fitting the Sheth-Tormen function (middle column) and residuals with respect to standard Sheth-Tormen function (right column). Rows 1,2,3 and 4 are for power-law indices $0.0$, $-0.5$, $-1.0$ and $-1.5$ respectively. Dotted orange and dashed red lines represent the Press-Schechter and the Sheth-Tormen mass function, respectively. Solid black lines represent the best-fit Sheth-Tormen mass function after $\chi^2$ analysis. Blue-green data points show the mass function calculated from the simulations. Data points color coded with respect to $r_{nl}$, color changes from blue to green as $r_{nl}$ increases as shown in adjacent color-bar. Gray filled area in the middle and right panel shows a standard deviation confidence interval for the fit.}
    \label{fig:mf_1}
\end{figure}

\begin{figure}
    \centering
    \includegraphics[width=0.99\textwidth]{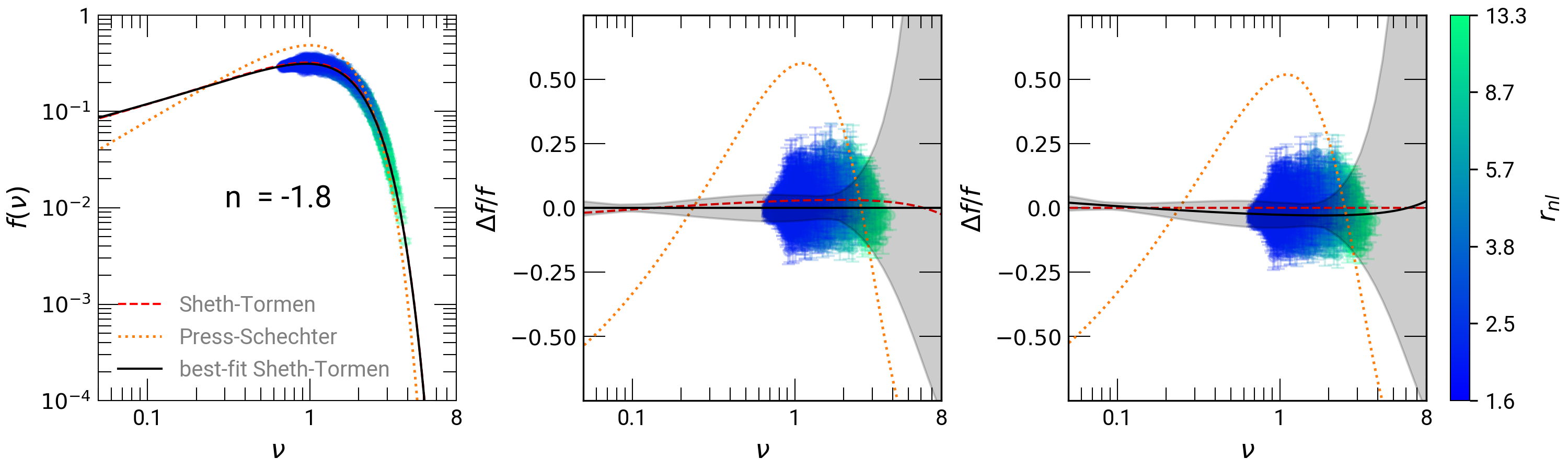}
    \includegraphics[width=0.99\textwidth]{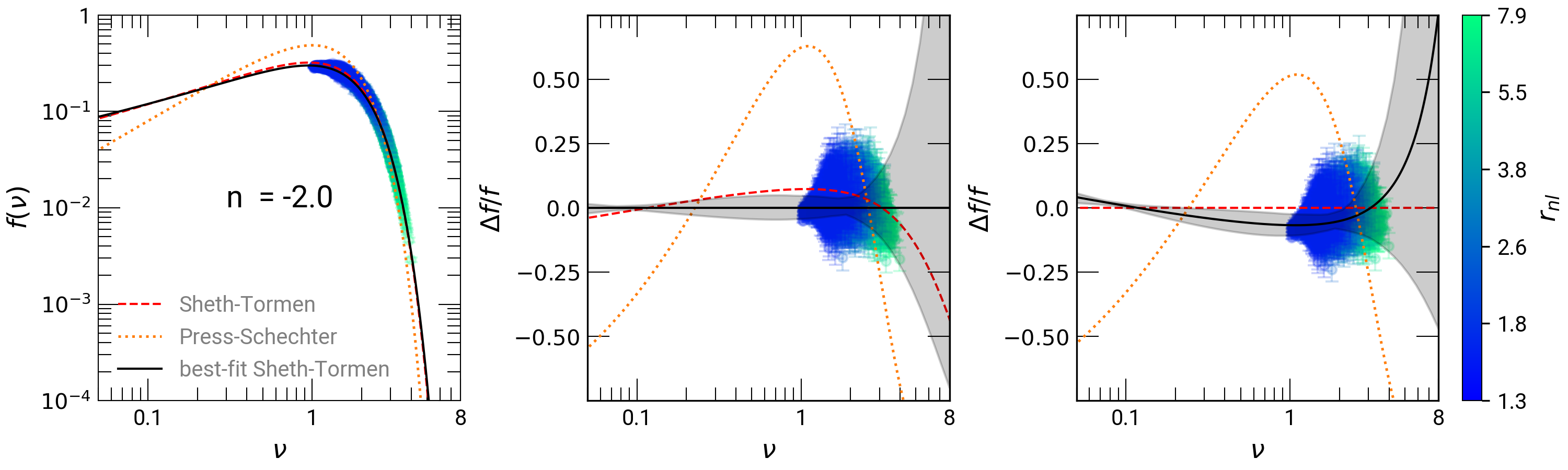}
    \includegraphics[width=0.99\textwidth]{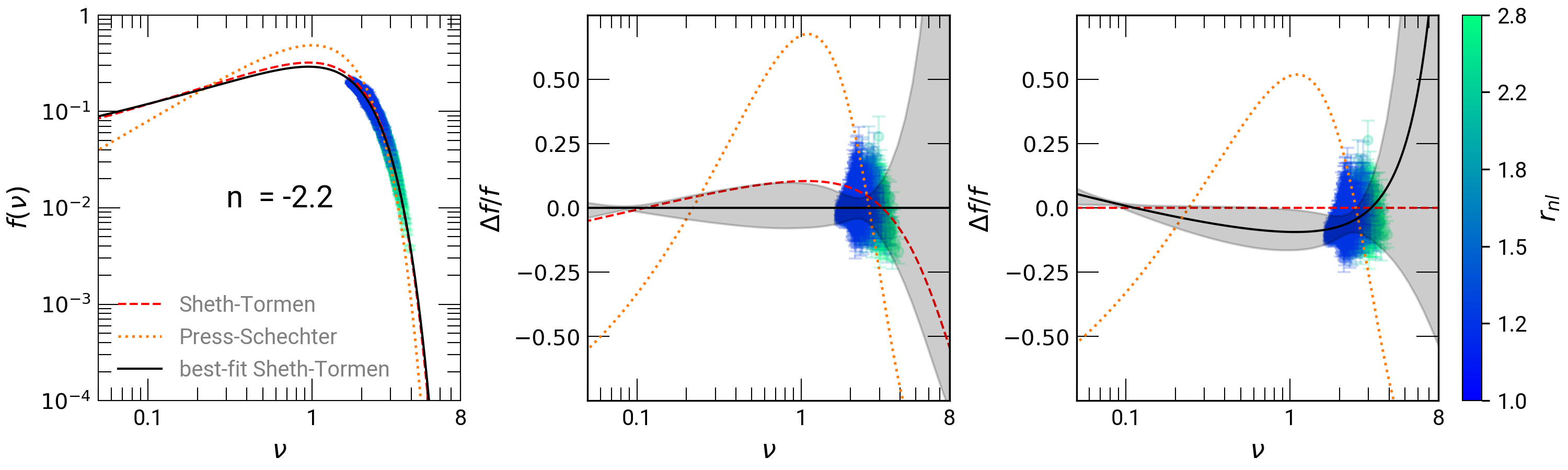}
    \caption[Mass function in scale-invariant models ($n=-1.8,-2.0,-2.2$)]{ \emph{Mass function:} The mass function from the simulations (left column), their residuals after fitting the Sheth-Tormen function (middle column) and residuals with respect to standard Sheth-Tormen function (right column). Rows 1,2 and 3 are for power-law indices $-1.8$, $-2.0$ and $-2.2$ respectively. Dotted orange and dashed red lines represent the Press-Schechter and the Sheth-Tormen mass function, respectively. Solid black lines represent the best-fit Sheth-Tormen mass function after $\chi^2$ analysis. Blue-green data points show the mass function calculated from the simulations. Data points color coded with respect to $r_{nl}$, color changes from blue to green as $r_{nl}$ increasesas shown in adjacent color-bar. Gray filled area in the middle and right panel shows a standard deviation confidence interval for the fit.}
    \label{fig:mf_2}
\end{figure}

We define spherical overdensity (SO) halo mass $M_{\Delta c}$ as the mass within a radius $r_{\Delta}$ inside which the average density is $\Delta$ times the critical density. We use masses with $\Delta=200$ to compute the mass function. $M_{\text{vir}}$ is the mass corresponding to the virial overdensity($\Delta=18\pi^2$). We only consider halos with more than $100$ particles to avoid discreteness noise. The fixed count for bins is chosen to be $300$ halos per bin. At the high mass end, the number counts of halos drop sharply and the fixed bin size requirement can create problems. The halo count at the high mass end is also affected strongly by cosmic variance. To remedy this, we remove the most massive $0.1\%$ halos during the computation of the mass function. We discuss our choice of these values in the \cref{sec:mf_tolerance}, along with variation in the best-fit values as we vary these thresholds/parameters.

\Cref{fig:mf_1} and \ref{fig:mf_2} show mass function results. The left panel shows mass function, while the middle and the right panel show the residual of the data with respect to best-fit Sheth-Tormen and standard Sheth-Tormen, respectively. Dotted orange and dashed red lines represent Press-Schechter and Sheth-Tormen mass functions, respectively. A solid black line represents the best-fit Sheth-Tormen mass function after $\chi^2$ analysis. Blue-green data points show mass function calculated from simulations. Data points are color coded for $r_{nl}$, and color changes from blue to green as $r_{nl}$ increases, as shown the in the adjacent color-bar. The gray filled area in the right panel shows a standard deviation confidence interval (throughout this chapter, we quote a single standard deviation confidence interval) around the best-fit. We find the following points noteworthy in these plots:
\begin{itemize}
  \item 
  The best-fit curve (black) in the left panels or deviation of residual Press-Schechter and Sheth-Tormen curves in the middle panels show a systematic pattern in the shape of the mass function. Its variation with the power-law power spectrum index $n$ can be seen.
  \item
  The right panels show data points, and the data point fit, moving away from standard Sheth-Tormen systematically. This points to the inadequacy of the Sheth-Tormen mass function in fully explaining the theory of mass function.
  \item
  We use adaptive binning (fixed number of halos per bin) to construct mass function. We verify that it gives better justification to the high mass end as fits are less biased to the low mass end here compared to fixed bin width in $\log_{10} (\nu)$. However, it is still insufficient, as data points show deviations at the high mass end, especially for $n=$ $0$, $-0.5$, $-1.0$. One probably needs additional theory parameters to explain the high mass end. Alternative methods, such as the maximum likelihood approach by \cite{Manera_2009}, can avoid binning entirely and provide a more direct comparison with theoretical predictions.
  \item 
  The range of the mass function is smaller as we move towards models with more negative power-law power spectrum indices. We have not evolved simulations with these indices over a broad range of epochs considering finite box size effects. A smaller range in $\nu$ for $n = 0.0$ compared to $n = -0.5$ is due to a smaller range of $r_{nl}$. Evolution in the non-linear regime becomes dramatically slow as $n$ approaches zero.
  \item 
  We notice $n=-1.5$ as an index around which deviation of the standard Sheth-Tormen function from the best-fit shifts sign. This may be related to the effective index in the simulations where the Sheth-Tormen function was calibrated.
  \item 
  As we progress towards late time, characteristic mass $m_*$ increases, leading to a $\nu$ interval shift towards high values (data points changing color from blue to green).
  \item 
  $\chi^2$ is biased by the low mass end as the data points are clustered there; we see a deviation of data points at the high mass end from the best-fit curve, especially for $n =$ $0.0$,$-0.5$ and $-1.0$.
  \item 
  Choosing a fixed number of halo counts per mass bin leads to equal Poisson error bars; however, we see some data points have smaller error bars at the low mass end. This is because there are more halos in a particular mass interval than the fixed count value.
  \item
  The low mass end behaviour of the Sheth-Tormen mass function is decided by $p$, while the high mass end exponential cutoff depends on $q$. This leads to a large variation in mass function at the low (high) mass end due to $q$ ($p$) change. In the transition region, the range of $p$ and $q$ in the fits is such that the Sheth-Tormen mass function does not vary much. Due to this, we see tapering of a standard deviation band around $\nu \approx 0.2$.
  
\end{itemize}

\subsection{Dependence of Sheth-Tormen parameters on power-law index and epoch}
\label{sec:n_rnl_dep}

\begin{figure}
    \centering
        \begin{subfigure}[b]{0.66\textwidth}
        \includegraphics[width=.99\textwidth]{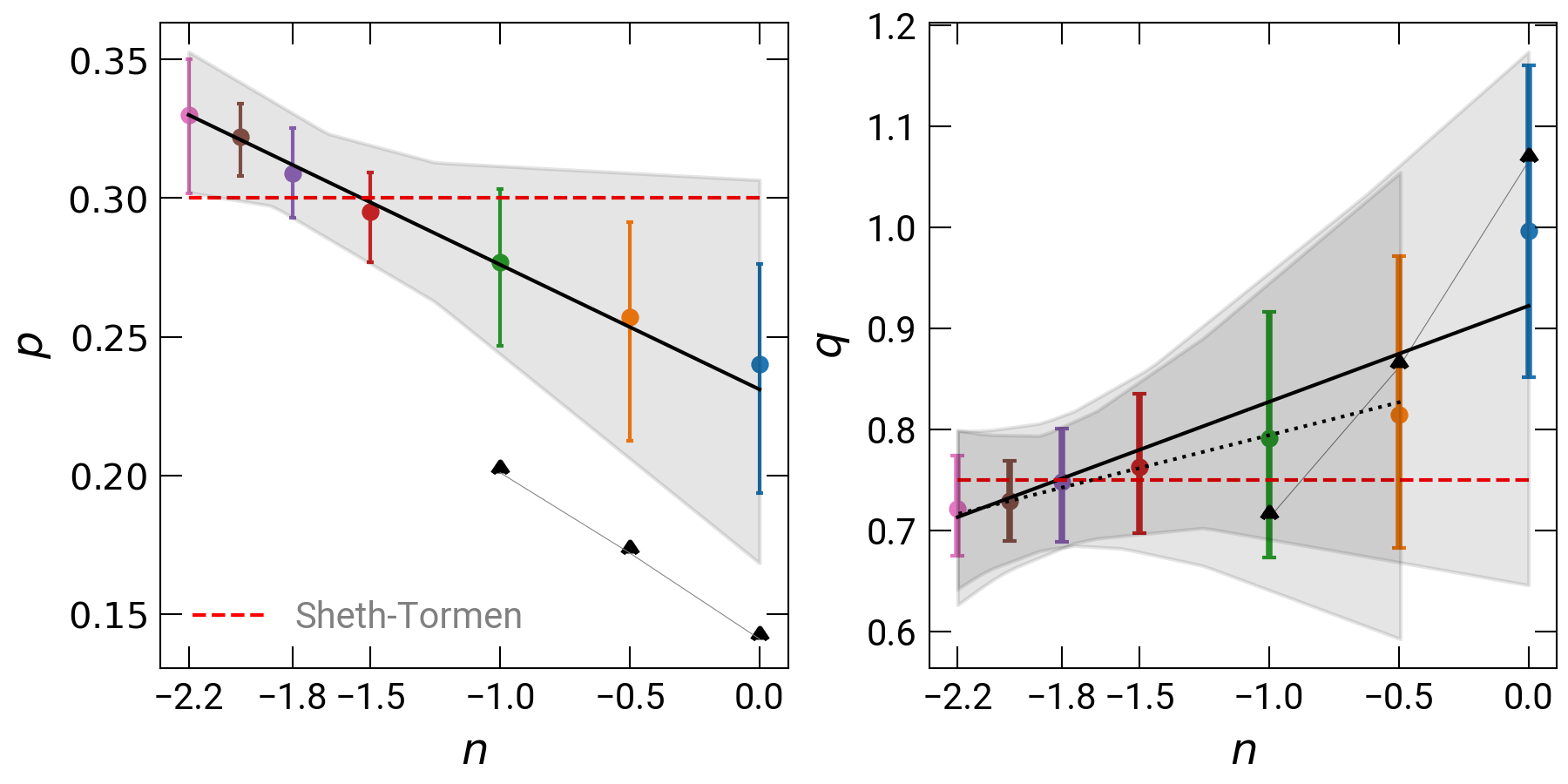}
        \end{subfigure}
        \begin{subfigure}[b]{0.33\textwidth}
        \includegraphics[width=.99\textwidth]{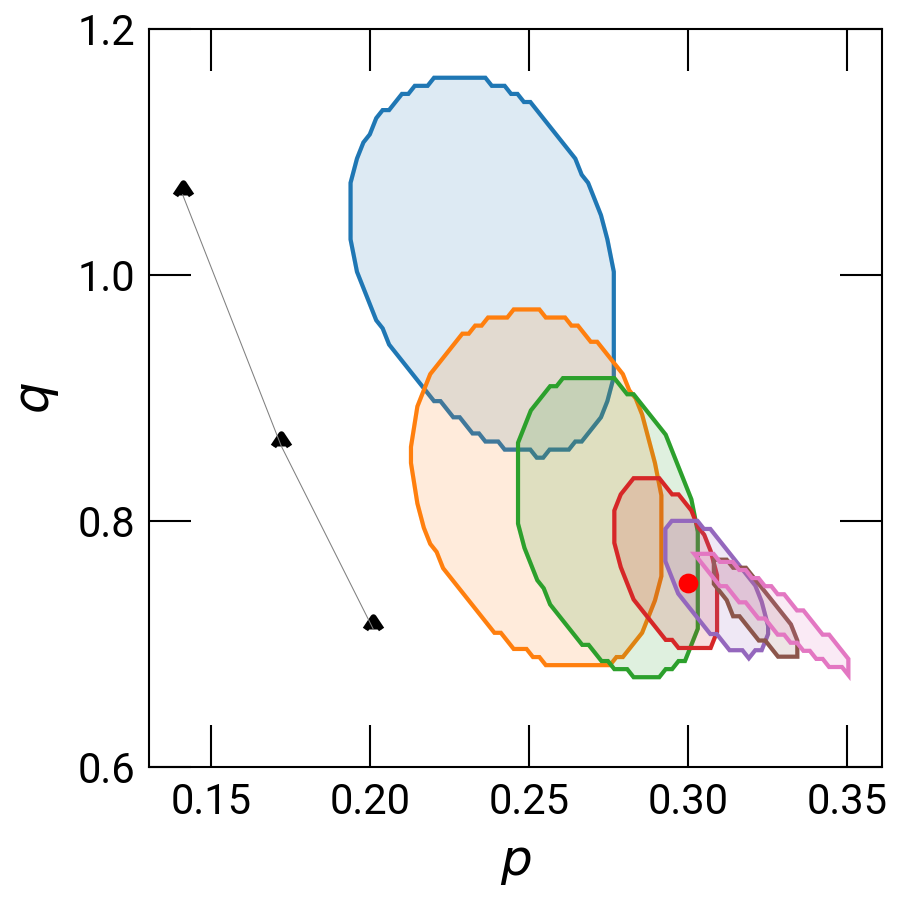}
        \end{subfigure}
    \caption[Power spectrum $n$ dependence of the Sheth-Tormen parameters]{\emph{Power spectrum dependence of the Sheth-Tormen parameters:} The variation of the Sheth-Tormen parameters with the power-law index ($p$: left panel, $q$: right panel). Errors in the $p$, $q$ corresponds to a standard deviation confidence in the fitting. A solid black line represents a linear fit to the scatter with the gray area as a standard deviation confidence of this fitting. The right panel shows a standard deviation contours of the mass function fittings in the $pq$-space. Dashed red lines in the left and middle panels show standard Sheth-Tormen parameters, the corresponding red marker is shown in the right panel. The dotted black line in the middle panel shows a revised linear fit to the scatter. Triangular data points show best-fit values from \protect\cite{Bagla_2009a}.}
    \label{fig:pq_n}
\end{figure}

We looked at the variation of best-fit Sheth-Tormen parameters $p$ and $q$ with power spectrum index $n$. We find the following approximate $n$ dependence of $p$ and $q$ after fitting straight lines. The choice of a linear relation appears to be a good first approximation, as seen in  \cref{fig:pq_n}.

\begin{equation}
  \begin{array}{l}
  p(n) \simeq -0.045\ n+0.231\\ 
  q(n) \simeq 0.095\ n+0.922
  \end{array}
    \label{eqn:pq_n}
\end{equation}
The left and middle panels of \cref{fig:pq_n} show the relation between Sheth-Tormen parameters and power-law power spectrum index $n$. Errors in $p$ and $q$ are derived from a standard deviation confidence in the fitting. Black lines show linear fit (\cref{eqn:pq_n}) to $p$ and $q$ scatter. Gray filled areas show a standard deviation confidence for the linear fits. Dashed red lines in the left and middle panels show standard Sheth-Tormen parameters. The right panel of \cref{fig:pq_n} shows a standard deviation contours of mass function in $pq$-space. Triangular data points show best-fit values from \cite{Bagla_2009a}. This shows a correlation between Sheth-Tormen parameters. The plot shows a clear trend of Sheth-Tormen parameters with power spectrum index $n$. Thus, we have clear evidence in these simulations of the dependence of the parameters of the mass function on the slope of the initial power spectrum. This indicates a departure from the universality of mass functions. This is a significant finding and requires theoretical investigations. Understanding this will lead to improved theoretical modelling of the collapse of halos and the theory of mass functions.

The propagation of errors from small to large scales can introduce transient features in N-body simulations. The sources of these errors can be mismatched mass and force resolution, discreteness effects, and stochastic noise. We study these in \cref{cha:5}, where we explore the impact of a finite shortest scale up to which the desired power spectrum is realised in simulations. We find that the impact of small-scale cutoff in the initial power spectrum and discreteness increases with $(n + 3)$. For the $n\ge-1.0$, resolving a halo with mass function errors within 1\% requires at least $300$ particles (see \cref{fig:th1}). However, in this analysis, we use a $100$-particle criterion. Consequently, for $n \geq -1.0$, spurious transients likely affect earlier epochs due to the lower particle threshold we use. Hence, we see $n=0$ as an outlier with the linear fit. If we eliminate this data point, we get a revised linear fit as in \cref{eqn:pq_n_revised}. Eliminating the $n=0$ data point does not affect the $p$ trend at all but affects the slope of the $q$ line significantly. Revised fit and corresponding standard deviation confidence interval is shown in the middle panel (black dashed line) of \cref{fig:pq_n}. It is a better predictor for $\Lambda$CDM models, as seen in \cref{sec:mf_lcdm}.

\begin{equation}
  \begin{array}{l}
  p(n) \simeq -0.045\ n+0.231\\ 
  q(n) \simeq 0.065\ n+0.859
  \end{array}
    \label{eqn:pq_n_revised}
\end{equation}

\begin{figure}
    \centering
    \includegraphics[width=.5\textwidth]{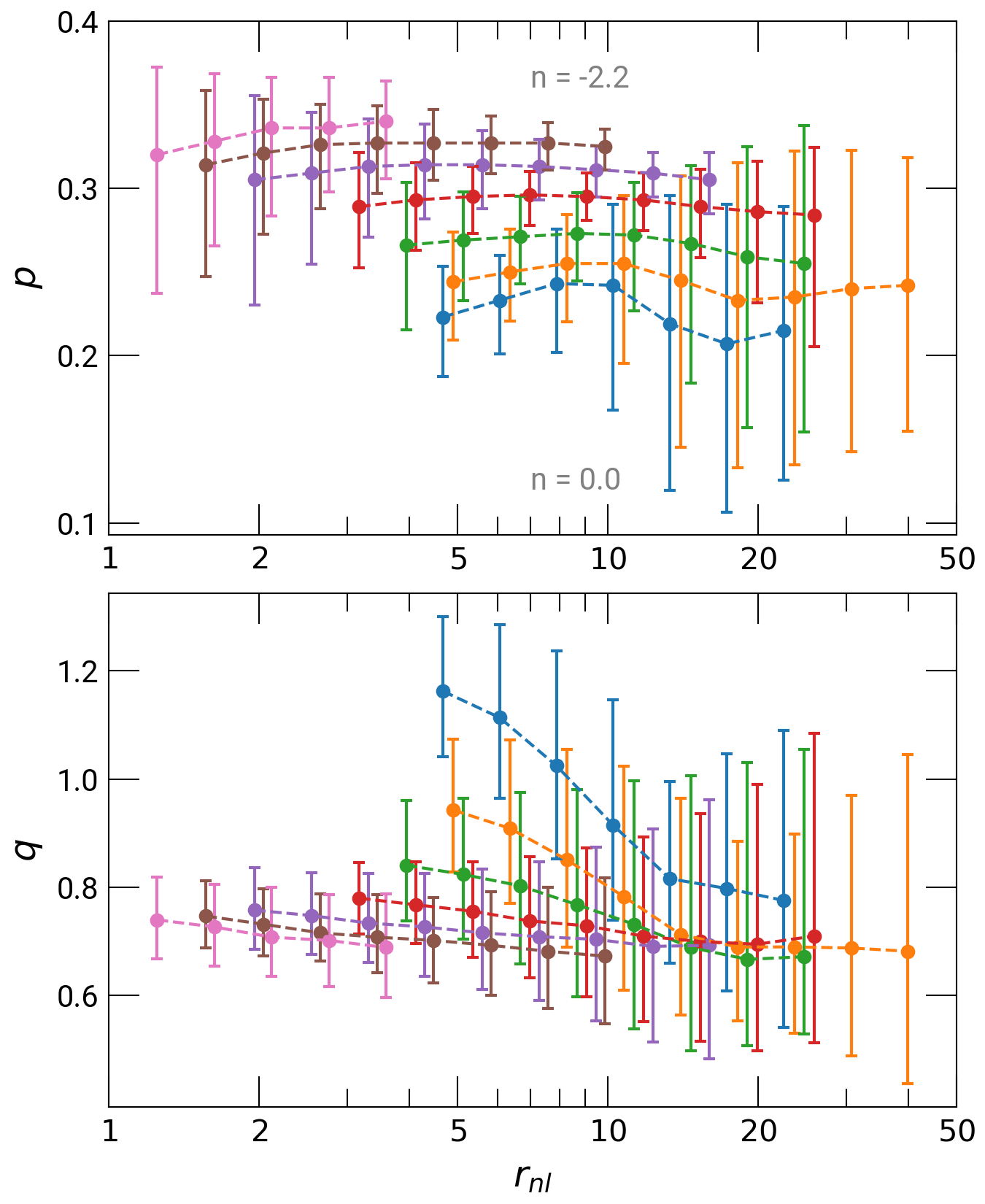}
    \caption[Epoch independence of the Sheth-Tormen parameters]{{\emph Epoch independence of the Sheth-Tormen parameters}: The variation of the Sheth-Tormen parameters ($p$: upper panel, $q$: lower panel) with the scale on non-linearity ($r_{nl}$). Errors in the $p$, $q$ corresponds to a standard deviation confidence in the fitting. Different colors used for the scatters represent different models with the power-law $n$ as shown in the legend of \cref{fig:sgm_err}.}
    \label{fig:pq_rnl}
\end{figure}

We separately fit epochs available for a model to investigate the time dependence of the shape of the mass function. \Cref{fig:pq_rnl} shows variation of $p$ and $q$ with $r_{nl}$. $p$ tend to remain stable with $r_{nl}$. $q$ varies with power-law index  $n$; variation in $q$ with $r_{nl}$ tends to increase as $n$ approaches zero. This indicates the presence of some transient effects. Thus, we conclude that the shape of the mass function does not vary with time in a significant manner for $n<-1$.

Trends in $p$ and $q$ presented by \cite{Bagla_2009a} match qualitatively with our findings. In the present study, we have explored more models and used larger simulations. We have also studied the tolerance towards various parameters used in the study, as discussed in the following section. 

We notice an $n$-dependent slope in the $p$-$q$ correlation. The possible reason could be the minimum mass cut in the simulations. At early times, halos above this cut mainly sample the exponential tail of the mass function, constraining $q$ better than $p$. As the halo mass fraction increases with time, the Sheth-Tormen mass function induces a correlation between $p$ and $q$ \cite{Manera_2009}. This correlation slope can vary with epoch due to fixed mass cuts used here. Hence, it is possibly contributing to the observed weak trends in $r_{\text{nl}}$ with $n$.

\subsection{Tolerance to parameters}
\label{sec:mf_tolerance}

\begin{figure}[t]
    \centering
    \includegraphics[width=.95\textwidth]{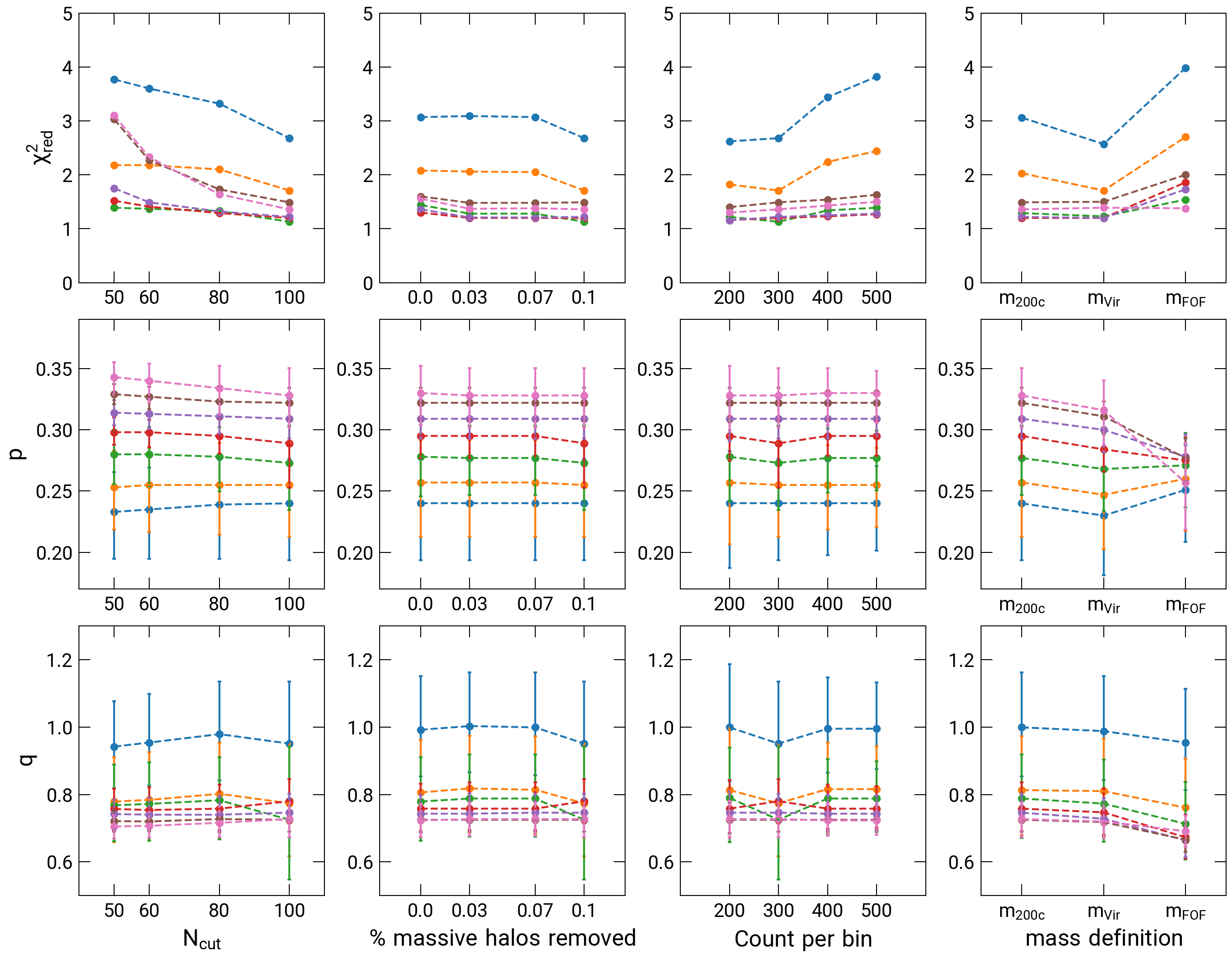}
    \caption[Mass function: tolerance to parameters]{\emph{Tolerance to parameters:} Effect of parameters like $N_{\text{cut}}$ (column 1), percentage of massive halos removed from halo catalog (column 2), number of halos in each bin (column 3), and mass definitions (column 4) on  $\chi^2_{\text{red}}$ (row 1) and fitting parameters ($p$: row 2, $q$: row3).
    Different curves in each panel represent different power-law models, colors used for models are the same as shown in \cref{fig:sgm_err}.}
    \label{fig:tol}
\end{figure}

This subsection presents how the criteria used to prepare the halo catalogue obtained from simulations affect the mass function and the best fit. We test the criteria: removal of small mass halos, removal of massive halos, number of halos in each bin, and halo mass definitions.

The parameters that we vary are:
\begin{itemize}
    \item 
    $N_{\text{cut}}$ is the minimum number of particles a halo must have for us to consider it in the halo catalogue used for fitting the mass function.
    \item
 Most massive halos are rare, and cosmic variance strongly affects the numbers.  Thus, we remove a small fraction of halos from the high mass end of the catalogue and fit the mass function in the remaining bins.  This is the second parameter used in the calculations.
    \item
 The number of halos in an individual bin affects the size of the error bars.  This is the third parameter that we have used. 
    \item
 There are multiple ways to define the mass of halos.  We consider three($m_{\text{200c}}$, $m_{\text{vir}}$, $m_{\text{FOF}}$) possibilities in this chapter.  We also study the tolerance of mass function to the choice of mass definition.

\end{itemize}

The fitting procedure gives us the best-fit values of the mass function parameters $p$ and $q$ and the reduced $\chi^2$. We study the variation of these three with our choices, and the results are shown in \cref{fig:tol}. We have plotted curves for each of the power-law models in each panel. Columns in this figure correspond to our choices, and rows correspond to the computed values of $\chi^2_{\text{red}}$, $p$ and $q$. We note that in most panels, the values of these three parameters are stable, indicating that our choices are not critical and the results we present do not depend significantly on the choices. Specific panels that require a discussion are as follows:

\begin{itemize}
    \item 
    There is an increase in $\chi^2_{\text{red}}$ with the number of halos per bin for models with a steeper power spectrum slope. This appears to be caused by a reduction in the number of bins. Interestingly, there is no variation in values of other parameters $p$ and $q$.
    
    As the number of counts in a bin becomes very large, statistical fluctuations in the data become small compared to the bin size. This led to a systematic shift in the last bin. Hence, we restrict our analysis to $500$ halos per bin.
       \item
    The estimated variation of $p$ with the power-law index is very small if we use $m_{\text{FOF}}$. Surprisingly, there is no corresponding effect in a variation of $q$ with the index $n$. Notably, trends in $p$ and $q$ are intact if the $n=-2.2$ model is ignored.

\end{itemize}

\section{Going beyond power-law and EdS}
\label{sec:mf_lcdm}

\begin{figure}
    \centering
\begin{subfigure}[b]{0.9\textwidth}
 \includegraphics[width=.95\textwidth]{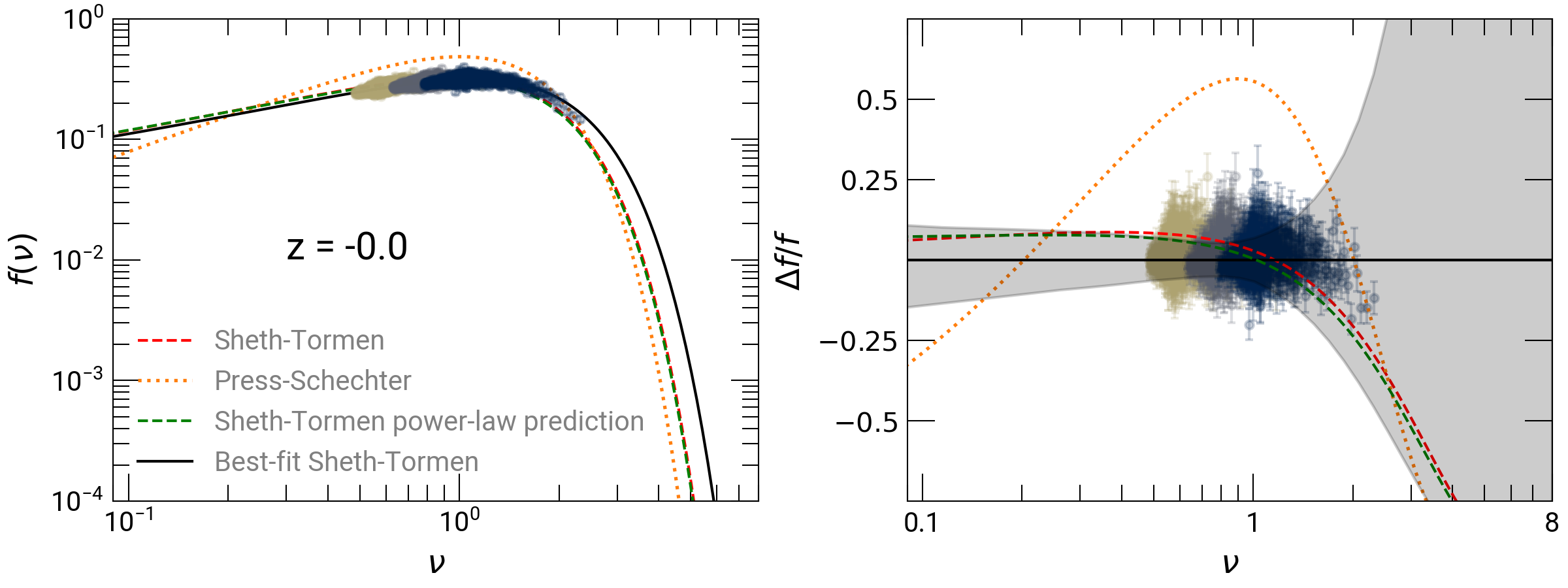}
  \end{subfigure}
      \begin{subfigure}[b]{0.9\textwidth}
 \includegraphics[width=.95\textwidth]{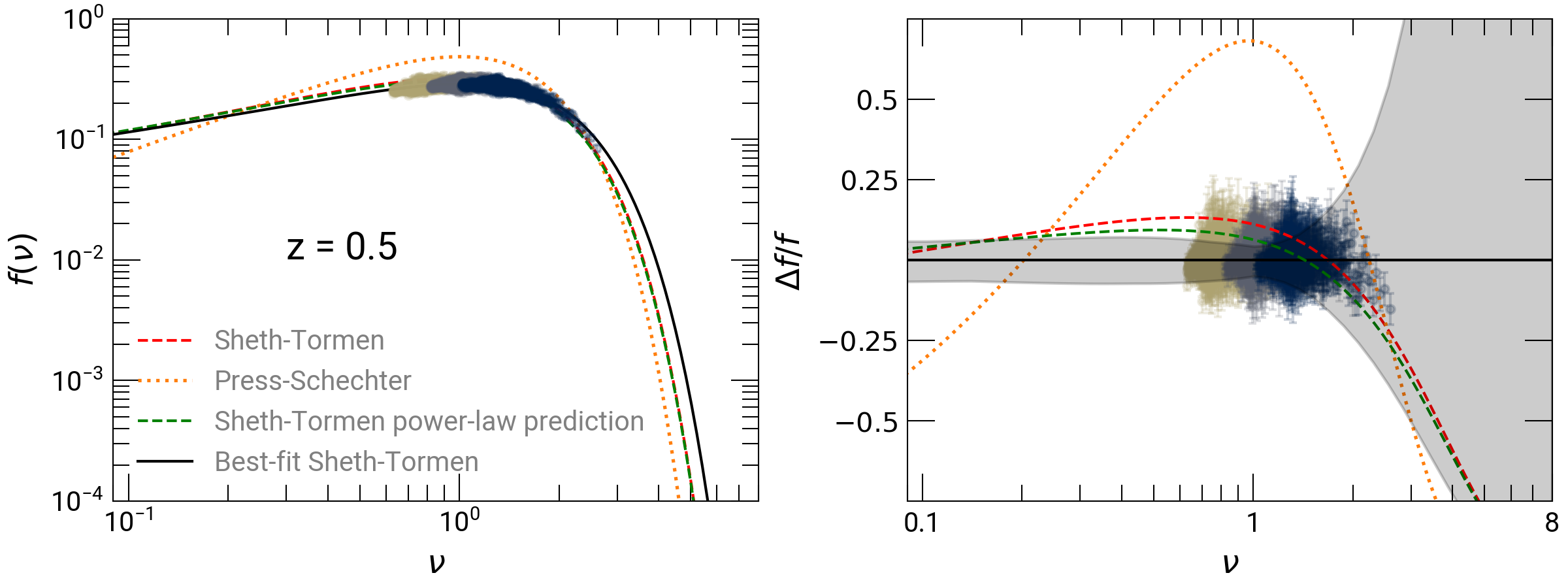}
  \end{subfigure} \\
      \begin{subfigure}[b]{0.9\textwidth}
 \includegraphics[width=.95\textwidth]{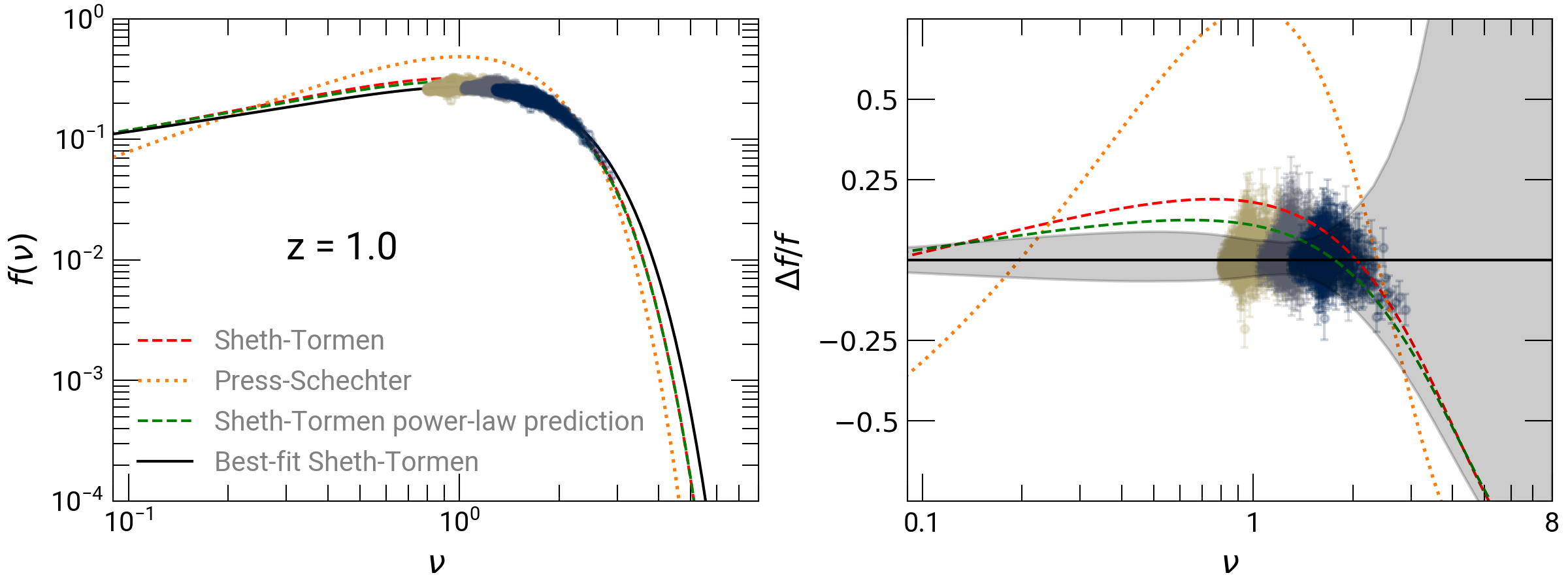}
  \end{subfigure}
      \begin{subfigure}[b]{0.9\textwidth}
 \includegraphics[width=.95\textwidth]{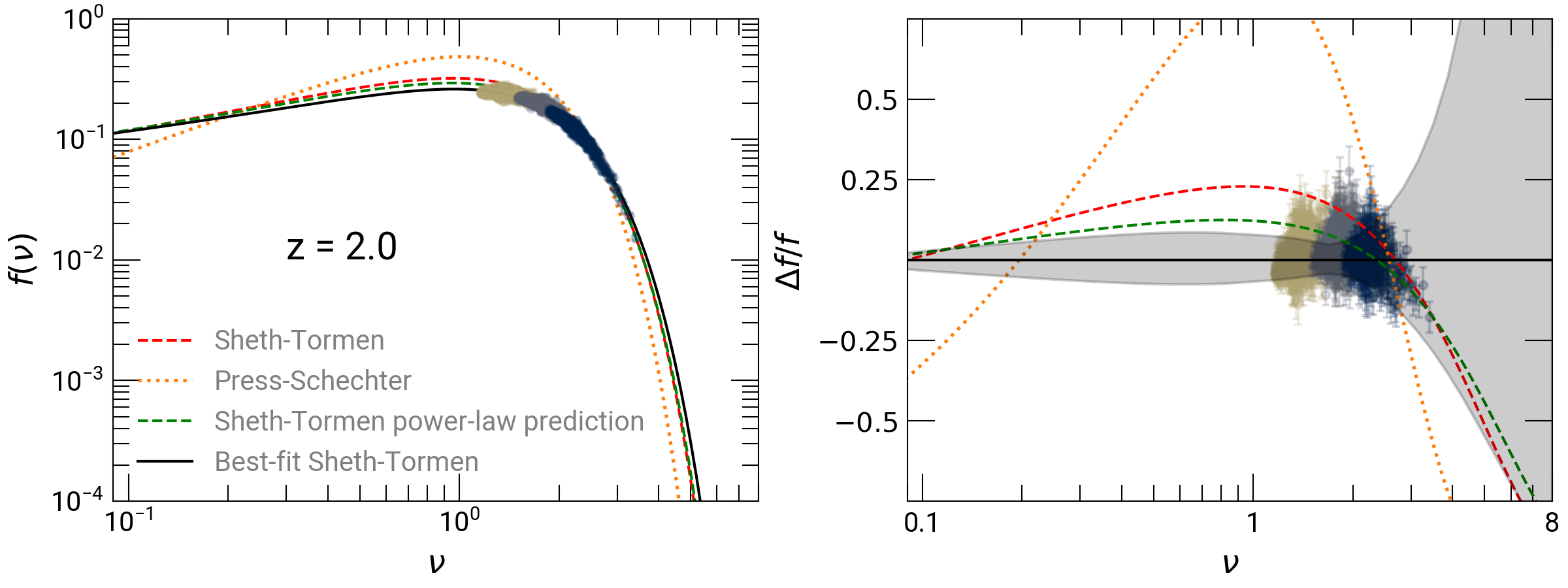}
  \end{subfigure}
  \caption[Mass function for $\Lambda$CDM models]{\emph{ Mass function for $\Lambda$CDM models: }  The mass function for $\Lambda$CDM simulations (left column), and their residuals after fitting the Sheth-Tormen function (right column). Rows 1,2,3 and 4 are for redshifts $0.0$, $0.5$, $1.0$ and $2.0$ respectively. Dotted orange, dashed red, and dashed green lines represent the Press-Schechter, the Sheth-Tormen mass function, and power-law prediction using \cref{eqn:pq_n_revised}. Solid black lines represent the best-fit Sheth-Tormen mass function after $\chi^2$ analysis. Brown-gray data points show the mass function calculated from the simulations. Three different colors of data points represent different simulations used for analysis. Gray filled area shows a standard deviation confidence interval for the fit.}
    \label{fig:mfl}
\end{figure}

\begin{figure}
    \centering
    \includegraphics[width=.7\textwidth]{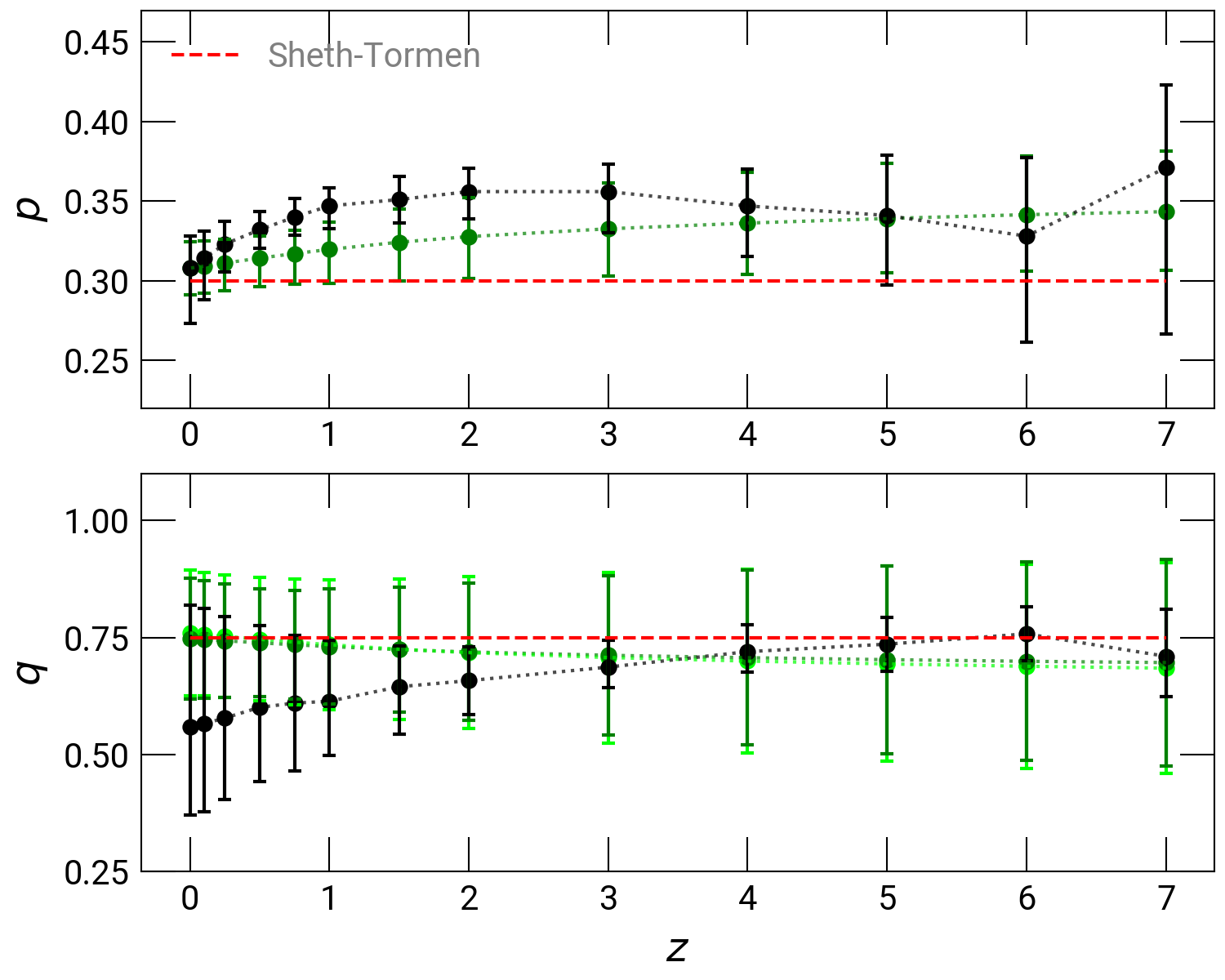}
    \caption[Sheth-Tormen parameters in $\Lambda$CDM cosmology]{\emph{ Sheth-Tormen parameters in $\Lambda$CDM cosmology:} The figure shows the variation of the Sheth-Tormen parameters ($p$: upper panel, $q$: lower panel) with the redshift for the $\Lambda$CDM model. Black scatter represents best-fit Sheth-Tormen values while the Green and light-green data points represent predictions from power-law models with respect to \cref{eqn:pq_n_revised} and \cref{eqn:pq_n} respectively. Errors in the $p$,$q$ correspond to a standard deviation confidence. Dashed red lines show standard Sheth-Tormen parameters.}
    \label{fig:lcdm_pq}
\end{figure}

\begin{figure}
    \centering
    \includegraphics[width=.7\textwidth]{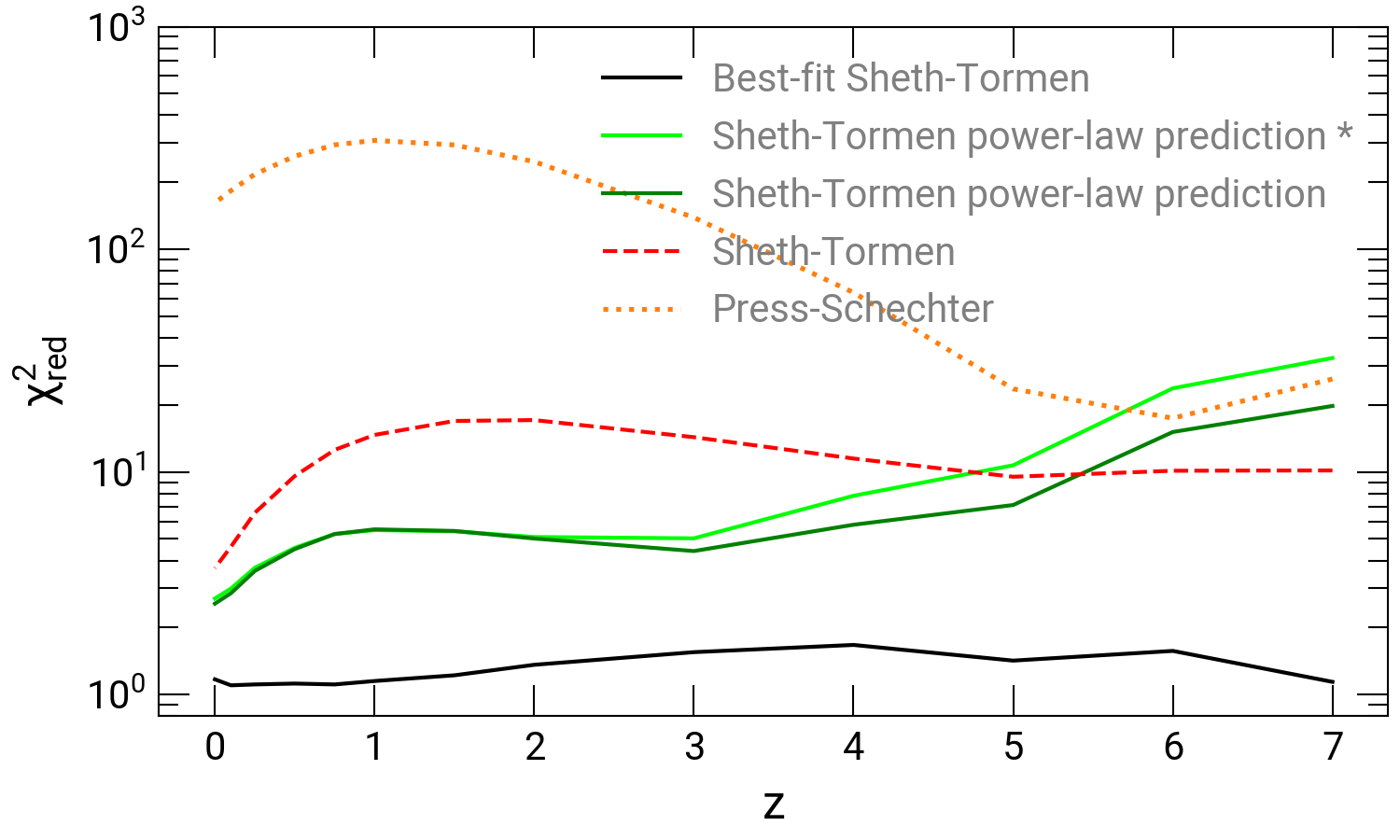}
    \caption[$\chi^2_{\text{red}}$ values for $\Lambda$CDM mass function fits]{\emph{ $\chi^2_{\text{red}}$ values for $\Lambda$CDM mass function fits:} Black curve is $\chi^2_{\text{red}}$ for best-fit Sheth-Tormen mass function, Green and light-green curve are power-law models prediction of Sheth-Tormen mass function (see text), dashed red curve is $\chi^2_{\text{red}}$ for standard Sheth-Tormen mass function, and the dotted orange curve is $\chi^2_{\text{red}}$ for  Press-Schechter mass function.}
    \label{fig:lcdm_chi}
\end{figure}

We have demonstrated conclusively that the mass functions for power-law models in an EdS background are not universal. Departure from universality is small but non-zero. The set of models used to demonstrate this departure from universality is simple; indeed, the simplicity allows us to demonstrate deviations from a universal behaviour. In this section, we seek to analyse how this departure from universality may affect the mass function in a more realistic model. 

Departures from universality in mass function can potentially arise from cosmology via the growth factor or the threshold for collapse, or it can come from the shape of the power spectrum. Our study has given us a quantitative handle on the variation in mass function with the slope of the power spectrum. In this section, we use the dependence of $p$ and $q$ on $n$ to estimate the expected variation of the mass function with the shape of the power spectrum. To map the varying slope of the $\Lambda$CDM power spectrum, we define an effective index at any redshift $z$ by computing
\begin{equation}
   {\frac{\d \log \sigma(m,z)}{\d \log m}} \bigg|_{\sigma=1} = - \frac{n_{\text{eff}}+3}{6}.
    \label{eqn:sigma_d}
\end{equation}

Then, we fit free Sheth-Tormen to $\Lambda$CDM simulations epoch-wise and compare the best-fit $p$ and $q$ with \cref{eqn:pq_n}. We use three Planck18 Cosmology \cite{Planck_2020} simulations with box sizes $150$, $300$ and $500$ Mpc/h with $1024^3$ particles, described in \cref{tab:lcdm_sim}. Small box simulations provide better mass resolution, essential to capture low mass halos. Further capturing low mass halos allows us to probe early epochs, as massive halos are rare. On the other hand, large box simulations provide us with high mass halos. We use one intermediate box to connect both mass ends. We calculate mass functions using the $M_{\text{200b}}$ halo mass definition, $300$ halos per bin, $N_{\text{cut}}=100$, and removing the top $0.1$\% of massive halos from each box. We verify that the mass function has converged with regard to box size by fitting individual boxes and comparing the results to a combined fit. \Cref{fig:mfl} shows the mass function for $\Lambda$CDM simulations (left column), and their residuals after fitting the Sheth-Tormen function (right column). Different panels are for different representative epochs. Dotted orange, dashed red, and dashed green lines represent the Press-Schechter, the Sheth-Tormen mass function, and power-law prediction using \cref{eqn:pq_n_revised}. Solid black lines represent the best-fit Sheth-Tormen mass function after $\chi^2$ analysis. Brown-gray data points show the mass function calculated from the simulations. Three different colors of data points represent different simulations used for analysis. Gray filled area shows a standard deviation confidence interval for the fit.

We show these fitted parameters, and in \cref{fig:lcdm_pq}, black data points show the best-fit Sheth-Tormen values of $p$ and $q$ for redshift. Green data points represent predictions from power-law models according to \cref{eqn:pq_n_revised}. Light-green scatter in the bottom panel shows \cref{eqn:pq_n} prediction. Errorbars represent a standard deviation errors. Dashed red lines show $p$ and $q$ values from standard Sheth-Tormen. We find that the trend in $p$ for both predicted and fitted values is similar. Their actual values are off by some factor, possibly because of cosmological contributions. $q$ in both cases is consistent within errorbars. As we increase redshift, the halo count decreases, so it is challenging to probe that part. It is not easy to probe mass function below $\nu<1$.

In \cref{fig:lcdm_chi} we present $\chi^2_{\text{red}}$ values of $\Lambda$CDM mass function fit with epoch. The black curve shows $\chi^2_{\text{red}}$ for the best-fit Sheth-Tormen mass function. Green and light-green curves represent power-law models prediction of Sheth-Tormen mass function from \cref{eqn:pq_n} and \cref{eqn:pq_n_revised} respectively. The dashed red and dotted orange curves show $\chi^2_{\text{red}}$ for standard Sheth-Tormen and Press-Schechter mass functions.

Power-law prediction better fits simulation data over standard Sheth-Tormen for $z \le 4.0$. For $z>4.0$, we have less number of data points resulting in large $\chi^2_{\text{red}}$ values. Using revised fits for $p$ and $q$ (\cref{eqn:pq_n_revised}), $\chi^2_{\text{red}}$ for power-law prediction improves further. This indicates the scope of improvement in Sheth-Tormen mass function.

\section{Conclusions}
\label{sec:mf_conclusion}

We have used power-law models in the Einstein-de Sitter background to investigate the mass function of collapsed halos in cosmological N-body simulations. We fit the mass function obtained from simulations with the Sheth-Tormen form to obtain the parameters $p$ and $q$. This is done for several models with different power-law indices in the $-2.2 \leq n \leq 0$. We summarise the key findings of this work below.
\begin{itemize}
    \item 
 We find clear evidence that the halo mass function is not universal and has explicit power spectrum dependence. 
    \item 
 The Sheth-Tormen parameter values that are obtained by fitting the mass function show a systematic trend with the power-law index (\cref{fig:mf_1} and \cref{fig:mf_2}. We also show a correlation between these parameters (\cref{fig:pq_n}).
    \item 
 The halo mass function has a weak epoch dependence: the best-fit value of Sheth-Tormen parameters varies with redshift. However, the values are consistent with a constant value for a given power-law model within a standard deviation limit.
    \item
 We have shown that these results are not sensitive to the process used for constructing the halo catalogue or subsequent analysis.
   \item    
 best-fit functions and the corresponding $\chi^2_{\text{red}}$ values indicate, as is also seen in the plots for residuals, that the Sheth-Tormen mass function form is inadequate. Better modelling of collapse with more parameters may be required. 
    \item 
 We show that the Sheth-Tormen mass function with parameter values derived from a matched power-law EdS cosmology provides a better fit to the $\Lambda$CDM mass function than the standard Sheth-Tormen mass function. We have not studied departures from universal mass function with cosmology, so we cannot rule out a similar contribution from such dependence, especially at low redshifts.
\end{itemize}

While we have not studied the cause of non-universality in mass functions, we can speculate about its origin. The variation we have seen is with the slope of the power spectrum, and in these models, the only contribution that can cause this can arise from the coupling of modes between the collapsing structures and the large scale density field. While such coupling is not significant for spherical collapse, it becomes an essential ingredient for ellipsoidal collapse. Recent studies have shown that shifting from density maxima to energy minima and positive-definite 'energy shear' tensors provides a better alignment with halo formation in simulations \cite{Musso_2021,Musso_2023,Musso_2024}. \cite{Castorina_2016} integrates excursion set theory with tidal effects to predict dark halo abundances and clustering, revealing nonlocal halo bias. The halo properties that correlate with the tidal environment are also examined \cite{Musso_2012, Ramakrishnan_2021}. A detailed coupling between halo formation and large scale tidal fields can be analysed in N-body simulations, and we are investigating it.



%% file: chapters/chap4.tex

\chapter[Dispersion in the Hubble-Lema\^{i}tre constant measurements from gravitational clustering]{Dispersion in the Hubble-Lema\^{i}tre constant measurements from gravitational clustering}
\label{cha:4}

{\small This chapter is based on the following manuscript: \fullcite{Gavas_2024}}\\

Hubble-Lema\^itre constant ($H_0$) is a key parameter in relativistic cosmology. It encapsulates the rate of expansion of the Universe at present \cite{Freedman_2010, Tully_2023, Verde_2023}. Being a dimensional quantity, it is also a measure of the age of the Universe in combination with other cosmological parameters. Measurements of $H_0$ at scales of a few to a few tens of megaparsecs are done using standard candles like Cepheid variables. Several distance indicators are used to extend these measurements to larger distances. Measurements of $H_0$ are also made using time delay measurements in gravitational lensed systems and CMBR anisotropies \cite{Brieden_2023}. There is, at present, a strong tension between the measured values of $H_0$ using local and distant sources \cite{Planck_2020, Riess_2021,Riess_2022, Brout_2022}. This has been given the name Hubble tension \cite{Shah_2021, Di_Valentino_2021, Verde_2023}. Many possible extensions of the standard cosmological model have been proposed to account for this discrepancy, though there appears to be no single solution that accounts for all observables \cite{Efstathiou_2021, Vagnozzi_2023}. A number of studies have also been carried out to determine whether the contribution of systematic errors can help alleviate tension \cite{Mortsell_2022, Huterer_2023, Majaess_2024, Shah_2024, Camilleri_2024}.

Local inhomogeneities, such as supervoids and peculiar velocities, significantly influence the measurement of the $H_0$ \cite{Wu_2017, Aluri_2023, Watkins_2023, Ruiz-lapuente_2023, Huang_2024, Kalbouneh_2024, Carreres_2024, Giani_2024}. Peculiar velocities, resulting from gravitational interactions between nearby galaxies, distort the smooth Hubble flow, affecting the observed recession velocities and subsequent distance measurements \cite{Maartens_2023}. Despite continuous efforts to improve distance measurement techniques and methodologies, discrepancies persist \cite{Anand_2024, Riess_2024,Verde_2023}, highlighting the importance of robust statistical analyses.

In this chapter, we present a study of the effect of inhomogeneities and gravitational clustering on the measurements of $H_0$. Our approach is inspired by ~\cite{Turner_1992}, where we use cosmological N-body simulations to estimate the effect of peculiar velocities on the measurement of $H_0$. Similar studies have been carried out by others \cite{Odderskov_2014, Wojtak_2014, Odderskov_2017}. In \cite{Odderskov_2014}, it was found that observers in random dark matter halos typically measure lower expansion rates than those in voids, with differences more pronounced in halo rest frames compared to the CMB frame. \cite{Wojtak_2014} explored how $H_0$ measurements are influenced by factors such as survey depth, size, sky coverage, and observer position. In \cite{Odderskov_2017}, results from linear perturbation theory were compared with non-linear velocity power spectra derived from N-body simulations, which typically show smaller variances. Although none of these approaches produces a variance large enough to explain the $H_0$ discrepancy fully, the differences are significant given the precision of local $H_0$ measurements. We use N-body simulations and consider galaxies to be embedded in dark matter halos. Considering observers to be located in galaxies, we compute the {\it measured} $H_0$ for each galaxy pair by considering the peculiar velocities. This allows us to construct and analyse the distribution for $H_0$ across different relative distance intervals for all observers. 

The chapter is structured as follows: In \cref{sec:ho_method}, we discuss the methodology to estimate local $H_0$ measurements. \Cref{sec:h0_simulations} outlines our simulations. In \cref{sec:ho_results}, we present results derived from our data analysis. Finally, we conclude our findings and discuss future directions in \cref{sec:ho_conclusion}.

\section{Methodology}
\label{sec:ho_method}

Assuming galaxies reside within dark matter halos, we use halos as a representative for both observers and observed galaxies. We use the distribution of halos identified in N-body simulations to create a group of hypothetical observers positioned on the halos. We assign a single galaxy or observer for each halo: this is clearly a simplification, especially for massive halos, and leads to an under-estimate the effect of peculiar velocities as we ignore the contribution of the intra-halo velocity distribution and the halo occupation distribution. We then derive the local Hubble-Lema\^itre constant ($H_L$) for observers in their neighbourhood volume by considering distances and peculiar velocities of halos within the volume. As a result, we get a set of local Hubble-Lema\^itre constant measurements, creating a distribution of measured values for different observers. This distribution is then compared with the global Hubble-Lema\^itre constant ($H_G$) value used in the simulation model. This approach closely follows the methodology used by \cite{Turner_1992}.

The contribution of cosmological expansion and peculiar velocities is combined to obtain redshifts that are then used to compute the local value of Hubble-Lema\^itre's constant for a given observer. To set the notation, we define $\mathbf{r}_i$ as the proper coordinate of the $i^{\text{th}}$ halo. This is related to the comoving coordinate as $\mathbf{r}_i = a(t) \mathbf{x}_i$. Hence, the velocities in the two coordinates are related as $\mathbf{V}_i = \dot{a} \mathbf{x}_i + a(t) \dot{\mathbf{x}}_i = H \mathbf{r}_i + a(t) \dot{\mathbf{x}}_i $. For the present epoch where we take $a(t_0) = 1$, we can write this as: $\mathbf{V}_i = H_0 \mathbf{r}_i + \dot{\mathbf{x}}_i$. The "observed" Hubble-Lema\^itre's by an observer at the origin is then $\mathbf{V}_i \cdot \mathbf{r}_i / r_i^2$. For a pair of halos $i$ and $k$, the measured Hubble-Lema\^itre's is given as,
\begin{equation}
  H_{ik} = H_0  + \frac{\left(\mathbf{x}_i - \mathbf{x}_k\right).\left(\dot{\mathbf{x}}_i - \dot{\mathbf{x}}_k\right)}{x_{ik}^2}.
  \end{equation}
Here, $x_{ik} = |\left(\mathbf{x}_i - \mathbf{x}_k\right)|$, $\mathbf{x}$ represents the comoving coordinates and $\dot{\mathbf{x}}$ is the peculiar velocity. Note that we have $a_0=1$ where $a_0$ is the present value of the scale factor. The measured Hubble-Lema\^itre's constant for the $k^{\text{th}}$ galaxy is the sum of this quantity over a set of galaxies (halos). The deviation of the measured Hubble-Lema\^itre's constant by the $k^{\text{th}}$ galaxy is then, 
\begin{equation}
    \label{eqn:delta_k}
  \delta_{H\, k} = \frac{1}{N} \sum_{i = 1}^{N} \frac{{\bf \hat{x}}_{ik} {\bf \cdot
      (\dot{\mathbf{x}}_i - \dot{\mathbf{x}}_k)}}{H_G |{\bf x_i} - {\bf x_k}|}.
  \end{equation}
The distribution of deviations from the reference value is described using the variable $\delta_H$ defined below in \cref{eqn:delta}. $\delta_H$ is a fractional deviation of the local $H_0$ value from the global $H_0$ value
\begin{equation}
    \label{eqn:delta}
    \delta_H \equiv \frac{H_L - H_G}{H_G}.
\end{equation}
Here, $N$ represents the total number of halos within volume $V$, ${\bf \hat{x}}_{ik}$ corresponds to the unit vector pointing in the direction of the $i^{\text{th}}$ halo from the observer's location. Averaging over all angles clearly reduces the dispersion in values, and hence $\delta_k$ is expected to be smaller than the corresponding estimation done in some patches in the sky for an observer. 

The size and the distance to the local volume affect the distribution $\delta$. We define this local volume as a spherical shell centred on the observer and use multiple concentric shells to explore how $\delta$ varies with distance from the observer. The lower radius ($r_{\text{min}}$) and upper radius ($r_{\text{max}}$) of the used volumes are displayed in the sixth column of \cref{tab:h0_sim}.  

While the estimation of Hubble-Lema\^itre's constant in shells is useful to understand the impact of peculiar velocities as a function of distance, real measurements are done over a wide range of scales using multiple distance indicators and cross-calibration. To mimic this, we also conduct mock observations over a wide range of shells to understand the impact of peculiar velocities on the measurement of $H_0$ in an ideal scenario. We restrict this part of the study to observers in Milky Way-size halos by mass.

\section{Simulations} 
\label{sec:h0_simulations}

We use three dark matter-only simulations with a different cubical comoving box size: $150$~Mpc/h, $500$~Mpc/h, and $1000$~Mpc/h, described in \cref{ssec:lcdm_cat}. These simulations are based on Planck18 cosmology \cite{Planck_2020}. We analyse data using the simulation output at redshift $z=0$ for this work. The large scale structure evolved in the simulations has significant cosmic variance on the scales comparable to box size and is also affected by discreteness errors at scales comparable to the grid size. Therefore, we use three box sizes to probe a sufficient range of distance scales with improved reliability. 

All three simulations use $1024^3$ particles and are subject to periodic boundary conditions. To ensure the consistency of results, we maintain overlap across the simulations in the spherical shells used as local volumes. Detailed simulation information is given in \cref{tab:h0_sim}. 

The simulations were carried out using the {\tt GADGET4} code \cite{Springel_2021}, and the halo catalogues were produced using its built-in {\tt SUBFIND} algorithm. The halo finder identifies Friends-Of-Friends (FOF) groups within the given distribution. It decomposes each detected object into substructures using the excursion set algorithm. A linking length of $0.2$ times the grid length is used, and all other default parameters of the code are used. Here, we focus on isolated halos, disregarding any halo part of a larger group or cluster. Substructures or sub-halos are not considered distinct entities; thus, each halo has only one observer, and additional galaxies/observers are not assigned to subhalos. Note that this leads to eliminating the finger of god effect, leading to an underestimate of the impact of peculiar velocities. Thus, our results are expected to underestimate the actual effect. We verify that including substructures increases observers by $\sim$30\% and increases dispersion by a few percent (see \cref{a4}).

\begin{table}
  \begin{center}
    \begin{tabular}{|>
      {\centering\arraybackslash}m{1.5cm} |>
      {\centering\arraybackslash}m{1.5cm} |>
      {\centering\arraybackslash}m{1.5cm} |>
      {\centering\arraybackslash}m{2cm}  |>
      {\centering\arraybackslash}m{5cm} |}
      \hline 
      $L_{\text{box}}$ & $N_{\text{part}}$  & ${N_{\text{halos}}}$ &  ${m_{\text{low}}}$ & Spherical shells: ($r_{\text{min}}$-$r_{\text{max}}$)\\
      (Mpc/h) & & & $(M_{\odot}$/h) &  (Mpc/h)\\ [0.5ex] 
      \hline 
      150 & $1024^3$ & 1,081,171 & $8.68 \times 10^9$ &  (15-20), (20-25), (25-30), (30-35), (35-40), (40-45), (45-50), (50-55) \\ 
      \hline
      500 & $1024^3$ & 1,599,064 & $3.22 \times 10^{11}$ & (40-55), (55-70), (70-85) (85-100), (100-115), (115-130), (130-145), (145-160) \\
      \hline
      1000 & $1024^3$ & 1,973,556 & $2.57 \times 10^{12}$ & (135-160) (160-185), (185-210), (210-235), (235-260), (260-285), (285-310), (310-335)\\
      \hline
    \end{tabular}
    \caption[$H_0$ estimates: Simulation details]{\emph{Simulation details: } Column 1: Side length of cubical simulation box, Column 2: Number of particles in the simulation, Column 3: Total number of halos in each simulation (also equal to the number of individual observers), Column 4: Lowest Halo mass, Column 5: Spherical shells used as a local volume.}
    \label{tab:h0_sim}
  \end{center}
\end{table}

One of the issues that can affect our results is the difference between a single time snapshot as compared to a light cone snapshot from simulations. We can estimate the variation of peculiar velocities at large scales using the Zel'dovich approximation \cite{Zeldovich_1970, Shandarin_1989}.  This is especially pertinent within our approach where we do not consider sub-halos.  It can be shown that the overall variation is that peculiar velocities come down over time \cite{Lahav_1991}, and hence, for the largest separations used here, we underestimate the effect of peculiar velocities by a few percent.  Note that the peculiar velocities are underestimated by a few percent, hence the overall effect on the dispersion in Hubble constant measurements is expected to be much smaller and can be ignored at this level. Similarly, we disregard the impact of baryonic processes on peculiar velocities, as studies have shown that this effect is well below 1\% at the scales considered in our analysis \cite{Dolag_2013,Hellwing_2016,Kuruvilla_2020}.

\section{Results}
\label{sec:ho_results}

\subsection{The \texorpdfstring{$\delta_H$}\ \ distributions}

\begin{figure}
    \centering
    \includegraphics[width=.88\textwidth]{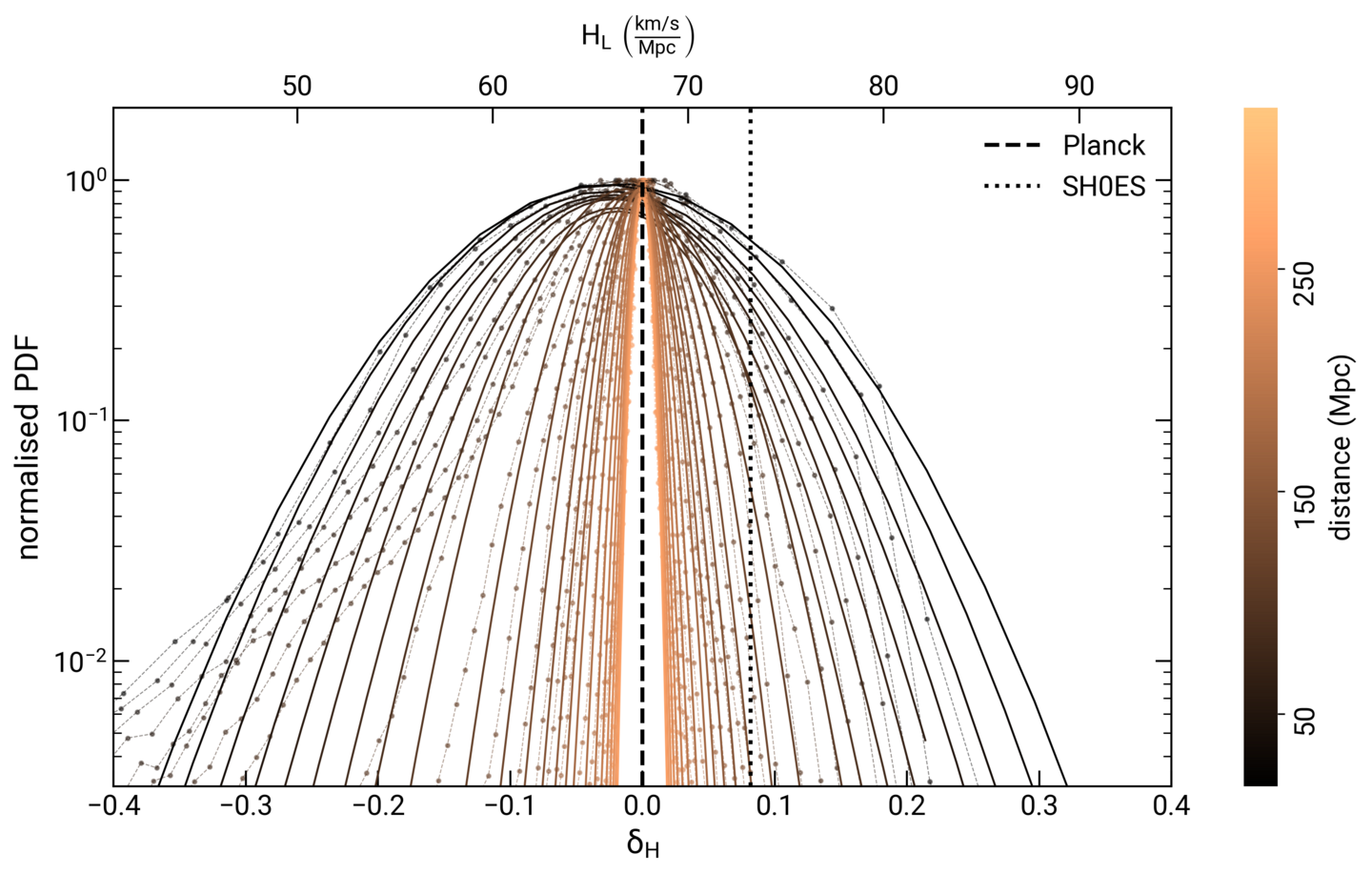}
    \caption[Local $H_0$ probability distribution functions]{\emph{Probability distribution functions:} normalised PDFs of the fractional deviation within a spherical local volume in redshift space. The scatter points represent the PDF obtained from the simulation data, while the solid lines represent Gaussian functions characterised by the first two moments of the PDF. The colors of lines and scatter points correspond to the distance to the local volume from the observer, as shown in the adjacent colorbar.}
    \label{fig:pdf1}
\end{figure} 

The distribution of $\delta_H$ obtained using \cref{eqn:delta} and \cref{eqn:delta_k}, considering different spherical shells in the three simulations, are illustrated in \cref{fig:pdf1}. Here, the distribution is obtained by averaging over all halos with a pairwise distance in redshift space corresponding to the given bin. The figure has points corresponding to the probability distribution in each bin and the Gaussian corresponding to the first two moments of the PDF. The colors of lines and scatter points correspond to the distance of local volume from the observer, as shown in the adjacent colorbar.

We observe a significant dispersion in the local $H_0$ values at all probed distances, with a maximum deviation on negative(positive) sides varying from $\sim 50$\%($70$\%) at $40$ Mpc/h to $\sim 5$\%($6$\%) at $335$ Mpc/h. As we approach larger distances, the PDFs become narrower, and the peaks shift to zero due to the increasing contribution of the Hubble flow to velocities.

PDFs show heavy negative halves, which is more evident for small-volume shells. Differences in peak location between PDFs (scatter) and Gaussian functions (solid lines) exhibit deviations of varying magnitude. These point towards a significant non-Gaussianity of the distributions.

\subsection{Statistics of distributions with distance}
\label{ssec:stat_with_d}

\begin{figure}[t]
  \centering
  \includegraphics[width=.82\textwidth]{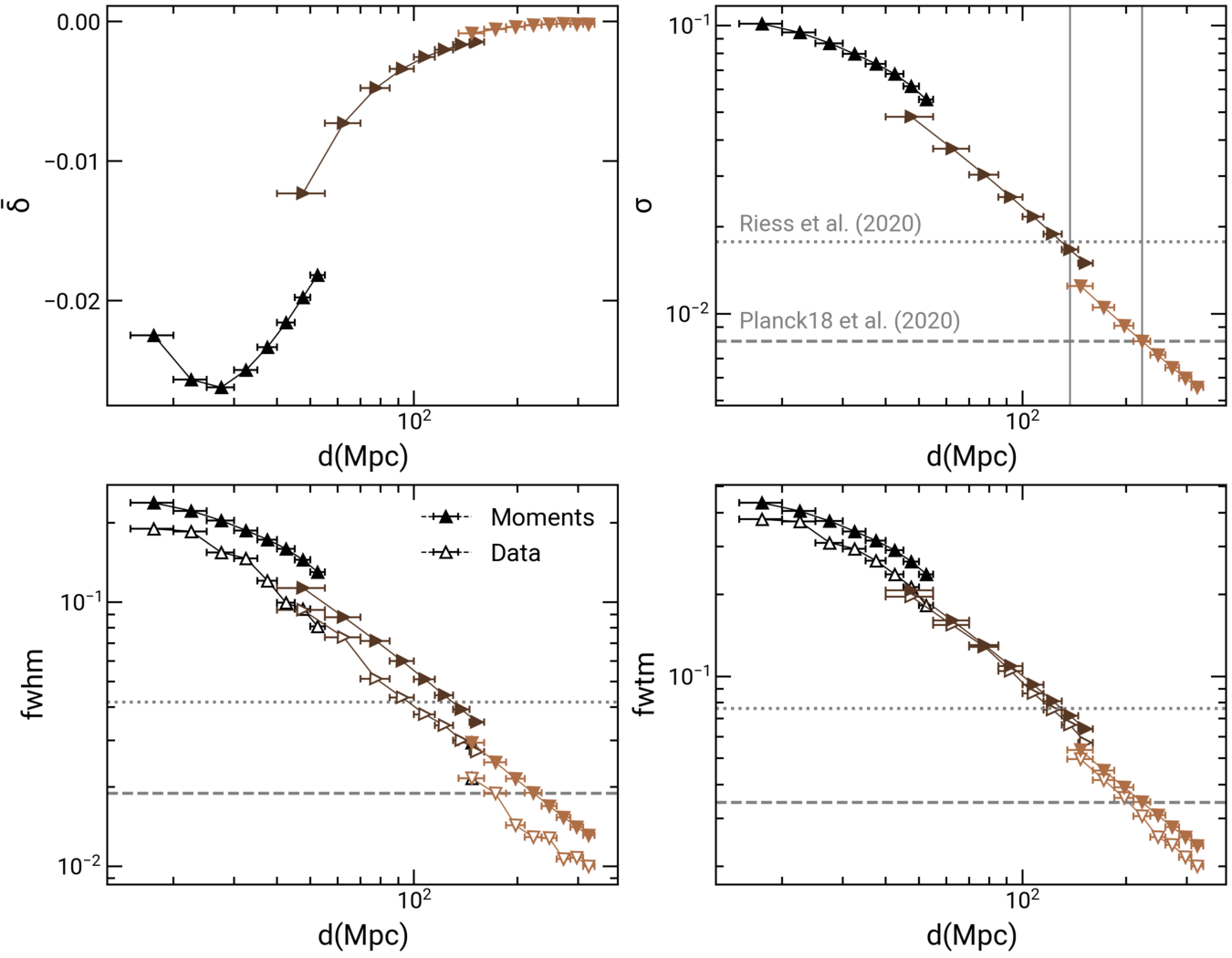}
  \caption[Statistics of $H_0$ distribution with distance]{\emph{Statistics of distribution with distance:} Mean ($\bar{\delta}$), standard deviation ($\sigma$), fwhm, and fwtm as indicated on the y-axis with distance in the x-axis. Filled markers represent quantities derived from the moments of PDF, and empty markers represent quantities computed from the discrete PDFs. The marker orientations up, right, and down correspond to simulations with box sizes of $150$ Mpc/h, $500$ Mpc/h, and $1000$ Mpc/h, respectively. The dotted and dashed gray lines depict quantities reported by \protect\cite{Riess_2021} and  \protect\cite{Planck_2020}.}
  \label{fig:stat_d}
\end{figure}

We analyse the $\delta_H$ distributions by calculating various statistical quantities such as mean ($\bar{\delta}$), standard deviation ($\sigma$), full-width half maximum (fwhm), and full-width tenth maximum (fwtm). In \cref{fig:stat_d}, we present these quantities as a function of distance in separate panels. In the figure, filled markers represent moments and quantities derived from the moments of PDF assuming a Gaussian distribution, while empty markers represent quantities directly computed from the discrete PDFs. Up, right, and down orientations of triangular markers correspond to simulations with box sizes of $150$~Mpc/h, $500$~Mpc/h, and $1000$~Mpc/h, respectively. Additionally, the dotted and dashed grey lines depict quantities reported by \cite{Riess_2021} and  \cite{Planck_2020}, providing a reference for comparison.

From the first panel of \cref{fig:stat_d}, we notice the mean approaches zero from the negative side with an increase in distance. This behaviour is expected as halos reside in denser environments than the average density of the Universe. As a result, the velocities of nearby halos are slowed, leading to an overall decrease in the measured value of $H_0$. This effect diminishes with an increase in the distance and size of the local volume.  A jump in the halo mass threshold causes the jump between the points from simulations of different sizes.

In the second panel of \cref{fig:stat_d}, we show the variation of $\sigma$, which consistently decreases as distance increases. The parameter $\sigma$, represents the expected statistical error in the measurement of the Hubble-Lema\^itre constant, originating from the varying positions of observers within the large scale structure. It indicates that deviations from the global value, greater than $\sigma(d)$, in the measured Hubble-Lema\^itre constant $H_L$ at a given distance $d$, are less likely to occur. As we investigate larger distances, this expected deviation diminishes, eventually leading to the convergence of the local ($H_L$) and global ($H_G$) Hubble-Lema\^itre constant values. The dotted and dashed gray lines in the plot represent the standard deviations quoted in observational measurements for local \cite{Riess_2021} and global \cite{Planck_2020} Hubble-Lema\^itre constant values, respectively. Given these factors and the existing errors in observational measurements, our analysis suggests a lower limit of $\sim 135$~Mpc/h for SH0ES and $\sim 220$~Mpc/h for Planck to get a robust convergence of these two Hubble-Lema\^itre constant values. 

Given the non-Gaussian nature of distributions, we use fwhm and fwtm in addition to mean and standard deviation for better description. Panels 3 and 4 of \cref{fig:stat_d} display the behavior of fwhm and fwtm. These plots reinforce our earlier conclusion regarding the lower limit for converging local and global Hubble constant values. Furthermore, they indicate that at a distance of $100$ Mpc/h, the probabilities (according to Gaussian) of encountering deviations exceeding 4\% and 10\% are $\sim$24\% and $\sim$3\%, respectively. These probabilities steeply decline as we move to distances beyond $100$ Mpc/h. Notably, the observed deviation of 8-9\% in observational measurements is rare in this context.

From the \cref{fig:stat_d}, we observe an offset in quantities as we transition across three simulations. The primary source of these offsets arises from the incomplete representation of the large-scale velocity field in simulations due to the finite box size. Smaller box sizes are more susceptible to this limitation, as the lack of large-scale information impacts peculiar velocities more significantly than the density field. The variation in the number of independent modes on small and large scales in simulations can be responsible for some of the contribution to the offsets in sigma between simulations. Additionally, finite mass resolution can be a secondary source of offsets across different simulations. We demonstrate this in \cref{a4}.

\subsection{Correlation with local density}
\label{ssec:corre}

\begin{figure}
    \centering
        \begin{subfigure}[b]{0.40\textwidth}
        \includegraphics[width=.99\textwidth]{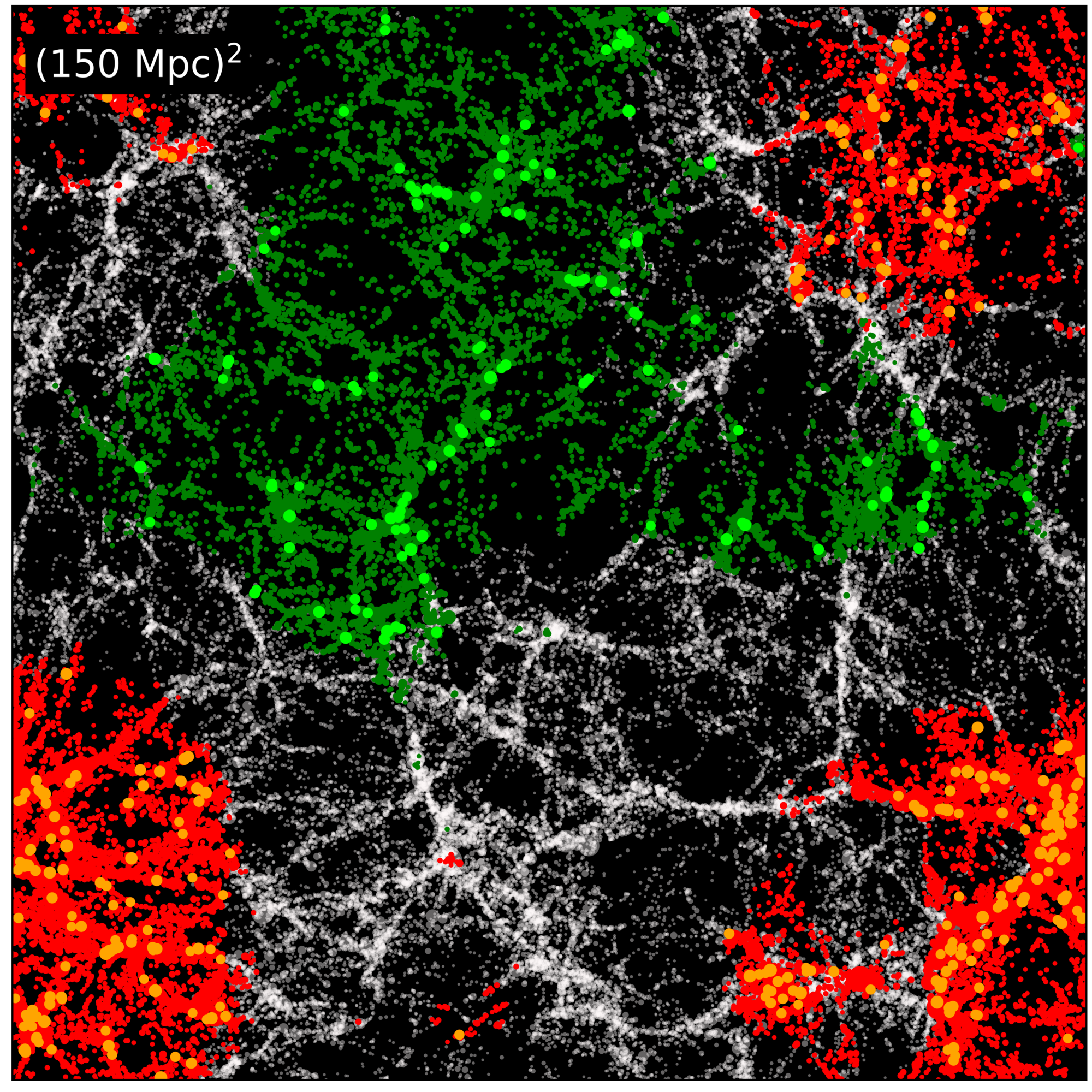}
        \end{subfigure}
        \begin{subfigure}[b]{0.40\textwidth}
        \includegraphics[width=.99\textwidth]{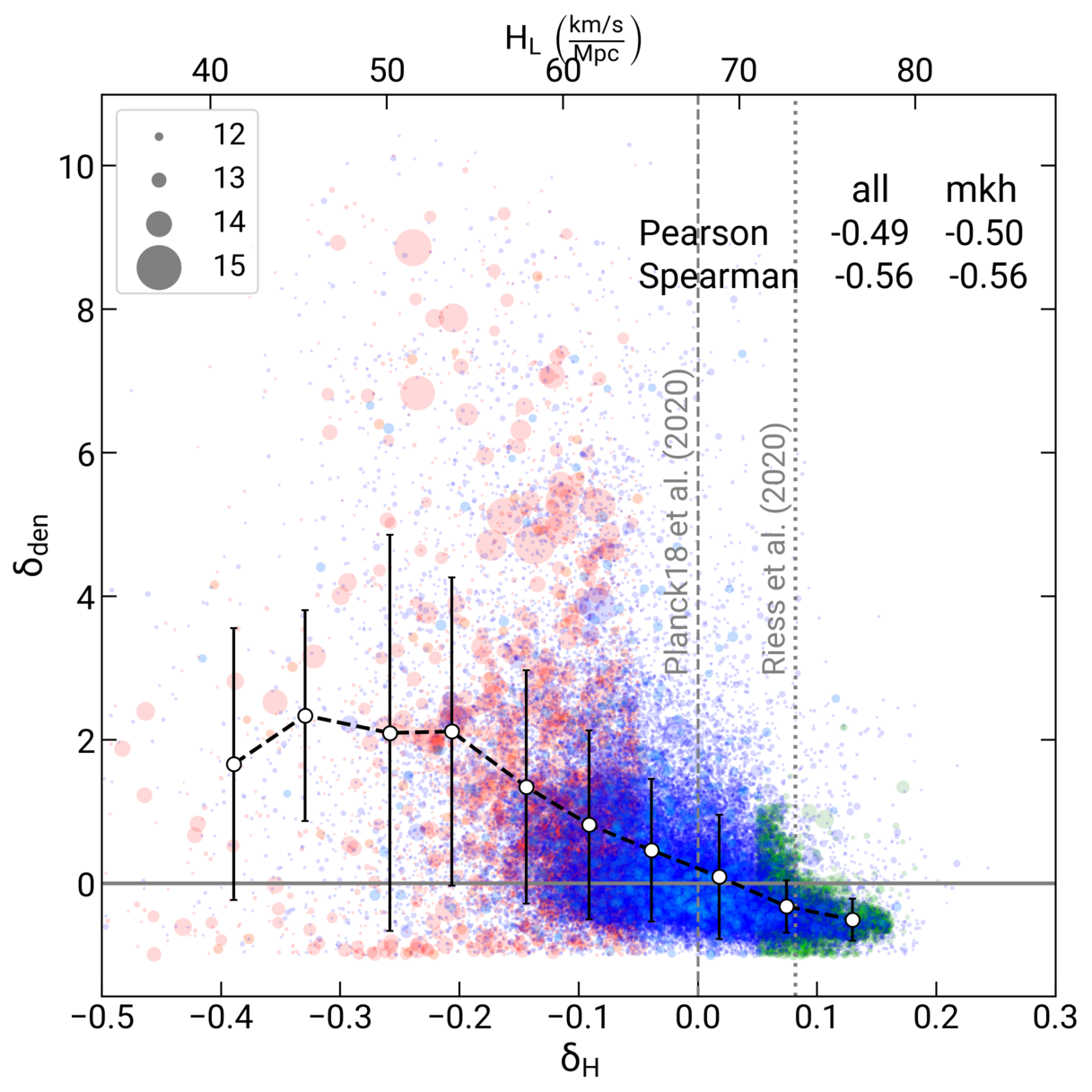}
        \end{subfigure}
        \begin{subfigure}[b]{0.40\textwidth}
        \includegraphics[width=.99\textwidth]{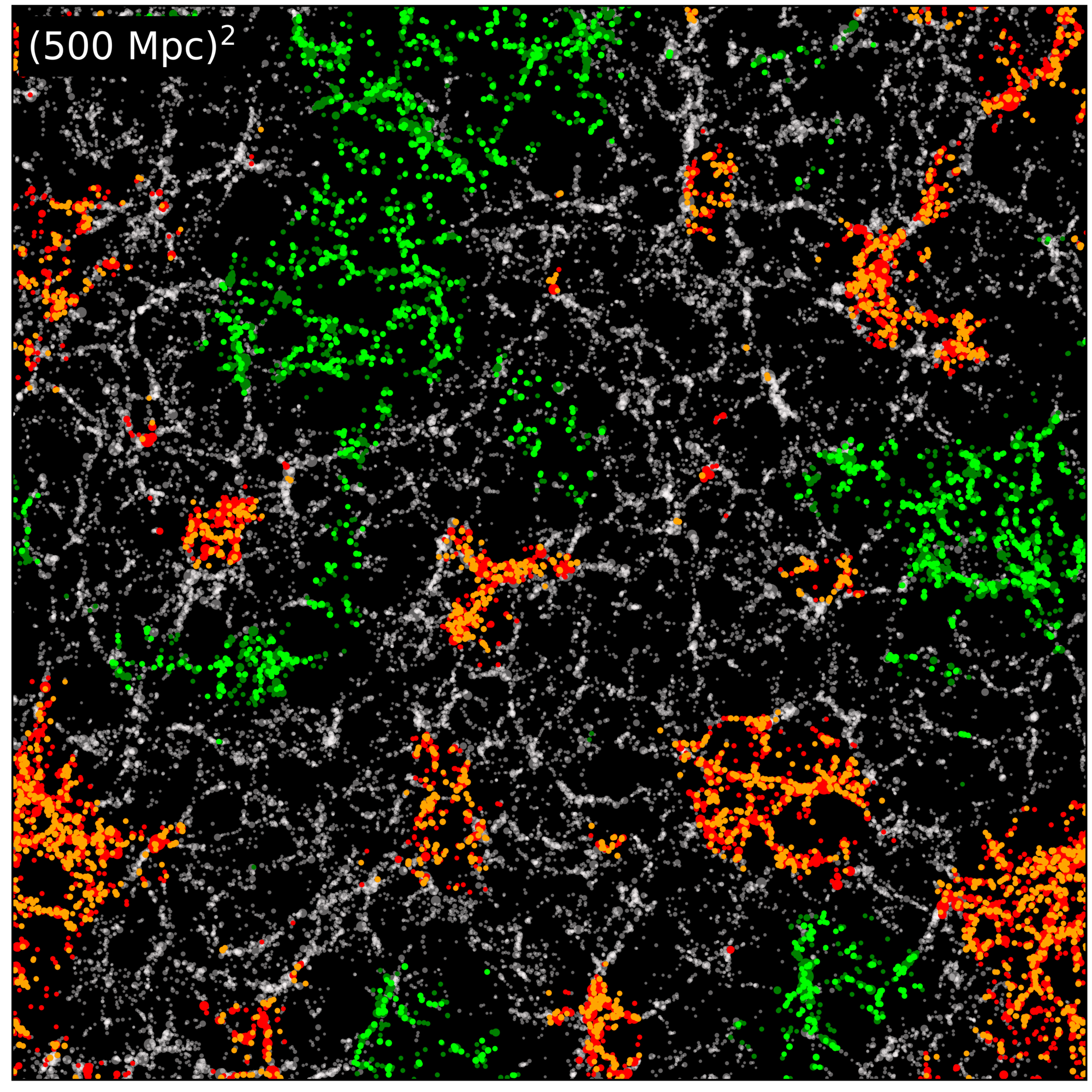}
        \end{subfigure}
        \begin{subfigure}[b]{0.40\textwidth}
        \includegraphics[width=.99\textwidth]{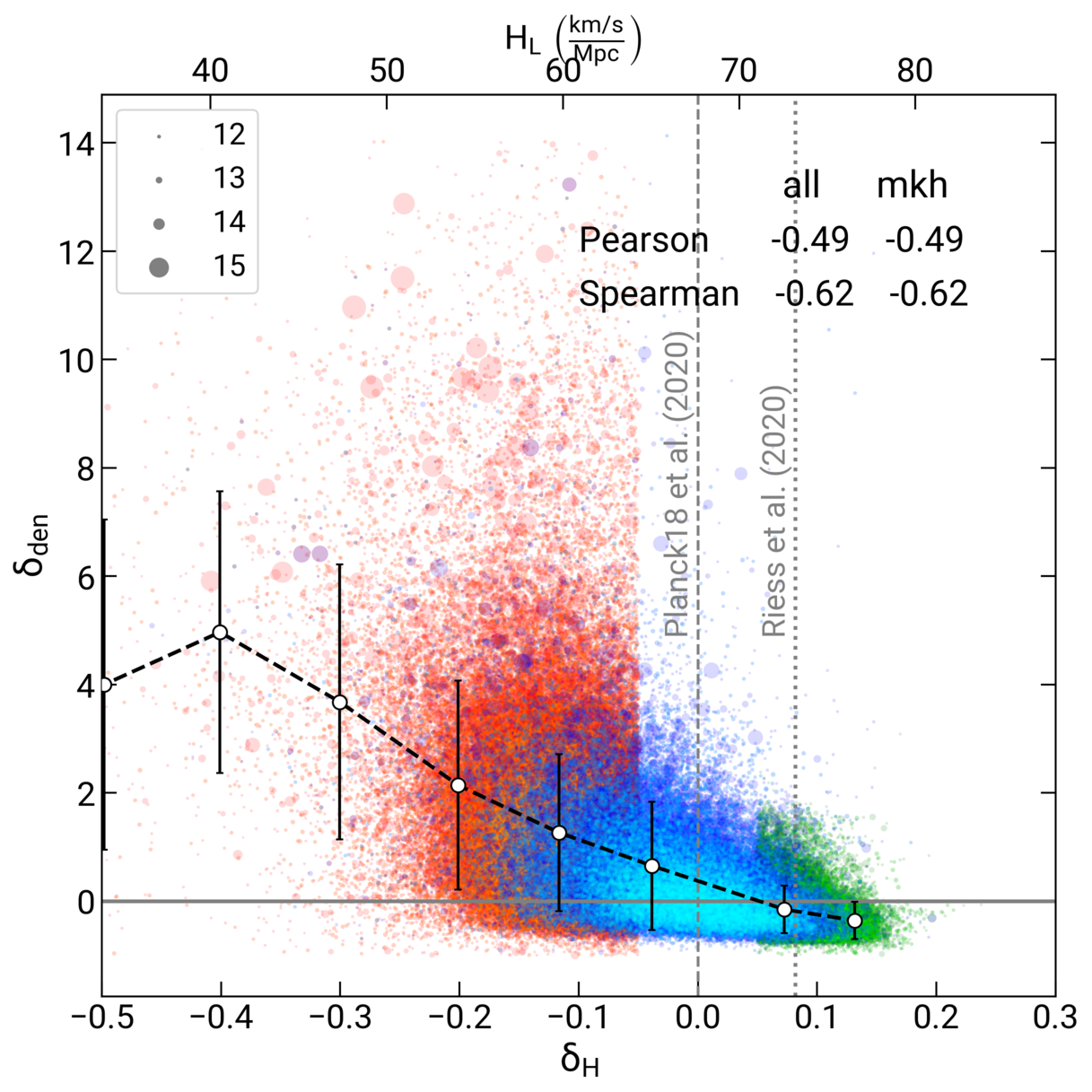}
        \end{subfigure}
        \begin{subfigure}[b]{0.40\textwidth}
        \includegraphics[width=.99\textwidth]{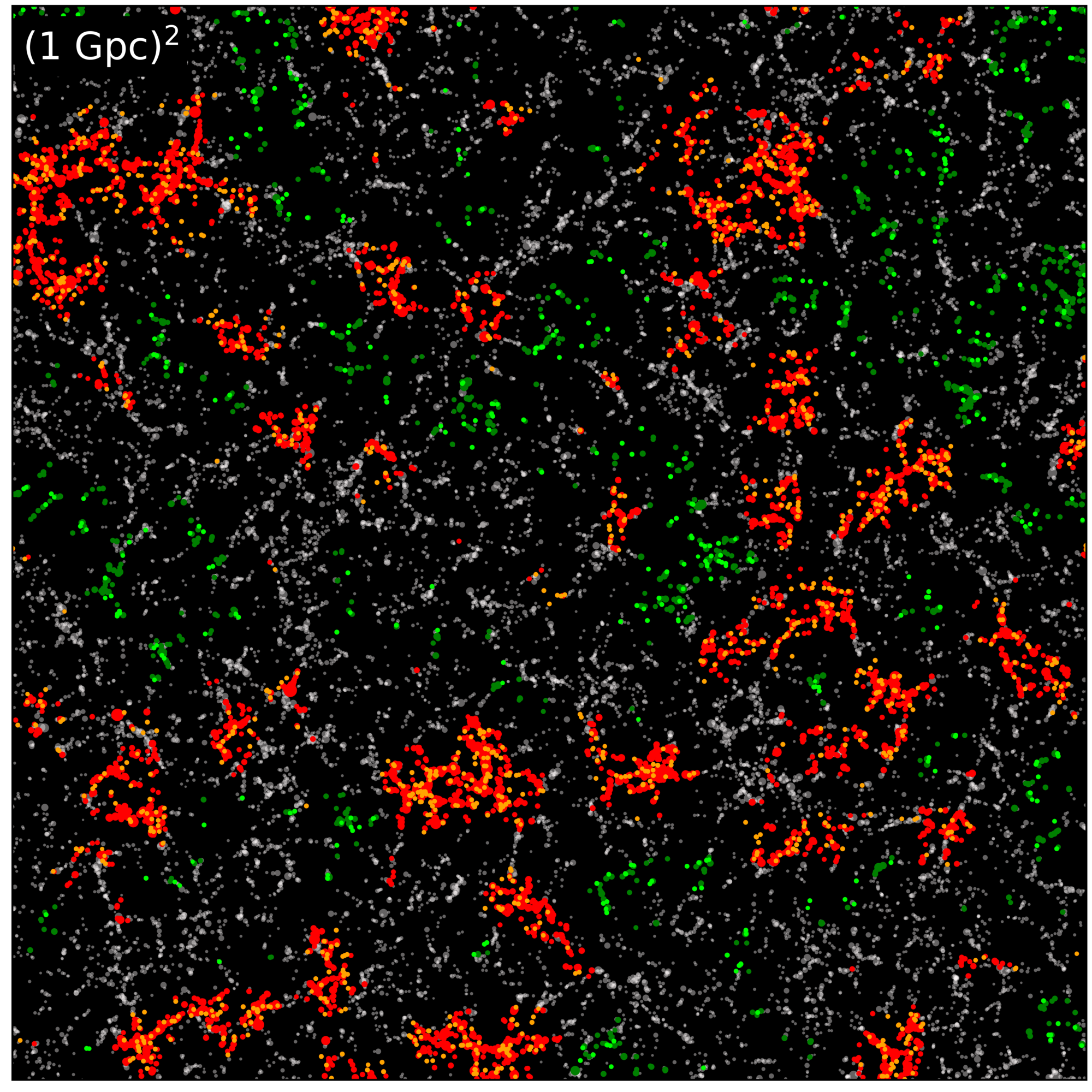}
        \end{subfigure}
        \begin{subfigure}[b]{0.40\textwidth}
        \includegraphics[width=.99\textwidth]{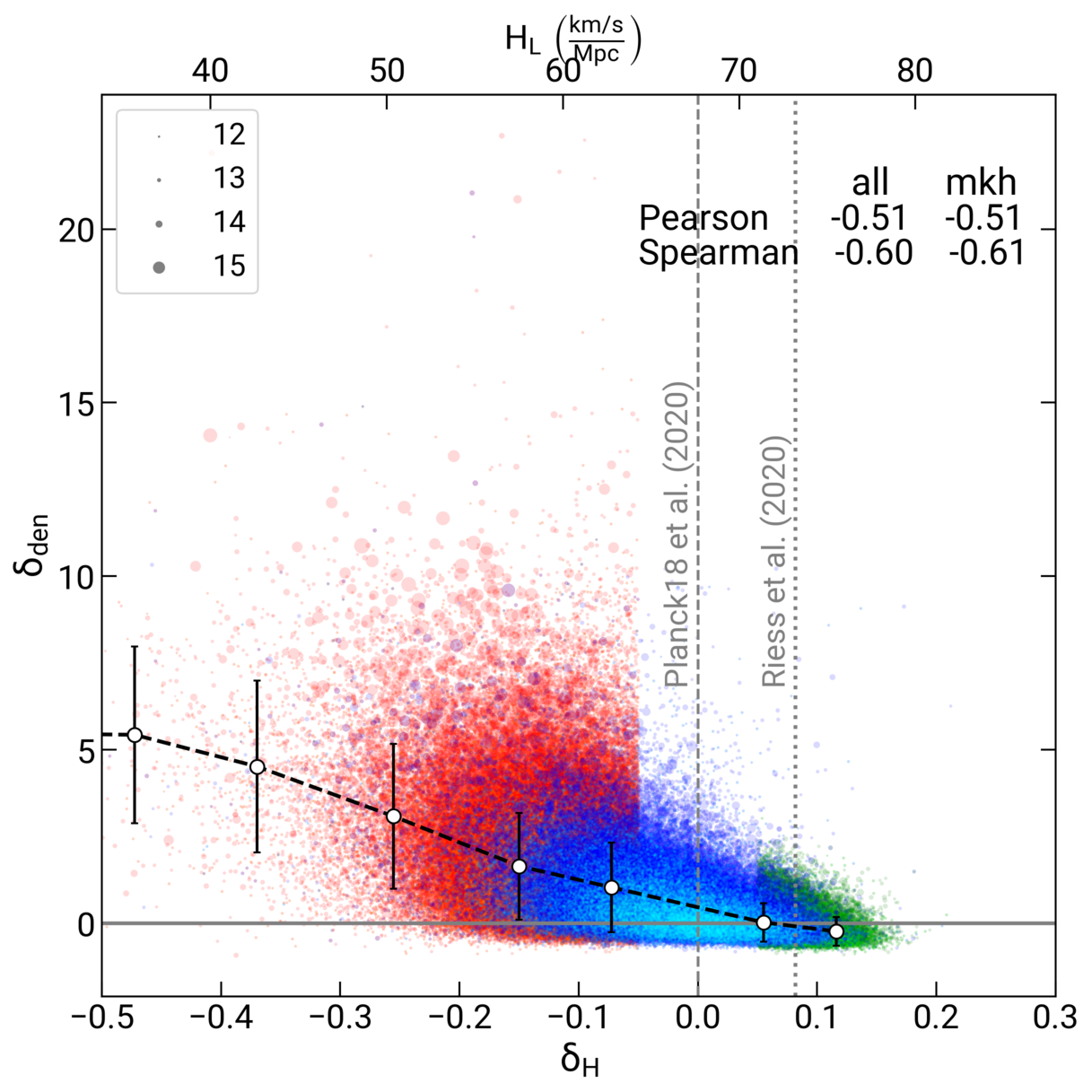}
        \end{subfigure}
    \caption[$H_0$ correlation with local overdensity]{\emph{Correlation with local overdensity}. Left panel: A $10$ Mpc/h slice of simulation box showing all halos in blue color, red and green correspond to halos with deviations exceeding $5\%$. The color red (green) for halos with negative (positive) deviation with respect to the global Hubble-Lema\^itre value. The sizes of the halos are proportional to their masses, with the $\text{log}_{10}$ exponent indicated in the legend of the left panel. Right panel: The plot shows the local overdensity of the halos within $10$ Mpc/h on the y-axis against the deviation in Hubble-Lema\^itre constant measurement for it on the x-axis. The mean and standard deviation in bins for three scatters(red+green, red, green) are plotted over scatters. The solid black line represents the combined scatter, while the dashed black lines, with red and green markers, represent the individual red and green scatter. The Pearson and Spearman correlation coefficient is displayed for combined scatter (\cref{ssec:corre}).}
    \label{fig:dev5p}
\end{figure}

We identify halos that exhibit a deviation of more than $5\%$ in their distributions between $25$-$40$ Mpc/h. Approximately $10$-$15\%$ of all observers show this characteristic. To visually represent these halos, we have included $10$~Mpc/h thick slices of the simulation box in the left panels of \cref{fig:dev5p}. The scatter points in white denote all halos within the slice, while those in red and green correspond to halos with deviations exceeding $5\%$. We use the color red for halos with negative deviation from the global Hubble-Lema\^itre value and green for those with negative deviation. We use orange and light green colors to highlight Milky Way-sized halos, with mass ranging from $6 \times 10^{11}$ to $3 \times 10^{12} M_{\odot}$/h. The sizes of the halo markers are proportional to the halo mass, with the $\text{log}_{10}$ exponent indicated in the legend of the right panels. 

We see regions with a lower than global Hubble-Lema\^itre constant, as determined above, lie in and around clusters, whereas regions with a higher determination are away from rich clusters. To quantify this, we examine the local surroundings of these identified halos to quantify their correlation. Specifically, we calculate the local over-density ($\delta_{\text{den}}$) of a halo within a $10$ Mpc/h radius and plot this against the deviation in Hubble-Lema\^itre constant measurement ($\delta_H$) for it within $25$-$40$ Mpc/h volume. The scatter from all halos is presented in the right panels of \cref{fig:dev5p}, with the same formatting as described for the left panels. We segment the data points for the all and Milky Way-sized halos into separate bins. We overplot the mean and standard deviation for each bin for the data sets. The dashed black line represents all the scatter. Further, we computed the Pearson and Spearman correlation coefficients for these. The results are tabulated in the right panel of \cref{fig:dev5p}, where columns `all' and `mkh' represent all and  Milky Way-sized halos. We find a moderate negative correlation between $\delta_H$ (deviation in Hubble-Lema\^itre constant) and $\delta_{\text{den}}$ (local over-density).

\subsection{Milky Way-sized halos}
\label{ssec:mkh}
\begin{figure}[t]
    \centering
        \begin{subfigure}[b]{0.45\textwidth}
        \includegraphics[width=.99\textwidth]{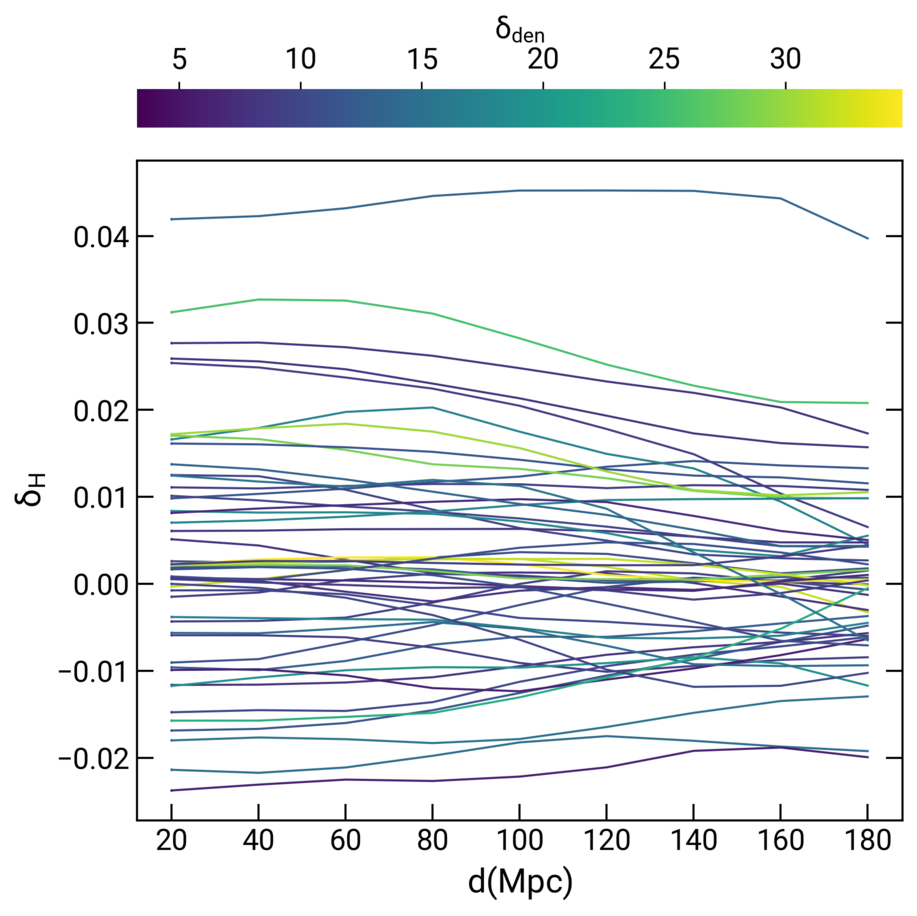}
        \end{subfigure}
        \begin{subfigure}[b]{0.45\textwidth}
        \includegraphics[width=.99\textwidth]{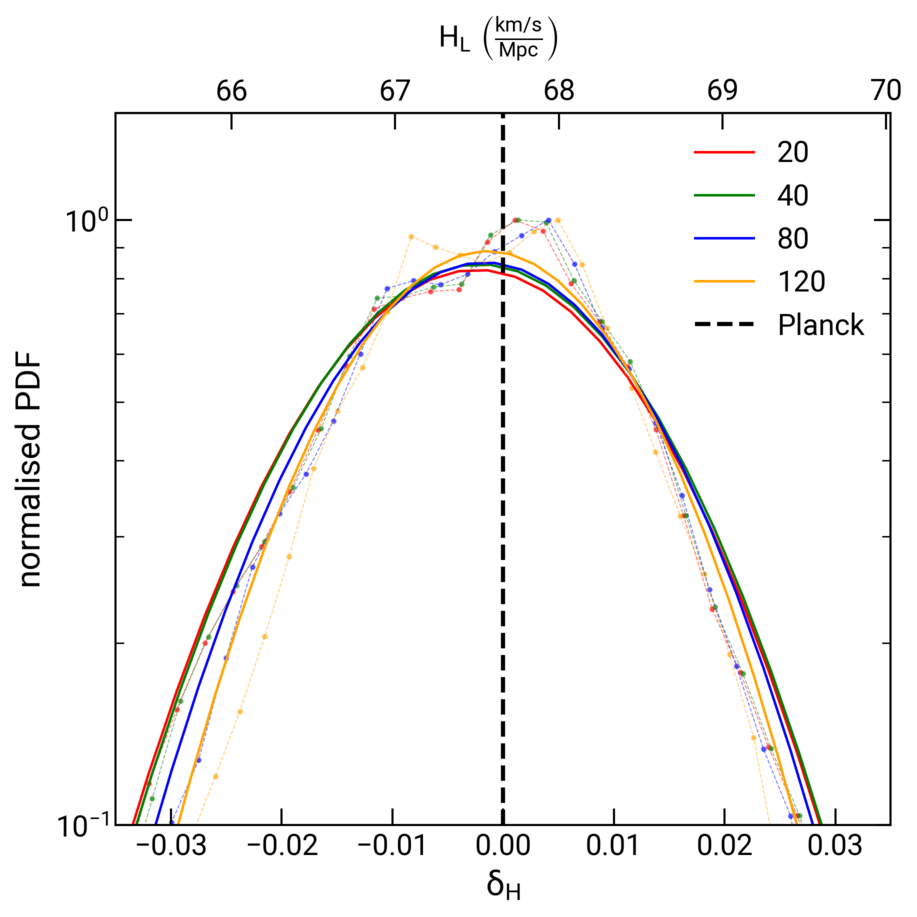}
        \end{subfigure}
    \caption[$H_0$ distributions for Milky way-sized halos]{\emph{Milky way-sized halos:} The left panel shows the deviations in local measurements for $50$ randomly selected Milky Way-sized halos across various spherical shells with lower radii ($r_{\text{min}}$) shown on the x-axis. The upper radius ($r_{\text{max}}$) for all measurements is fixed at $200$ Mpc/h. The colors of the lines represent the local overdensity of these halos, as shown in the colorbar. The right panel show normalised PDFs for Milky-sized halos within spherical shells ($20$-$200$), ($40$-$200$), ($80$-$200$), and ($120$-$200$).}
    \label{fig:mkh}
\end{figure}

\begin{table}[t]

  \begin{center}

    \begin{tabular}{|c |c |c |c |c|}
      \hline 
      Shell($r_{\text{min}}$-$r_{\text{max}}$) & (20-200) & (40-200)  & (80-200) &  (120-200) \\ [.1cm]
      \hline 
      1$\sigma$ & [-0.045, 0.037] & [-0.044, 0.037] & [-0.042, 0.038] & [-0.046, 0.036]\\ [.1cm]
      \hline
      2$\sigma$ & [-0.042, 0.034] & [-0.042, 0.034] & [-0.040, 0.033] & [-0.039, 0.034]\\ [.1cm]
      \hline
       3$\sigma$ & [-0.042, 0.034] & [-0.042, 0.034] & [-0.040, 0.033] & [-0.037, 0.032]\\ [.1cm]
      \hline
      
    \end{tabular}
    \caption[Standard deviation of PDF for Milky Way-sized halos]{1$\sigma$, 2$\sigma$ and 3$\sigma$ confidence intervals for PDFs shown in \cref{fig:mkh}}
    \label{tab:mkh}
  \end{center}
\end{table}

The analysis presented above is limited to spherical shells. However, the observational determination of Hubble-Lema\^itre's constant relies on observations across a wide range of scales, often using multiple distance indicators. To emulate this, we randomly select $50$ Milky Way-sized halos within our $500$ Mpc/h simulation box. We ensure that these halos are not located close to each other to ensure the statistical independence of our analysis. The left panel of \cref{fig:mkh} depicts the deviations in local measurements across various spherical shells observed by these individual observers. On the y-axis, we represent these deviations, and the x-axis denotes the lower radius ($r_{\text{min}}$) of the spherical shell. The upper radius ($r_{\text{max}}$) for all measurements is fixed at $200$ Mpc/h. We color the trajectories of these observers according to their local overdensity, as shown in the colorbar at the top.

In the right panel of \cref{fig:mkh}, using a $500$ Mpc/h simulation box, we show normalised PDFs for Milky-sized halos within spherical shells ($20$-$200$), ($40$-$200$), ($80$-$200$), and ($120$-$200$). 1$\sigma$, 2$\sigma$ and 3$\sigma$ intervals for these PDFs is tabulated in \cref{tab:mkh}. We note that the standard deviation of the distribution of $\delta_H$ is about a factor $2$ higher when compared to the stated standard deviation in SH0ES. We also note that a low probability tail extends to fairly large deviations from the global value.

\section{Conclusion and Discussion}
\label{sec:ho_conclusion}

In this work, we have considered the effect of peculiar velocities on the error budget of the measurement of Hubble-Lema\^itre's constant in the local Universe. We have worked with a model where we consider observers and galaxies to be located in halos. The halo distribution and velocities are obtained from N-body simulations. We make a simplifying assumption of one observer and one target galaxy per halo. This is an oversimplification as it ignores subhalos, however this is expected to lead to a conservative estimates of the contribution of peculiar velocities as we ignore internal velocities within halos and the finger of god effect.  

Our key conclusions may be summarised as follows.
\begin{itemize}
\item 
We reproduce the result that the local measurements of the Hubble-Lema\^itre constant exhibit significant and systematic differences from the global value due to gravitational clustering and peculiar motions. 
\item 
We observe a negative correlation between the deviation of Hubble-Lema\^itre constant measurements from the global value and the local halo population. Observers in high-density regions measure smaller $H_0$ values, while those in low-density regions measure larger values. While earlier studies indicated such a correlation \cite{Odderskov_2014, Wojtak_2014, Odderskov_2017}, this work provides a novel result by explicitly quantifying it.
\item 
The study suggests that, regardless of how accurately we measure the local value of the Hubble-Lema\^itre constant, it cannot be directly interpreted as a measure of the global value and deviations of the order of $3$-$5\%$ may be expected. 
\end{itemize}

We would like to note that observations have also revealed coherent bulk flow at scales in excess of $100$~Mpc/h in the region around the Milky Way galaxy. This puts our region as an outlier in the expected distribution of galaxies and flows in the $\Lambda$CDM model \cite{Whitford_2023, Hoffman_2024, Lopes_2024}. An assessment of how these flows may contribute to the measurement of Hubble-Lema\^itre's constant requires improved simulations as we will need to simulate a much larger physical region than we have done so far while retaining high resolution. We plan to address this issue in a follow-up study while noting that if large-scale bulk flows indicate that we are away from the average, then it may very well be that we are in the tail of the distribution for observed $H_0$. Whether this can help alleviate the Hubble tension is something that can only be answered with more detailed studies. 

In summary, we would like to recommend that the impact of peculiar velocities and large scale structure needs to be studied in greater detail while interpreting results from observations.


%% file: chapters/chap5.tex
\chapter[On the origin of transient features in cosmological N-body Simulations]{On the origin of transient features in cosmological N-body Simulations}
\label{cha:5}
{\small This chapter is based on the following manuscript: \fullcite{Bagla_2024}} \\

It is believed that structures like galaxies and clusters of galaxies formed by the growth of small perturbations due to gravitational clustering \cite{Peebles_1981, Shandarin_1989, Peacock_1998, Padmanabhan_2002, Bernardeau_2002}. The same processes are thought to be responsible for the formation of the large scale structure, i.e., the clustered distribution of galaxies in space. Dark matter is believed to be the dominant form of matter in galaxies and clusters of galaxies: dark matter is generally assumed to be made up of non-relativistic, weakly interacting massive particles (WIMPs) \cite{Trimble_1987, Spergel_2003, Planck_2020}. In view of the weak interactions within the dark matter sector, it responds mainly to gravitational forces. As dark matter dominates over baryonic matter in terms of total matter content, assembly of matter into halos and the large scale structure is driven mainly by gravitational amplification of initial perturbations. In the standard cosmological model, it is believed that galaxies form when gas in highly over-dense halos forms stars due to cooling, collapse and fragmentation of gas \cite{Hoyle_1953, Rees_1977, Silk_1977, Binney_1977, Benson_2010, Somerville_2015, Wechsler_2018}. Thus, the evolution of density perturbations due to gravitational clustering in an expanding universe is a key process for the study of large scale structure and its evolution. 
The basic equations for this can be derived from first principles \cite{Peebles_1974a}. These equations can be solved easily when the amplitude of perturbations is small. At this stage, perturbations at each scale evolve independently, and mode coupling is sub-dominant. Once the amplitude of perturbations at relevant scales becomes large, the coupling with perturbations at other scales becomes important and cannot be ignored, though a few results have been obtained. 
The equation for the evolution of density perturbations cannot be solved for generic perturbations in the non-linear regime. One can use dynamical approximations for studying mildly non-linear perturbations \cite{Zeldovich_1970, Gurbatov_1989, Matarrese_1992, Brainerd_1993, Bagla_1994, Sahni_1995, Hui_1996, Bernardeau_2002,Crocce_2006a, Kitaura_2013,  Pixius_2022, Ota_2023, Garny_2023, Garny_2023a, Rampf_2023}. Scaling relations can be used if the aim is to describe the system using statistical descriptors \cite{Davis_1977, Hamilton_1991, Jain_1995, Kanekar_2000, Ma_1998, Nityananda_1994, Padmanabhan_1996a, Padmanabhan_1996, Peacock_1996, Smith_2003, Carrasco_2012,Baumann_2012}. We require cosmological N-body simulations \cite{Bagla_1994, Bertschinger_1998, Bagla_2004, Angulo_2022} to follow the detailed evolution of generic systems.

Cosmological N-body simulations work with a representative region of the universe. This region is typically large but finite. Perturbations at scales smaller than the mass resolution of the simulation, and at scales larger than the box cannot be represented or taken into account. Indeed, perturbations at scales comparable to the box are under-sampled with only a handful of independent modes. Perturbations at scales much larger than the simulation volume can affect the results of N-body simulations \cite{Gelb_1994, Gelb_1994a, Tormen_1996, Cole_1997, Bagla_2005, Bagla_2006, Power_2006, Takahashi_2008, Bagla_2009, Angulo_2016, Michaux_2020}. It is possible to estimate whether a given simulation volume is large enough to be representative or not \cite{Bagla_2005, Bagla_2006}. It has also been shown that for gravitational dynamics in an expanding universe, perturbations at small scales do not influence the collapse of large scale perturbations in a significant manner if the scale of non-linearity is larger than the smaller scales under consideration \cite{Peebles_1974a, Peebles_1985, Little_1991, Bagla_1997a, Couchman_1998}. This is certainly valid for the correlation function or power spectrum at large scales. This has led to a belief that ignoring perturbations at scales much smaller than the scales of interest does not affect the results of N-body simulations. In earlier work, we have shown that if large scale collapse is highly symmetric, then the presence of perturbations at much smaller scales affects the evolution of density perturbations at large scales \cite{Bagla_2005a}. However, such effects are not obvious when we study the evolution of perturbations with generic initial conditions.

Substructure is known to play an important role in the dynamical relaxation of halos. It can induce mixing in phase space \cite{Lynden-bell_1967, Weinberg_2001}, or change halo profiles by introducing transverse motions \cite{Peebles_1990, Subramanian_2000}. Further, gravitational interactions between substructures can introduce an effective collisionality even for a collisionless system \cite{Ma_2004, Ma_2003}. It is, therefore, important to understand the role played by substructure in gravitational clustering and dynamical relaxation in the cosmological context.

The key mechanism for perturbations across scales affecting each other is mode coupling. We know the following from earlier work. 
\begin{itemize}
\item
Perturbations at large scales influence perturbations at small scales in a significant manner. If the initial conditions are modified by filtering out perturbations at small scales or other modifications restricted to small scales, then mode coupling generates small scale power. If the scale of filtration is smaller than the scale of non-linearity at the final epoch, then the non-linear power spectrum as well as the appearance of large scale structure is similar to the original initial conditions without modification \cite{Peebles_1985, Little_1991, Bagla_1997a, Couchman_1998}.
\item
Non-linear evolution of density perturbations {\it drives} every model towards a weak attractor ($P(k)\simeq k^{-1}$) in the mildly non-linear regime ($1 \leq \bar\xi \leq 200$) \cite{Klypin_1992, Bagla_1997a}. 
\item
If there are no initial perturbations at large scales, mode coupling generates power with ($P(k) \simeq k^4$) that grows very rapidly at early times \cite{Bagla_1997a}. This can be explained in many ways. The second-order perturbation theory as well as momentum and mass conserving motion of a group of particles provide an adequate explanation. The $k^4$ tail can also be derived from the full non-linear equation for density \cite{Peebles_1974a, Peebles_1981, Zeldovich_1965}.
\item
In case the large scale perturbations are highly symmetric, e.g. planar, then small scale fluctuations play an important role in the non-linear evolution of perturbations at large scales \cite{Bagla_2005a}. 
\end{itemize}

The effect of large scales on small scales is clearly significant, particularly if the larger scales are comparable to the scale of non-linearity. On the other hand, the effect of small scales on larger scales is known to be small in most situations. Though this effect has not been studied in detail, many tools have been developed that exploit the presumed smallness of the influence of small scales on large scales \cite{Bond_1996, Monaco_2002, Monaco_2002a}. 

\emph{Pre-initial} conditions refer to the distribution of particles that is laid out before imposing initial density and velocity perturbations. The choice of pre-initial conditions is known to impact the results \cite{Baertschiger_2007, Baertschiger_2007a, Baertschiger_2007b, Bagla_1997, Gabrielli_2006, Joyce_2007, Joyce_2007a, Joyce_2020, Joyce_2005, Marcos_2006, Huillier_2014, Michaux_2020, zhang_2021}. Pre-initial conditions are expected to have no density perturbations or symmetry. It is clear, however, that at least one of these requirements must be relaxed in practice. 
This leads to growth of some modes deviating from expectations in perturbation theory. The work presented here allows us to estimate the effect such discrepant modes can have on the non-linear evolution of clustering across scales. 

The representation and evolution of perturbations at small scales depend strongly on the mass and force resolution in the simulation. A high force resolution without a matching mass resolution can lead to better modelling of dense halos but gives rise to two body collisions and misleading results in some situations  \cite{Kuhlman_1996,Splinter_1998, Binney_2002, Diemand_2004, Binney_2004, El-zant2006, Romeo_2008, Mansfield_2020}. Further, discreteness and stochasticity limit our ability to measure physical quantities in simulations \cite{Thiebaut_2008, Romeo_2008,Angulo_2022}. The question we address here is: can these errors remain significant at late times, and, can these errors propagate to larger scales through mode coupling? Propagation of these errors can potentially introduce transient features in N-body simulations that are likely to remain dominant till the mass scale of non-linearity is well resolved. 

We present our analysis in two parts. In \cref{sec:mc}, we discuss mode coupling and the expected effect of collapsed halos at small scales on perturbations at much larger scales. In \cref{sec:mc_err}, we present an analysis of the effect of unrepresented perturbations at small scales. We also present an analysis of the same using N-body simulations. This is followed by discussion and summary in \cref{sec:mc_summ}.

\section{Mode Coupling}
\label{sec:mc}

The evolution of density perturbations in a system of particles interacting gravitationally in an expanding Universe \cite{Peebles_1981} can be written as,
\begin{eqnarray}
{\ddot\delta}_{\mathbf k} + 2 \frac{\dot{a}}{a} {\dot\delta}_{\mathbf k} &=&
A_{\mathbf k} - B_{\mathbf k}, \\
A_{\mathbf k} &=& \frac{1}{M}\sum\limits_{j} m_j 
\left [ \iota {\mathbf k}. \left(- \frac{\mathbf \nabla \phi_j}{a^2} \right)
\exp\left(\iota{\mathbf  k}.{\mathbf x}_j \right) \right ], \\
B_{\mathbf k} &=& \frac{1}{M} \sum\limits_{j} m_j \left({\mathbf k}.{\dot
    {\mathbf x}_j } \right)^2   \exp\left[\iota{\mathbf  k}.{\mathbf x}_j \right] .
\end{eqnarray}
The usual linear term in $\delta_{\mathbf k}$ is absorbed in $A_{\mathbf k}$. It has been shown that a single cluster of particles in virial equilibrium does not contribute to $A_{\mathbf k} - B_{\mathbf k}$ at wave numbers much smaller than the inverse of the cluster size. Here, we consider the effect of interaction between two clusters in order to estimate the effect of mode coupling and transfer of power from small scales to large scales.

The first mode coupling term can be written as,
\begin{equation}
A_{\mathbf k} = \frac{1}{M}\sum\limits_{j} m_j 
\left [ \iota {\mathbf k}. \left( - \frac{\mathbf \nabla \phi_j}{a^2} \right)
\exp\left(\iota{\mathbf  k}.{\mathbf x}_j \right) \right ] =
\frac{\iota}{M} \sum\limits_{j} \left({\mathbf k}.{\mathbf f}_j \right) 
\exp\left(\iota{\mathbf  k}.{\mathbf x}_j \right)  .
\end{equation}
If there are two clusters $C_1$ and $C_2$ then the sum in RHS can be written separately for particles in two clusters,
\begin{equation}
A_{\mathbf k} = \frac{\iota}{M} \sum\limits_{j\in C_1} 
 \left({\mathbf k}.{\mathbf f}_j \right) \exp\left(\iota{\mathbf  k}.{\mathbf
     x}_j \right)  + \frac{\iota}{M} \sum\limits_{j \in C_2}  \left({\mathbf
     k}.{\mathbf f}_j \right) \exp\left(\iota{\mathbf  k}.{\mathbf x}_j \right) ,
\end{equation}
representing the force acting on particles in halo '1' due to halo '2' by ${\mathbf f}^{12}$, and the force due to particles within the same halo by ${\mathbf f}^{11}$, we can rewrite the sum\footnote{Note that there is no  self-interaction implied, each particle experiences a force due to all the others.} as,   
\begin{align}
A_{\mathbf k} &= \frac{\iota}{M} \sum\limits_{j \in C_1} \left({\mathbf k}.{\mathbf
    f}_j^{11} \right) \exp\left(\iota{\mathbf k}.{\mathbf x}_j \right) +
\frac{\iota}{M} \sum\limits_{j \in C_1} \left({\mathbf k}.{\mathbf f}_j^{12}
\right) \exp\left(\iota{\mathbf k}.{\mathbf x}_j \right)  
\nonumber \\
& \quad + \frac{\iota}{M} \sum\limits_{j \in C_2} \left({\mathbf k}.{\mathbf f}_j^{22}
\right) \exp\left(\iota{\mathbf k}.{\mathbf x}_j \right) +  \frac{\iota}{M}
\sum\limits_{j \in C_2} \left({\mathbf k}.{\mathbf f}_j^{21} \right)
\exp\left(\iota{\mathbf k}.{\mathbf x}_j \right)  .
\end{align}
If the wave numbers of interest are such that $\left|{\mathbf k}. {\mathbf r} \right|  \ll 1$, then we can expand the exponential in each term,
\begin{align}
A_{\mathbf k} &= \frac{\iota}{M} \sum\limits_{j \in C_1}  \left({\mathbf k}.{\mathbf
    f}_j^{11} \right) \left[ 1 + \iota \left({\mathbf k}.{\mathbf x}_j \right) +
  \mathcal{O}(k^2) \right] 
+ \frac{\iota}{M} \sum\limits_{j \in C_1}  \left({\mathbf k}.{\mathbf
    f}_j^{12} \right) \left[ 1 + \iota \left({\mathbf k}.{\mathbf x}_j \right) +
  \mathcal{O}(k^2) \right] \nonumber \\
& \quad + \frac{\iota}{M} \sum\limits_{j \in C_2}  \left({\mathbf k}.{\mathbf
    f}_j^{22} \right) \left[ 1 + \iota \left({\mathbf k}.{\mathbf x}_j \right) +
  \mathcal{O}(k^2) \right] 
+ \frac{\iota}{M} \sum\limits_{j \in C_2}  \left({\mathbf k}.{\mathbf
    f}_j^{21} \right) \left[ 1 + \iota \left({\mathbf k}.{\mathbf x}_j \right) +
  \mathcal{O}(k^2) \right] \nonumber \\
&= \frac{\iota}{M} {\mathbf k}. \left[ \sum\limits_{j \in C_1} {\mathbf f}_j^{11}
+ \sum\limits_{j \in C_1} {\mathbf f}_j^{12}
+ \sum\limits_{j \in C_2} {\mathbf f}_j^{22}
+ \sum\limits_{j \in C_2} {\mathbf f}_j^{21}
           \right] \nonumber \\
& \quad -  \frac{1}{M} {\mathbf k}. \left[ \sum\limits_{j \in C_1} {\mathbf
    f}_j^{11} \left( {\mathbf k}.{\mathbf x}_j \right)
+ \sum\limits_{j \in C_1} {\mathbf f}_j^{12} \left( {\mathbf k}.{\mathbf x}_j
\right) 
+ \sum\limits_{j \in C_2} {\mathbf f}_j^{22} \left( {\mathbf k}.{\mathbf x}_j
\right) 
+ \sum\limits_{j \in C_2} {\mathbf f}_j^{21} \left( {\mathbf k}.{\mathbf x}_j
\right)  \right] \nonumber \\
& \quad + \mathcal{O}(k^3) .
\end{align}
The first and the third term at $\mathcal{O}(k)$ are zero as there is no net internal force. Second and fourth terms at the same order cancel by virtue of Newton's third law. Thus, there is no contribution at this order, and the lowest order contribution is at $\mathcal{O}(k^2)$ from this term. The external force may be assumed to be constant across the cluster\footnote{In principle, we can evaluate the tidal interaction term as well, but that is subdominant in most situations, and we choose to ignore it here.}. This allows us to simplify the second and fourth terms,
\begin{align}
A_{\mathbf k} &\simeq
- \frac{1}{M} {\mathbf k}. \left[ \sum\limits_{j \in C_1} {\mathbf
    f}_j^{11} \left( {\mathbf k}.{\mathbf x}_j \right)
+ \sum\limits_{j \in C_1} {\mathbf f}_j^{12} \left( {\mathbf k}.{\mathbf x}_j
\right) 
+ \sum\limits_{j \in C_2} {\mathbf f}_j^{22} \left( {\mathbf k}.{\mathbf x}_j
\right) 
+ \sum\limits_{j \in C_2} {\mathbf f}_j^{21} \left( {\mathbf k}.{\mathbf x}_j
\right)  \right] \nonumber \\
&\simeq
- \frac{1}{M} {\mathbf k}. \left[ \sum\limits_{j \in C_1} {\mathbf
    f}_j^{11} \left( {\mathbf k}.{\mathbf x}_j \right)
+  {\mathbf F}^{12} \left( {\mathbf k}.{\mathbf X}_1
\right) 
+ \sum\limits_{j \in C_2} {\mathbf f}_j^{22} \left( {\mathbf k}.{\mathbf x}_j
\right) 
+ {\mathbf F}^{21} \left( {\mathbf k}.{\mathbf X}_2
\right)  \right]   \nonumber \\
&=
- \frac{1}{M} {\mathbf k}. \left[ \sum\limits_{j \in C_1} {\mathbf
    f}_j^{11} \left( {\mathbf k}.{\mathbf x}_j \right)
+ \sum\limits_{j \in C_2} {\mathbf f}_j^{22} \left( {\mathbf k}.{\mathbf x}_j
\right)  
\right] 
-  \frac{1}{M} {\mathbf k}.{\mathbf F}^{12} \left( {\mathbf k}.\left({\mathbf
      X}_1 - {\mathbf  X}_2 \right) \right) .
\end{align}
Here ${\mathbf F}^{ij}$ is the force due to the $j^{\text{th}}$ cluster in the $i^{\text{th}}$ cluster, and ${\mathbf X}_i$ is the centre of mass of the $i^{\text{th}}$ cluster. It is easy to see that the combination of reduced terms scales inversely with the separation of the two clusters. Terms that refer to the internal forces can also be rewritten by switching to the centre of mass frame of each cluster. We use centre of mass coordinates ${\mathbf y}_j^i = {\mathbf x}_j - {\mathbf X}_i$ for the $j^{\text{th}}$ particle in the $i^{\text{th}}$ cluster. We will drop the superscript as it is obvious from the context,
\begin{align}
A_{\mathbf k} &=
- \frac{1}{M} {\mathbf k}. \left[ \sum\limits_{j \in C_1} {\mathbf
    f}_j^{11} \left( {\mathbf k}.\left({\mathbf y}_j + {\mathbf X}_1\right)
  \right) 
+ \sum\limits_{j \in C_2} {\mathbf f}_j^{22} \left( {\mathbf k}.\left({\mathbf
      y}_j + {\mathbf X}_2\right) \right)  \right] \nonumber \\
& \quad
-  \frac{1}{M} {\mathbf k}.{\mathbf F}^{12} \left( {\mathbf k}.\left({\mathbf
      X}_1 - {\mathbf  X}_2 \right) \right) 
\nonumber \\
&=
- \frac{1}{M} {\mathbf k}. \left[ \sum\limits_{j \in C_1} {\mathbf
    f}_j^{11} \left( {\mathbf k}.{\mathbf y}_j  \right) 
+ \sum\limits_{j \in C_2} {\mathbf f}_j^{22} \left( {\mathbf k}.{\mathbf
      y}_j \right)  \right] \nonumber \\
&
 \quad -  \frac{1}{M} {\mathbf k}.{\mathbf F}^{12} \left( {\mathbf k}.\left({\mathbf
      X}_1 - {\mathbf  X}_2 \right) \right) + \mathcal{O}(k^3) .
\end{align}
The second step follows as the internal forces cancel out for each cluster. This is the final expression for $A_{\mathbf k}$, ignoring the tidal interaction terms between the two clusters.

Now, we consider the second mode coupling term,
\begin{equation}
B_{\mathbf k} = \frac{1}{M} \sum\limits_{j} m_j \left({\mathbf k}.{\dot
    {\mathbf x}_j } \right)^2   \exp\left[\iota{\mathbf  k}.{\mathbf x}_j \right] .
\end{equation}
It is evident that this term will contribute at $\mathcal{O}(k^2)$ or higher orders to the evolution of $\delta_{\mathbf k}$. We can again write the sum in two parts, one for each cluster,
\begin{align}
B_{\mathbf k} &= \frac{1}{M} \sum\limits_{j} m_j \left({\mathbf k}.{\dot
    {\mathbf x}_j } \right)^2   \exp\left[\iota{\mathbf  k}.{\mathbf x}_j \right] 
\nonumber \\
  &= \frac{1}{M} \sum\limits_{j \in C_1}  m_j \left({\mathbf k}.{\dot
    {\mathbf x}_j } \right)^2   \exp\left[\iota{\mathbf  k}.{\mathbf x}_j \right]
+ \frac{1}{M}  \sum\limits_{j \in C_2}  m_j \left({\mathbf k}.{\dot
    {\mathbf x}_j } \right)^2   \exp\left[\iota{\mathbf  k}.{\mathbf x}_j \right]
\nonumber \\
&\simeq
\frac{1}{M}  \sum\limits_{j \in C_1}  m_j \left({\mathbf k}.{\dot
    {\mathbf x}_j } \right)^2
+ \frac{1}{M} \sum\limits_{j \in C_2}  m_j \left({\mathbf k}.{\dot
    {\mathbf x}_j } \right)^2  + \mathcal{O}(k^3) .
\end{align}
We can now switch to the centre of mass frame for each cluster. Let ${\mathbf V}$ be the velocity of the centre of mass and $\dot{{\mathbf y}}_j$ the velocity of the $j^{\text{th}}$ particle in this frame,
\begin{align}
B_{\mathbf k} &= \frac{1}{M}  \sum\limits_{j \in C_1}  m_j \left[{\mathbf
    k}.\left({\dot {\mathbf y}_j} + {\mathbf V}_1\right) \right]^2
+ \frac{1}{M} \sum\limits_{j \in C_2}  m_j \left[{\mathbf k}.\left({\dot
      {\mathbf y}_j} + {\mathbf V}_2\right)\right]^2  + \mathcal{O}(k^3) 
\nonumber \\
&= \frac{1}{M}  \sum\limits_{j \in C_1}  m_j \left[\left({\mathbf
    k}.{\dot {\mathbf y}_j}\right)^2 + \left({\mathbf k}.{\mathbf
    V}_1\right)^2 + 2 \left({\mathbf k}.{\dot {\mathbf y}_j}\right)
   \left({\mathbf k}.{\mathbf  V}_1\right) \right]
\nonumber \\
& \quad +  \frac{1}{M}  \sum\limits_{j \in C_2}  m_j \left[\left({\mathbf
    k}.{\dot {\mathbf y}_j}\right)^2 + \left({\mathbf k}.{\mathbf
    V}_2\right)^2 + 2 \left({\mathbf k}.{\dot {\mathbf y}_j}\right)
   \left({\mathbf k}.{\mathbf  V}_2\right) \right]  + \mathcal{O}(k^3)  
\nonumber \\
&= \frac{1}{M}  \sum\limits_{j \in C_1}   m_j \left({\mathbf
    k}.{\dot {\mathbf y}_j}\right)^2 
    +  \frac{1}{M}  \sum\limits_{j \in C_2}   m_j \left({\mathbf
    k}.{\dot {\mathbf y}_j}\right)^2 
\nonumber \\
& \quad  + \frac{1}{M}  \left[ M_1 \left({\mathbf k}.{\mathbf V}_1\right)^2 + M_2
      \left({\mathbf k}.{\mathbf V}_2\right)^2 \right]  + \mathcal{O}(k^3)  .
\end{align}
The total contribution to mode coupling can be written as,
\begin{align}
A_{\mathbf k} - B_{\mathbf k} &= 
- \frac{1}{M} {\mathbf k}. \left[ \sum\limits_{j \in C_1} {\mathbf
    f}_j^{11} \left( {\mathbf k}.{\mathbf y}_j  \right) 
+ \sum\limits_{j \in C_2} {\mathbf f}_j^{22} \left( {\mathbf k}.{\mathbf
      y}_j \right)  \right] 
\nonumber \\
& \quad +  \frac{1}{M}  \sum\limits_{j \in C_1}   m_j \left({\mathbf
    k}.{\dot {\mathbf y}_j}\right)^2 
    +  \frac{1}{M}  \sum\limits_{j \in C_2}   m_j \left({\mathbf
    k}.{\dot {\mathbf y}_j}\right)^2 
\nonumber \\
& \quad -  \frac{1}{M} {\mathbf k}.{\mathbf F}^{12} \left( {\mathbf k}.\left({\mathbf
      X}_1 - {\mathbf  X}_2 \right) \right)
+ \frac{1}{M}  \left[ M_1 \left({\mathbf k}.{\mathbf V}_1\right)^2 + M_2
  \left({\mathbf k}.{\mathbf V}_2\right)^2 \right]  
\nonumber \\
& \quad + \mathcal{O}(k^3) .
\end{align}
Using the fact that $A_{\mathbf k} - B_{\mathbf k}$ for a virialised cluster, computed in its centre of mass is zero at this order, we find,
\begin{align}
A_{\mathbf k} - B_{\mathbf k} &= - \frac{1}{M} {\mathbf k}.{\mathbf F}^{12} \left( {\mathbf k}.\left({\mathbf
      X}_1 - {\mathbf  X}_2 \right) \right)
+ \frac{1}{M}  \left[ M_1 \left({\mathbf k}.{\mathbf V}_1\right)^2 + M_2
  \left({\mathbf k}.{\mathbf V}_2\right)^2 \right]  + \mathcal{O}(k^3) .
\end{align}
Thus, the contribution of interacting clusters to mode coupling leads to the influence of small scales on larger scales, which can be thought of in terms of clusters acting as point masses to the leading order. Here, we have ignored the subdominant tidal interactions between the clusters. Both the leading and subleading terms lead to the generation of a power spectrum that goes as $k^4$ at small $k$. This expression encodes the fact that the internal structure of virialised interacting clusters at small scales does not affect the evolution of perturbations at much larger scales to the leading order. This equation can be generalised to an arbitrary number of interacting clusters. 

Going further with this analysis, especially to put these in the context of specific models is non-trivial. However, the message to take home from here is that there is a weak coupling between non-linear structures at small scales and perturbations at very large scales. This coupling can be represented correctly in cosmological N-body simulations once halos are resolved at small scales. Therefore, we can expect simulations to match our expectations once the scale of non-linearity exceeds the mass resolution scale by more than an order of magnitude. At earlier times, there are likely to be errors in the resulting distribution of particles, and these may impact different indicators like power spectrum, mass function, etc., differently. 

These expressions provide a firm footing for the ideas of the renormalisability of gravitational clustering in an expanding universe \cite{Carrasco_2012, Couchman_1998, Crocce_2006a, Hui_1996, Nishimichi_2016, Rampf_2023}.

\section{Sources of errors at small scales}
\label{sec:mc_err}

\begin{figure}[t]
  \begin{center}
      \includegraphics[width=0.95\textwidth]{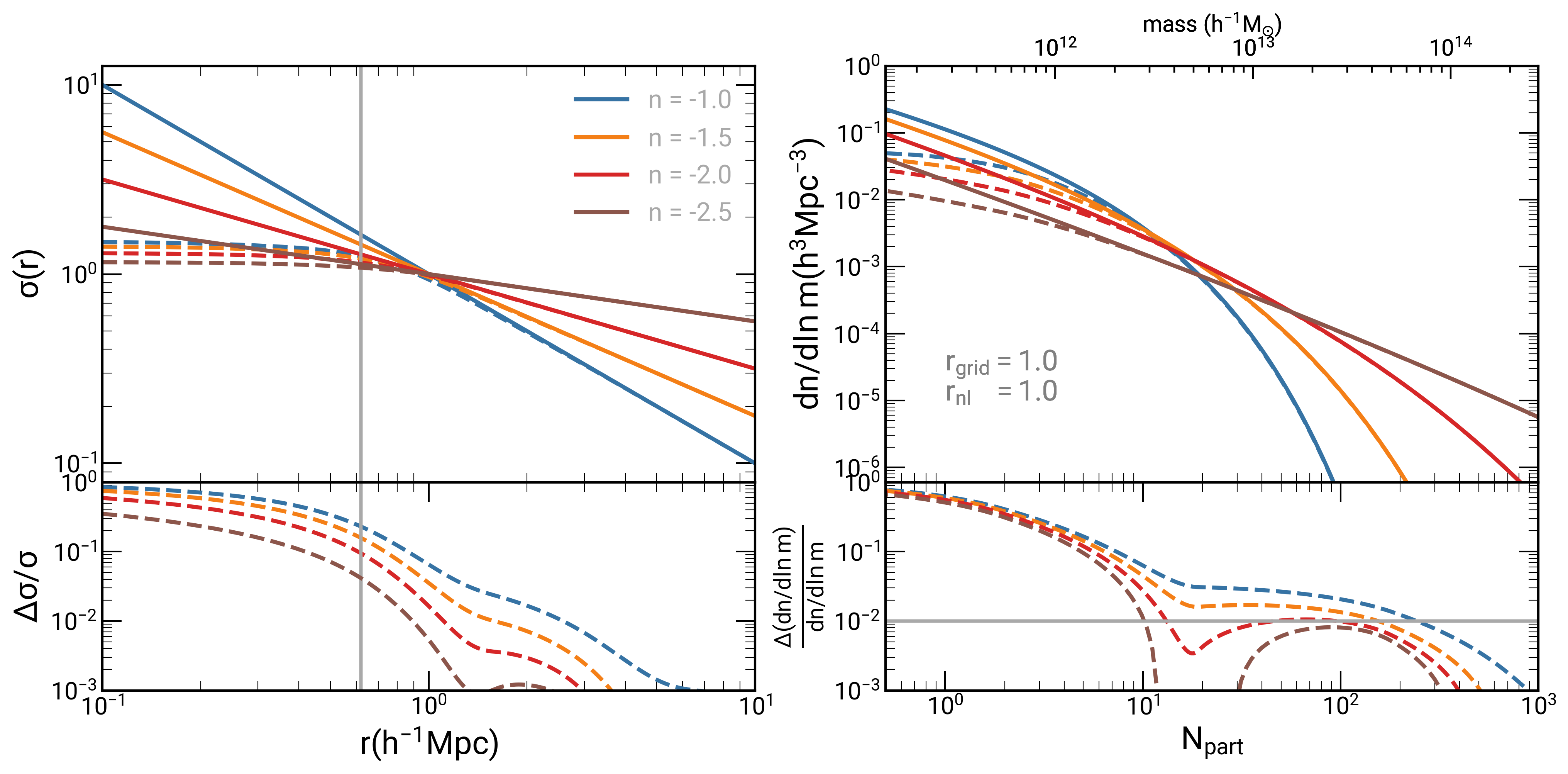}\\
      \caption[Effect of missing small scale power on root-mean-square fluctuations and mass function]{\emph{Effect of missing small scale power on root-mean-square fluctuations and mass function:} Top left panel show mass variance for the power-law power spectrum of indices $n$. Solid lines are computed using full power spectrum, while dashed lines include constrained power spectrum up to Nyquist frequency. The bottom left panel show fractional errors in mass variance. The vertical grey line shows the size of a single particle. The top right panel shows the halo mass function at a time when the scale of non-linearity is equal to the unit grid length. Solid lines involve a full power spectrum, while dashed lines include a constrained power spectrum. The bottom right panel show fractional errors in mass function. Horizontal grey lines represent the $1$\% mark.}
      \label{fig:th1}
  \end{center}
  \end{figure}

  \begin{figure}
    \begin{center}
        \includegraphics[width=0.95\textwidth]{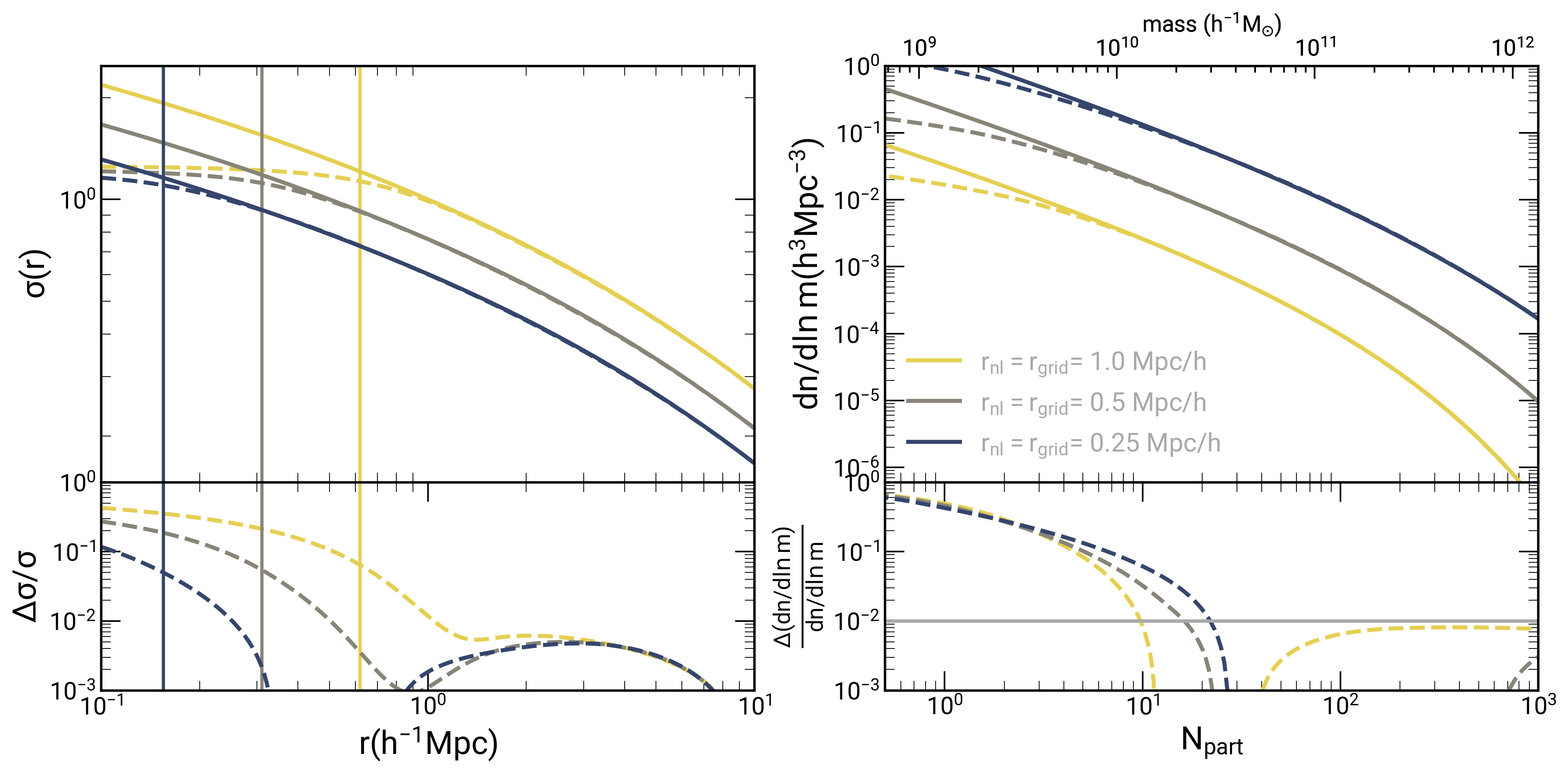}\\
        \caption[Effect of missing small scale power for $\Lambda$CDM models]{\emph{Effect of missing small scale power for $\Lambda$CDM models} The top panels show mass variance and mass function in the $\Lambda$CDM models with three values of $r_{\text{nl}} = r_{\text{grid}}=1.0$, $0.5$, $0.25$ Mpc/h. Solid lines are computed using full power spectrum, while dashed lines include constrained power spectrum up to Nyquist frequency. The bottom panels show fractional errors in mass variance and mass function. The vertical lines in the left panels mark the size of a single particle. Horizontal grey lines represent the $1$\% errors.}
        \label{fig:th_lcdm}
    \end{center}
\end{figure}

Cosmological N-body simulations have many potential sources of errors in the results at small scales. We have seen an estimate of the mode coupling between small scale virialised structures and large scale density fluctuations. Any errors or unrealistic representation of density fluctuations and halos at small scales will lead to such errors at large scales via mode coupling. The foremost source of errors at small scales is due to discreteness: particles in a simulation have a finite mass and it is not possible to use these to describe the density field at mass scales that are comparable to or smaller than this mass. Indeed, the density field can only be defined at mass scales much larger than the mass of an individual N-body particle. Further, simulations cannot resolve collapsed and virialised halos of masses lower than (at least) tens of N-body particles.  

To be realistic, we also need to worry about errors in inter-particle interaction force, but we drop this aspect as it depends on the specific code being used. 

A corollary of the inability to represent the density field at small mass scales is that no initial perturbation is imprinted in the density distribution at these scales. This missing power translates into modified root-mean-square fluctuations at scales comparable to the grid scale in simulations. We can estimate the effect of this missing power on the mass function of halos using the approach where the root-mean-square fluctuations computed from the power spectrum realised in the initial conditions is used \cite{Bagla_2006} and compared with the expected mass function without this limitation. \Cref{fig:th1} shows the expected outcome of such a calculation. It shows the shift in root-mean-square fluctuations for power law models in the left panel and the corresponding shift in the mass function in the right panel. Here, we have chosen to plot these for the epoch where the scale of non-linearity is one grid length. The lower panels show fractional error as a result of the missing power. We notice that the root-mean-square fluctuations have an error that is larger for models with a larger $(n+3)$, with $n$ being the index of the power spectrum. For some models, the errors in the root-mean-square exceed $1\%$ out to many grid cells. The corresponding errors in the mass function are also significant, and it is clear that for some models, the error exceeds $1\%$ for halos up to a few hundred particles. \Cref{fig:th_lcdm} displays the case $\Lambda$CDM cosmology. The plots are presented for the epoch at which the non-linearity scale equals one grid length. As the effective spectral index is such that $(n_{\text{eff}}+3)$ is small at very small scales in this model, the errors drop to 1\%  very quickly and hence the missing power is not a serious problem for $\Lambda$CDM models.

\begin{figure}
\begin{center}
\begin{subfigure}[b]{0.32\textwidth}
    \includegraphics[width=0.99\textwidth]{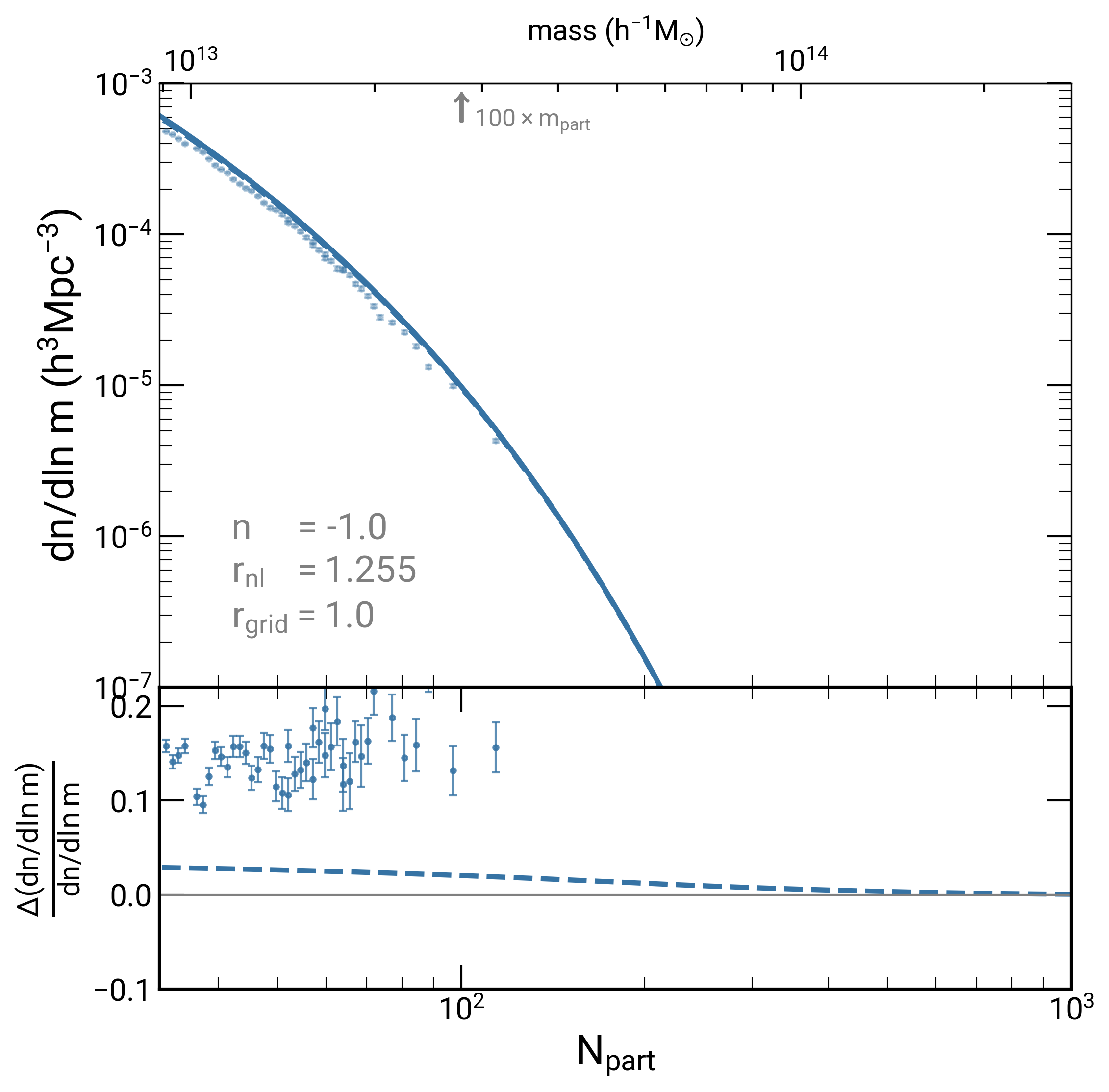}
\end{subfigure} 
\begin{subfigure}[b]{0.32\textwidth}
    \includegraphics[width=0.99\textwidth]{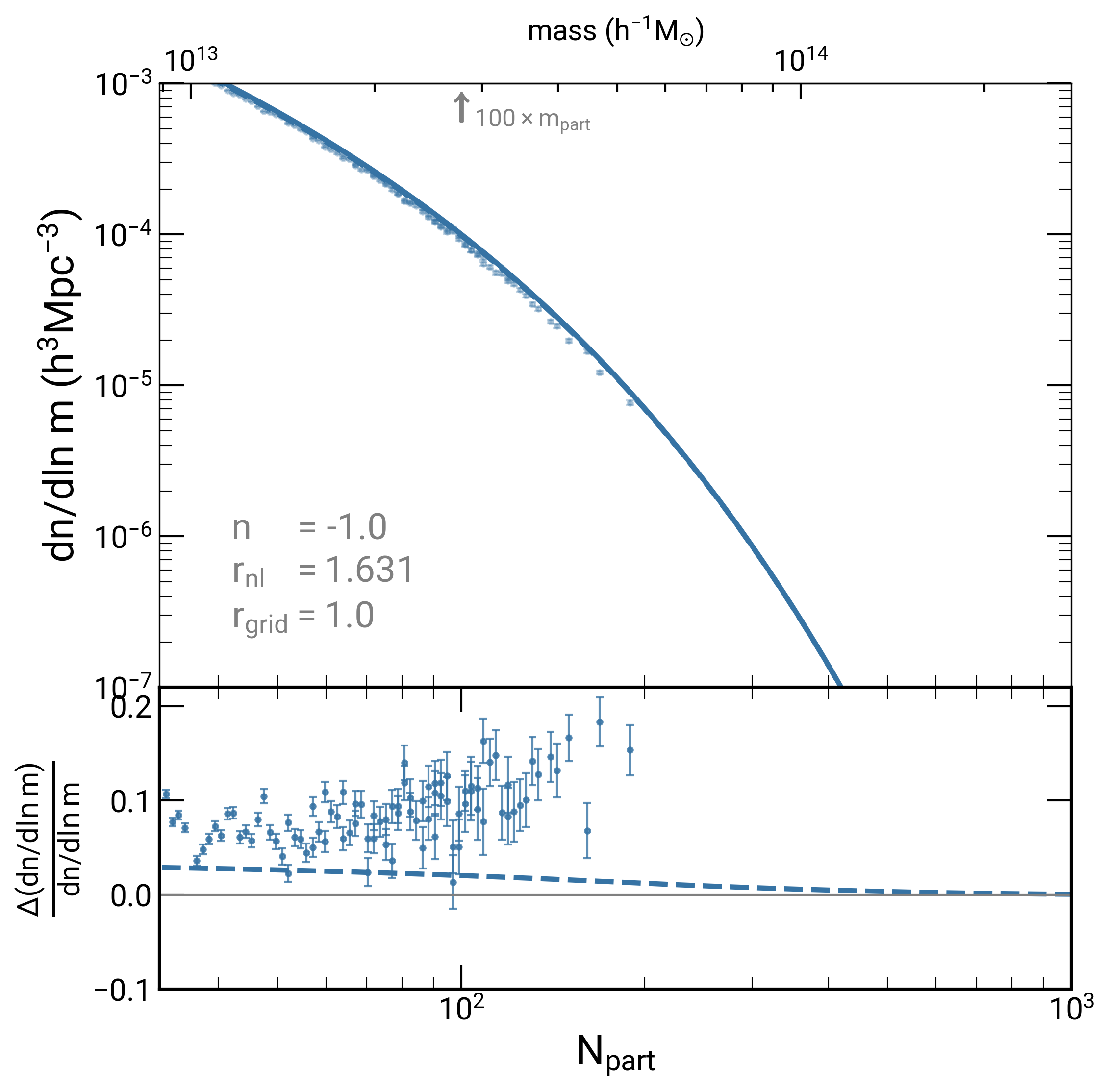}
\end{subfigure} 
\begin{subfigure}[b]{0.32\textwidth}
    \includegraphics[width=0.99\textwidth]{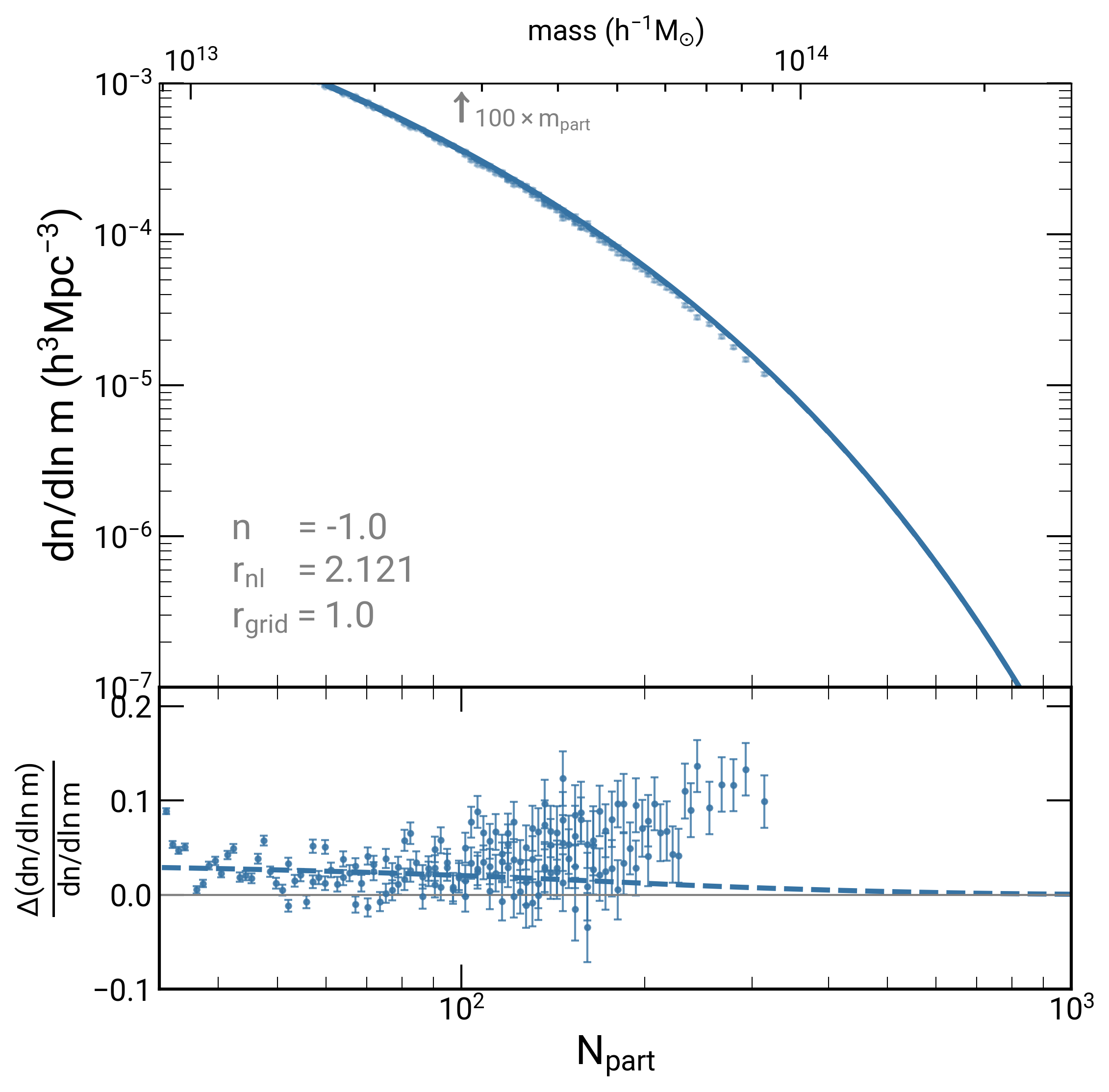}
\end{subfigure} \\
\begin{subfigure}[b]{0.32\textwidth}
    \includegraphics[width=0.99\textwidth]{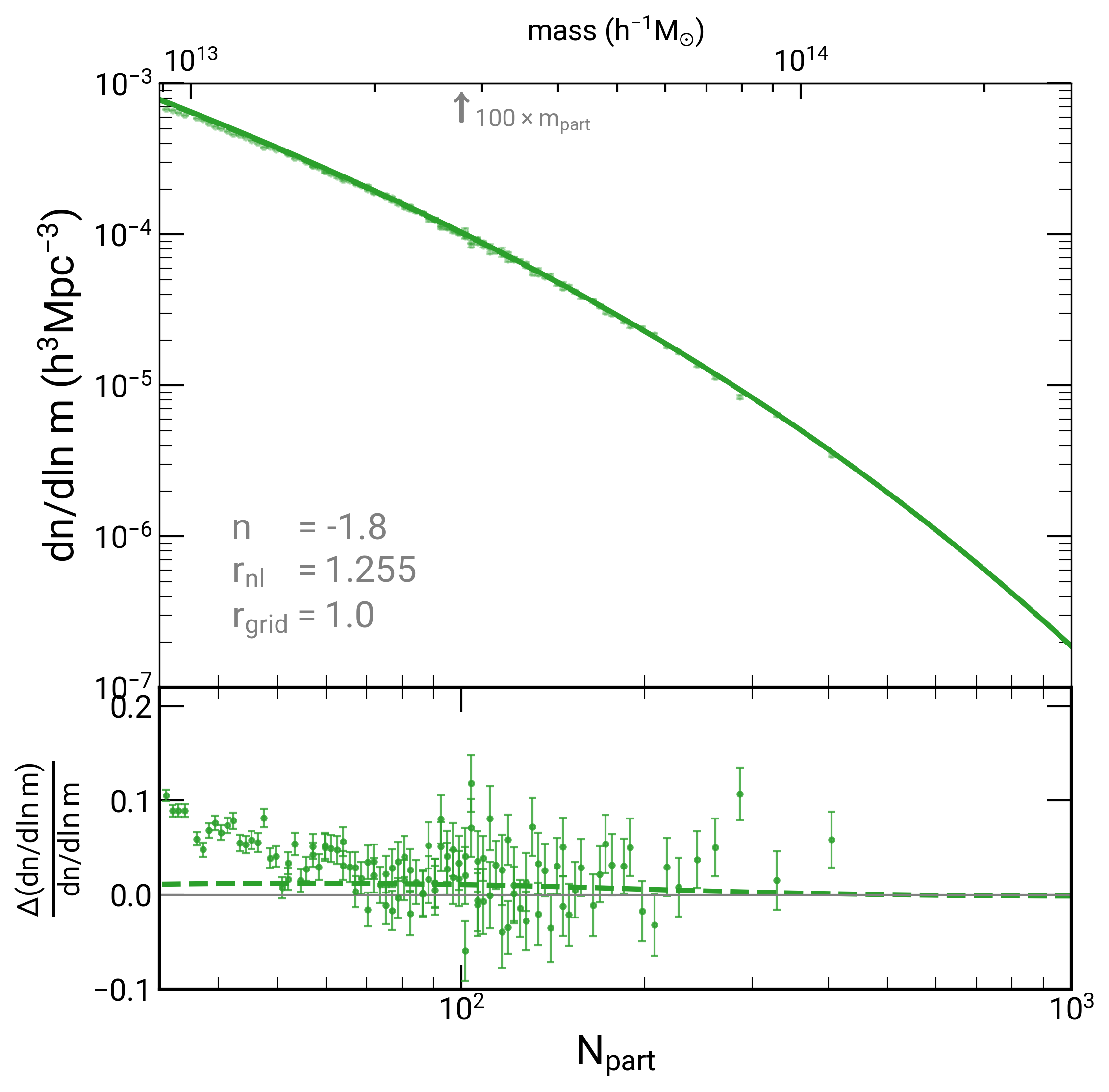}
\end{subfigure} 
\begin{subfigure}[b]{0.32\textwidth}
    \includegraphics[width=0.99\textwidth]{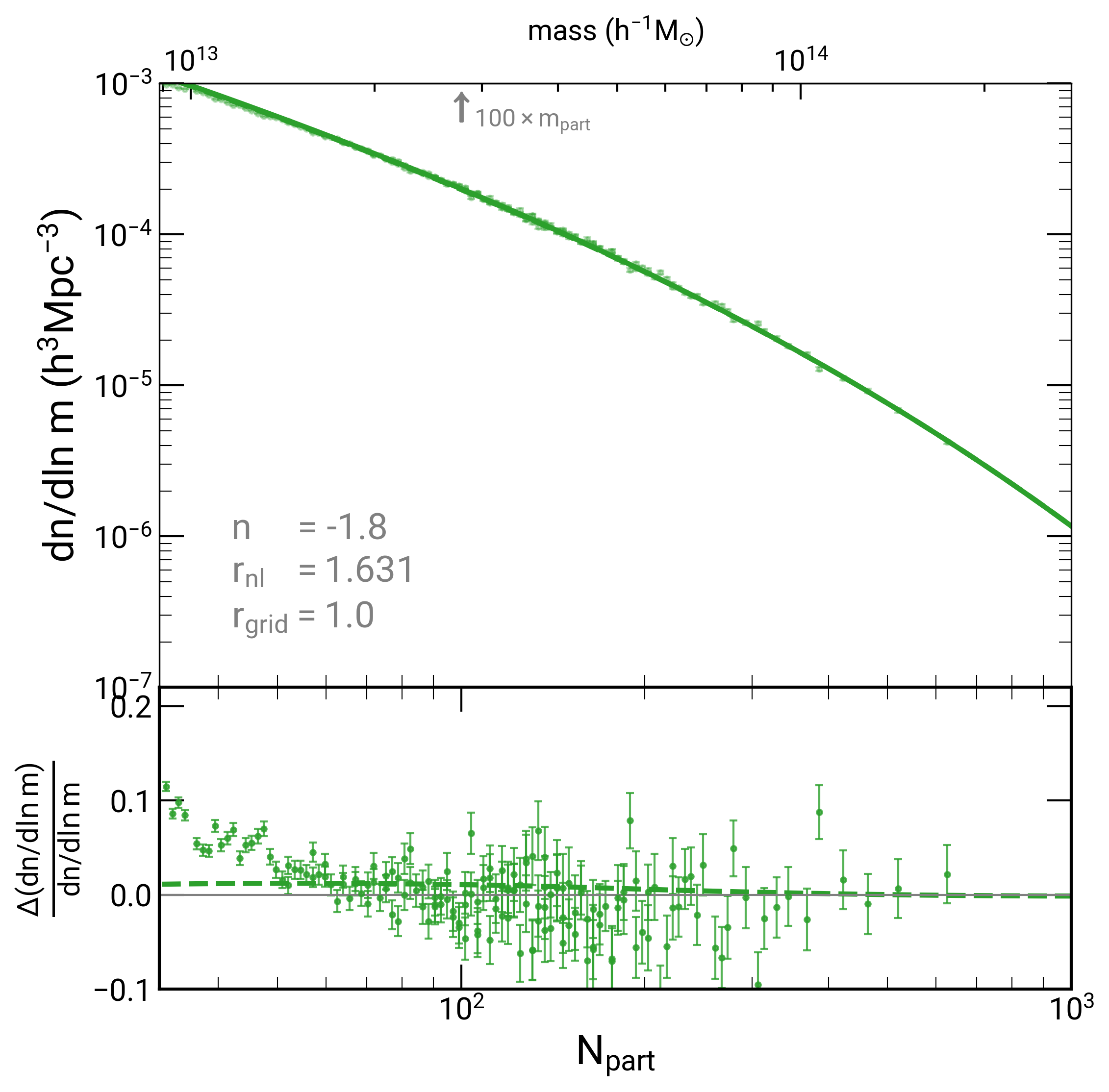}
\end{subfigure} 
\begin{subfigure}[b]{0.32\textwidth}
    \includegraphics[width=0.99\textwidth]{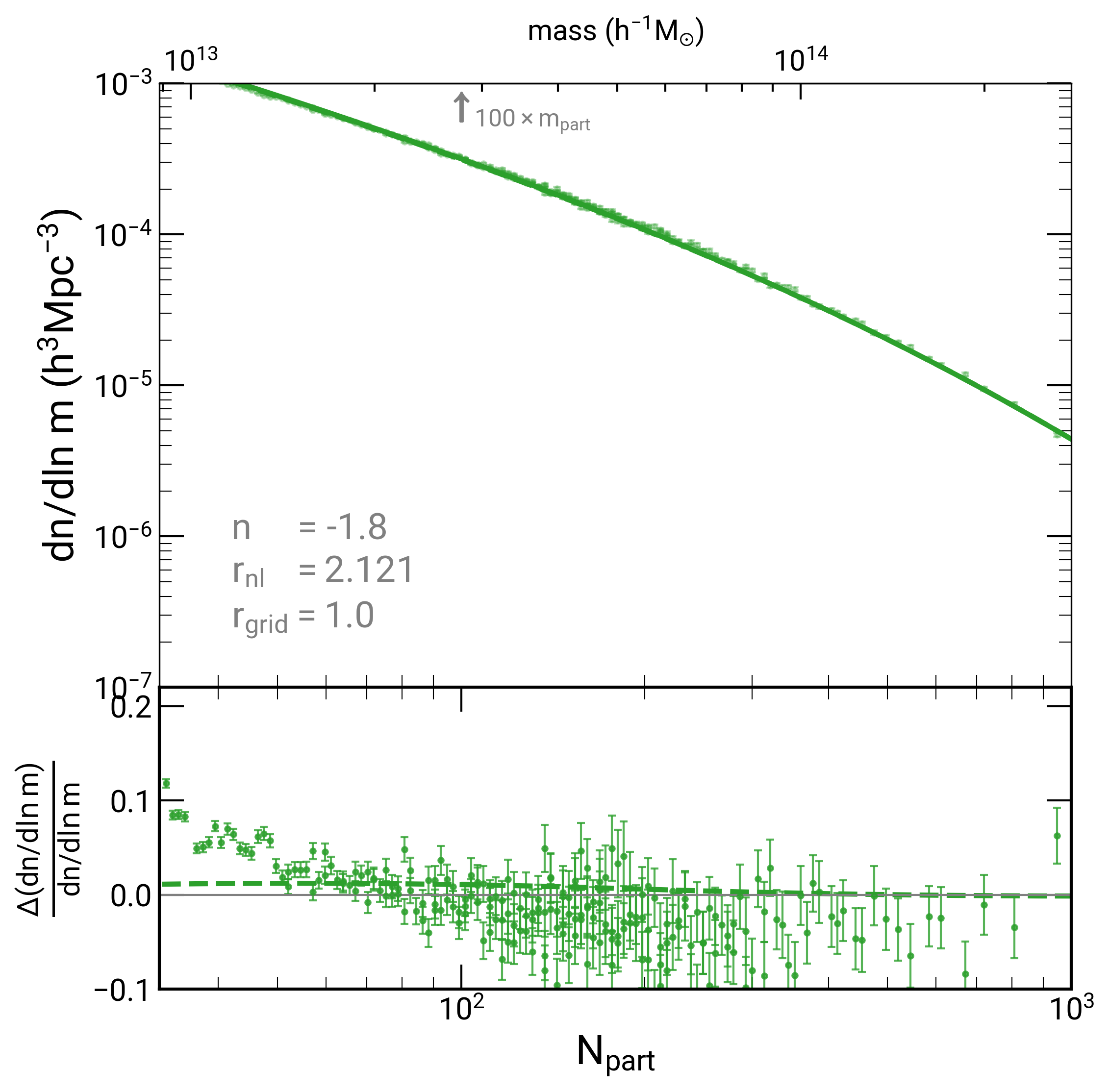}
\end{subfigure} \\
\begin{subfigure}[b]{0.32\textwidth}
    \includegraphics[width=0.99\textwidth]{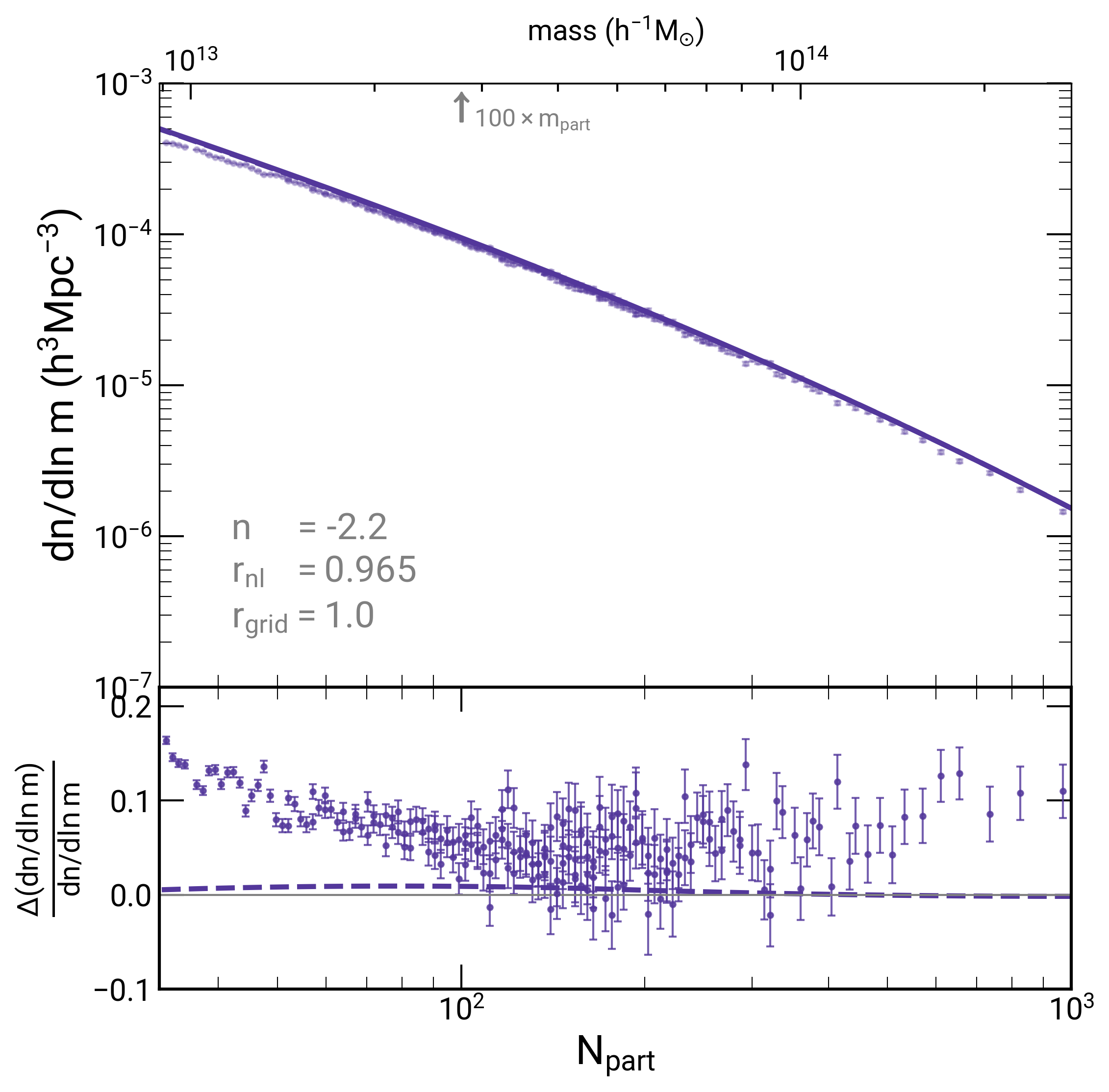}
\end{subfigure} 
\begin{subfigure}[b]{0.32\textwidth}
    \includegraphics[width=0.99\textwidth]{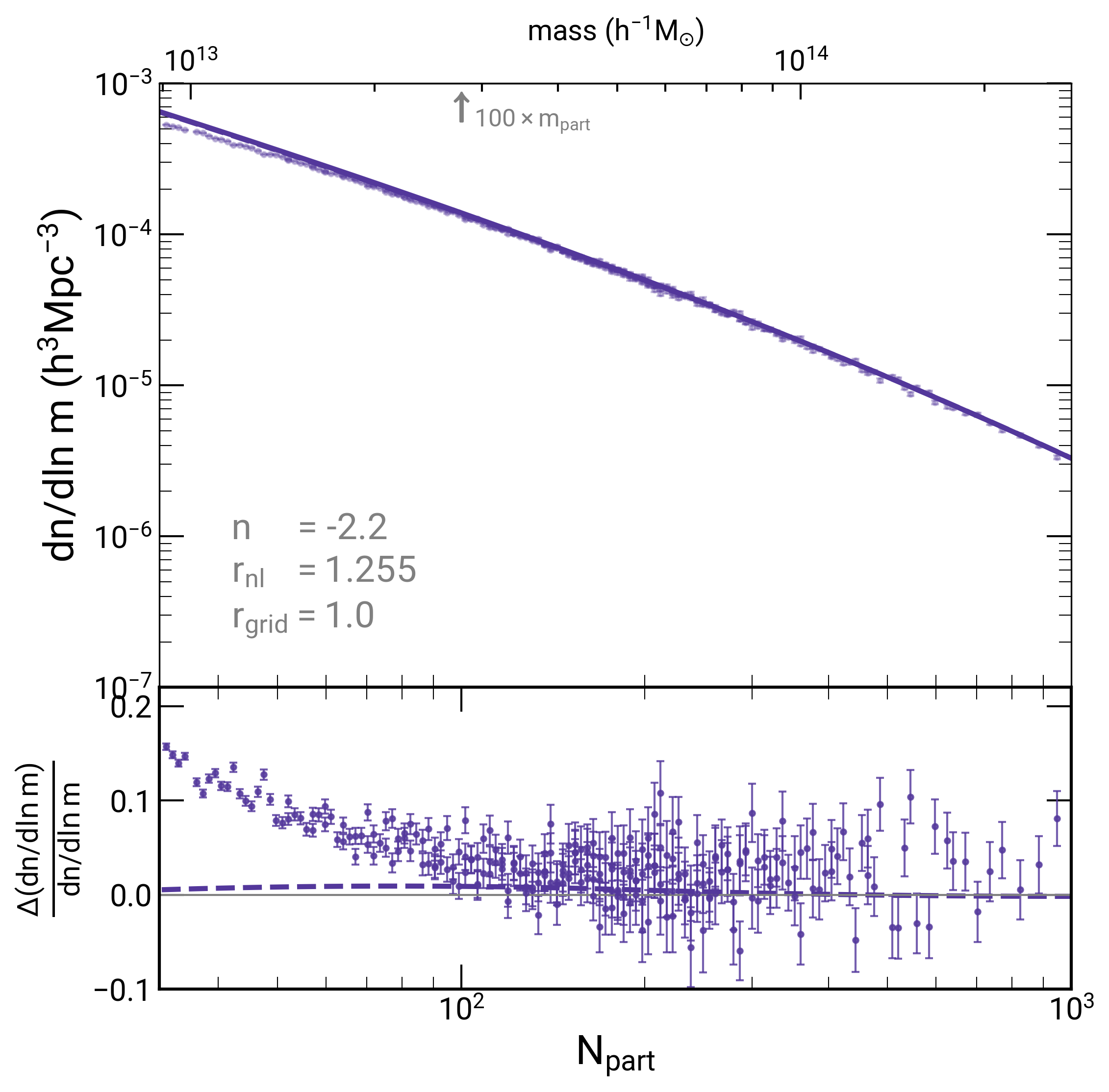}
\end{subfigure} 
\begin{subfigure}[b]{0.32\textwidth}
    \includegraphics[width=0.99\textwidth]{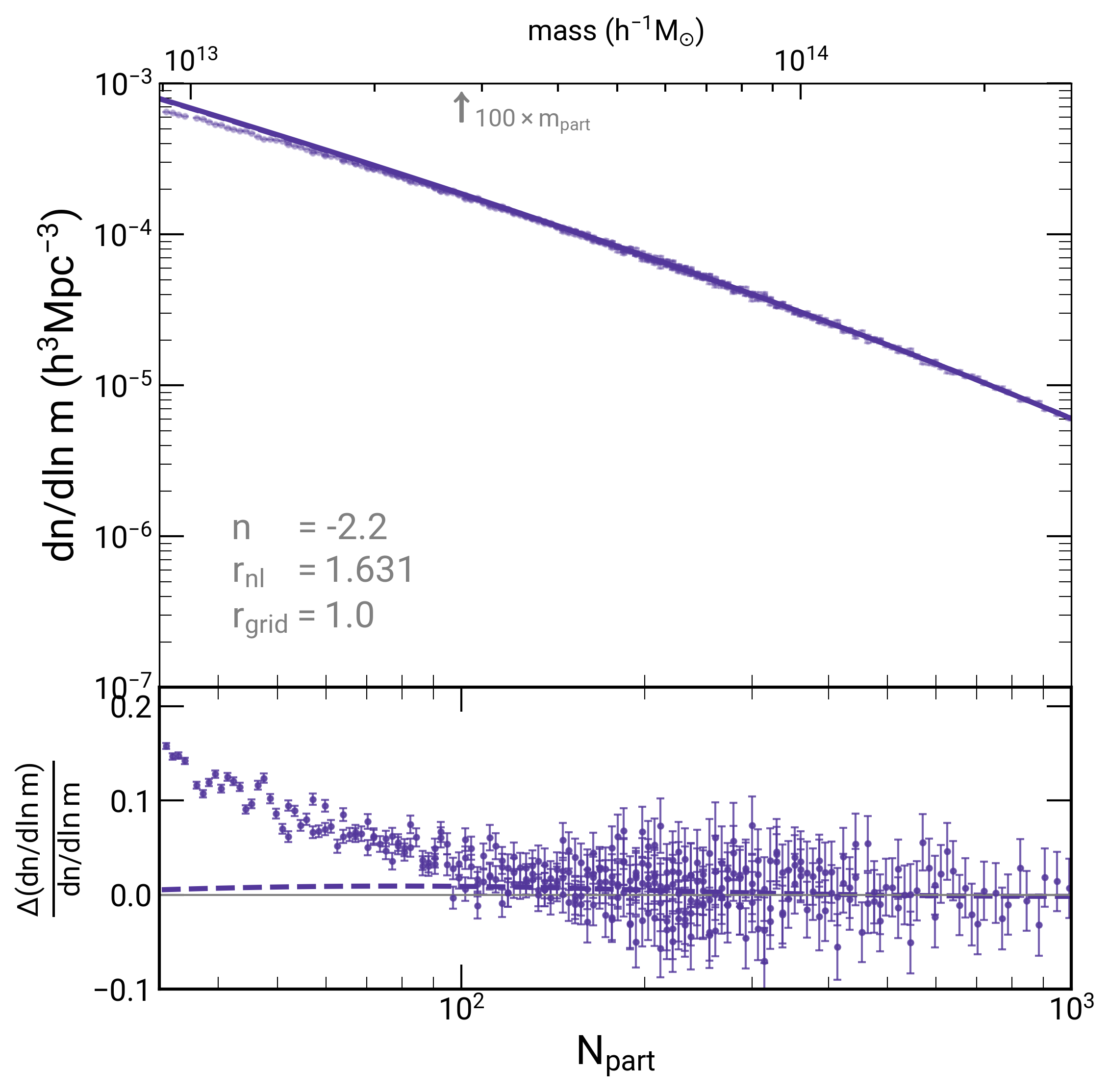}
\end{subfigure}
\caption[Errors in the simulation mass function]{\emph{Errors in the simulation mass function: }Each panel show the halo mass function and fractional errors in mass function as a function of a number of particles. Data points represent simulation, whereas solid lines show theoretical mass function. The top, middle and bottom rows represent the power-law power spectrum index $n$ values $-1.0$, $-1.8$ and $-2.2$, respectively. Columns represent different scales of non-linearity ($r_{nl}$). }
\label{fig:mc_sim}
\end{center}
\end{figure}

\Cref{fig:mc_sim} shows the mass function for three models at three epochs each.  Each row is for a given index: the top row is for $n=-1$, the middle row is for $n=-1.8$, and the bottom row is for $n=-2.2$. Each row has plots for three epochs corresponding to an increasing scale of non-linearity as we go from left to right. Each panel shows the expected mass function shown in \cref{eqn:mc_mf} and the mass function from N-body simulations. The lower part of each panel shows the fractional error in the mass function. We note that the error is indeed highest at early epochs for $n=-1$, which corresponds to the largest $(n+3)$ in this set. This is in agreement with the impact of the missing power at small scales estimated above. We see that the errors come down over a substantial range in mass at later epochs as a number of data points distributed around the zero-deviation mark show a systematic shift with the epoch. Some of the remaining disagreement results from a poor fitting function for the theoretical estimate, discussed in \cref{cha:3}. \Cref{fig:mf_1}, \cref{fig:mf_2}, and \cref{fig:pq_n} illustrate the poor fit of the halo mass function to scale-free simulations.
\begin{subequations}
\label{eqn:mc_mf}
\begin{align}
\frac{\d n}{\d \ln m} &= \frac{\bar{\rho}}{m} \frac{\d \ln \sigma^{-1}}{\d \ln m} f(\nu), \\
f(\nu) &= A(p) \nu \sqrt{\frac{2q}{\pi}} [1+(q\nu^2)^{-p}]  \exp(-q\nu^2 /2),\\
  p(n) &= -0.045 \ n+0.231 ,\\ 
  q(n) &= 0.095 \ n+0.922,
\end{align}
\end{subequations}
where, $\nu$ is peak height $\nu \equiv \delta_c /(\mathcal{G}(z)\sigma(m))$. Here $\delta_c$ is the linearly extrapolated critical over density required for a spherically symmetric perturbation. $\mathcal{G}$ is the growth factor. $\sigma(m)$ is the variance in the initial density fluctuation field, linearly extrapolated to the present epoch and smoothed with a top hat filter $W(k,m)$ of scale R= $(3m/4\pi\bar{\rho})^{1/3}$.
  
\section{Discussion and Summary}
\label{sec:mc_summ}

We have seen in the last two sections that the mode coupling between small scales and large scales has a component arising from interactions of halos. This mode coupling arises from the departure of shapes of halos from spherical shapes and tidal interactions. These components clearly require the halos to be resolved in the simulations with a large number of particles. Thus, we expect that this is a source of errors in cosmology N-body simulations and that it arises from discreteness. It is reasonable to expect that the error will diminish at late times once the scale of non-linearity is sufficiently larger than the average inter-particle separation. This is motivated by observations that clustering at small scales is dominated by the power spectrum around the scale of non-linearity \cite{Little_1991, Bagla_1997a}.

The second effect on the limitation of N-body simulations at small scales and at early times arises from missing power. It is not possible to represent density perturbations below the Nyquist frequency, and hence, there is a sharp departure from the scenario being simulated. We show that this results in an underestimation of the root-mean-square fluctuations at small scales in the simulation. Obviously, this affects models with steep power spectra much more than other models. We quantify this by computing the root-mean-square amplitude of fluctuations and the corresponding mass function of collapsed halos. We find that the errors percolate to an under-estimated root-mean-square amplitude of fluctuations out to several grid lengths and mass function for halos with more than $10^2$ particle halos at early times. 

We test this deficit in the mass function using N-body simulations and find that this is seen clearly at early times for the $n=-1$ power spectrum. We also note that the error diminishes at later times. The error is smaller for other models for halos with $50$ or more particles, even at early times. 

The summary of our findings is that there are significant errors in cosmological N-body simulations at small scales at early times. In general, the results of simulations are not reliable till the scale of non-linearity is larger than one grid length. For steeper power spectra with an index where $(n+3) > 1$, it may be better to only use data once the scale of non-linearity is much larger than a grid length. Note that we have not considered any errors that may arise due to errors in force or numerical implementation, e.g., \cite{Bouchet_1985, Bagla_1997}. These limitations on cosmological N-body simulations are complementary to the limitations arising from a finite box size \cite{Bagla_2006}.

A comparison with some of the other estimates of expected transients is also in order.  \cite{Scoccimarro_1998} estimated transients arising from the deviations of initial conditions generated using the Zel'dovich approximation as compared to the expected evolution.  Note that the Zel'dovich approximation gives correct evolution at linear order and hence any deviations arise from its use to set up initial conditions directly in the mildly non-linear regime.  This estimation was done using higher order Lagrangian perturbation theory.  We would like to point out that we have assumed that the generation of initial conditions as well as dynamical evolution is {\sl ideal} within the limitations of mass resolution and the range of wave modes available in the simulation. Interestingly, analytical calculations using a gravitational dynamics approach also indicate a spectrum dependence in the influence of small scales on larger scales \cite{Bharadwaj_1996} where a similar factor of $(n+3)$ makes an appearance\footnote{See end of \S5 in \cite{Bharadwaj_1996}.}.

The primary concern raised in this work is that N-Body simulations are plagued by transient features introduced by two factors: discreteness induced errors in mode coupling and, missing power at small scales.  We do not dwell on any other {\sl non-ideal} aspects of evolution of perturbations in cosmological N-Body simulations.  As we have shown, the errors arising out of these factors are significant at small scales and at early epochs where the scale of non-linearity is smaller than a few grid lengths.  Once the scale of non-linearity is such that halos with tens of particles have collapsed and virialised, these errors become insignificant.  Thus we are pointing out that the reliability of cosmological N-Body simulations is limited at smaller scales.  When combined with the limitations arising from box size (See, e.g., \cite{Bagla_2006}), we see that the dynamic range of cosmological simulations in length scales is about an order of magnitude smaller than the box size (in terms of number of cells).  As these limitations cannot be overcome within the context of simulations, other approaches that help enhance the dynamic range of simulations must be studied. These results imply that the use of very large simulations to generate mock catalogues needs to be evaluated carefully and suitable corrections devised in order to ensure that these are reliable, e.g. \cite{Ramakrishnan_2021}. Of course, as all such methods are calibrated with simulations, sufficient care and caution needs to be exercised.  Approaches that suffice for some quantities of interest may fail for other observables and hence the validation of these methods has to be carried out with some care before deploying these for real applications.

One may wonder whether such errors arising at small scales will contribute to larger scales at late time. However, we know from many numerical experiments that once larger scales become non-linear these can erase differences at smaller scales \cite{Little_1991,Bagla_1997a}.

We may summarise the lessons from this work as follows: N-body simulations can be trusted for generic CDM/$\Lambda$CDM type models once the scale of non-linearity exceeds the mass resolution by at least an order of magnitude. For such situations the {\it correctness} of cosmological N-body simulations and renormalisability of gravitational clustering can be justified.


%% file: chapters/chap6.tex

\chapter{Conclusion}
\label{cha:6}

The linear theory predictions of structure formation are based on the large scale homogeneity and isotropy of the Universe, which allows for simpler mathematical descriptions of perturbation growth. As structures grow and the amplitude of perturbations becomes large at small scales, the limitations of such assumptions become noticeable, necessitating non-linear investigations of structure formation \cite{Peebles_1981, Dodelson_2020, Baumann_2022, Buchert_1994, Baumann_2012, Padmanabhan_1996}. The non-universality of halo mass functions that depend on the underlying power spectrum indicates a need to go beyond the isolated ellipsoidal gravitational collapse assumption. Further, The systematic biases introduced by peculiar velocities in local Hubble-Lema\^itre constant measurements can arise due to inhomogeneities in the Universe. These findings in this thesis reassure that simplifying assumptions needs to be improved in capturing the details of large scale structure. Non-linear processes, such as mode coupling driven by halo interactions and the influence of large scale bulk flows, can play a role in modifying observables both on small and large scales. In simulations, discreteness and resolution errors can introduce significant biases on small scales. As large scale surveys and precision cosmology push for tighter constraints, it becomes evident that a more fine approach that accounts for mode coupling, tidal interactions, and local environmental factors is essential for accurate theoretical predictions \cite{Lopes_2024, Delos_2024, Angrick_2010, Castorina_2016}.

In this chapter, we provide a summary of our results and discuss their implications.

\section*{Summary}

We first investigate the power spectrum dependence of the halo mass function using a suite of dark-matter-only N-body simulations for seven power-law models with an Einstein-de Sitter cosmology. Our findings demonstrate a clear non-universality in the mass function. We provide fits for the parameters of the Sheth-Tormen mass function across a range of power-law power-spectrum indices, observing a mild evolution in the overall shape of the mass function over epochs. We extend these results to a $\Lambda$CDM cosmology and show that the Sheth-Tormen mass function with parameters derived from a matched power-law EdS cosmology more accurately fits the $\Lambda$CDM mass function than the standard Sheth-Tormen model. These results suggest that an improved analytical theory is necessary to describe the mass function better.
 
We further quantify the impact of peculiar velocities on estimating the Hubble-Lema\^itre constant $H_0$. By considering observers located within dark matter halos, we compute the distribution of the estimated $H_0$ across all such observers. We find that the dispersion of this distribution is significant at small scales but diminishes with increasing separations, approaching the level of quoted statistical errors in SH0ES and Planck measurements at distances beyond  ~135 Mpc/h and ~220 Mpc/h, respectively. A negative correlation is observed between the local over-density around an observer and the deviation between the local and global $H_0$ values. Deviations exceeding 5\% of the global value occur frequently on scales up to ~40 Mpc/h, which is considerably larger than in the range of distance scales where majority of large distance indicators like supernovae of type Ia are calibrated. We also analyse the cumulative impact of these errors on mock measurements of $H_0$ as deduced from Milky Way-sized halos. Our results show that this error is sensitive to the minimum distance used for measurements. The distribution of $H_0$ in mock measurements exhibits a large tail, suggesting that a deviation of a few percent from the global value is possible in the atypical regions of the Universe.
  
Lastly, we investigate the impact of small scale gravitational clustering on large scales by analysing mode coupling between virialised halos. Building on Peebles' (1974) \cite{Peebles_1974a} calculation, which shows that a virialised halo does not contribute mode coupling terms at small wave numbers $k$, we use a perturbative expansion in wave numbers to demonstrate that this effect is minor. We relate these findings to the limitations of finite mass resolution in cosmological N-body simulations on the evolution of perturbations at early times. We explore the effects of a finite shortest scale up to which the desired power spectrum is realised in simulations. We provide basic estimates of the magnitude of these effects and their dependence on the power spectrum. Our results indicate that the influence of small-scale cutoffs in the initial power spectrum and discreteness effects increases with $(n + 3)$, where $n$ is the power spectrum index. Cosmological simulation data should be utilised only if the non-linearity scale is sufficiently larger than the average inter-particle separation.

\section*{Implications and outlook}

\Cref{cha:3} explores the test of the accuracy of theoretical modelling of the halo mass function, focusing on its dependence on the input power spectrum. The widely used assumption that the halo mass function is universal across different models has often been questioned \cite{Bhattacharya_2011, Ondaro_2022, Watson_2013, Despali_2015, Euclid_2023}. This challenges when predicting the abundance of halos at high masses and early redshifts, further impact galaxy formation models. The halo mass function is also used to make mock catalogues to estimate the efficacy of future observational programs and a covariance matrix to estimate errors. In the precision cosmology era, we need to ensure that underlying assumptions do not introduce errors in the system. Our long-term aim is to identify these sources of deviations, which will improve our understanding of underlying processes and enhance the precision of theoretical calculations. The observed non-universality in the mass functions may originate from mode coupling between collapsing structures and the large scale density field, particularly in non-spherical collapse scenarios. It is important to note that the physical scenario relevant to excursion set theory is far more complex than the assumptions of the ellipsoidal collapse \cite{Delos_2024, Angrick_2010, Castorina_2016}. We are investigating this using N-body simulations.

The effect of peculiar velocities on measuring the Hubble-Lema\^itre constant $H_0$ in the local Universe has essential implications for understanding the Hubble tension \cite{Maartens_2023, Odderskov_2014, Odderskov_2017}. Study in \cref{cha:4} supports the idea that local measurements of $H_0$ are systematically different from the global value due to gravitational clustering. It adds to the evidence that the Hubble tension may not only be due to observational systematics or unknown physics but could also be influenced by local arrangements of structures. Further, one needs to investigate how it impacts the calibrations involved in the observations. Observed large scale bulk flows around the Milky Way \cite{Lopes_2024} suggest that our region might be an outlier in the expected distribution of galaxies and flows according to the concordance $ \Lambda $CDM model. This implies that local measurements might be largely biased if we live in a region with atypical flows. However, more detailed studies are needed to evaluate this possibility.

N-body simulations are essential tools in cosmology and galaxy formation studies. However, understanding their limitations is necessary to avoid misinterpreting simulation data. The finite resolution of N-body simulations leads to missing power at small scales. In \cref{cha:5}, we see that this results in errors in the root-mean-square fluctuations and the corresponding mass function. These errors are most significant at early times and on small scales. Analytical calculations suggest that the mode coupling between small and large scales is tied to the resolution of collapsed halos. Thus, accurate mode coupling estimates require sufficient halos in the simulation.

The thesis highlights the need to refine existing theoretical models, such as the halo mass function. Such refinement is essential for improving galaxy abundance and their distribution predictions, impacting the large scale surveys like Euclid ~\cite{Euclid_2011, Euclid_2023} and LSST  \cite{lsst_collaboration_2009, Ivezi__2019}. The recent discovery of a very large number of massive galaxies at high redshifts in JWST \cite{Labbe_2023, Finkelstein_2022, Naidu_2022} data demands a closer look at galaxy abundances. Identifying sources of deviations in halo mass function can lead to more accurate theory-observation alignment. Moreover, the role of peculiar velocities in local Hubble-Lema\^itre constant ($H_0$) measurements needs to be reconsidered while analysing the Hubble tension. Linking local density to variations in $H_0$ can provide a statistical framework to contextualise discrepancies. Examining cosmological simulation limitations helps bridge the gaps in modelling the structure formation within theoretical frameworks. Resolving such errors is necessary to ensure simulations can reliably inform future observations.


%% file: chapters/appendix1.tex

\chapter[Observables in the expanding Universe]{Consequence of expanding Universe on observables}
\label{a1}

The expansion of the Universe influences the measurement of physical quantities in various ways. Here, we discuss the Hubble scale, redshift and redshift-space distortions.

\section{The Hubble scale/radius}
\label{a1:hs}

The Hubble parameter sets a characteristic time scale for the expansion of the Universe, which in turn defines a corresponding length scale over which physical processes function coherently~\cite{Peebles_1981, padmanabhan_1993}. Hubble scale $R_H$ is defined as,

\begin{equation}
   R_H = cH^{-1},
\end{equation}
where, $c$ denotes the speed of light, and $H$ is the Hubble parameter which signifies the rate of cosmic expansion. As the Universe expands, the Hubble scale marks the threshold at which the curvature of spacetime, as governed by general relativity, begins to impact cosmic dynamics. On scales smaller than this, Newtonian gravity typically suffices.

\section{Redshift}
\label{a1:r}

Redshift denotes the increase in the wavelength and a corresponding decrease in the frequency of light emitted from distant objects as a consequence of the expansion of the Universe. The redshift $z$ can be measured as,

\begin{equation}
   1+z = \frac{\lambda_{\text{observerd}}}{\lambda_{\text{emitted}}},
\end{equation}
where, $\lambda_{\text{observerd}}$ is the observed wavelength of light and $\lambda_{\text{emitted}}$ is  the wavelength of light emitted by the source. It can be shown that 

\begin{equation}
   1+z(t) = \frac{a(t_{\text{observerd}})}{a(t_{\text{emitted}})}.
\end{equation}
The redshift function $z(t)$ and the scale factor $a(t)$ serve as interchangeable descriptors of cosmic expansion and can be used as time variables. At small redshifts ($z \ll 1$), the relationship between redshift and distance can be established using the Hubble's law ($v \approx cz = H_0 d$).

\section{Redshift-space distortions}
\label{a1:rsd}

Redshift-space distortions occur due to the peculiar velocities of galaxies within the large scale structure of the Universe. Velocities of galaxies along the line of sight result in distortions in their observed redshifts. This effect introduces anisotropy in the distribution of galaxies, particularly in galaxy surveys where redshift is used as a proxy for distance~\cite{Kaiser_1987, Peacock_2001, Hawkins_2003, Guzzo_2008}.

Two distinct effects contribute to redshift distortions. The first, known as the \emph{Fingers of God}, results from the velocity dispersion of galaxies within rich clusters, elongating the cluster in redshift space. This effect influences only redshift, not sky position, so the elongation occurs solely along the radial direction. Hence, the term `fingers' points to the observer. The other significant redshift distortion, the \emph{Kaiser effect}~\cite{Kaiser_1987}, arises from galaxies bound to a central mass and undergoing infall. Unlike the Fingers of God, the peculiar velocities in the Kaiser effect are coherent, not random, and directed towards the central mass, although the effect is more nuanced.

In general, redshift has two components: the cosmological component ($z_{\text{H}}$) due to the expansion of the Universe, as discussed above, and the Doppler effect of peculiar velocities ($z_{\text{peculiar}}$). Therefore, if $r$ and $v_{\text{peculiar}}$ are the distance and the peculiar velocity of the source galaxy,  the total observed redshift is

\begin{equation}
   (1+z_{\text{observed}}) = (1+z_{\text{H}}) (1+z_{\text{peculiar}}),
\end{equation}
where, $z_{\text{peculiar}} \approx v_{\text{peculiar}} /c$ for $v_{\text{peculiar}} \ll c$. Hence, for small distances ($cz_{\text{H}}\approx H_0 r$)

\begin{equation}
    z_{\text{observed}} \approx \frac{1}{c} \left[ rH_0 + v_{\text{peculiar}}\right].
\end{equation}

We can define a redshift space $\mathbf{s} \equiv (s_1,s_2,s_3)$ which is a transform of the real space $\mathbf{r} \equiv (r_1,r_2,r_3)$, as 

\begin{equation}
   s_1 = r_1 = \frac{cz}{H_0} \theta_1,\text{~~~~~}s_2 = r_2 = \frac{cz}{H_0} \theta_2,\text{~and~~~~~} s_3 = r_3 + \frac{v_3}{H_0},
\end{equation}
where, $\theta_1$ and $\theta_1$ are two angles on the sky and $v_3$ is peculiar velocity along the line of sight.

The radial axis is influenced by Doppler effects resulting from peculiar velocities. Further, these peculiar velocities arise from the clustering itself, introducing complexity. Consequently, the apparent clustering pattern observed in redshift space deviates systematically from that in real space. As a result, the spatial correlation function of galaxies, typically isotropic in real space, becomes anisotropic in redshift space.


%% file: chapters/appendix2.tex
\chapter{Fourier transform}
\label{a2}

The Fourier transform proves crucial in large-scale structure analyses, given that differentiation operations translate to simple multiplication in Fourier space. Here, we explore the Fourier transform of density contrast, the Vlasov equation, correlation functions, and various window functions. We will use to the following Fourier convention:

\begin{equation}
   F(\mathbf{x}) = \int \frac{\d^3 \mathbf{k}}{(2\pi)^3} F(\mathbf{k}) e^{\iota \mathbf{k} \cdot \mathbf{x}}.
\end{equation}

\section{Density contrast}
\label{a2:del}

The Fourier component $\delta_{\mathbf{k}}$ of the density contrast $\delta_{\mathbf{x}} \equiv \delta(\mathbf{x})$ (\cref{eq:d_contrast}) are related through a Fourier transform

\begin{equation}
   \delta_{\mathbf{x}} = \int \frac{\d^3 \mathbf{k}}{(2\pi)^3} \delta_{\mathbf{k}} e^{\iota \mathbf{k} \cdot \mathbf{x}}, \\
   \delta_{\mathbf{k}} = \int \frac{\d^3 \mathbf{k}}{V} \delta_{\mathbf{x}} e^{- \iota \mathbf{k} \cdot \mathbf{x}}.
\end{equation}
Since $\delta_{\mathbf{x}}$ is a real quantity $\delta_{\mathbf{k}}$ satisfies the reality condition, i,e. $\delta_{- \mathbf{k}} = \delta_{\mathbf{k}}^*$.

\section{Vlasov equation}
\label{a2:vl}

The Vlasov~\cref{eq:de_delta} dictates the dynamics of densitiy contrast \cref{eq:d_contrast} in expanding Universe as discussed in~\cref{ssec:lss_equations}. Assuming Universe as a periodic volume $V$, , comprised of particles, with $M$ representing the total mass within $V$, Fourier transform Vlasov equation is given as (refer \S 27 of ~\cite{Peebles_1981})

 

\begin{subequations}
\begin{align}
  {\ddot\delta}_{\mathbf k} + 2 \frac{\dot{a}}{a} {\dot\delta}_{\mathbf k} &=
  A_{\mathbf k} - B_{\mathbf k} ,\\
  A_{\mathbf k} &= \frac{1}{M}\sum\limits_{j} m_j 
  \left[ \iota {\mathbf k}. \left( - \frac{\mathbf \nabla \phi_j}{a^2} \right)
  e^{\iota{\mathbf  k}.{\mathbf x}_j} \right] ,\\
  B_{\mathbf k} &= \frac{1}{M} \sum\limits_{j} m_j \left({\mathbf k}.{\dot
      {\mathbf x}_j } \right)^2   e^{\iota{\mathbf  k}.{\mathbf x}_j } .
  \end{align}
\end{subequations}
The terms $A_{\mathbf{k}}$ and $B_{\mathbf{k}}$ denote nonlinear couplings between different modes. Specifically, $B_{\mathbf{k}}$ couples density contrasts indirectly via the peculiar velocities of particles.

\section{$n$-point Fourier-space correlations}
\label{a2:cf}

The correlation functions~\cref{eq:1-4pcf} give information about the statistic of spatial arrangements of density contrast $\delta$, as discussed in~\cref{ssec:npcf}. Their Fourier space counterpart are called Fourier-space correlation functions are given as,

\begin{subequations}
  \begin{align}
    \label{eq:f1-4pcf}
    \langle \delta(\mathbf{k_1}) \delta(\mathbf{k_2}) \rangle &=  (2\pi)^3 \delta_D(\mathbf{k_1} + \mathbf{k_2}) P(k_1),\\
    \langle \delta(\mathbf{k_1}) \delta(\mathbf{k_2}) \delta(\mathbf{k_3}) \rangle &=  (2\pi)^3 \delta_D(\mathbf{k_1} + \mathbf{k_2} + \mathbf{k_3}) B(\mathbf{k_1}, \mathbf{k_2}, \mathbf{k_3}),\\
    \langle \delta(\mathbf{k_1}) \delta(\mathbf{k_2}) \delta(\mathbf{k_3}) \delta(\mathbf{k_4}) \rangle_c &= (2\pi)^3 \delta_D(\mathbf{k_1} + \mathbf{k_2} + \mathbf{k_3} + \mathbf{k_4}) T(\mathbf{k_1}, \mathbf{k_2}, \mathbf{k_3}, \mathbf{k_4}),
\end{align}
\end{subequations}
where, $\delta_D$ is the Dirac delta function, the quantities $P$, $B$ and $T$ are known as the power spectrum, bispectrum and trispectrum, respectively.

\section{Filter/window functions}
\label{a2:wf}

A common concept in manipulating cosmological density fields involves filtering, achieved through the convolution of the density field with a filter/window function: $\delta \rightarrow \delta * W$ or in Fourier space $\delta_k \rightarrow \delta_k * W_k$. For example, the filtered power spectrum is  $P(k) |W_k|^2$. Below are three standard filter functions. Each filter allows the definition of a mass  $M = f_W r^3 \rho_m$, where $f_W$ is a constant dependent on the shape of the filter. Thus, a filter can be characterised by size $R$ or mass $M$.

\subsection*{Gaussian filter:}
The Gaussian window function $W(r,R)$ for characteristic scale $R$ or $M$, its Fourier transform $\tilde{W} (k, R)$, and corresponding volume factor $f_W$ is given as

\begin{subequations}
  \begin{align}
    \label{eq:fg}
    W (r, R) &= \frac{1}{(2\pi)^{3/2} R^3} e^{-\frac{r^2}{2R^2}},\\
    \tilde{W} (k, R) &=  e^{-\frac{k^2R^2}{2}},\\
    f_W &= (2\pi)^{3/2}.
  \end{align}
\end{subequations}

\subsection*{Sharp $k$-space Filter:}
The Sharp $k$-space window function $W(r,R)$ for characteristic scale $R$ or $M$, its Fourier transform $\tilde{W} (k, R)$, and corresponding volume factor $f_W$ is given as

\begin{subequations}
\begin{align}
    \label{eq:fk}
    W (r, R) &= \frac{1}{2\pi^2 r^3} \left[ \sin \left( \frac{r}{R} \right) -  \left(\frac{r}{R}\right) \cos \left(\frac{r}{R}\right) \right],\\
    \tilde{W} (k, R) &= \begin{cases} 1 & \text{~~} k \le \frac{1}{R} \\
                                      0 & \text{~~} k > \frac{1}{R}
                        \end{cases},\\
    f_W &= 6\pi^2.
  \end{align}
\end{subequations}

\subsection*{Top-hat Filter:}
The top-hat window function $W(r,R)$ for characteristic scale $R$ or $M$, its Fourier transform $\tilde{W} (k, R)$, and corresponding volume factor $f_W$ is given as

\begin{subequations}
  \begin{align}
      \label{eq:ft}
      W (r, R) &= \begin{cases}\frac{3}{4\pi R^3} & r \le R\\
                               0 & r> R
      \end{cases},\\
      \tilde{W} (k, R) &=  \frac{3}{k^3R^3} \left[ \sin(kR) - (kR) \cos (kR) \right] ,\\
      f_W &= \frac{4\pi}{3}.
    \end{align}
\end{subequations}


%% file: chapters/appendix3.tex
\chapter{Mass function: supplementary analysis}
\label{a3}

We present an additional examination of the halo mass function, testing the robustness of our computed mass function for scale-free models discussed in \cref{sec:mf_result}. We explore its sensitivity to the choice of initial redshift (\cref{a3:1}), box size (\cref{a3:2}), and error types (\cref{a3:3}). In \cref{a3:4}, we provide box-wise fits for the mass function in the $\Lambda$CDM model, elaborated in \cref{sec:mf_lcdm}. The color formatting for figures in this chapter follows the conventions specified in \cref{cha:3}, specifically referencing \cref{fig:mf_1}, \ref{fig:pq_n}, and \ref{fig:mfl}. For all the mass functions variations presented in this chapter, we use the $m_{\text{200c}}$ halo mass definition, with $300$ halos per bin (a bin size of $>500$ halos is found to produce systematic effects), $N_{\text{cut}}=100$, and excluded the top 0.1\% of massive halos.

\section{Effect of initial redshift}
\label{a3:1}

\begin{figure}[h]
    \centering
          \begin{subfigure}[b]{0.32\textwidth}
        \includegraphics[width=.99\textwidth]{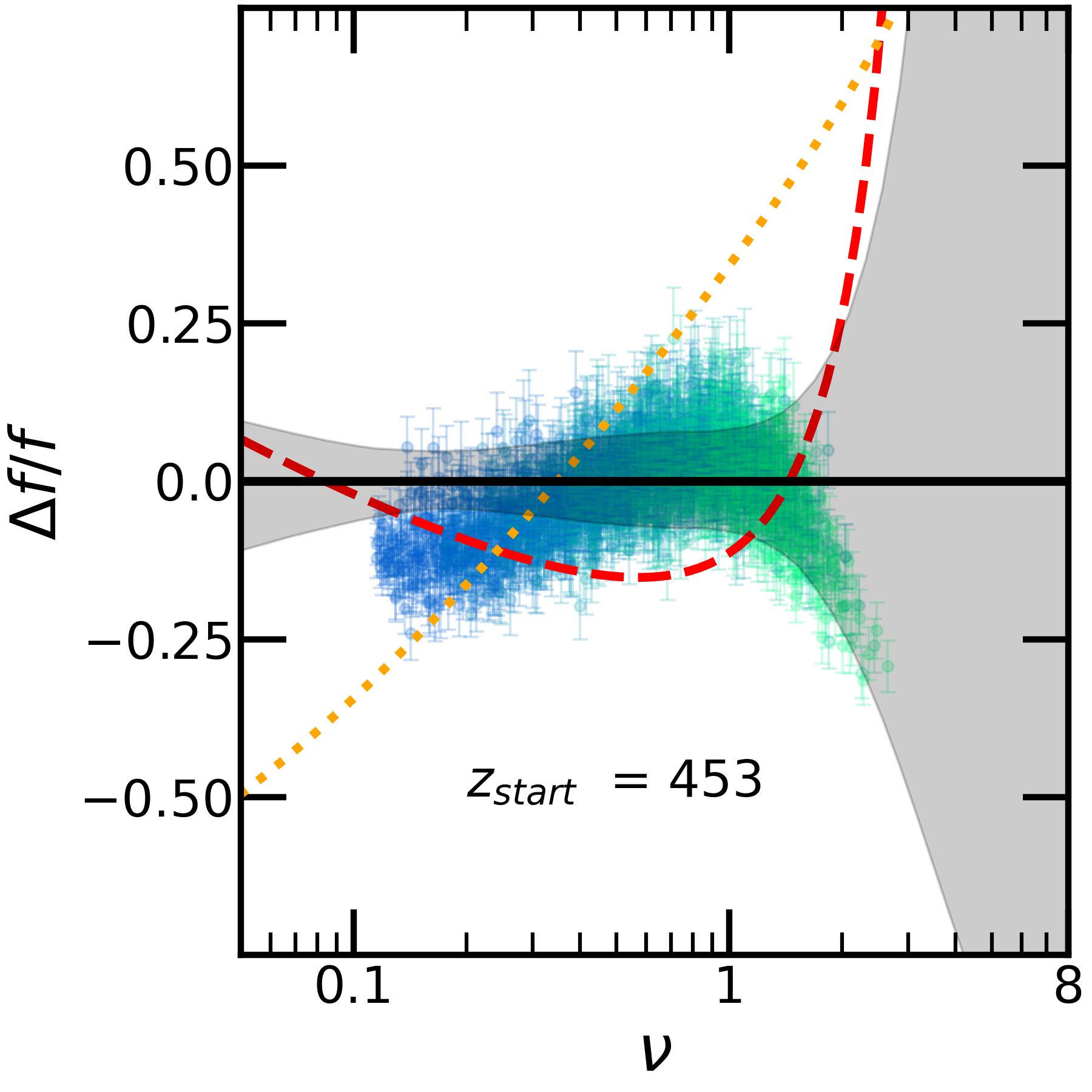}
        \end{subfigure}
        \begin{subfigure}[b]{0.32\textwidth}
        \includegraphics[width=.99\textwidth]{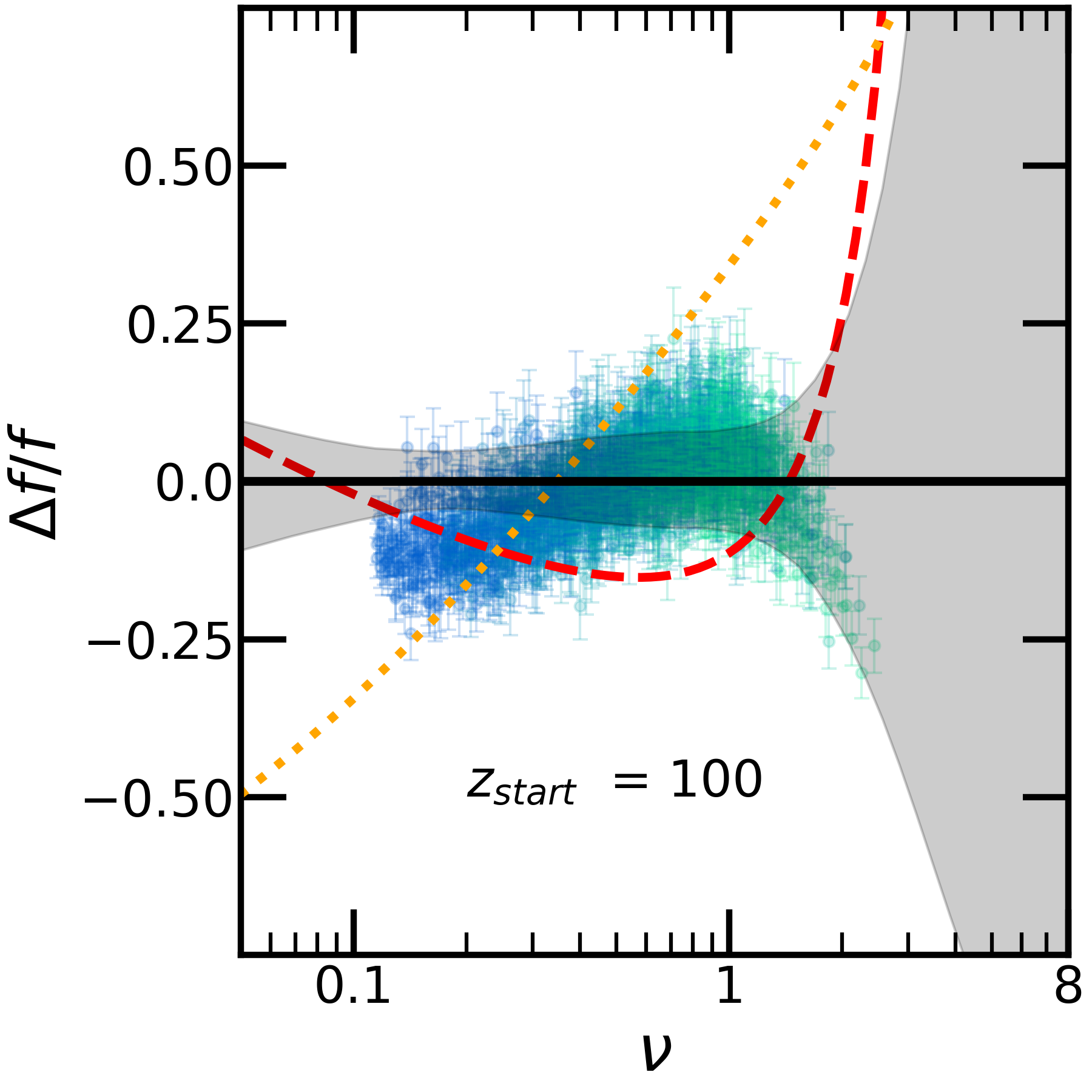}
        \end{subfigure}
        \begin{subfigure}[b]{0.32\textwidth}
        \includegraphics[width=.99\textwidth]{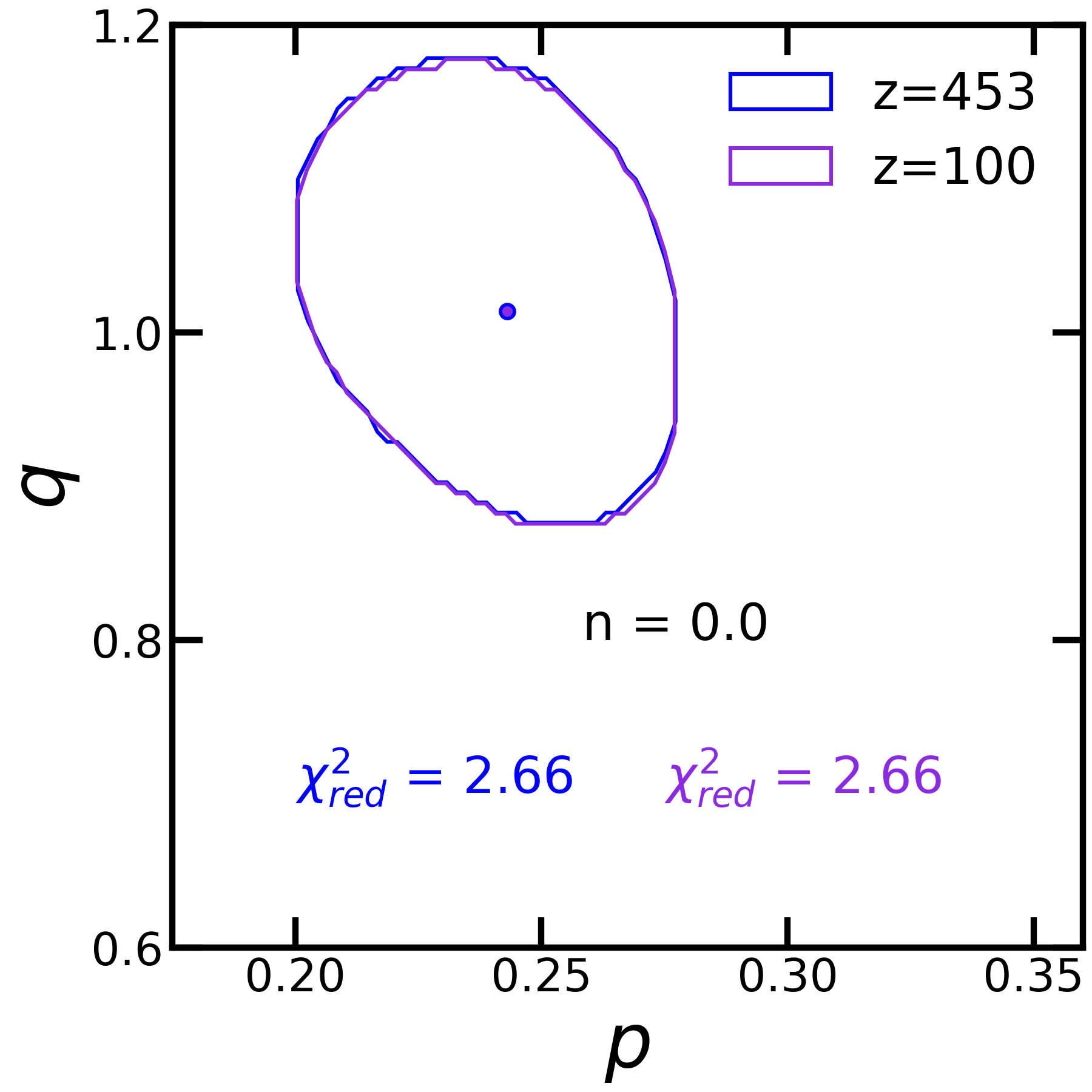}
        \end{subfigure}
    \caption[Mass function: Effect of initial redshift]{\emph{Effect of initial redshift on mass funtion in scale-free model $n=0.0$} }
    \label{fig:a3:1}
\end{figure}
Generally, varying the initial redshifts can introduce significant systematic errors in the mass function \cite{Michaux_2020}. To address this, we used simulations with $n=0$, starting at $z_{\text{start}} = 100$, and compared the resulting mass function with that from a simulation starting at $z_{\text{start}} = 453$, keeping all other parameters identical. The results are presented in \cref{fig:a3:1}. The left and middle panels show the residuals of the mass function data relative to the best-fit Sheth-Tormen model, while the right panel shows a standard deviation contours of the mass function for the two cases. The solid dots within the contours represent the best-fit values for the two simulations. We observe no change in the best-fit Sheth-Tormen parameters or a standard deviation confidence levels. Additionally, both fits show no deviation from the best fit in the mass function computed using ten realisations.

\section{Effect of Box size}
\label{a3:2}

\begin{figure}[h]
    \centering
          \begin{subfigure}[b]{0.32\textwidth}
        \includegraphics[width=.99\textwidth]{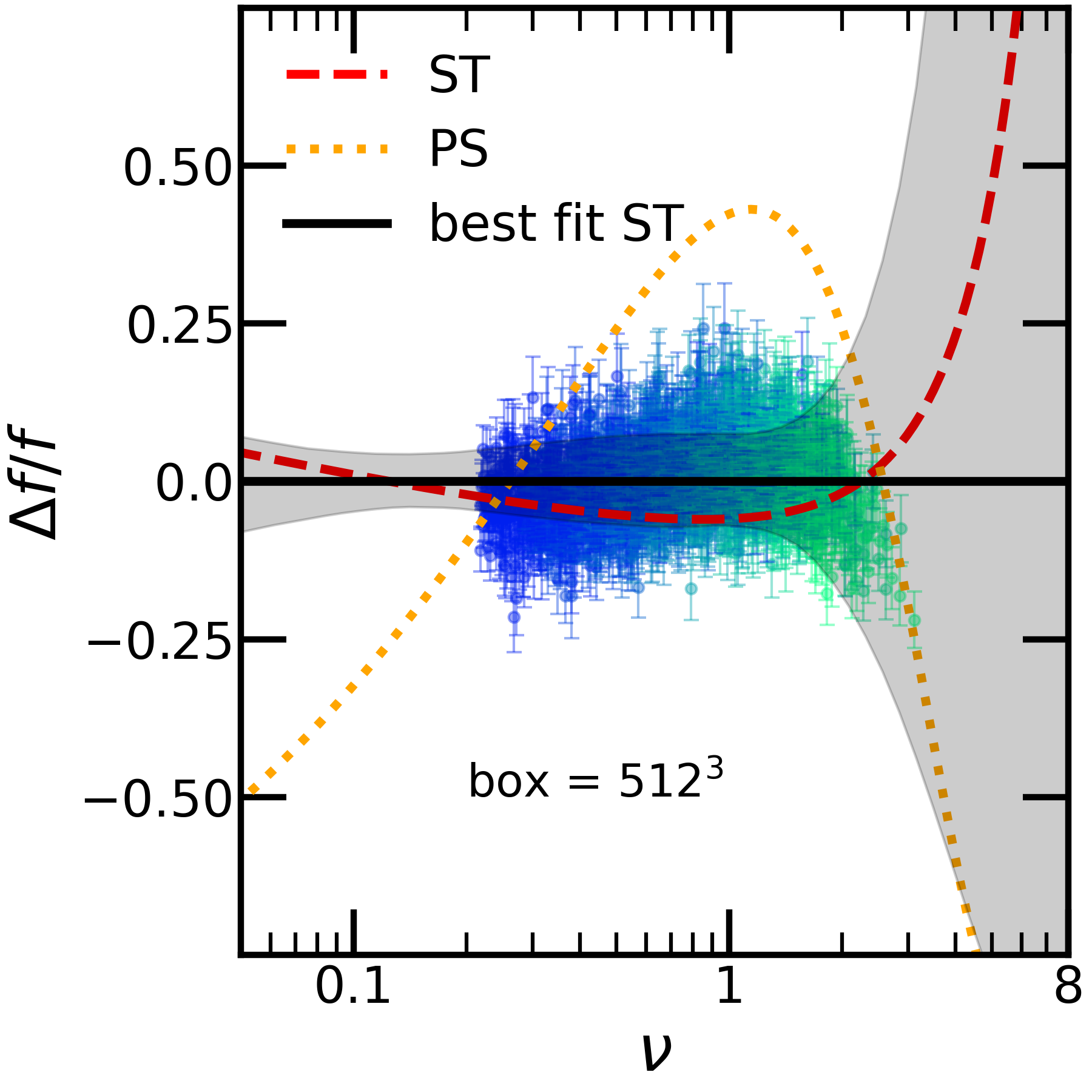}
        \end{subfigure}
        \begin{subfigure}[b]{0.32\textwidth}
        \includegraphics[width=.99\textwidth]{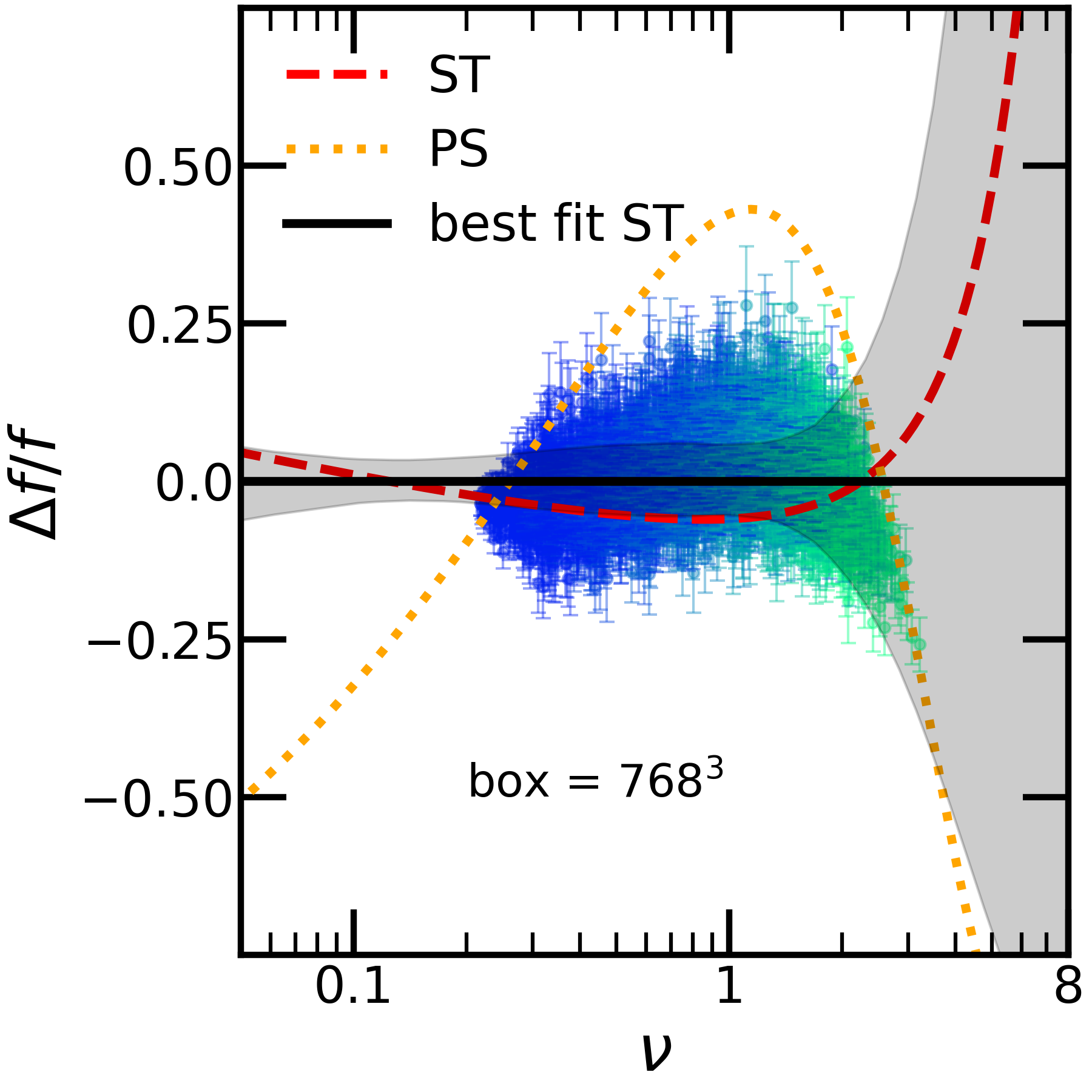}
        \end{subfigure}
        \begin{subfigure}[b]{0.32\textwidth}
        \includegraphics[width=.99\textwidth]{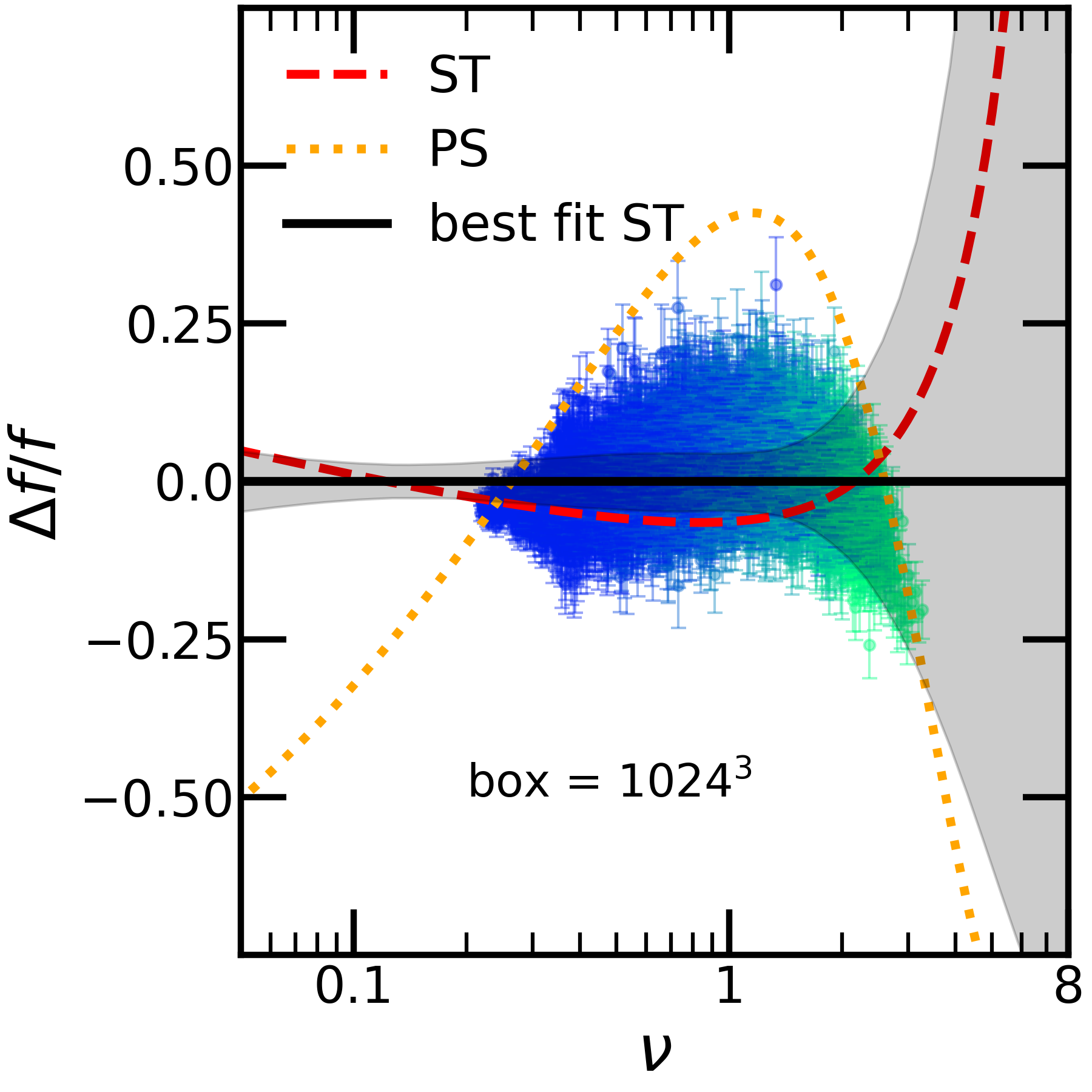}
        \end{subfigure}
        \begin{subfigure}[b]{0.32\textwidth}
        \includegraphics[width=.99\textwidth]{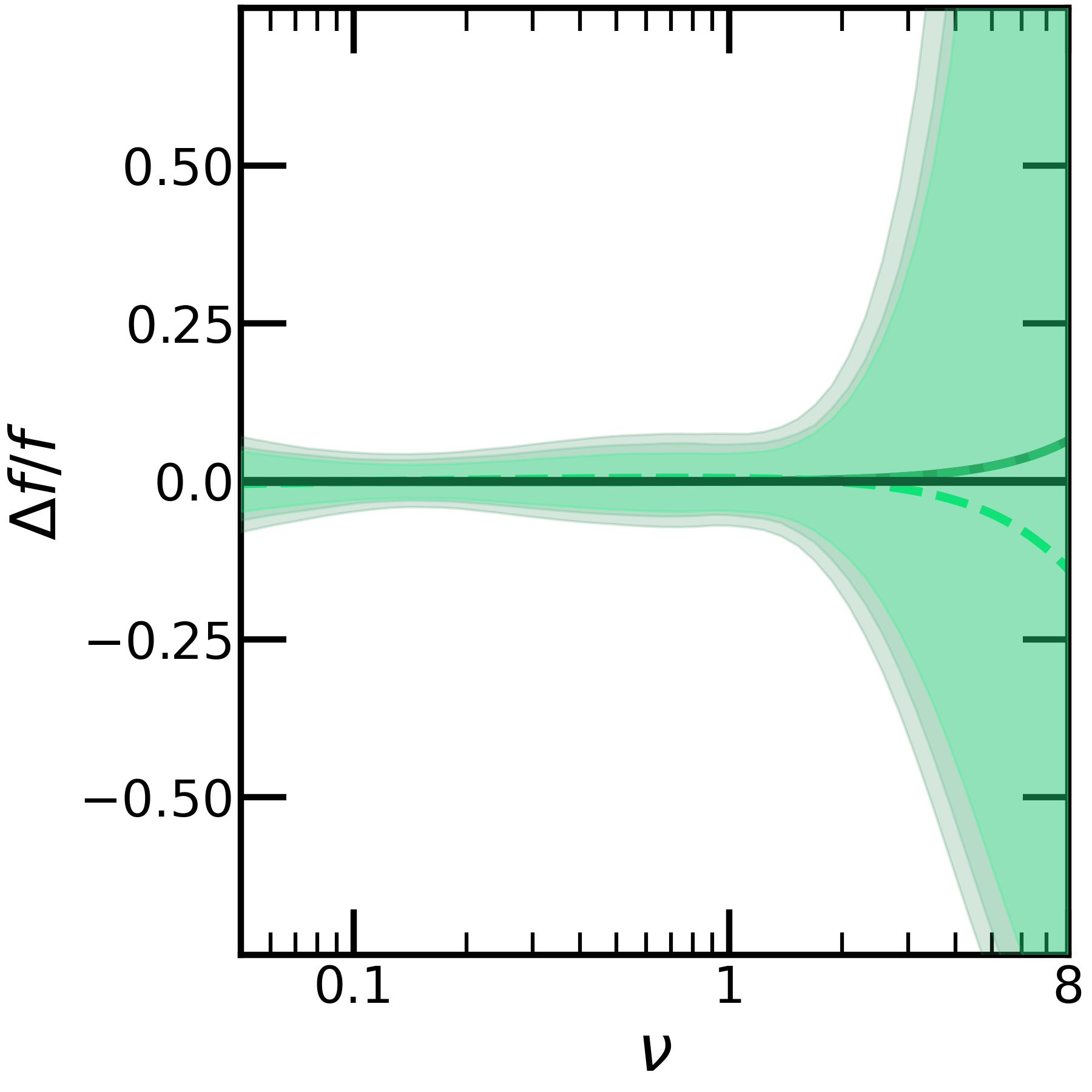}
        \end{subfigure}
        \begin{subfigure}[b]{0.32\textwidth}
        \includegraphics[width=.99\textwidth]{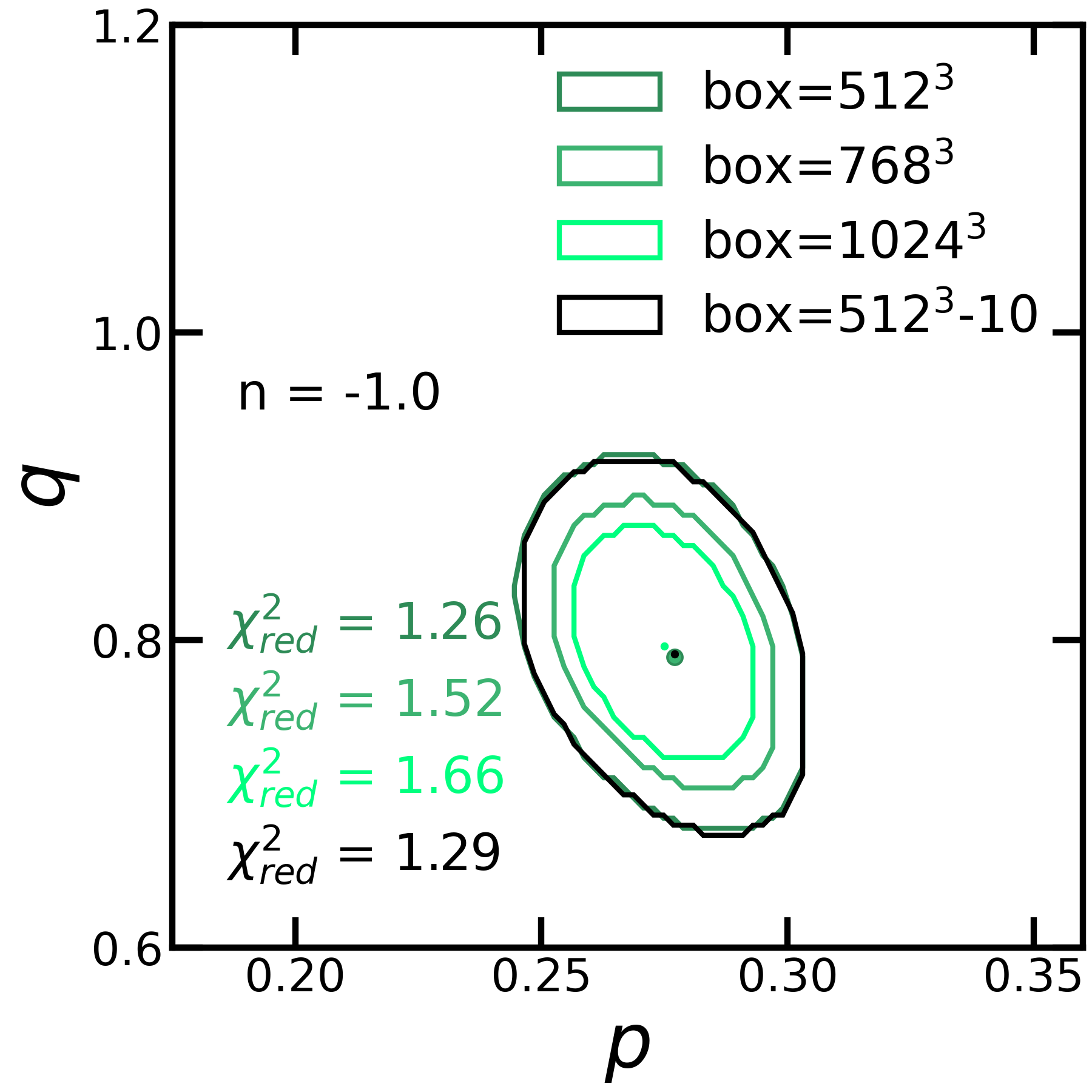}
        \end{subfigure}
        \begin{subfigure}[b]{0.32\textwidth}
        \includegraphics[width=.99\textwidth]{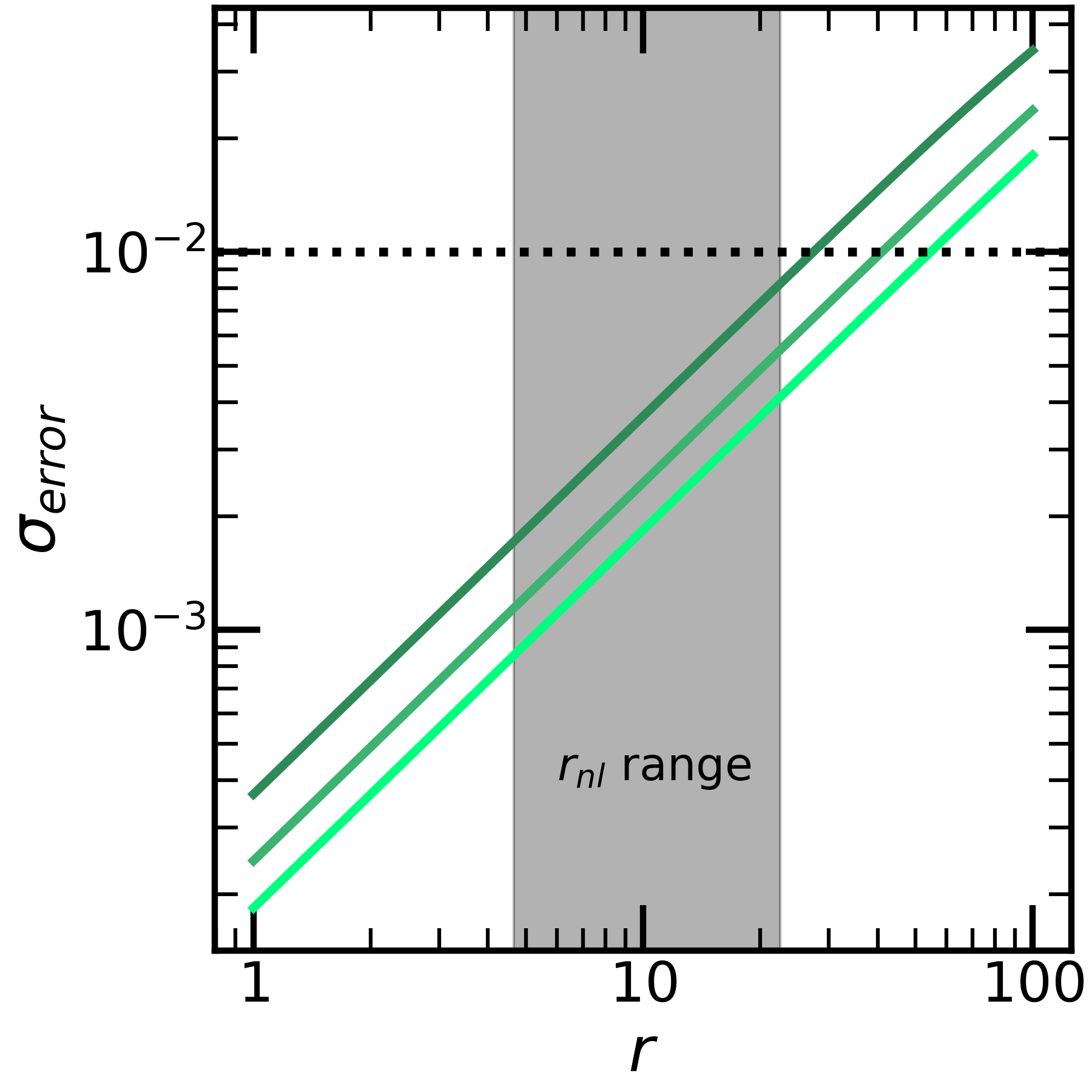}
        \end{subfigure}
    \caption[Mass function: Effect of Box size]{\emph{Effect of Box size on mass funtion in scale-free model $n=-1.0$} }
    \label{fig:a3:2}
\end{figure}
The small box sizes used for some power-law models might introduce systematic errors. However, because the power-law slopes used in these smaller boxes are relatively smaller, the impact of finite box effects is expected to be minimal \cite{Bagla_2006, Klypin_2019}. To ensure these effects are negligible, we compared three simulations with $n=-1$, using box sizes of $512^3$, $768^3$, and $1024^3$. All three simulations share the same parameters, differing only in box size and the number of particles ($N_{\text{part}} = \text{box size}$). The results are shown in \cref{fig:a3:2}. The top row of panels displays the residuals of the mass function data relative to the best-fit Sheth-Tormen model for box sizes $512^3$ (left), $768^3$ (middle), and $1024^3$ (right). The bottom left panel illustrates the deviations of the three fits relative to the best-fit (solid black) presented in \cref{sec:mf_result} which is based on ten realisations. The bottom middle panel shows a standard deviation contours for the fits, with the solid dots representing the best-fit values for the corresponding fits. We observe no significant change in the best-fit values, though a standard deviation interval narrows as the box size increases.

The decrease in a standard deviation interval is due to including larger scales in simulations with larger box sizes. The theoretical errors in the mass variance for the three simulations are shown in the bottom right panel of \cref{fig:a3:2}. This plot determines the upper limit of $r_{nl}$ for calculating the mass function. The gray area in the plot represents the $r_{nl}$ range based on a $512^3$ box. The same $r_{nl}$ range is applied across the three simulations. The upper limit of $r_{nl}$ is set to ensure that $\sigma_{\text{error}}$ remains below 1\%, while the lower limit is determined by technical considerations, ensuring that each $r_{nl}$ value includes at least 10 data points. The plot confirms that the chosen box sizes are sufficient for obtaining reliable results within the specified $r_{nl}$ range outlined in \cref{tab:sim_mf}.

\section{Effect of error-types} 
\label{a3:3}

\begin{figure}[h]
    \centering
          \begin{subfigure}[b]{0.32\textwidth}
        \includegraphics[width=.99\textwidth]{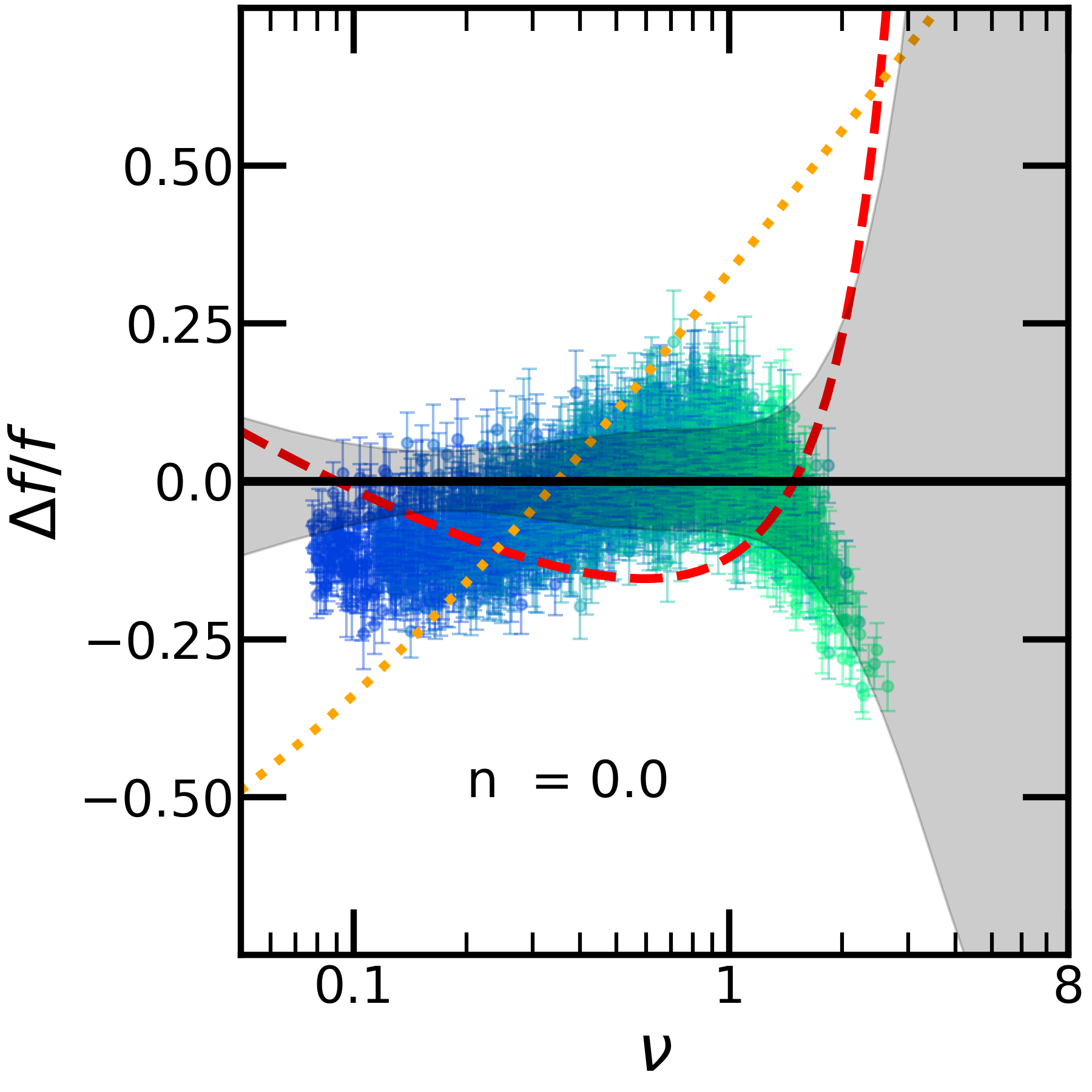}
        \end{subfigure}
        \begin{subfigure}[b]{0.32\textwidth}
        \includegraphics[width=.99\textwidth]{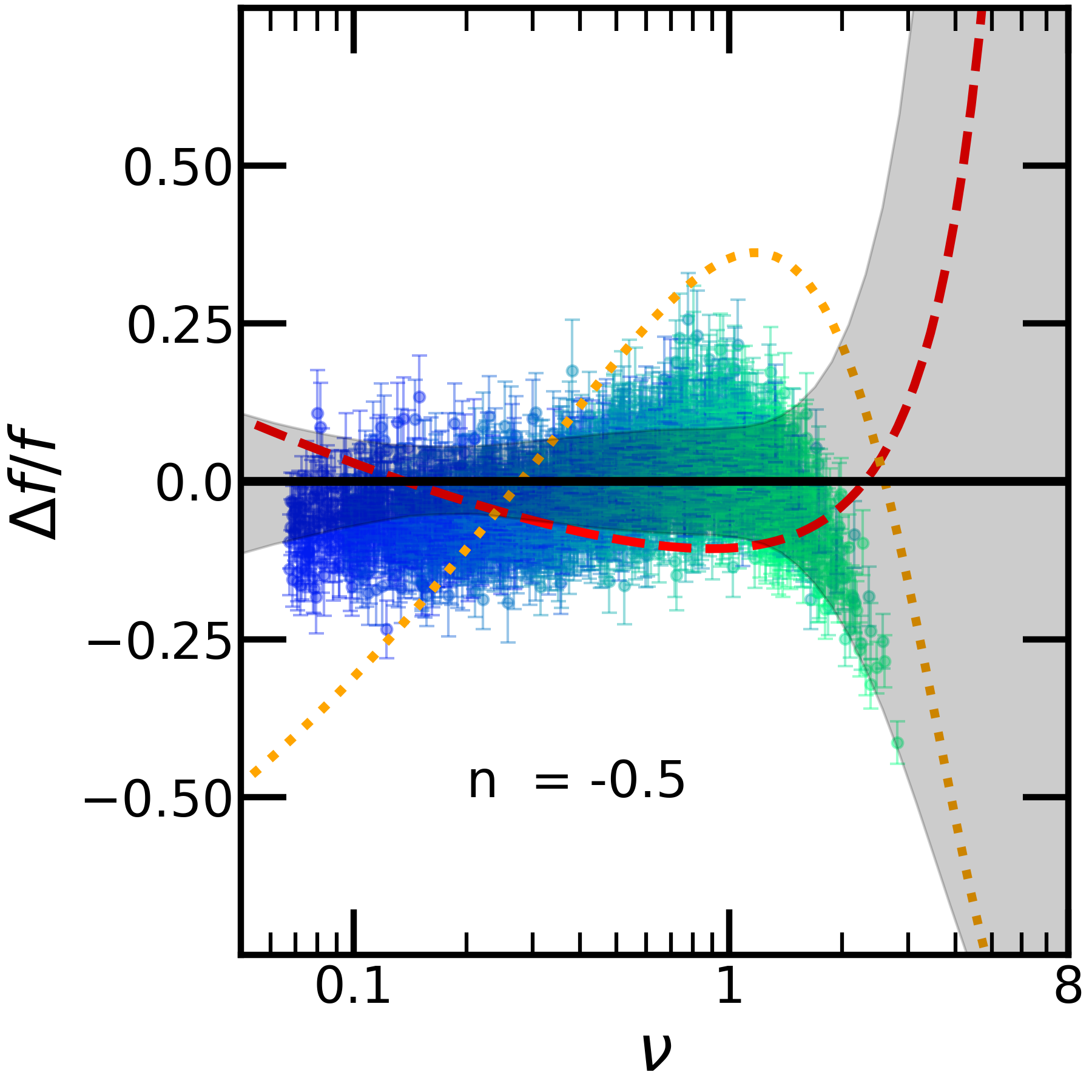}
        \end{subfigure}
        \begin{subfigure}[b]{0.32\textwidth}
        \includegraphics[width=.99\textwidth]{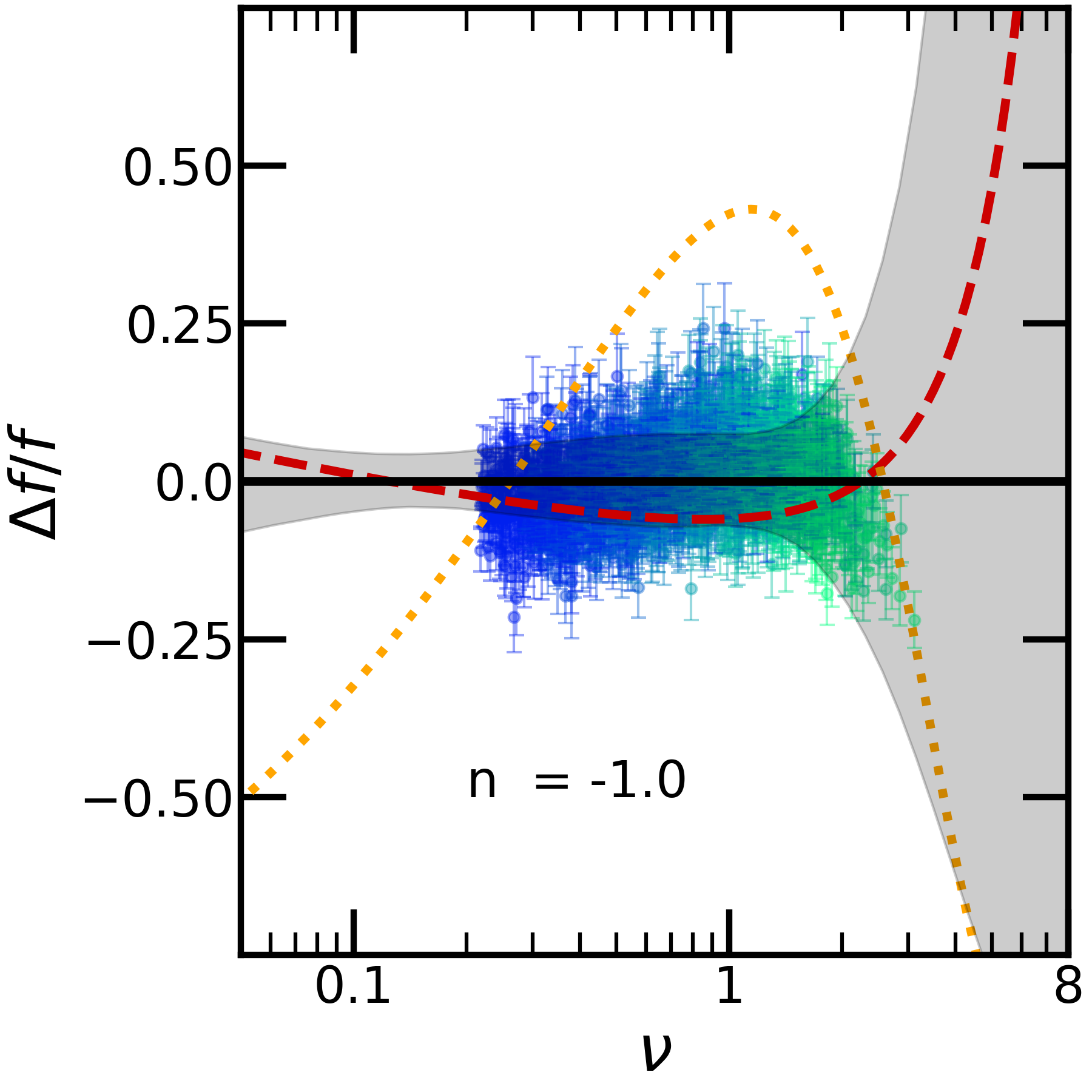}
        \end{subfigure}
    \caption[Mass function: Poisson error]{\emph{Mass function: Poisson error in scale-free models $n=0.0$,$-0.5$,$-1.0$}}
    \label{ps}
\end{figure}  

\begin{figure}[h]
    \centering
          \begin{subfigure}[b]{0.32\textwidth}
        \includegraphics[width=.99\textwidth]{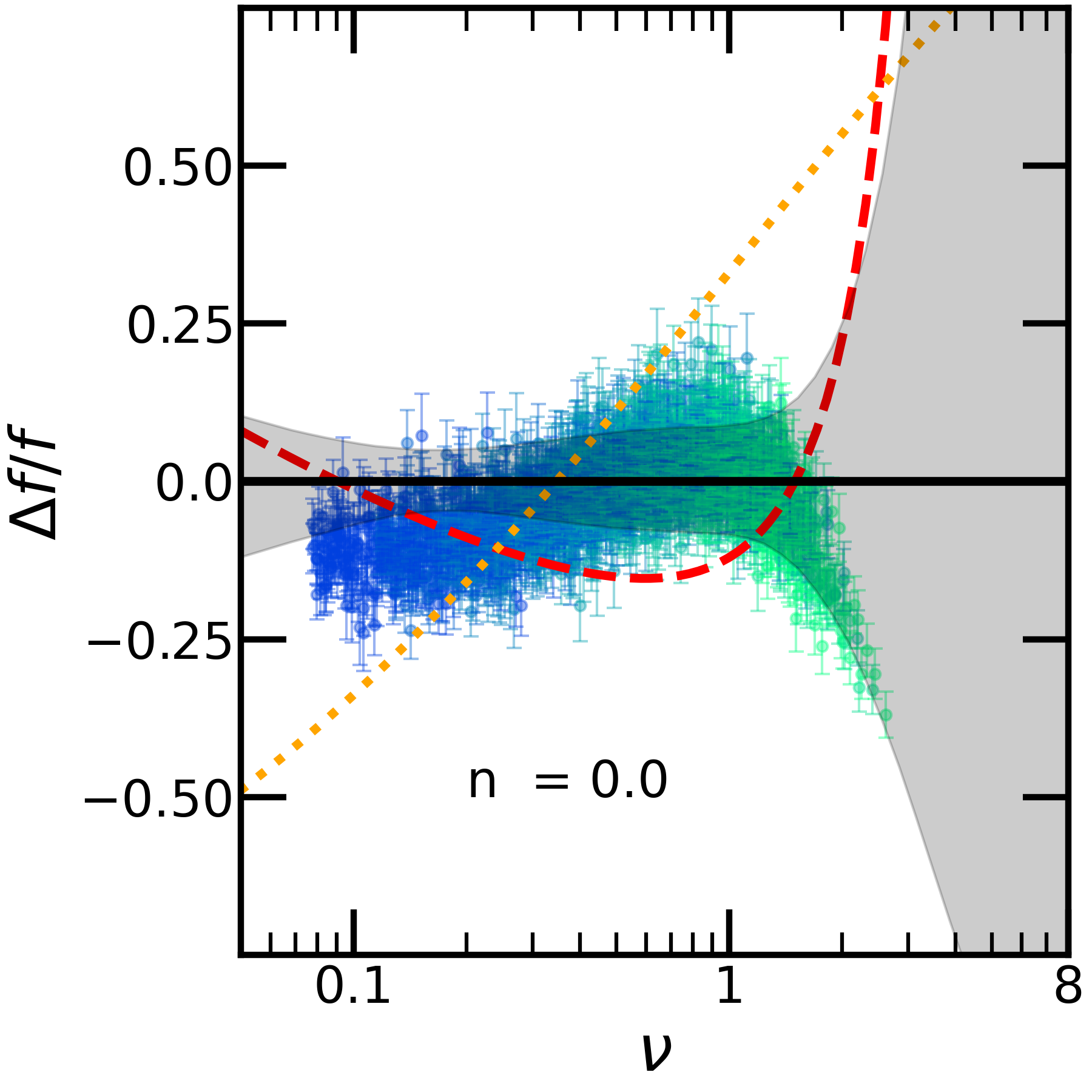}
        \end{subfigure}
        \begin{subfigure}[b]{0.32\textwidth}
        \includegraphics[width=.99\textwidth]{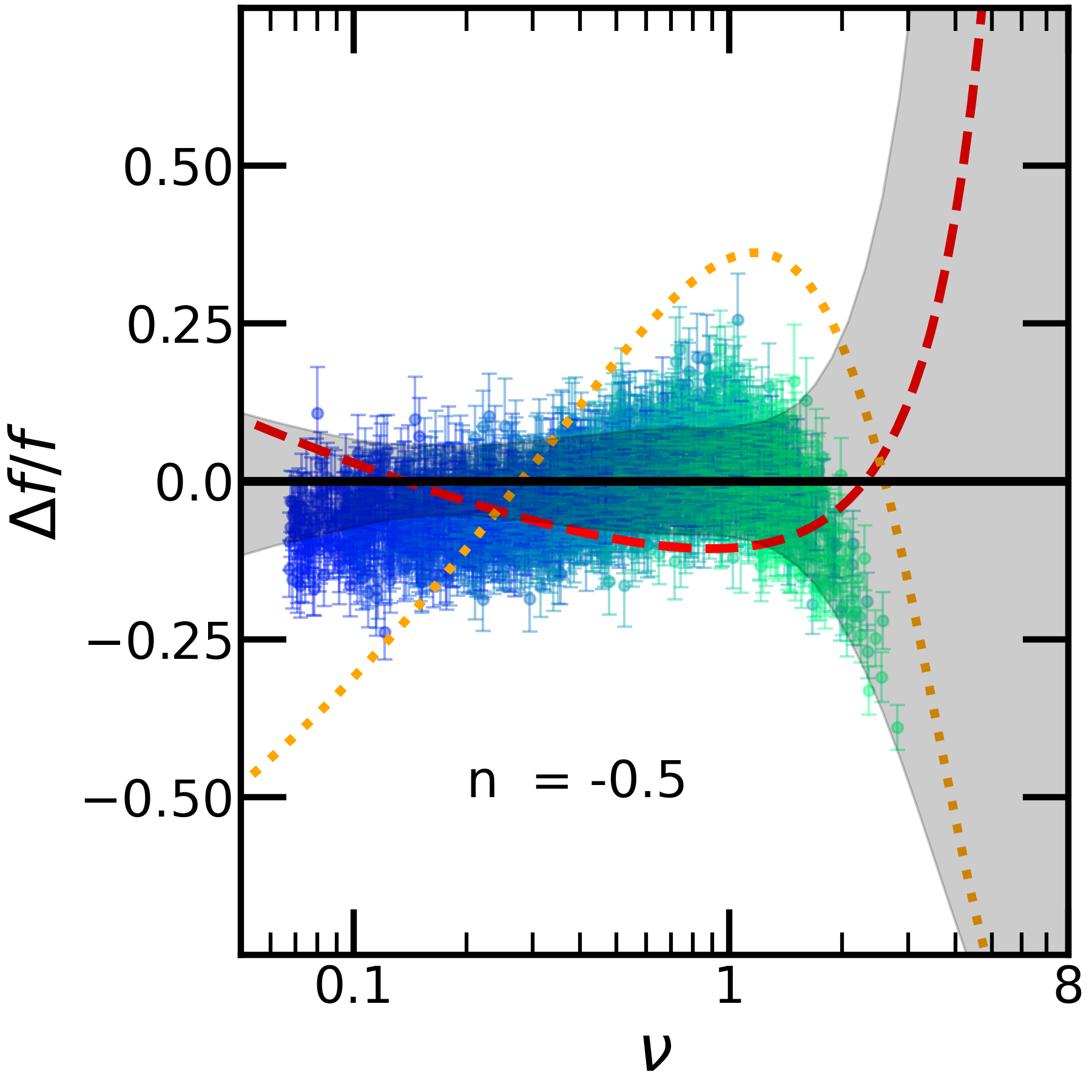}
        \end{subfigure}
        \begin{subfigure}[b]{0.32\textwidth}
        \includegraphics[width=.99\textwidth]{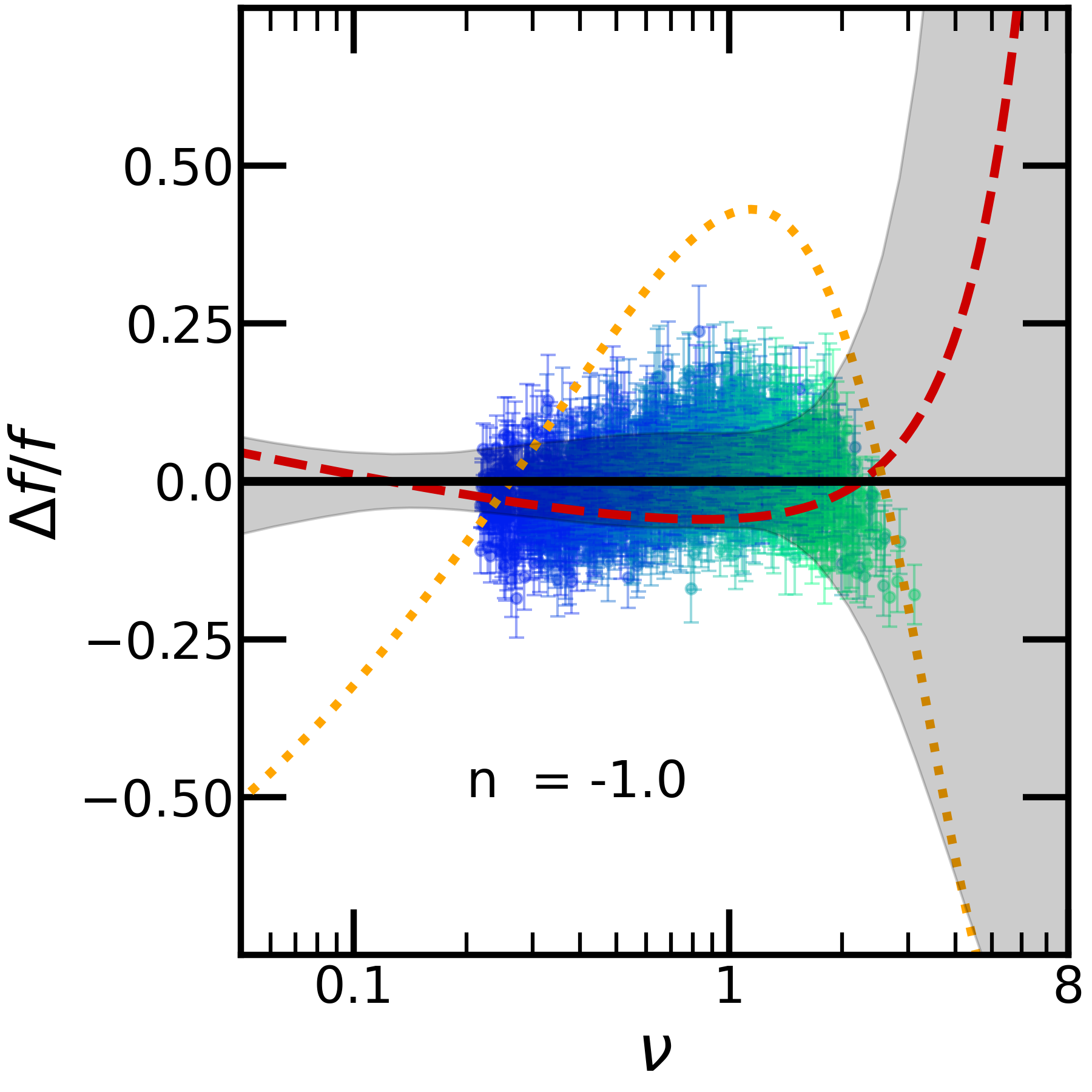}
        \end{subfigure}
    \caption[Mass function: Jackknife error]{\emph{Mass function: Jackknife error in scale-free models $n=0.0$,$-0.5$,$-1.0$}}
    \label{jk}
\end{figure}   

    \begin{figure}[h]
    \centering
          \begin{subfigure}[b]{0.32\textwidth}
        \includegraphics[width=.99\textwidth]{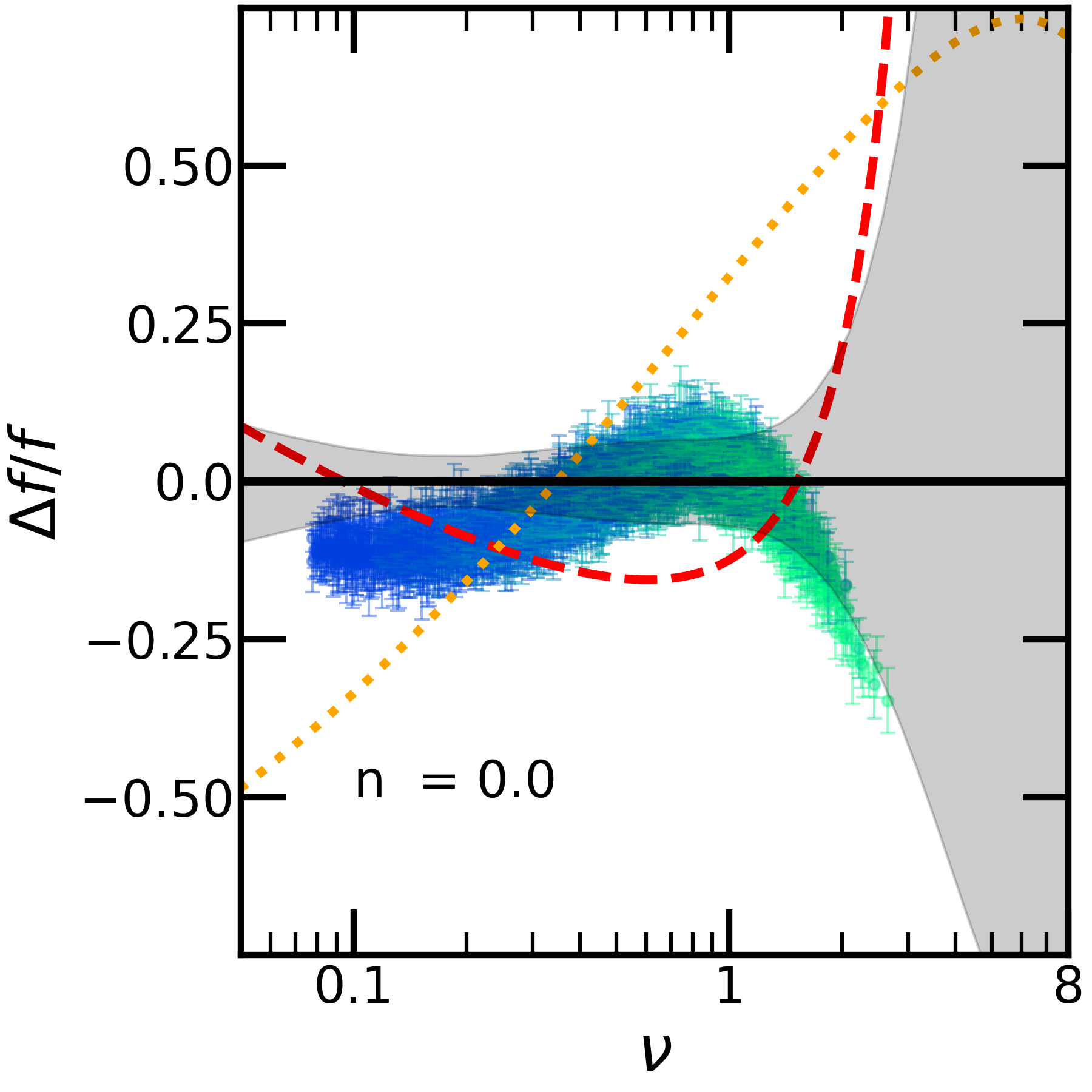}
        \end{subfigure}
        \begin{subfigure}[b]{0.32\textwidth}
        \includegraphics[width=.99\textwidth]{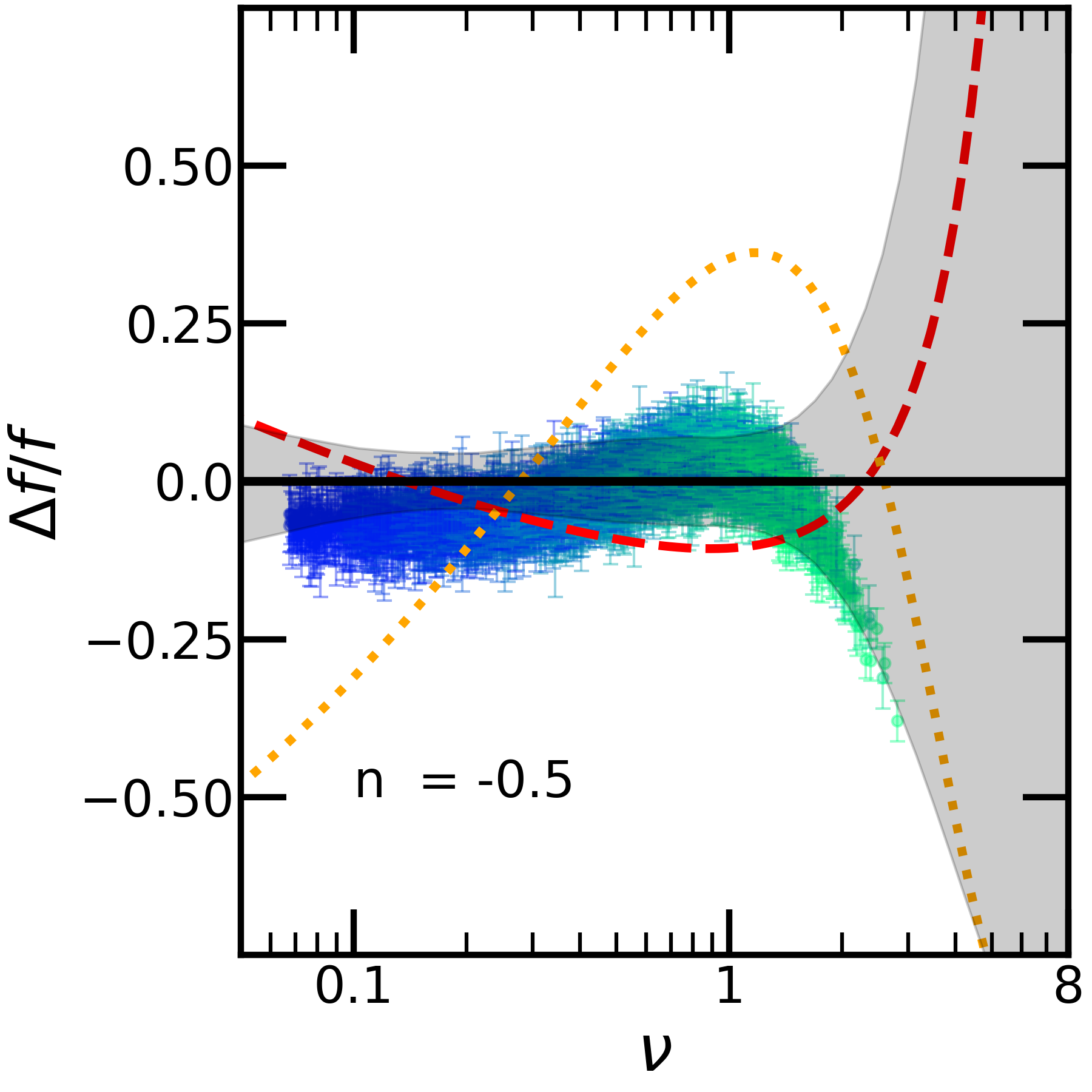}
        \end{subfigure}
        \begin{subfigure}[b]{0.32\textwidth}
        \includegraphics[width=.99\textwidth]{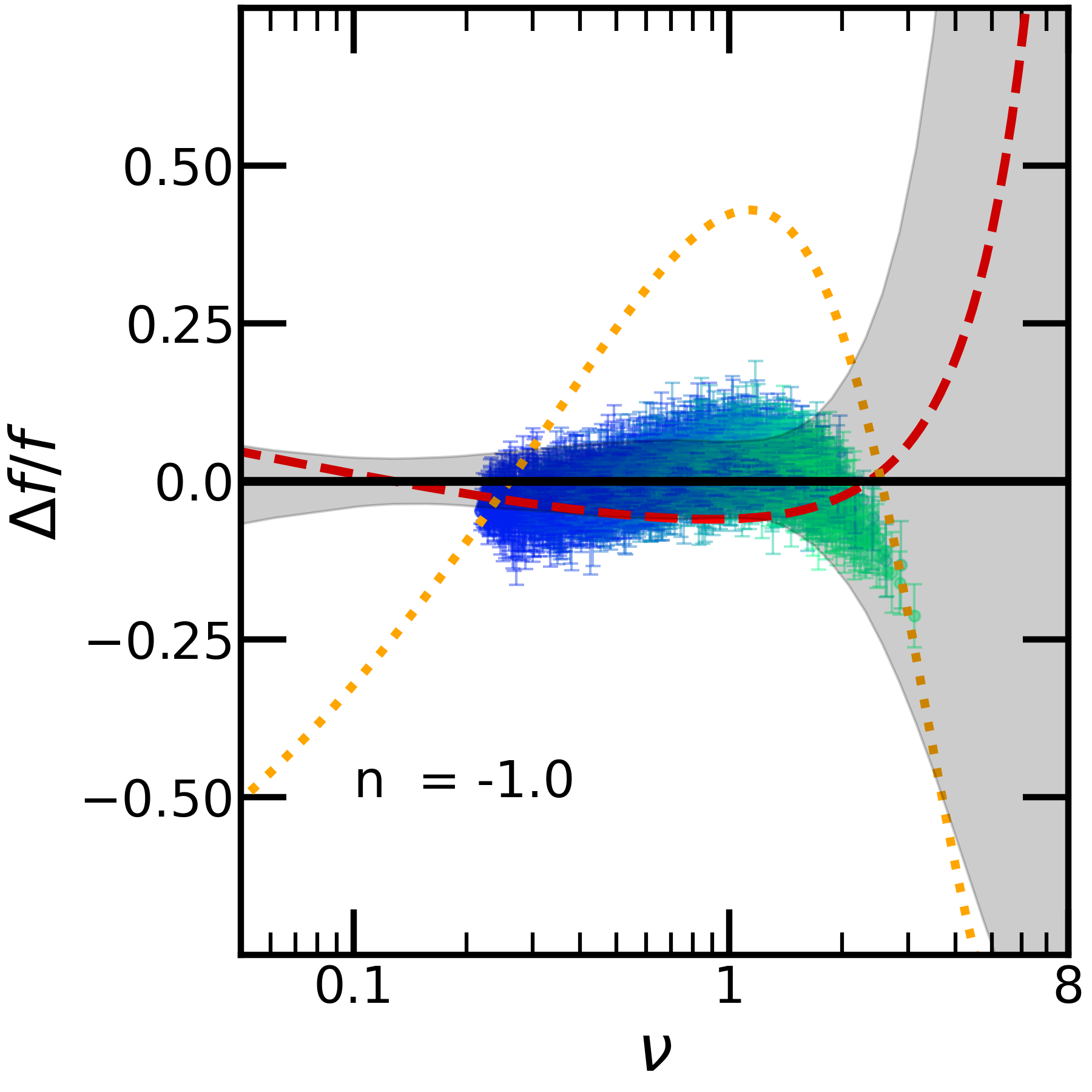}
        \end{subfigure}
    \caption[Mass function: Sample variance error]{\emph{Mass function: Sample variance error in scale-free models $n=0.0$,$-0.5$,$-1.0$}}
    \label{sv}
\end{figure}  

Poisson errors for halo counts in bins may under-estimate the errors at the low mass end of the halo mass function. An alternate estimate can be obtained using jackknife errors. We have conducted Jackknife error estimates for scale-free models. We divide a simulation volume into eight equal subvolumes to generate jackknife samples. Furthermore, as we have multiple realisations for three models with power law index $n=0$, $-0.5$, $-1.0$, we also estimated sample variance error and included them here. \Cref{jk}, \ref{sv}, and \ref{ps} depict the residue of the mass function data with the best-fit Sheth-Tormen for Poisson errors, Jackknife, and sample variance, respectively.

We have found that Poisson errors and Jackknife errors are comparable and do not offer any benefit over one another. Sample variance is more effective as it uses information from multiple realisations. However, we do not implement it because the additional computational time outweighs the fit improvement.

\begin{figure}[h]
    \centering
          \begin{subfigure}[b]{0.64\textwidth}
        \includegraphics[width=.99\textwidth]{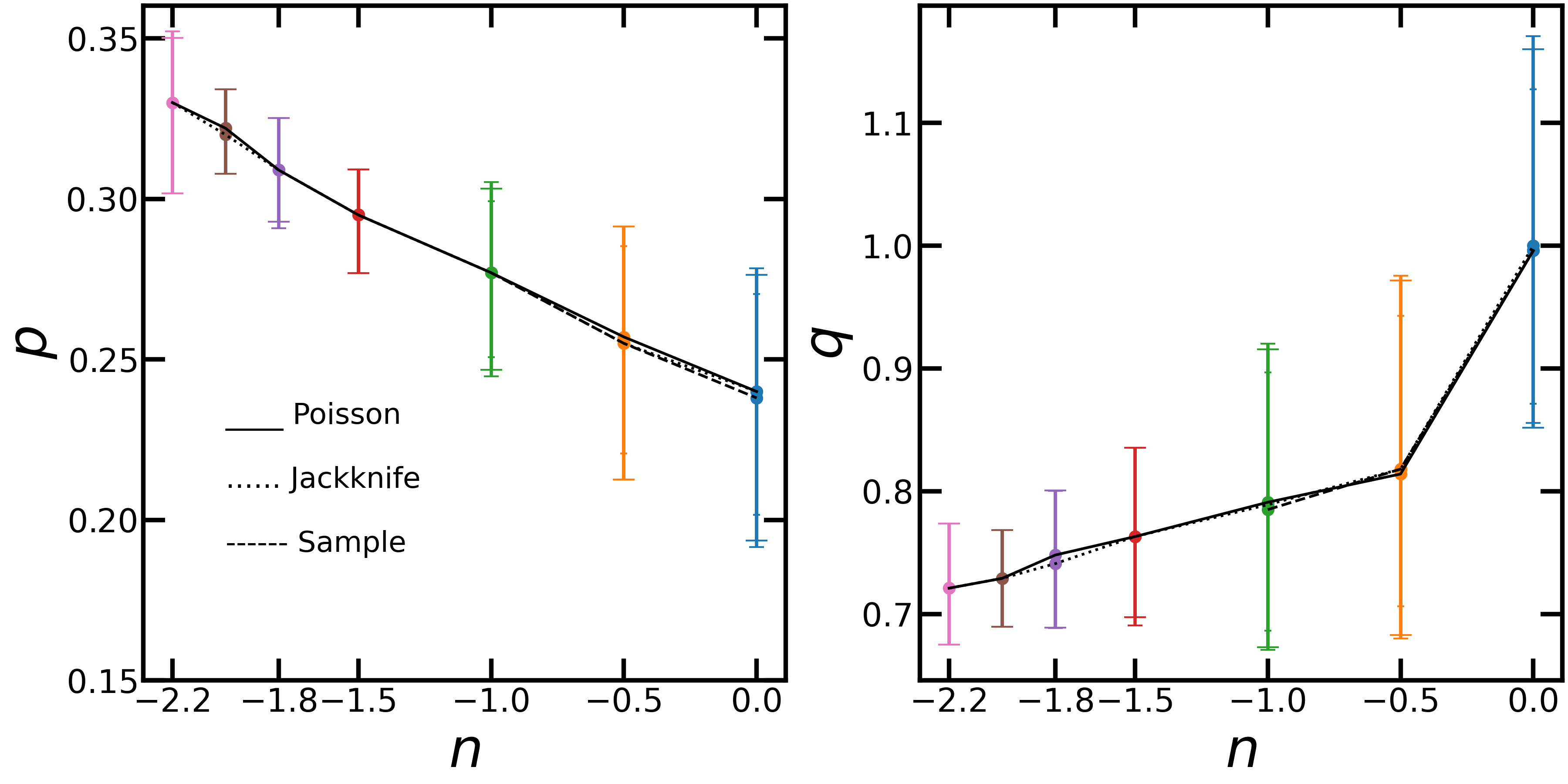}
        \end{subfigure}
        \begin{subfigure}[b]{0.32\textwidth}
        \includegraphics[width=.99\textwidth]{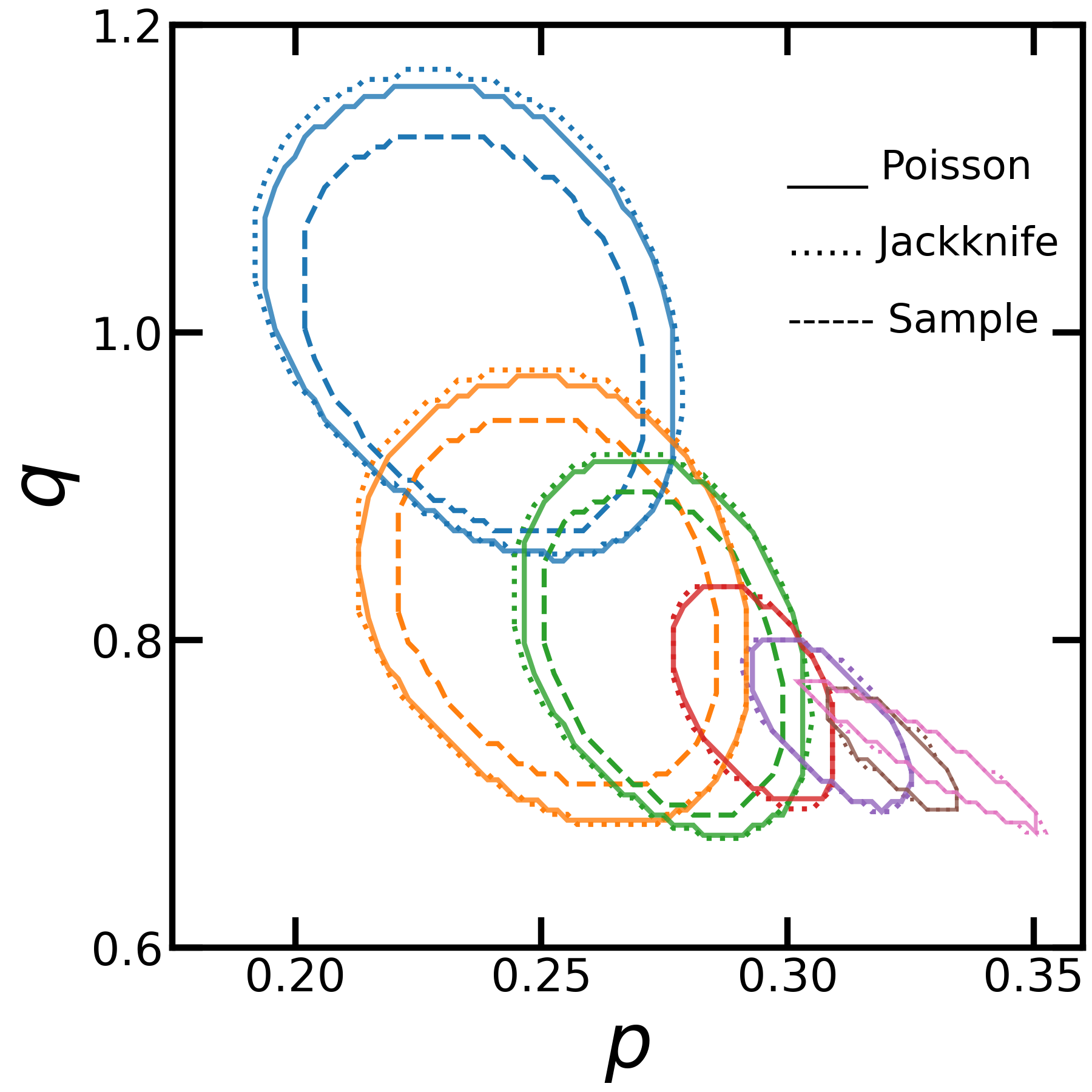}
        \end{subfigure}
    \caption[Mass function: Effect of error-types]{\emph{Mass function: Effect of error-types for all scale-free models}}
    \label{pqer}
\end{figure}

The \cref{pqer} shows the best-fit parameters of $p$ (left panel), $q$ (middle panel), and power law index $n$ in three cases, with the right panel displaying a standard deviation contours. Solid, dotted, and dashed lines represent Poisson, Jackknife, and Sample Variance errors. The analysis reveals no substantial change in the best-fit values of $p$ and $q$. A standard deviation contours for Poisson and Jackknife errors are relatively unchanged, but the Sample Variance contours are comparatively tighter. As a result, using Poisson errors for our purposes is sufficient.

\newpage
\section{Box-wise \texorpdfstring{$\Lambda$CDM}\ \ mass function fits}
\label{a3:4}

\begin{figure}[h]
    \centering
        \includegraphics[width=.99\textwidth]{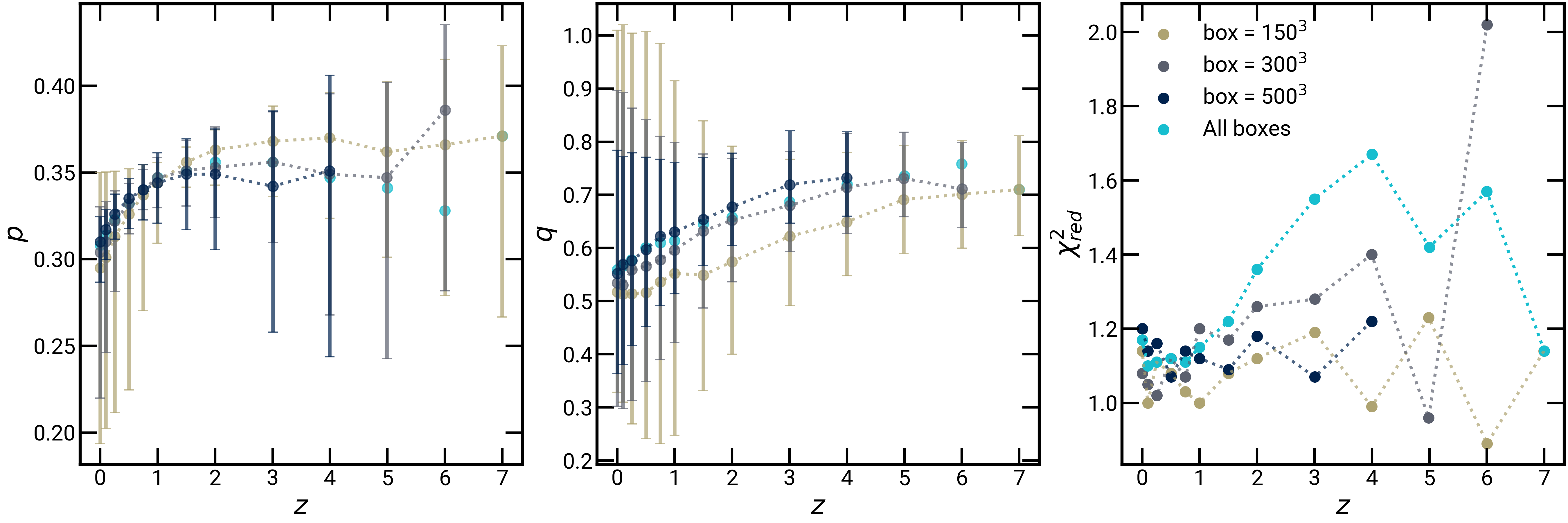}
    \caption[Box-wise $\Lambda$CDM mass function]{\emph{Box-wise $\Lambda$CDM mass function}}
    \label{box_lcdm}
    \end{figure}  
    
We present the box-wise mass function fit for $\Lambda$CDM models in \cref{box_lcdm}. We confirm that the mass function remains consistent within the error bars across different box sizes by fitting each box individually and comparing the results with a combined fit.


%% file: chapters/appendix4.tex
\chapter{$H_0$ measurements: supplementary analysis}
\label{a4}
In this chapter, we further analyse $H_0$ measurements using simulations referencing the effects of substructure and halo mass cuts on the mean and standard deviation of the distributions.

\section{Effect of substructure}
\begin{figure}[h]
    \centering
         \begin{subfigure}[b]{0.3\textwidth}
        \includegraphics[width=0.99\textwidth]{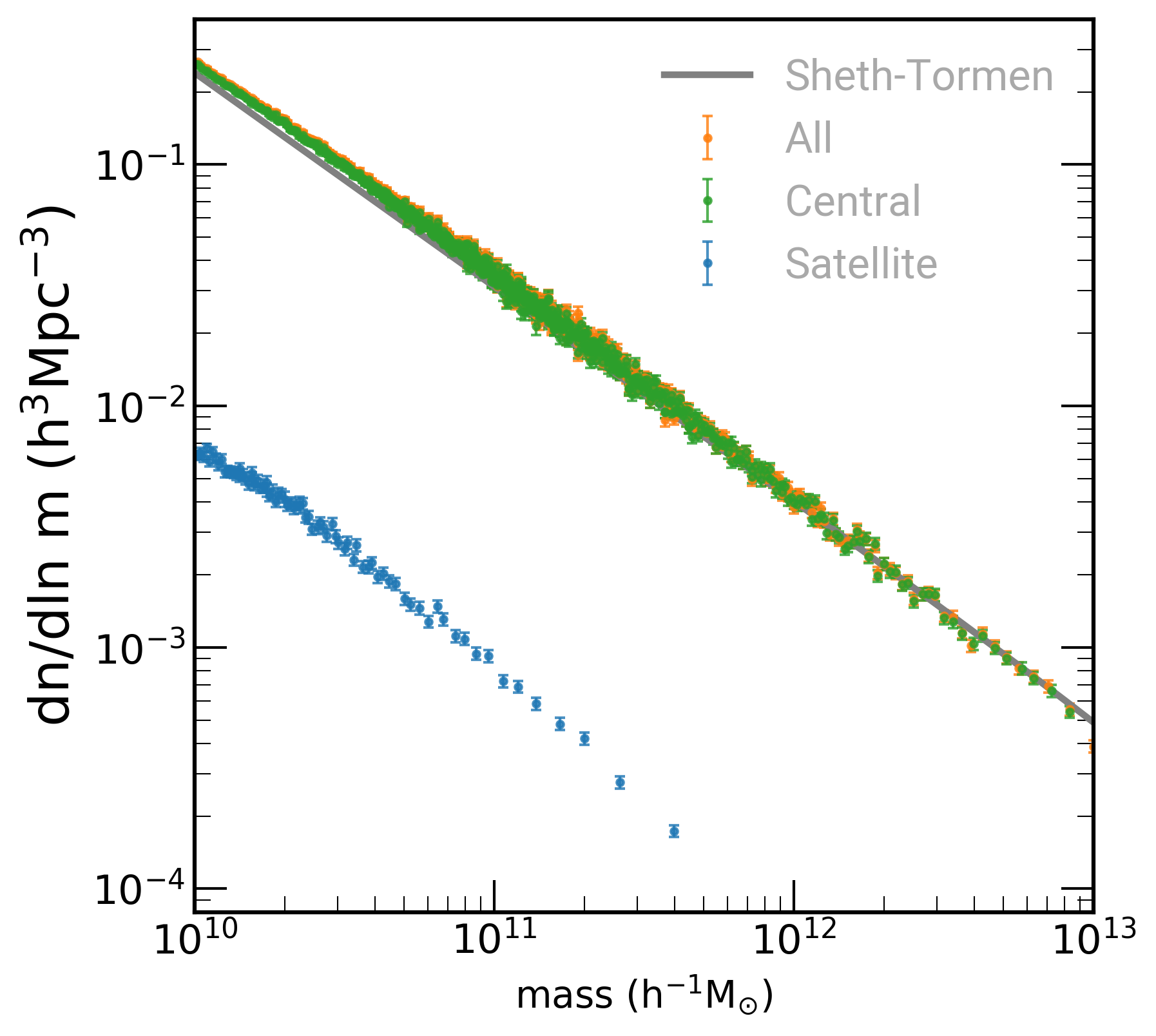}
        \end{subfigure}
         \begin{subfigure}[b]{0.68\textwidth}
        \includegraphics[width=0.99\textwidth]{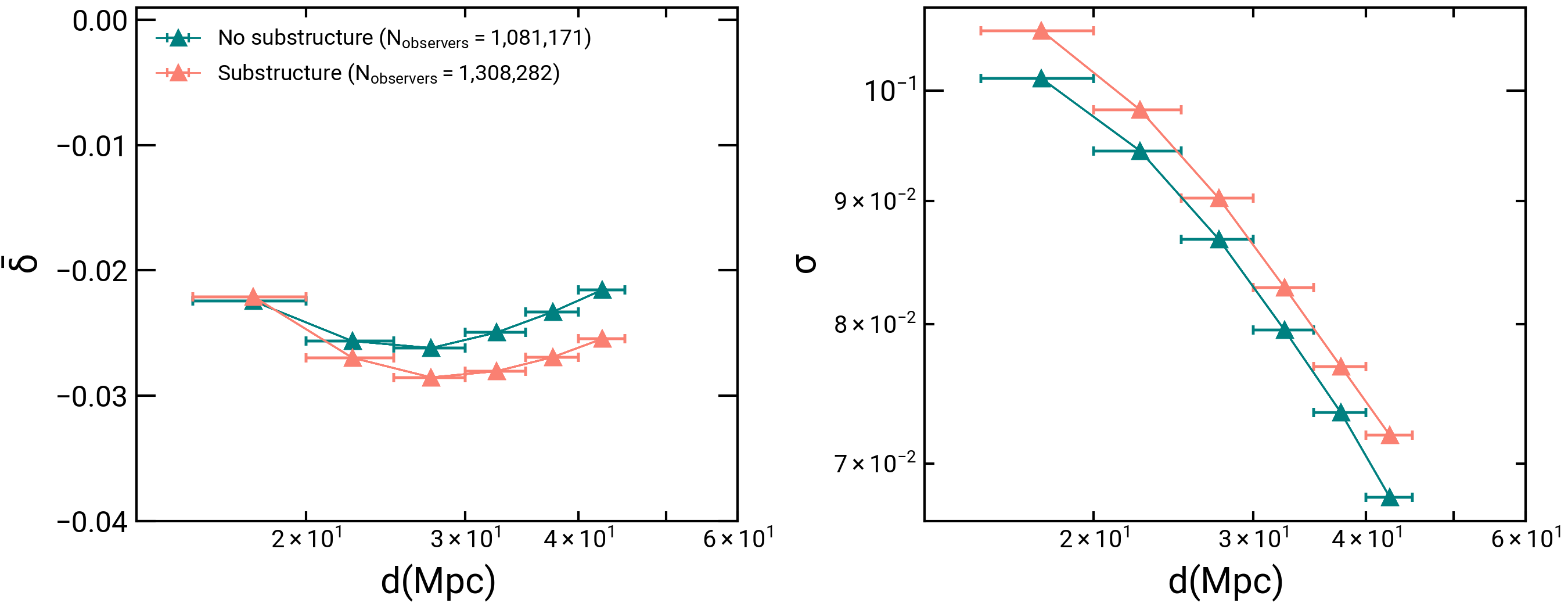}
        \end{subfigure}
    \caption[$H_0$ PDF: Effect of substructure]{\emph{$H_0$ PDF: Effect of substructure}}
\label{substr}
\end{figure}

We analyse the distributions by including substructures, using the $L_{box}$(Mpc/h) /$N_{part}$: 150/1024$^3$ simulation. The first panel of \cref{substr} shows the halo mass function for central and satellite halos in our halo catalogue. The substructure fraction generally increases with host halo mass but remains a minority ($\sim$10-30\%) of the total halo population in simulations \cite{Santos_2022,Gao_2010,Springel_2008}. The second and third panels show mean and standard deviations for five different shells with and without substructure. Including substructures increases the number of observers by 30\% compared to when substructures are not considered. This leads to a slight increase in the dispersion by a few percent.

\section{Convergence in the simulations}
\begin{figure}[h]
    \centering
        \begin{subfigure}[b]{0.9\textwidth}
        \includegraphics[width=0.99\textwidth]{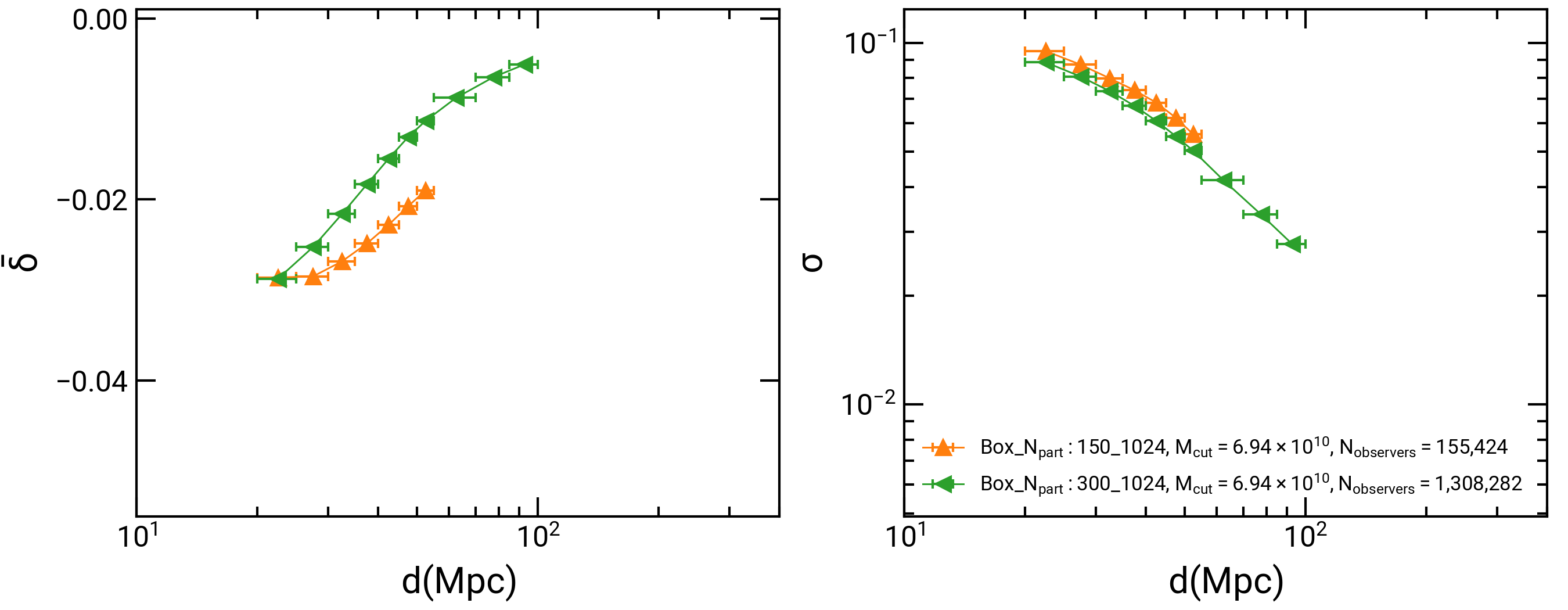}
        \end{subfigure}\\
        \begin{subfigure}[b]{0.9\textwidth}
        \includegraphics[width=0.99\textwidth]{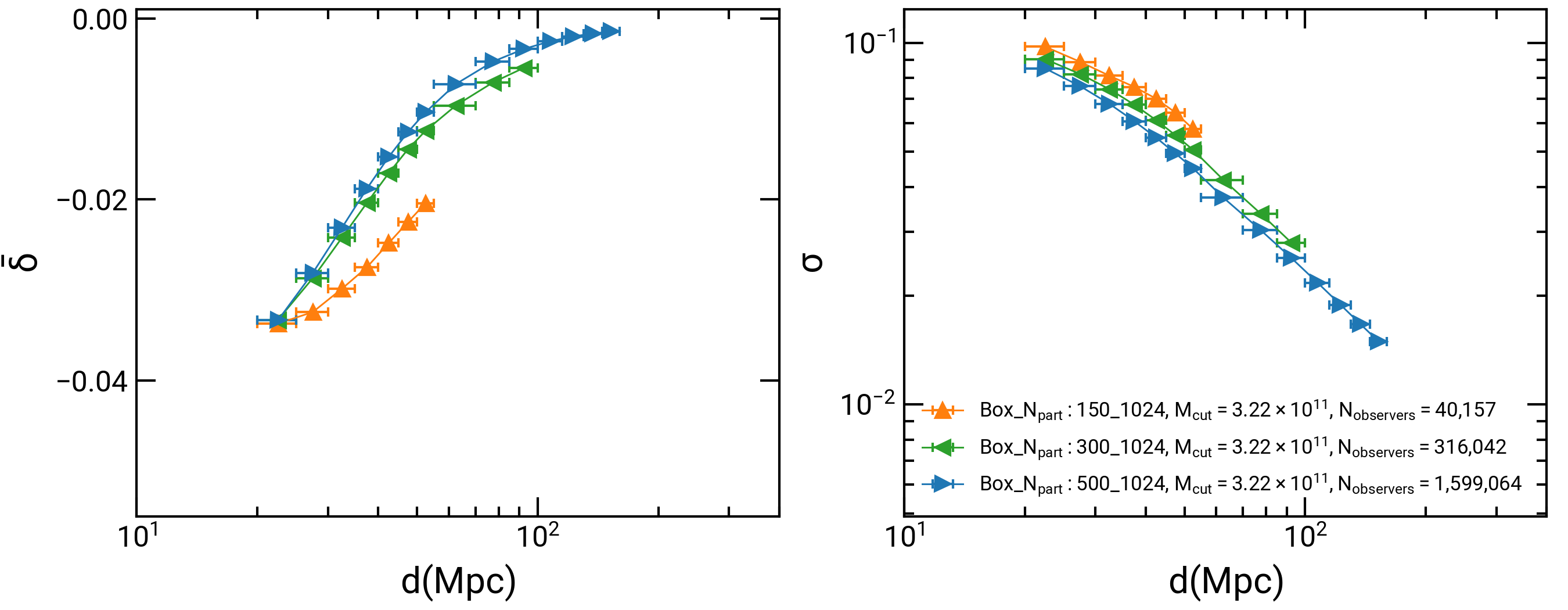}
        \end{subfigure}\\
        \begin{subfigure}[b]{0.9\textwidth}
        \includegraphics[width=0.99\textwidth]{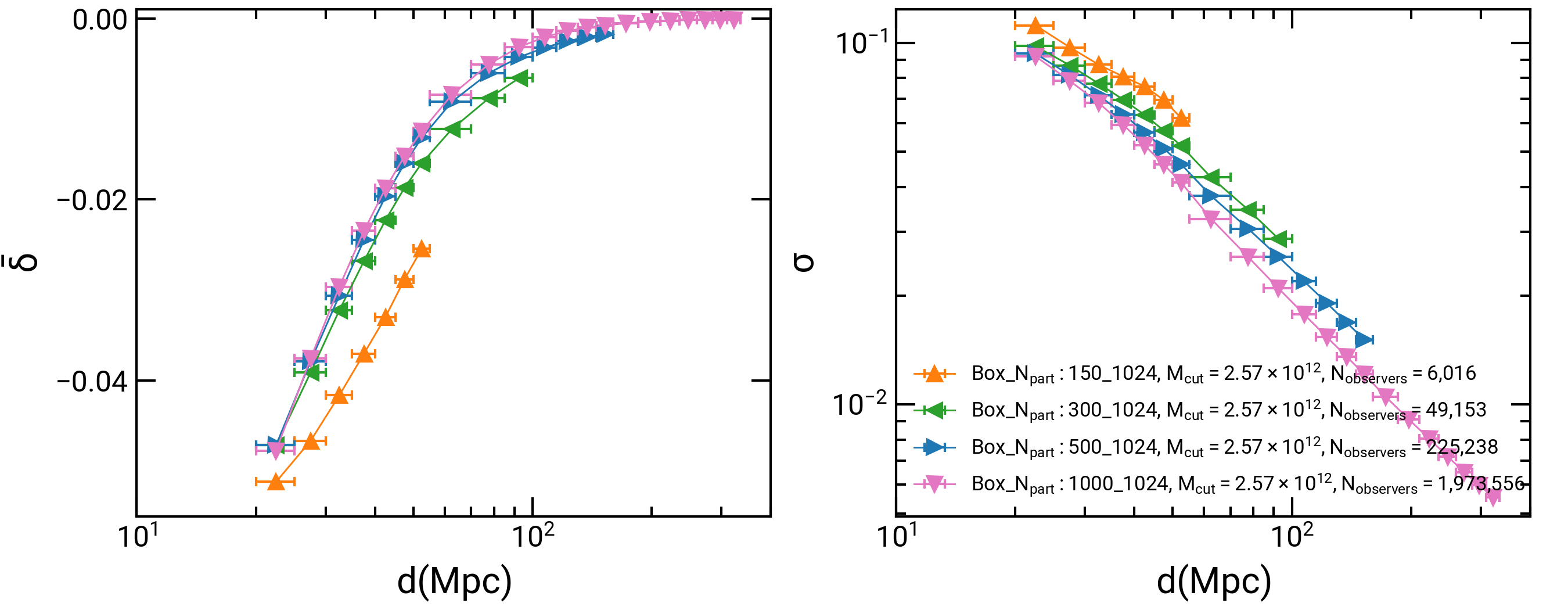}
        \end{subfigure}
        \caption[$H_0$ PDF: Convegence in the simulations]{\emph{$H_0$ PDF: Convegence in the simulations}}
        \label{stat_conv}
\end{figure}

We analyse the distributions by applying different mass cuts as a proxy for convergence study. We use four simulations with $L_{box}$(Mpc/h)/$N_{part}$: 150/1024$^3$, 300/1024$^3$, 500/1024$^3$ and 1000/1024$^3$, with corresponding lowest halo masses $(M_{\odot}$/h) of 8.68 $\times$ 10$^9$, 6.94 $\times$ 10$^{10}$, 3.22 $\times$ 10$^{11}$ and 2.57 $\times$ 10$^{12}$. These lowest halo masses are used as mass cuts to compute the distributions, which are shown in the \cref{stat_conv}. 

In the first row, we use a mass cut of $300$ Mpc/h box simulation to the $300$ and $150$ Mpc/h box simulations, considering distance bins up to one-third of the box size. Similarly, the second row compares a mass cut from the $500$ Mpc/h box simulation to the $500$, $300$, and $150$ Mpc/h box simulations. The last row shows the mass cut from the $1000$ Mpc/h box simulation, compared to the $1000$, $500$, $300$, and $150$ Mpc/h box simulations. The figure demonstrates that a larger box size reduces the mean and standard deviations offset significantly for the same mass cut. Additionally, the offset in both quantities increases with distance. We verify that the sample size of a number of observers ($N_{\text{observers}}$) does not alter the results. 

%% file: chapters/appendix5.tex
\chapter{Simulations in \cref{cha:5}}
\label{a5}

In this chapter, we describe the simulations used in \cref{cha:5} and \cref{fig:th1}. We also demonstrate the convergence of the mass function computed in these simulations.

\begin{table}[h]
	\begin{center}
	 \begin{tabular}{|c c c c c c c|} 
	 \hline 
	 n & $z_{\text{start}}$ & $N_{\text{box}}$ & $\sqrt[3]{N_{\text{part}}}$  & $r_{nl}^{\text{sim}}$ & $r_{nl}^{\text{mf}}$ & $r_{nl}^{\text{max}}$ \\ [0.5ex] 
	 \hline
	 -1.0 & 160 & 512 & 512  & 1.25-49.4& 1.265, 1.631, 2.121 & 27.4 \\[0.25ex]
	 -1.0 & 160 & 768 & 768  & 1.25-49.4& 1.265, 1.631, 2.121 & 27.4 \\[0.25ex]
	 -1.0 & 160 & 1024 & 768  & 1.25-49.4& 1.631, 2.121, 2.757 & 27.4 \\[0.25ex]
	 -1.0 & 160 & 1024 & 1024  & 1.25-49.4& 1.265, 1.631, 2.121 & 27.4 \\[0.25ex]
	 -1.8 & 70 & 1024 & 1024  & 1.25-49.4& 1.265, 1.631, 2.121 & 14.5 \\[0.25ex]
	 -2.2 & 46 & 1536 & 1536  & 0.2-10.24& 0.965, 1.265, 1.631 & 4.4 \\[0.25ex]
	 \hline
	 \end{tabular}
	 \caption[Simulations in \cref{cha:5}]{{\emph{Simulation Setup}:} Column 1: Power law power spectrum index for the model, Column 2: Initial redshift used to start the simulation, Column 3: Side length of the cubical simulation box, Column 4: Cube root of the total number of particles put in the simulation, Column 5: Range of the scale of non-linearity($r_{nl}$) covered in the simulation, Column 6: $r_{nl}$s used to compute the mass function, Column 8: Maximum limit on $r_{nl}$ considering the finite box size effect.}
     \label{tab:sim}
	\end{center}
\end{table}
    
We use six dark-matter-only simulations initialised with three power-law power spectrum conditions. Additional details about the simulations are provided in~\cref{tab:sim} and \cref{ssec:pl_cat}.

\begin{figure}
\begin{center}
\begin{subfigure}[b]{0.32\textwidth}
    \includegraphics[width=0.99\textwidth]{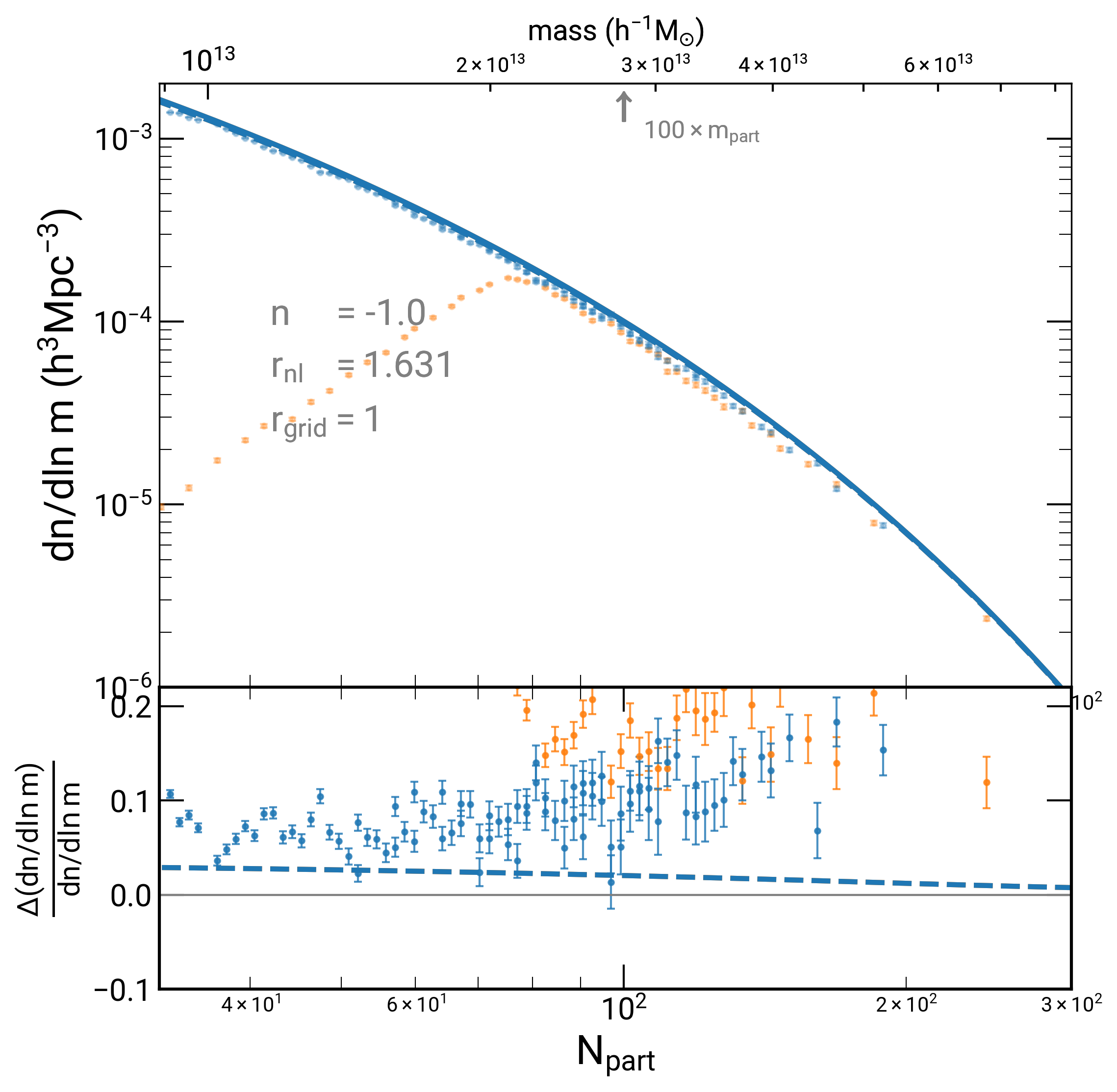}
\end{subfigure} 
\begin{subfigure}[b]{0.32\textwidth}
    \includegraphics[width=0.99\textwidth]{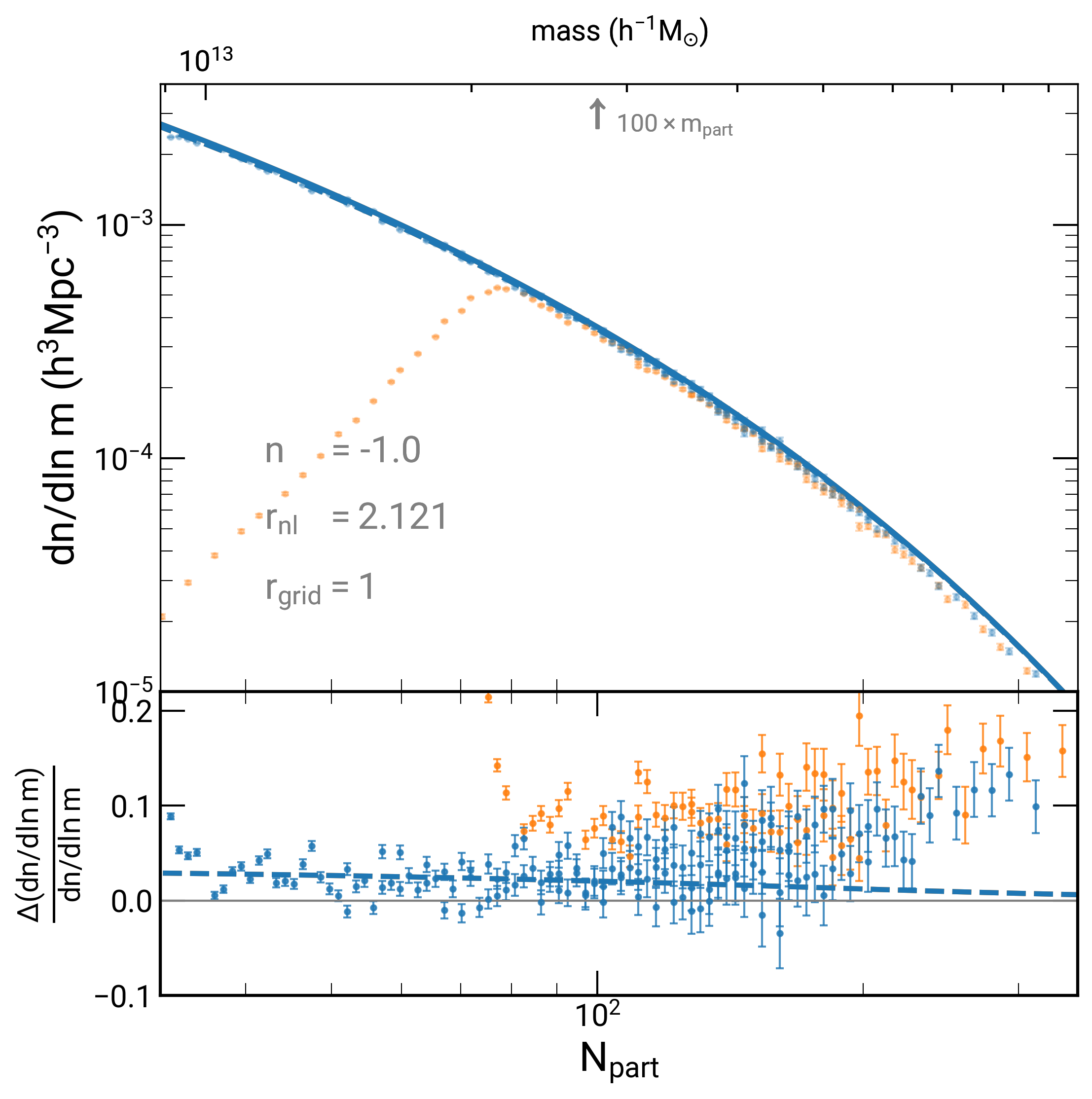}
\end{subfigure} 
\begin{subfigure}[b]{0.32\textwidth}
    \includegraphics[width=0.99\textwidth]{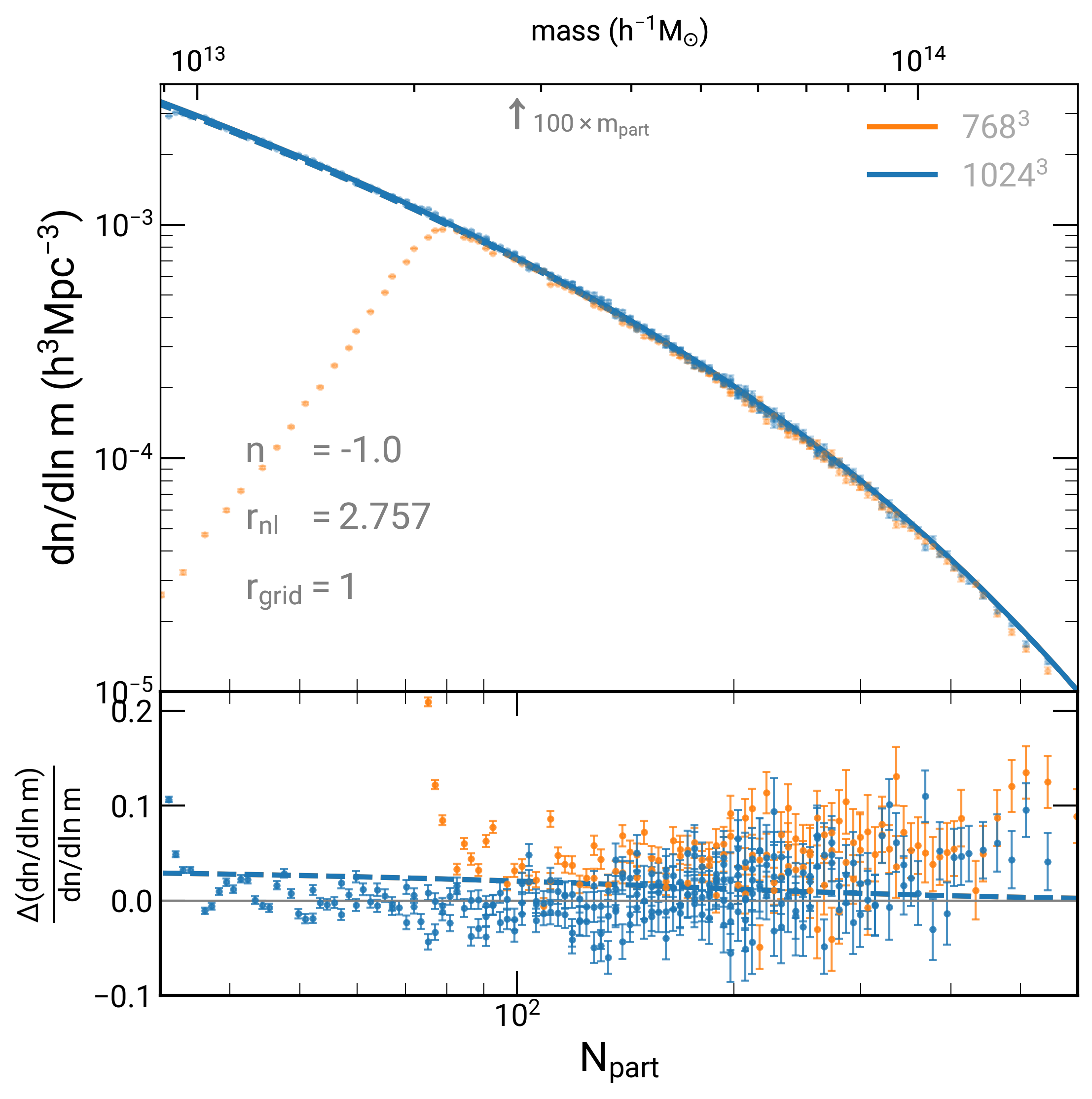}
\end{subfigure} 
\caption[Mass function: Effect of finite mass resolution]{ \emph{Mass function: Effect of finite mass resolution:} Each panel shows the halo mass function and fractional errors in mass function in the simulation for the $n=-1.0$ model with two values of $N_{\text{part}}= 768^3$ (orange), and $1024^3$ (blue) of box size $1024^3$ shown in Table \ref{tab:sim}. Data points represent simulation, whereas solid lines show theoretical mass function. Three panels represent different scales on non-linearity ($r_{nl}$).}
\label{fig:sim_conv1}
\end{center}
\end{figure}

\begin{figure}
\begin{center}
\begin{subfigure}[b]{0.32\textwidth}
    \includegraphics[width=0.99\textwidth]{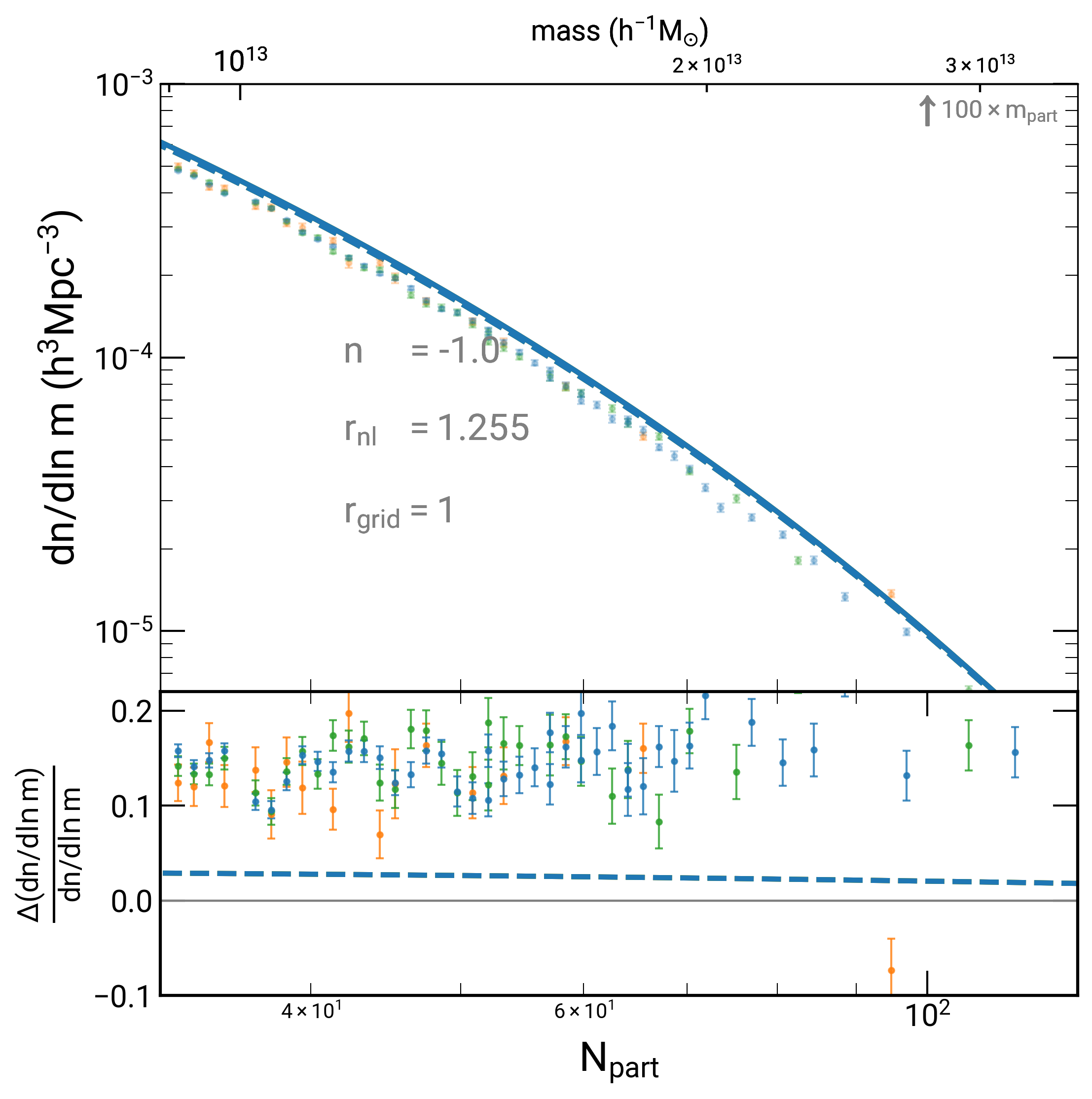}
\end{subfigure} 
\begin{subfigure}[b]{0.32\textwidth}
    \includegraphics[width=0.99\textwidth]{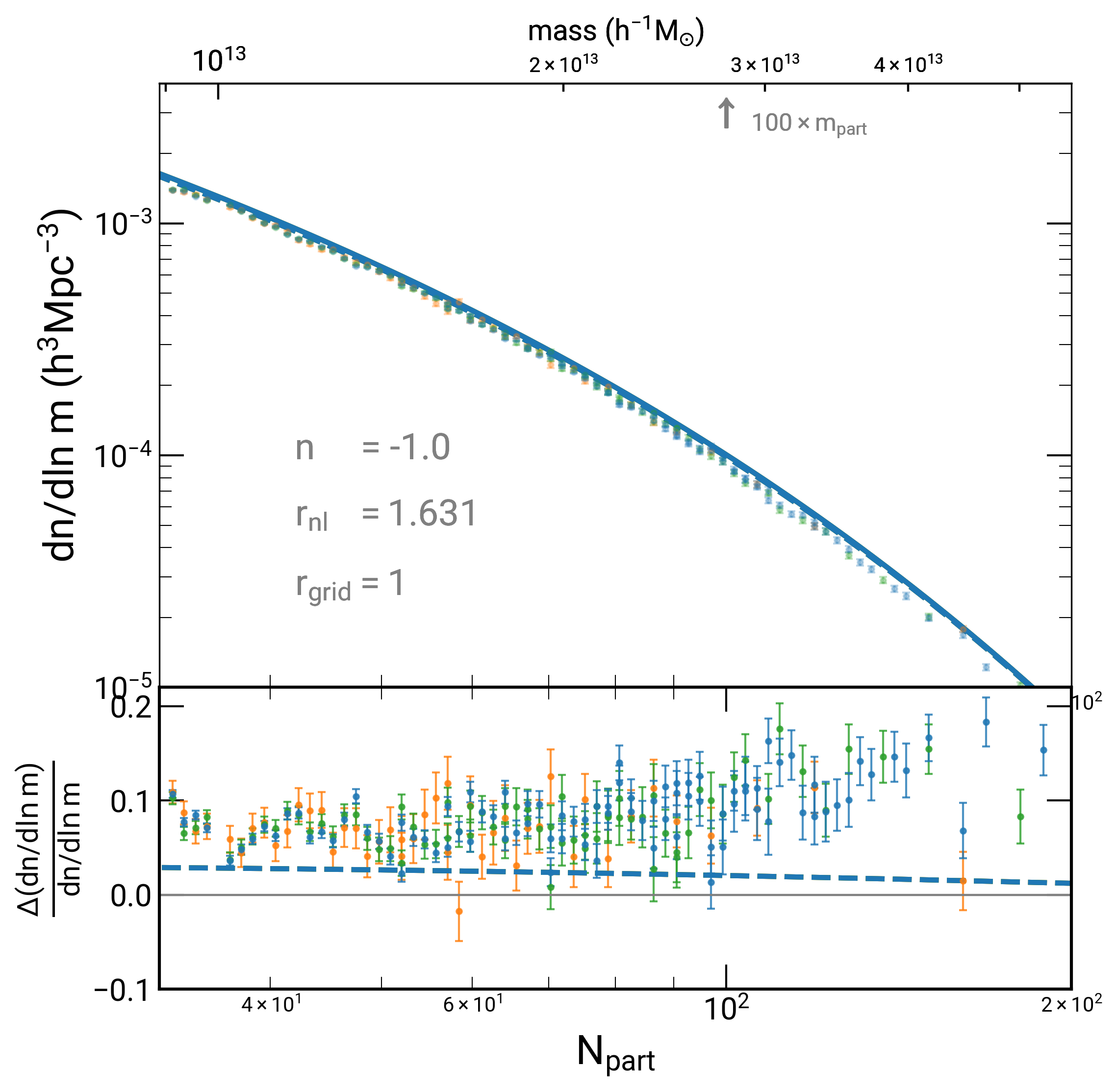}
\end{subfigure} 
\begin{subfigure}[b]{0.32\textwidth}
    \includegraphics[width=0.99\textwidth]{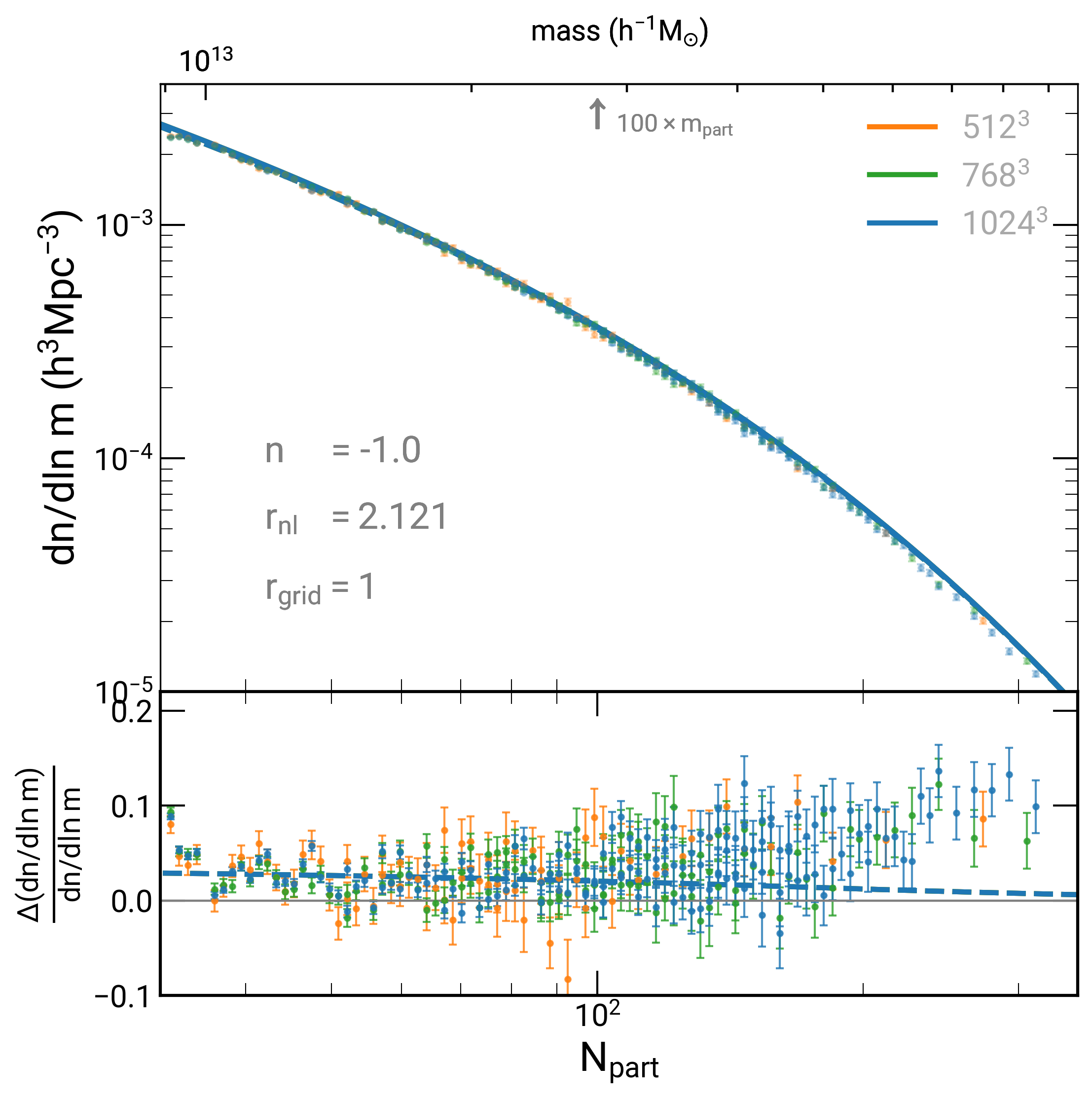}
\end{subfigure} 
\caption[Mass function: Effect of finite box size]{\emph{Mass function: Effect of finite box size:} Each panel shows the halo mass function and fractional errors in mass function in the simulation for three variations of $N_{\text{part}}=N_{\text{box}}^3=$ $512^3$ (orange), $768^3$ (green), $1024^3$ (blue) for the $n=-1.0$ model shown in Table \ref{tab:sim}. Data points represent simulation, whereas solid lines show theoretical mass function. Three panels represent different scales on non-linearity ($r_{nl}$).}
\label{fig:sim_conv2}
\end{center}
\end{figure}

\Cref{fig:sim_conv1} shows a conversion of the mass function computed in the simulation with respect to finite mass resolution. The lower-resolution scatters show higher fractional errors in mass function at a given $r_{nl}$. However, these errors settle similarly with the $r_{nl}$. Likewise, \cref{fig:sim_conv2} displays the impact of finite box size on the mass function. The box size affects the range of masses that can be probed, with larger box sizes allowing for better coverage of the higher-mass end.

